\RequirePackage{silence} 
\documentclass[ twoside,openright,titlepage,numbers=noenddot,
                headinclude,footinclude,cleardoublepage=empty,abstract=on,
                BCOR=5mm,paper=a4,fontsize=11pt,
                ]{scrreprt}

\PassOptionsToPackage{utf8}{inputenc}
  \usepackage{inputenc}

\PassOptionsToPackage{T1}{fontenc} 
  \usepackage{fontenc}

\PassOptionsToPackage{
  drafting=false,    
  tocaligned=false, 
  dottedtoc=true,  
  eulerchapternumbers=true, 
  linedheaders=true,       
  floatperchapter=true,     
  eulermath=false,  
  beramono=true,    
  palatino=true,    
  style=classicthesis 
}{classicthesis}

\newcommand{\myTitle}{Gravitational Waves and the Galactic Potential\xspace}
\newcommand{\mySubtitle}{\xspace}

\newcommand{\myName}{Francisco Miguel Batista Duque\xspace}

\newcommand{\myFaculty}{Put data here\xspace}

\newcommand{\myUni}{Instituto Superior Técnico\xspace}
\newcommand{\myLocation}{Lisbon\xspace}
\newcommand{\myTime}{June 2023\xspace}
\newcommand{\myVersion}{draft v1}

\providecommand{\mLyX}{L\kern-.1667em\lower.25em\hbox{Y}\kern-.125emX\@}
\newcommand{\ie}{i.\,e.}

\newcommand{\eg}{e.\,g.}

\newcommand{\fd}[1]{{\textcolor{cyan}{\textbf{[Francisco: #1]}} }}

\PassOptionsToPackage{ngerman,american}{babel} 
    \usepackage{babel}

\usepackage{csquotes}
\PassOptionsToPackage{%
  backend=bibtex8,bibencoding=ascii,%
  language=auto,%
  style=numeric-comp,%
  sorting=none, 
  maxbibnames=10, 
  natbib=true 
}{biblatex}
    \usepackage{biblatex}

\PassOptionsToPackage{fleqn}{amsmath}       
  \usepackage{amsmath,amssymb,lmodern,amsthm}

\usepackage{graphicx,psfrag,mathrsfs} %
\usepackage{adjustbox}
\usepackage{csquotes}
\usepackage{multirow}
\usepackage{bold-extra}
\usepackage{bm}
\usepackage{scrhack} 
\usepackage{xspace} 
\PassOptionsToPackage{printonlyused,smaller}{acronym}
  \usepackage{acronym} 

\usepackage{tabularx} 
\usepackage{subfig}

\usepackage{listings}
\usepackage{classicthesis}

\hypersetup{%
  colorlinks=true, linktocpage=true, pdfstartpage=3, pdfstartview=FitV,%
  breaklinks=true, pageanchor=true,%
  pdfpagemode=UseNone, %
  plainpages=false, bookmarksnumbered, bookmarksopen=true, bookmarksopenlevel=1,%
  hypertexnames=true, pdfhighlight=/O,
  urlcolor=CTurl, linkcolor=CTlink, citecolor=CTcitation, 
  pdftitle={\myTitle},%
  pdfauthor={\textcopyright\ \myName, \myUni, \myFaculty},%
  pdfsubject={},%
  pdfkeywords={},%
  pdfcreator={pdfLaTeX},%
  pdfproducer={LaTeX with hyperref and classicthesis}%
}

\makeatletter
\@ifpackageloaded{babel}%
  {%
    \addto\extrasamerican{%
    }%
    \addto\extrasngerman{%
    }%
    }{\relax}
\makeatother

\def\nn{\nonumber}

\def\be{\begin{equation}}
\def\ee{\end{equation}}
\def\beq{\begin{eqnarray}}
\def\eeq{\end{eqnarray}}

\begin{document}
\frenchspacing
\raggedbottom
\selectlanguage{american} 
\pagenumbering{roman}
\pagestyle{plain}
\thispagestyle{empty}
\pdfbookmark[1]{Title}{title}
%
%
%

\begin{addmargin}[-1cm]{-3cm}
\begin{flushleft} ~\\ \vspace{-30mm} \hspace{-12mm}  \includegraphics[width=8cm]{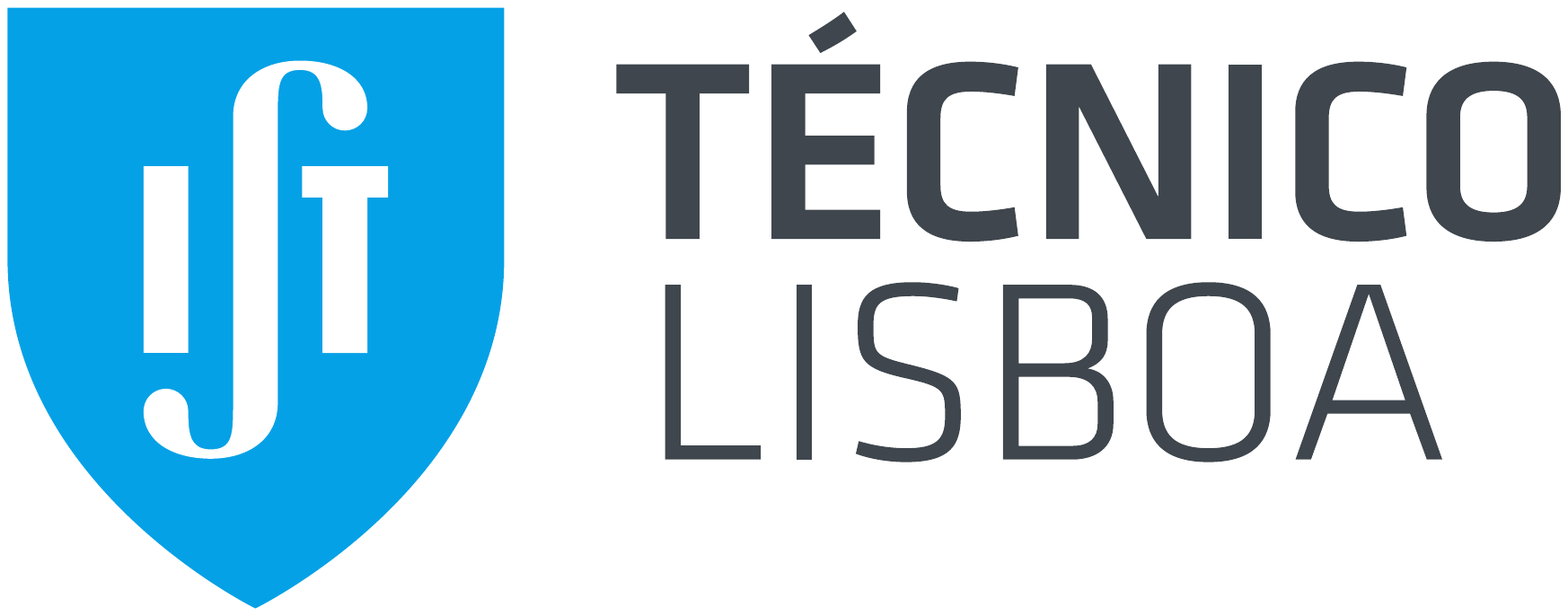} 
	
	\centering
	\LARGE \textbf{\spacedallcaps{UNIVERSIDADE DE LISBOA \\ INSTITUTO SUPERIOR TÉCNICO}}
	\\ \vspace{33mm}

	
	\centering
	\LARGE \spacedallcaps{\myTitle}
	\\ \vspace{5mm}
	\LARGE \spacedlowsmallcaps{\myName} 
	\vspace{2.5cm}
	
	\begin{minipage}{\textwidth}
		\begin{tabularx}{\textwidth}{ l @{ } l }
			\hspace{-7.5mm} \large \spacedlowsmallcaps{\textbf{\uppercase{S}upervisor}} : & \large Doctor Vítor Manuel dos Santos Cardoso\\
		\end{tabularx}
		
	\end{minipage}
	\\ \vspace{26mm}
	\centering
	\large \spacedlowsmallcaps{\textbf{\uppercase{T}hesis approved in public session to obtain the \uppercase{P}h\uppercase{D} degree in}}\\
	\large \spacedallcaps{PHYSICS}\\
	\vspace{12mm}
	\spacedlowsmallcaps{\textbf{\uppercase{J}ury final classification}}:
	\large \spacedlowsmallcaps{Pass with distinction and honour}

	\vspace{8mm}
	
	\Large \textbf{2023} \\
	\let\thepage\relax
\end{flushleft}
\end{addmargin}
\pagebreak

\thispagestyle{empty}
\pdfbookmark[1]{Title}{title}
%
%
%

\begin{titlepage}
\begin{addmargin}[-1cm]{-3cm}
\begin{flushleft} ~\\ \vspace{-30mm} \hspace{-12mm}  \includegraphics[width=8cm]{gfx/IST_A_CMYK_POS} 
	
	\centering
	\LARGE \textbf{\spacedallcaps{UNIVERSIDADE DE LISBOA \\ INSTITUTO SUPERIOR TÉCNICO}}
	\\ \vspace{10mm}

	
	\centering
	\Large \spacedallcaps{\myTitle}
	\\ \vspace{3mm}
	\Large \spacedlowsmallcaps{\myName} 
	\vspace{0.5cm}
	
	\begin{minipage}{\textwidth}
		\begin{tabularx}{\textwidth}{ l @{ } l }
			\hspace{-14.5 mm} \normalsize \spacedlowsmallcaps{\textbf{\uppercase{S}upervisor}} : & \normalsize Doctor Vitor Manuel dos Santos Cardoso\\
		\end{tabularx}
		
	\end{minipage}
	\\ \vspace{10mm}
	\centering
	\normalsize \spacedlowsmallcaps{\textbf{\uppercase{T}hesis approved in public session to obtain the \uppercase{P}h\uppercase{D} degree in}}\\
	\normalsize \spacedallcaps{Physics}\\
	\vspace{2mm}
	\spacedlowsmallcaps{\textbf{\uppercase{J}ury final classification}}:
	\large \spacedlowsmallcaps{Pass with distinction and honour}
	\vspace{10mm}
	
	\normalsize \spacedallcaps{Jury}
	
	\vspace{2mm}

			\normalsize \spacedlowsmallcaps{Chairperson} :  \\
			\vspace{1mm}  
			\hspace{-8.5 mm} Doctor Ilídio Pereira Lopes\small, Instituto Superior Técnico da Universidade de Lisboa\\
			\vspace{2mm}
			\normalsize \spacedlowsmallcaps{Members of the Committee} :  \\
			\vspace{1mm}
			\hspace{-15.5mm} \normalsize Doctor Scott Alexander Hughes\small, Massachusetts Institute of Technology, EUA\\ \vspace{1.5mm}
			\hspace{-40.0 mm} \normalsize Doctor Gaurav Khanna\small, The University of Rhode Island, EUA\\ \vspace{1.5mm}
			\hspace{-8.0 mm} \normalsize Doctor Ilídio Pereira Lopes\small, Instituto Superior Técnico da Universidade de Lisboa\\ \vspace{1.5mm} 
			\hspace{5.0 mm} \normalsize Doctor Vítor Manuel dos Santos Cardoso\small, Instituto Superior Técnico da Universidade de Lisboa\\ \vspace{1.5mm}
			\hspace{-45.0 mm} \normalsize Doctor Andrea Maselli\small, Gran Sasso Science Institute, Itália\\ \vspace{1.5mm}\normalsize 
			\hspace{2.0 mm}\normalsize Doctor David Matthew Hilditch\small, Instituto Superior Técnico da Universidade de Lisboa\\ \vspace{1.5mm}

	\vspace{5mm}
	

	\vspace{5mm}

	{\normalsize \spacedlowsmallcaps{Funding institution:}}\\ \vspace{1mm}	
	{\large Fundação para a Ciência e Tecnologia} 
	
	\vspace{5mm}

	\Large \textbf{2023} \\
	\let\thepage\relax
\end{flushleft}
\end{addmargin}
\pagebreak
\end{titlepage}
\begin{titlepage}
    \begin{addmargin}[-1cm]{-3cm}
    \begin{center}
        \large

        \hfill

        \vfill

        \begingroup
            \LARGE \color{CTtitle}\spacedallcaps{\myTitle} \\  \bigskip
            \Large \mySubtitle
        \endgroup
        
        \vfill

        \Large \spacedlowsmallcaps{\myName}

        \vfill

        \includegraphics[width=8cm]{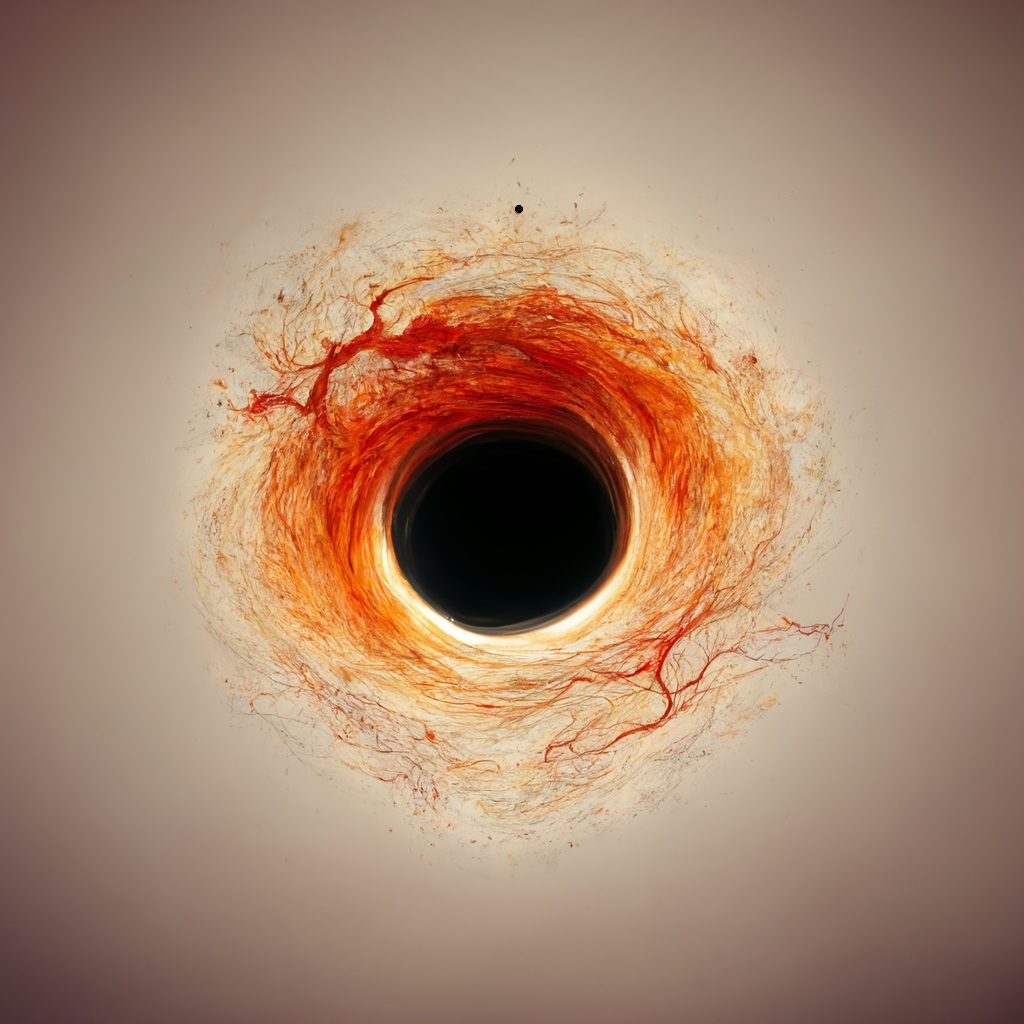} \\ \medskip

        \vfill 
        
        \spacedlowsmallcaps{\myTime}

    \end{center}
  \end{addmargin}
\end{titlepage}

\thispagestyle{empty}

\hfill

\vfill

\noindent\myName, \textit{\myTitle:} \mySubtitle, 
\textcopyright\ \myTime

%
%
%
%
%

\cleardoublepage
\thispagestyle{empty}
\phantomsection
\pdfbookmark[1]{Dedication}{Dedication}

\vspace*{1cm}
\spacedlowsmallcaps{Estilo} \vspace*{0.5cm}  \\
	\textit{
	 Se eu quisesse, enlouquecia. Sei uma quantidade de histórias terríveis. Vi muita coisa, contaram-me casos extraordinários, eu próprio… Enfim, às vezes já não consigo arrumar tudo isso. Porque, sabe?, acorda-se às quatro da manhã num quarto vazio, acende-se um cigarro… Está a ver? A pequena luz do fósforo levanta de repente a massa das sombras, a camisa caída sobre a cadeira ganha um volume impossível, a nossa vida… (...) Há felizmente o estilo. Não calcula o que seja? Vejamos: o estilo é o modo subtil de transferir a confusão e a violência da vida para um plano mental de uma unidade de significação. (...) É forçoso encontrar um estilo. Seria bom colocar grandes cartazes nas ruas, fazer avisos na televisão e cinemas. Procure o seu estilo, se não quer dar em pantanas. Arranjei o meu estilo estudando matemática e ouvindo um pouco de música. - João Sebastião Bach. Conhece o Concerto Brandeburguês n. º 5? Conhece com certeza essa coisa tão simples, tão harmoniosa e definitiva que é um sistema de três equações a três incógnitas. Primário, rudimentar. Resolvi milhares de equações. Depois ouvia Bach. Consegui um estilo. Aplico-o à noite, quando acordo às quatro da madrugada (...)
	} 
	\\ \medskip \\
	--- Herberto Helder \emph{em} ''Os Passos em Volta'' 
\vspace{1cm}



\begin{center}
    {Dedicado à minha avó Maria e à minha avó Peta.} \\ \smallskip
\end{center}

\cleardoublepage
\pdfbookmark[1]{Abstract}{Abstract}

\chapter*{Resumo}

O progresso rápido e recente da astronomia de ondas gravitacionais tornou necessário modelar fontes cada vez mais complexas. Durante a próxima década, interferómetros de terceira geração e a missão espacial LISA observarão binárias em centros galácticos envolvendo buracos negros supermassivos com milhões de massas solares. O seu sinal diferá substancialmente das mais ``comuns'' binárias de buracos negros de igual massa que têm dominado as detecções de ondas gravitacionais. Medições mais precisas de eventos mais extremos que excitam campos gravitacionais mais fortes podem ter um impacto tremendo em física fundamental, astrofísica e cosmologia. Porém, à escala galáctica, discos de acreção, halos de matéria escura e densas populações de objectos compactos podem interagir gravitacionalmente com corpos em coalescência. O papel que estas estruturas astrofísicas desempenham na evolução e respectiva assinatura de ondas gravitacionais de sistemas binários continua por explorar e estudos prévios  dependeram com frequência de aproximações Newtonianas ad-hoc. Nesta tese, pretendemos melhorar este panorama e responder a questões como: podem ambientes de não-vácuo comprometer testes de Relatividade Geral e da natureza de buracos negros? Podemos colocar constrangimentos nas propriedades de ambientes astrofísicos com futuras observações de ondas gravitacionais? Em particular, estudamos como deformações de maré de matéria presente à volta de buracos negros podem mascarar desvios a Relatividade Geral, ou destruir estruturas ``cabeludas'' que poderiam sinalizar a existência de candidatos de matéria escura como campos bosónicos ultraleves. Também exploramos a conexão profunda entre anéis de luz - órbitas fechadas de partículas sem massa - e os modos próprios de oscilação de objectos compactos. Mostramos que, independentemente da presença de um ambiente, o anel de luz controla como observadores distantes vêem matéria a cair num buraco negro, ou como o buraco negro final formado numa colisão relaxa para estacionariedade. Finalmente, desenvolvemos o primeiro \textit{framework} completamente relativista capaz de estudar emissão de ondas gravitacionais em ambientes de não-vácuo. Aplicamo-lo a binárias de buracos negros galácticas rodeadas por um halo de matéria escura e observamos a conversão entre ondas de matéria e ondas gravitacionais. Este acoplamento resulta em diferenças significativas no fluxo de energia emitido, que poderão ajudar a constrangir as propriedades de distribuições galácticas de matéria. Os nossos métodos podem tornar-se a ferramenta de referência para estudos de efeitos de ambiente em astronomia de ondas gravitacionais e ser implementado em \textit{pipelines} de análise de dados de colaborações futuras.

\vspace{0.7cm}
\noindent \spacedlowsmallcaps{Palavras-chave:} relatividade geral; buracos negros; ondas gravitacionais, efeitos de ambiente, matéria escura. 

\vfill

\newpage

\begin{otherlanguage}{ngerman}
\pdfbookmark[1]{Zusammenfassung}{Zusammenfassung}
\chapter*{Abstract}

The rapid progress of gravitational-wave astronomy in recent years has raised the need to model increasingly more complex sources. Over the next decade, third-generation interferometers and the space-based LISA mission will observe binaries in galactic centers involving supermassive black holes with millions of solar masses. Their signal will differ substantially from the more ``standard'' equal-mass black-hole binaries that have dominated gravitational-wave detections. More precise measurements of more extreme events that probe stronger gravitational fields can have a tremendous impact on fundamental physics, astrophysics, and cosmology. However, at the galactic scale, accretion disks, dark matter halos, and dense populations of compact objects can interact gravitationally with coalescing bodies. The role these astrophysical structures play in the evolution and gravitational-wave signature of binary systems remains largely unexplored and previous studies have often relied on ad-hoc Newtonian approximations. In this thesis, we aim to improve this picture and answer questions like: Can non-vacuum environments jeopardize tests of General Relativity and of the nature of black holes? Can we constrain the properties of astrophysical environments from future gravitational-wave observations? In particular, we study how tidal deformations of matter surrounding black holes can mask off deviations from General Relativity, or destroy ``hairy'' structures that could signal the existence of dark matter candidates like ultralight bosonic fields.
We also explore the deep connection between light rings - closed orbits of massless particles - and the proper oscillation modes of compact objects. We show that independently of the presence of an environment, the light ring controls the late-time appearance of infalling matter to distant observers and how the final black hole formed in a collision relaxes to stationarity. Finally, we develop the first fully-relativistic framework capable of studying gravitational wave emission in non-vacuum environments. We apply it to galactic black-hole binaries surrounded by a dark matter halo and observe the conversion between matter and gravitational waves. This coupling results in significant changes in the energy flux emitted, which could help constraining the properties of galactic matter distributions. Our methods can become the benchmarking tool for studies of environmental effects in gravitational-wave astronomy and be implemented in the data analysis pipelines of future collaborations. 

\vspace{0.7cm}
\noindent \spacedlowsmallcaps{Key-words:} general relativity; black holes; gravitational waves; environmental effects, dark matter;
\end{otherlanguage}


\vfill

\cleardoublepage
\pdfbookmark[1]{Publications}{publications}
\chapter*{Publications}

Most of this doctoral thesis is based on the following publications:\\\\

{\small
\noindent V. Cardoso, K. Destounis, \underline{F. Duque}, R. A. Macedo, A. Maselli, \textit{Gravitational waves from extreme-mass-ratio systems in astrophysical environments}, Phys. Rev. Lett. 129 (2022) 24, 241103; \href{https://arxiv.org/abs/2210.01133}{\footnotesize arXiv:2210.01133 [gr-qc]}\\

\noindent E. Berti, V. Cardoso, M. Ho-Yeuk Cheung, F. Di Filippo, \underline{F. Duque}, P. Martens, S. Mukohyama, \textit{Stability of the fundamental quasinormal mode in time-domain observations against small perturbations}, Phys. Rev. D106 (2022) 8, 084011; \href{https://arxiv.org/abs/2205.08547}{\footnotesize arXiv:2205.08547 [gr-qc]}.\\

\noindent V. Cardoso, \underline{F. Duque}, \textit{Resonances, black hole mimickers, and the greenhouse effect: Consequences for gravitational-wave physics}, Phys. Rev. D105 (2022) 10, 104023; \href{https://arxiv.org/abs/2204.05315}{\footnotesize arXiv:2204.05315 [gr-qc]}.\\

\noindent V. Cardoso, K. Destounis, \underline{F. Duque}, R. A. Macedo, A. Maselli, \textit{Black holes in galaxies: Environmental impact on gravitational-wave generation and propagation}, Phys. Rev. D105 (2022) 6, L061501; \href{https://arxiv.org/abs/2109.00005}{\footnotesize arXiv:2109.00005 [gr-qc]}.\\

\noindent V. Cardoso, \underline{F. Duque}, A. Foschi, \textit{Light ring and the appearance of matter accreted by black holes}, Phys. Rev. D103 (2021) 10, 104044; \href{https://arxiv.org/abs/2102.07784}{\footnotesize arXiv:2102.07784 [gr-qc]}.\\

\noindent V. Cardoso, \underline{F. Duque}, G. Khanna, \textit{Gravitational tuning forks and hierarchical triple systems}, Phys. Rev. D103 (2021) 8, L081501; \href{https://arxiv.org/abs/2101.01186}{\footnotesize arXiv:2101.01186 [gr-qc]}.\\

\noindent V. Cardoso, \underline{F. Duque}, Taishi Ikeda, \textit{Tidal effects and disruption in superradiant clouds: a numerical investigation}, Phys. Rev. D101 (2020) 6, 064054; \href{https://arxiv.org/abs/2001.01729}{\footnotesize arXiv:2001.01729 [gr-qc]}.\\

\noindent V. Cardoso, \underline{F. Duque}, \textit{Environmental effects in gravitational-wave physics: Tidal deformability of black holes immersed in matter}, Phys. Rev. D101 (2020) 6, 064028; \href{https://arxiv.org/abs/1912.07616}{\footnotesize arXiv:1912.07616 [gr-qc]}. \\\\

\pagebreak

\noindent During the years of my doctoral program, I also coauthored the following works (not discussed in this thesis):\\\\

\noindent V. Cardoso, \underline{F. Duque}, A. Maselli, D. Pereniguez, \textit{The dipolar death of massive gravity}; \href{https://arxiv.org/abs/2304.01252}{\footnotesize arXiv:2304.01252 [gr-qc]}\\

\noindent V. Baibhav, M. Ho-Yeuk Cheung, E. Berti, V. Cardoso, G. Carullo, R. Cotesta, W. Del Pozzo, \underline{F. Duque}, \textit{Agnostic black hole spectroscopy: quasinormal mode content of numerical relativity waveforms and limits of validity of linear perturbation theory}; \href{https://arxiv.org/abs/2304.01252}{\footnotesize arXiv:2302.03050 [gr-qc]}\\

\noindent V. Baibhav, M. Ho-Yeuk Cheung, E. Berti, V. Cardoso, G. Carullo, R. Cotesta, W. Del Pozzo, \underline{F. Duque}, \textit{Nonlinear Effects in Black Hole Ringdown}, Phys. Rev. Lett. 130 (2023) 8, 8; \href{https://arxiv.org/abs/2210.01133}{\footnotesize arXiv:2210.01133 [gr-qc]}}

\cleardoublepage
\pdfbookmark[1]{Acknowledgments}{acknowledgments}

\bigskip

\begingroup
\let\clearpage\relax
\let\cleardoublepage\relax
\let\cleardoublepage\relax
\chapter*{Agradecimentos}

É difícil escapar aos lugares comuns e desfile sentimentalista que se reservam aos agradecimentos de uma tese. Porém, ao fim de quatro anos de labuta sinto-me na ousadia de arriscar algo que destoe da minha natureza lacónica.

Começo por agradecer à Fundação para a Ciência e Tecnologia pelo apoio financeiro através da bolsa SFRH/BD/143657/2019. Também agradeço o suporte da European Union's 2020 ERC Consolidator Grant "Matter and strong-field gravity: New frontiers in Einstein’s theory" grant agreement no. MaGRaTh-646597, European Union’s H2020 ERC Advanced Grant “Black holes: gravitational engines of discovery” grant agreement no. Gravitas–101052587, e ajuda financeira para networking da GWverse COST Action CA16104, “Black holes, gravitational waves and fundamental physics.” e da bolsa Marie Sklodowska-Curie No. 101007855. 

Tratado o parágrafo burocrata, agradeço aos colegas com quem tive a oportunidade de trabalhar e me divertir ao longo deste doutoramento. A todo o GRIT, em especial aos meus ``irmãos" científicos  Arianna, David, Lorenzo e Rodrigo, pelos momentos de intenso debate e ainda mais intenso deboche. Quero também agradecer aqueles que durante as minhas visitas mais longas ao estrangeiro me faziam sentir em casa com a sua hospitalidade: aos \textit{strongers} que fazem de Copenhaga uma cidade latina; ao \textit{bon vivant} Andrea Maselli pelos conselhos académicos e gastronómicos; ao Gaurav Khanna, ao Nur Rifat, à Asia Haque e ao restante grupo de gravidade da UMass Darthmouth; ao Scott Hughes e ao seu grupo no MIT; e ao Emanuele Berti e a sua \textit{mafia} da Johns Hopkins University, em particular ao Andrea Antonelli e ao Nick Speeney. Um obrigado também ao João Paulo e à Rita - que também são colegas - por tornarem toda a logística não-científica tão fácil. 

Canta a sabedoria popular que um bom aluno é obra de um bom professor, e eu fui apaparicado com o \textit{filet mignon} da orientação científica. Muitas vezes me perguntam como é que trabalhar com \textit{o} Vitor Cardoso, esse Sansão da gravitação. Talvez a gravidade quântica esteja escondida nos seus cabelos. Costumo responder que  ``é uma experiência''. Daqui a alguns anos, gostava de não ter saudades das chamadas de Skype às 7 da manhã ao fim-de-semana, pois será sinal de que continuaremos a trabalhar juntos e elas lá acontecerão invariavelmente. Podia florir mais algumas palavras pomposas, mas sei que a melhor forma de lhe agradecer toda a ajuda que me dá é ir trabalhar no próximo artigo. Para lá caminho. 

Quero também deixar um obrigado às minhas professoras Adelaide e Clara, por me ensinarem a importância da visão crítica. Curioso que um aluno de Física recorde como mais marcantes as suas aulas de Português e Filosofia. Penso que não há maior elogio que lhes possa fazer. Numa mão a pena e noutra o ábaco. 

Tenho também que  agradecer ao meu primo Ruben pelo volume impossível de conselhos que me oferece desde que tenho idade para me lembrar de algo. É forçoso encontrar um estilo, e ele ajudou-me a encontrar o meu. 

Ensaiei várias vezes um parágrafo dedicado a todos os meus amigos. Iguais vezes tive de o apagar. Sei que munidos de vaidade, só abriram esta tese para verem o vosso nome esparrado nos agradecimentos e desistirem ao fim da terceira página da Introdução. \textit{Kaputt}. Infelizmente para vós, tenho a sorte de poder dizer que se tentasse agradacer a cada um, este parágrafo rapidamente se transformaria num capítulo por si só. Não preciso de vos nomear \textit{a la} memorial da guerra para saberem que são merecedores de aqui estar. Enfim, é a minha queda para a iconoclastia. Um poema de João Luís Barreto Guimarães diz que "Segundo Albert Einstein o tempo/ não passa de igual modo para mim/ (que estou aqui sentado) e para aquele que/ ali vai a correr./ Tanto pior para mim. A Física só traz problemas. O sol/ (dentro de instantes) passará exactamente/ por entre os/ gargalos das garrafas que estivemos/ a beber. É o solstício da amizade./ A qualquer dia/ do ano." A Física está correcta. Espero que o resto também se verifique e o resto são gafanhotos.

Falta-me o obrigado aqueles que tiveram que lidar no dia a dia com a frustração, por vezes ira, inerente a uma ocupação que consiste em estar metade do tempo perdido sem saber bem o que fazer. Primeiro, à Carolina por me ensinar que a bondade é mais bela que o binómio de Newton. Falta escrevermos um artigo juntos. 

Quis o acaso que em 2014 me sentasse ao lado da Beatriz numa aula de Relatividade de uma escola de Verão de Física da Universidade do Porto. Desde então tem sido a minha binária dançante, que por fim coalesceu. O que seria de uns agradecimentos de uma tese de Física sem um jogo de palavras carrascão de cariz romântico. Equações de Eins tein? Metafísica? Prova da existência de Deus? \textit{Pathos}? Passarola Voadora? Tudo isto existe, tudo isto é Fado. 

Por fim os Pais, de que sou Filho e Espírito Santo. Um obrigado à minha Mãe pela infindável paciência e carinho com que garante a minha sobrevivência. Deus quer, o Homem sonha, a Mãe cuida, a Obra nasce. E findado este gongórico devaneio, falta-me rasgo para honrar com palavras aquilo que o meu Pai tanto merece. Como o corte de um \textit{haiku}, amo com um amor mais forte e mais profundo, aquela praia extasiada e nua, e a cadeira ao teu lado na mesa de jantar. 

\endgroup

\cleardoublepage
\pagestyle{scrheadings}
\pdfbookmark[1]{\contentsname}{tableofcontents}
\setcounter{tocdepth}{2} 
\setcounter{secnumdepth}{3} 
\manualmark
\markboth{\spacedlowsmallcaps{\contentsname}}{\spacedlowsmallcaps{\contentsname}}
\tableofcontents
\automark[section]{chapter}
\renewcommand{\chaptermark}[1]{\markboth{\spacedlowsmallcaps{#1}}{\spacedlowsmallcaps{#1}}}
\renewcommand{\sectionmark}[1]{\markright{\textsc{\thesection}\enspace\spacedlowsmallcaps{#1}}}
\clearpage
\begingroup
    \let\clearpage\relax
    \let\cleardoublepage\relax
    \pdfbookmark[1]{\listfigurename}{lof}
    \listoffigures

    \vspace{8ex}

    \pdfbookmark[1]{\listtablename}{lot}
    \listoftables

    \vspace{8ex}

    \pdfbookmark[1]{Conventions and Units}{conventions}
    \markboth{\spacedlowsmallcaps{Conventions and Units}}{\spacedlowsmallcaps{Conventions and Units}}
	\chapter*{Conventions, Notation and Units}
	In this thesis I follow the conventions of Refs.~\cite{Misner:1974qy,Wald:1984rg}.
	Unless stated otherwise, I use geometrized units~$(c=G=1)$ and work with the~\emph{mostly positive} metric signature~$(-\, +\, +\; +)$. \\
	
	\begin{tabular}{lll}
		&$\alpha,\beta,\gamma, ...$          &  spacetime indices (from~$0$ to~$3$) \\
		&$i,j, k, ...$     & 3-spatial indices (from~$1$ to~$3$) \\
		&$V_\alpha W^\alpha \equiv \sum_{\alpha=0}^{3} V_\alpha W^\alpha$ & Einstein's notation\\
		&$T_{(\alpha_1\,...\,\alpha_l)}\equiv= \frac{1}{l!}\sum_{\,\sigma} T_{\alpha_{\sigma(1)}\,...\,\alpha_{\sigma(l)}}$ & symmet. over all permutat.~$\sigma$\\
		&$g_{\alpha \beta} \,, \eta_{\alpha \beta}$       & curved, flat spacetime metric  \\
		&$(\,\cdot\,)_{,\alpha}=\partial_\alpha(\cdot)=\frac{\partial}{\partial x^\alpha}(\cdot)$ &  coord. derivative \\
		&$(\,\cdot\,)_{;\alpha}= \nabla_\alpha(\cdot)$  & Levi-Civita derivative \\
		&$\Box(\cdot) \equiv \nabla_\alpha \nabla^\alpha(\cdot)$ &  Levi-Civita d'Alembertian 
	\end{tabular}
	
    \vspace{8ex}

    \pdfbookmark[1]{Acronyms}{acronyms}
    \markboth{\spacedlowsmallcaps{Acronyms}}{\spacedlowsmallcaps{Acronyms}}
    \chapter*{Acronyms}
    \begin{acronym}[UMLX]
        \acro{AGN}{active galactic nucleus/nuclei}
    	\acro{BH}{black hole}
    	\acro{BHB}{black hole binary}
    	\acro{CDM}{cold dark matter}
        \acro{CM}{center of mass}
    	\acro{DF}{dynamical friction}
    	\acro{DM}{dark matter}
        \acro{ECO}{exotic compact object}
    	\acro{EMRI}{extreme-mass-ratio inspiral}
    	\acro{EOM}{equation(s) of motion}
        \acro{GR}{General Relativity}
    	\acro{GW}{gravitational wave}
        \acro{IMRI}{intermediate-mass-ratio inspiral}
        \acro{ISCO}{innermost stable circular orbit}
        \acro{LIGO}{Laser Interferometer Gravitational-Wave Observatory}
        \acro{LISA}{Laser Interferometer Space  Antenna}
        \acro{LR}{light ring}
        \acro{LVK}{LIGO-Virgo-Kagra (collaboration)}
    	\acro{MACHO}{massive compact halo object}
    	\acro{QNM}{quasinormal mode}
        \acro{SMBH}{supermassive black hole}
        \acro{SNR}{signal-to-noise ratio}
        \acro{TLN}{tidal Love number}
    	\acro{WIMP}{weakly interacting massive particle}
    \end{acronym}

\endgroup

\cleardoublepage
\pagestyle{scrheadings}
\pagenumbering{arabic}
\cleardoublepage
\part*{Gravitational Waves and the Galactic Potential}\label{pt:introduction}
\chapter{Introduction}\label{ch:introduction}

\section{The GWzoic}

The theory of General Relativity (\acs{GR}) is often regarded as one of humanity's most outstanding intellectual achievements. Its mathematical elegance and ability to pass experimental tests remain unmatched in the history of science. For over a century, tests were limited to the weak gravity regime. Yet, the excitement about the theory stemmed from its puzzling predictions when a large amount of mass concentrates in a small region of space. According to \acs{GR}, if stars become too massive, they can collapse under their gravity, forming black holes (\acs{BH}s) - regions of space where the gravitational attraction is so strong that not even light can escape. They are deceivingly simple. The ``no-hair" theorem states that any isolated \acs{BH} in the Universe belongs to the Kerr family and is fully characterized by just two numbers: its mass and spin (how fast it rotates)~\cite{PhysRevLett.26.331}.

\acs{GR} also predicts that when accelerated, \acs{BH}s, or any other compact objects, create tidal deformations of the gravitational field that propagate at the speed of light: gravitational waves (\acs{GW}s). In $2015$, the Laser Interferometer Gravitational-Wave Observatory (\acs{LIGO}), in the U.S.A., detected a \acs{GW} signal from the coalescence of two \acs{BH}s with masses of $36$ and $24 \, M_\odot$~\footnote{$M_\odot = 1.9891 \times 10^{31} \, \text{kg}$ is the mass of the Sun}. This event became known as GW$150914$~\footnote{GW events are named as GW-YEAR-MONTH-DAY}~\cite{PhysRevLett.116.241102}. It originated billions of years ago, around $\sim 410 \, \text{Mpc}$ away from the Earth~\footnote{$1\, \text{pc} \approx 3.1 \times 10^{16} \, \text{m}$}, and lasted for $0.2\, \text{s}$ causing a disturbance of $10^{-21} \, \text{m}$ on its passage, smaller than the size of a proton. Until this point, we based our observations of the outer Universe almost solely on detecting light, but since then, we can also "hear" it through \acs{GW}s.

Unlike their electromagnetic relatives, \acs{GW}s interact weakly with matter and remain mostly unaffected during their journey. They offer an almost perfect fingerprint of some of the most violent events in the Universe, where objects with masses exceeding that of the Sun are compressed to a few kilometers and collide at speeds surpassing half the speed of light. GW$150914$ reached a peak power emission rate of $\sim 10^{49} \, \text{W}$ in \acs{GW}s, more than the power radiated by all the known stars in the Universe. In this regime, \acs{GR} is pushed to its limits, and GW$150914$ sparked what many have called the dawn of the \textit{gravitational-wave astronomy era}. Welcome to the \textit{GWzoic}.

Since then, two new detectors, Virgo (Italy) and KAGRA (Japan), were added to the network, and more than 90 \acs{GW}-events were confirmed by the LIGO-Virgo-KAGRA (\acs{LVK}) collaboration~\cite{LIGOScientific:2018mvr, LIGOScientific:2021usb, LIGOScientific:2021djp}. These have included stellar mass \acs{BH} binaries (\acs{BHB}s) as well as neutron star binaries with an electromagnetic counterpart
~\cite{LIGOScientific:2017vwq, LIGOScientific:2017ync, LIGOScientific:2017zic}, neutron star-\acs{BH} binaries
~\cite{LIGOScientific:2021qlt}, asymmetric binaries~\cite{LIGOScientific:2020stg}, a curious object which is either the lightest \acs{BH} or the heaviest neutron star ever observed (or perhaps, even ``something else")~\cite{LIGOScientific:2020zkf}, and
\acs{BH}s that are too massive to be formed via standard stellar collapse~\cite{LIGOScientific:2020iuh}. In the latter event, GW$190521$, the first intermediate-mass \acs{BH} with $\sim 142\, M_\odot$, whose existence in the Universe was often questioned, was also discovered~\cite{LIGOScientific:2020iuh}. This vast catalog has provided opportunities to test \acs{GR} with unprecedented levels of precision, impose constraints and exclude possible modifications of it, and gain insights about dense matter~\cite{Chatziioannou:2020pqz}, the formation of heavy elements in nature, and the astrophysical populations of \acs{BH}s, i.e. their masses, spins, or distance to the Earth~\cite{PhysRevX.13.011048, LIGOScientific:2021sio}. 

And the future \textit{sounds} even brighter. The \acs{LVK} is starting a new run with increased sensivity~\cite{KAGRA:2013rdx}, and it is planned that LIGO-India will join the network by the end of the decade. There are also ambitious plans to build third-generation detectors, such as the Einstein Telescope in Europe~\cite{Maggiore:2019uih} and Cosmic Explorer in the U.S.A.~\cite{Evans:2021gyd}. These advanced detectors will feature longer arms and better technology, boosting their sensitivity by one order of magnitude beyond that of \acs{LVK}~\cite{Kalogera:2021bya}. We expect to detect $\sim10^{6}$ \acs{GW} events per year with signal-to-noise ratios (\acs{SNR}) on the order of $\sim 10^{3}$, far surpassing GW$150914$'s \acs{SNR} of $24$. These \textit{louder} sources will decrease the relative error in parameter estimation by several orders of magnitude, improve the sky localization of the sources ($\sim 1 \, \text{deg}^2$ \textit{vs} the current $\sim 10^2 \, \text{deg}^2$), and observe events originated in the early Universe. 

In the late 2030s, the European Space Agency will also launch the Laser Interferometer Space  Antenna (\acs{LISA})~\cite{2017arXiv170200786A}. This space-based \acs{GW} detector will consist of three spacecrafts separated by approximately $2.5\times 10^6 \, \text{km}$, working together like an interferometer. While ground-based detectors are sensitive to \acs{GW}s between $\left[1, 10^3\right] \, \text{Hz}$, \acs{LISA} will operate in the much lower frequency range of $\left[10^{-4}, 10^{-1}\right] \, \text{Hz}$. It will \textit{hear} a very different class of sources compared to ground-based detectors, from millions of white dwarf binaries lurking in our galactic center to binaries involving supermassive black holes (\acs{SMBH}) with masses of $10^5-10^{10} M_\odot$, possibly at very high redshift ($z \sim 20$)~\cite{LISA:2022yao}. The latter may appear as monstruous \acs{SMBH}-\acs{SMBH} binaries~\cite{Klein:2015hvg, 1980Natur.287..307B}, or in extreme-mass-ratio inspirals (\acs{EMRI}s), where a stellar-mass object such as a \acs{BH} or a neutron star orbits around the central \acs{SMBH}~\cite{Babak:2017tow, Barack:2018yly}. \acs{EMRI}s perform around~$10^4-10^5$ orbital cycles while in the \acs{LISA} band~\cite{Hinderer:2008dm}, allowing the measurement of parameters like the masses, the \acs{SMBH} spin, and orbital eccentricity with statistical errors of order $10^{-4}-10^{-6}$~\cite{Babak:2017tow}.  The scientific potential of \acs{EMRI}s in both astrophysics~\cite{LISA:2022yao} cosmology~\cite{LISACosmologyWorkingGroup:2022jok}, and fundamental physics~\cite{LISA:2022kgy} is almost immeasurable. However, we need extremely accurate models to follow the waveform for months, possibly years, with errors below $1$ radian in the \acs{GW} phase~\cite{Hughes:2021exa, Wardell:2021fyy, Albertini:2022rfe}.

Finally, the International Pulsar Timing Array has very recently reported evidence for a stochastic \acs{GW} background in the nanoHertz range, due to a cosmic population of \acs{SMBH}s with billions of solar masses~\cite{NANOGrav:2023gor, Antoniadis:2023rey, NANOGrav:2023hfp, Antoniadis:2023zhi}. These observations are based on correlated shifts on the period of milisecond pulsars, caused by the passing of GWs. Together with the ground and space-based interferometers, they allow for the detection of GWs across a frequency band spanning almost $10$ orders of magnitude.

\section{Galactic Centers: gravity's potluck}

It is common belief that most \acs{SMBH}s grow in the center of galaxies~\cite{10.1093/mnras/152.4.461, 10.1093/mnras/200.1.115, 2010RvMP...82.3121G, Do:2019txf, 2018A&A...615L..15G, Volonteri_2003, 10.1111/j.1365-2966.2006.10467.x}. These are far from being simple \textit{ecosystems}, as perfectly illustrated by the \acs{BH} images captured by the Event-Horizon Telescope~\cite{1642263, 2019ApJ...875L...1E} and the pictures taken by the James Webb Telescope. The \acs{SMBH} residing there can efficiently accrete matter, forming active galactic nuclei (\acs{AGN}s), which are the brightest sources of electromagnetic radiation in the Universe. \acs{AGN}s consist of gas orbiting in a thin accretion disk, with surface densities between $10^3-10^6 \, \text{g} \cdot \text{cm}^{-2}$ and temperatures as high as $10^6 \, \text{K}$~ \cite{Shakura:1972te, Abramowicz:2011xu, Derdzinski:2022ltb}. Galaxies are also thought to be surrounded by dark-matter (\acs{DM}) distributions, which play a key role in their formation and evolution~\cite{1970ApJ...160..811F, 1980ApJ...238..471R, Navarro:1995iw, Wechsler:2018pic}. Finally, the various stars and other compact objects orbiting close to the central \acs{SMBH} attract each other gravitationally, making the motion of binary systems more complex. We need very accurate models for their orbital evolution to do proper precision physics with \acs{EMRI}s. The question to make is whether non-vacuum environments in galactic centers lead to non-negligible effects in an \acs{EMRI} that are detectable with future \acs{GW} experiments. Or can we ignore the environment and safely model these systems as isolated binaries in vacuum \acs{GR}?

\subsection{How many? Many... }

Before trying to answer this question, we should first assess how many \acs{EMRI}s in \acs{AGN}s is \acs{LISA} expected to observe. In the standard ``dry'' formation channel, stellar-mass compact objects evolve through gravitational scattering between each other until they are close enough to the central \acs{SMBH} to become gravitationally bound to it. \acs{EMRI} population rates will therefore depend on the typical number of stars and stellar mass \acs{BH}s in galactic centers, whose density distribution can reach $10^6\, M_\odot \cdot \text{pc}^{-3}$~\cite{2014CQGra..31x4007S}, their initial positions and velocities, the distribution of \acs{SMBH}s in the Universe, and the fraction of these which is present in dense stellar clusters. It should come as no surprise that forecasts on the number of detectable \acs{EMRI}s formed via this multibody scattering mechanism ranges between $1-10^3$~\cite{Babak:2017tow, Vazquez-Aceves:2021xwl}.

Despite the uncertainties, there is astrophysical evidence for binaries evolving in galactic centers. For instance, OJ$287$~\cite{2010ApJ...709..725V, Pihajoki_2013} is a binary system made up of two \acs{SMBH}s, with masses of $10^{10}\, M_\odot$ and $10^{8} \, M_\odot$, orbiting in an accretion disk with $10^{2} M_\odot$. Their orbit is slightly inclined with respect to the plane of the disk. Hence, as the lighter \acs{BH} (often referred to as the \textit{secondary}) crosses the disk, periodic electromagnetic flares are emitted~\cite{2010MNRAS.402.1614D}. Additionally, the large masses of the \acs{BH}s coalescing in GW$195021$~\cite{LIGOScientific:2020iuh}, $85-66 \, M_\odot$, suggest they are second generation \acs{BH}s that formed in previous mergers, which typically requires these objects to be in dense environments~\cite{2017ApJ...835..165B, Tagawa_2020, 2021ApJ...908..194T, 2021ApJ...920L..42G, Derdzinski:2022ltb}. Moreover, the Zwicky Transient Facility observed an electromagnetic flare $34$ days after GW$195021$, which has been proposed to be due to the movement of the remnant \acs{BH} in an \acs{AGN} at $200\, \text{km} \cdot \text{s}^{-1}$~\cite{Graham:2020gwr, Bustillo:2021tga}.

However, the estimates mentioned above for detectable \acs{EMRI} population rates ignore the interaction of the small compact object with the environment around it. Just as a ball moving through a fluid experiences drag and pressure differentials that lead to interesting phenomena like the Magnus effect~\cite{2018PhRvD..98b4026C}, a compact object moving through gas-rich environments will experience several dynamical effects related to accretion flows. The same applies to \acs{DM} distributions. Typically, these effects accelerate the inspiral towards the central \acs{BH}. Recent estimates point out that these ``wet'' \acs{EMRI}s may actually dominate the fraction of detectable \acs{EMRI}s by \acs{LISA} and be several orders of magnitude more common than the ones born through the ``dry'' formation channel~\cite{Pan:2021ksp, Pan:2021oob} (see Fig.~1 in Ref.~\cite{Pan:2021oob}).

\subsection{Surfing the waves}

An \acs{EMRI} evolving in an accretion disk experiences various dynamical effects not present in vacuum. Both the \acs{SMBH} and the stellar-mass compact object accrete matter, which increases their masses through the inspiral~\cite{1944MNRAS.104..273B, Edgar:2004mk}. The disk's self-gravity modifies the gravitational potential, altering the acceleration of the secondary object and potentially inducing orbital precession. Additionally, as the compact object pierces through the gas, it generates density wakes that trail behind and interact with itself, exerting additional torques and drag. These effects are broadly referred to as dynamical friction~\cite{1943ApJ....97..255C, Ostriker:1998fa, 1980ApJ...241..425G,  Tanaka_2002, 10.1046/j.1365-8711.1999.02623.x, 10.1093/mnras/277.3.758, Chakrabarti:1995dw}\footnote{The astrophysics community dubs them differently depending on the origin of the interaction, e.g. hydrodynamical drag, migration torques...}. \acs{GW}s tend to circularize orbital motion, but the dynamical effects created by the environment compete with this and may increase the orbit's eccentricity~\cite{Cardoso:2020iji}.

A systematic study of the effects mentioned above, focusing on geometrically thin, radiatively efficient, stationary accretion disks described by the Shakura-Sunayev model~\cite{Shakura:1972te}, found that dynamical friction from \textit{planetary-migration}-like torques results in the largest deviation from vacuum waveform templates for circular, non-inclined \acs{EMRI}s targeted by \acs{LISA}~\cite{Kocsis:2011dr, Yunes:2011ws} (see Fig.~$1$ of Ref.~\cite{Kocsis:2011dr}). This study predicts a dephasing in the \acs{GW} signal as large as $\sim 10^2\, \text{rad}$, which could severely constrain the binary environmental density.

Further order-of-magnitude estimates confirm this picture~\cite{Barausse:2014tra, Cardoso:2019rou}. A Bayesian analysis using the state-of-the-art waveform models for circular-equatorial \acs{EMRI}s suggests that \acs{GW} observations will be able to distinguish between different accretion models~\cite{Chua:2020stf, Hughes:2021exa, Katz:2021yft}. If environmental effects are not included in waveform templates, the estimation of the mass and spin of the objects could be biased and compromise tests of fundamental physics~\cite{Speri:2022upm}. In addition, environmental effects are expected to dominate over still unmodelled vacuum corrections in the waveform~\cite{Zwick:2022dih}.

Unfortunately, most literature on environmental effects in \acs{EMRI}s has only modeled the environment at the Newtonian level or utilized the lowest-order quadrupole formula to estimate \acs{GW} emission and dynamics. Neither of these approximations is expected to hold for \acs{EMRI}s, whose large number of orbital cycles in the \acs{LISA} band naturally probe the strong-field regime of gravity. This has been corroborated by $2$D general-relativistic-hydrodynamic simulations of intermediate-mass-ratio inspirals (\acs{IMRI}s)~\footnote{EMRIs have mass ratios between the secondary and the SMBH $\lesssim 10^{-5}$ while IMRIS have $10^{4}-10^{-3}$.}~\cite{Derdzinski:2020wlw}. The gas torques exerted on the inspiralling body depend on the disc parameters and the mass ratio of the binary. For the early inspiral, the analytical estimates from planetary migration~\cite{Tanaka_2002} capture the correct evolution of the torque with the inspiral rate. However, their order of magnitude may differ (check Fig.~$3$ and $5$ of Ref.~\cite{Derdzinski:2020wlw}). In some configurations, the torques can even become positive, acting as a thrust instead of a drag, which slows down the inspiral. In the later stages, the torque evolution is highly oscillatory because the inspiral timescale becomes comparable to the relaxation timescale of the disk, which depends on its viscosity and temperature. However, this work still uses simplified disc models, focuses only on circular orbits, and considers Newtonian gravity and the quadrupole formula for the binary's evolution, so further investigations are necessary. Gas torques also exhibit stochastic fluctuations that may result in non-zero secular phase shifts when accumulated over many orbits and introduce high-frequency glitches in the waveform. These can also appear as a stochastic background in the mHz band~\cite{Zwick:2021dlg}.  

Detecting an \acs{EMRI} evolving in an \acs{AGN} is also interesting from the perspective of electromagnetic observations. The angular resolution to which these systems can be localized should highly reduce the number of possible \acs{AGN}s hosting the \acs{EMRI}. The combination of \acs{GW} and electromagnetic observations could provide a direct measurement of the disk viscosity~\cite{Speri:2022upm} and allow for independent tests of cosmology through distance measurements of the system using electromagnetic redshift and \acs{GW} luminosity~\cite{Schutz:1986gp, Holz:2005df}. 

\subsection{The unbearable darkness of...}

We already mentioned that another important \textit{player} at the galactic scale is \acs{DM}. Extensive experimental evidence supports its existence, including the flatness of rotation curves of stars in spiral galaxies~\cite{1970ApJ...160..811F, 1980ApJ...238..471R, Navarro:1995iw, 1990ApJ...356..359H, Wechsler:2018pic} and gravitational lensing~\cite{Massey:2010hh}. According to the standard model of cosmology, \acs{DM} behaves as a collisionless, non-relativistic fluid, hence the name cold dark matter (\acs{CDM}). Its local density and dispersion velocity are approximately $\rho_\text{DM}\sim 10^{-2} \, M_\odot \cdot \, \text{pc}^{-3}$ and $v_\text{DM} \sim 150 \, \text{km} \cdot \text{s}^{-1}$, respectively. \acs{CDM} tends to clump in halos, where baryonic matter can become gravitationally bound and form large-scale structures like galaxies. Our own Milky Way is expected to be immersed in a \acs{DM} halo with $\sim 10^{12} M_\odot$, in comparison with the $\sim 10^{11} M_\odot$ of luminous matter, and a radius of $\sim 10^{2}\, \text{kpc}$~\cite{Guo:2009fn, Nesti:2013uwa}. 

However, \acs{CDM} suffers from several pathologies at subgalactic scales~\cite{Weinberg:2013aya, DelPopolo:2016emo}. Two illustrative examples are the \textit{cuspy-halo} problem~\cite{1994Natur.370..629M, Navarro:1996gj}, where $N$-body simulations favor ``cuspy'' halos with a steep increase in the density of \acs{DM} at their center, while galaxy rotation curves indicate more flat profiles; and the \textit{dwarf-galaxy} problem, which arises from the conflict between the small number of dwarf galaxies observed and the larger number of subhalos predicted by simulations. 

Many of these problems are due to our limited understanding of some astrophysical processes and the actual fundamental nature of \acs{DM}. There is an entire \textit{zoo} of possible answers~\cite{RevModPhys.90.045002} ranging from macroscopic objects with a few solar masses, such as primordial \acs{BH}s formed in the early universe due to curvature fluctuations~\cite{10.1093/mnras/168.2.399, Carr:2009jm}, to new ultralight bosonic fields with masses as light as $\mu \sim10^{-22}\, \text{eV}$~\cite{PhysRevLett.38.1440, PhysRevD.81.123530, Brito:2015oca, Hui:2021tkt}. The latter provides a natural solution to the cuspy-halo problem, as they admit stable, self-gravitating compact configurations, known as \textit{boson stars}, which are a good description of flat \acs{DM} halo cores~\cite{Liebling:2012fv, Hui:2016ltb, Annulli:2020lyc, Cardoso:2022nzc}. If different \acs{DM} models have a particular signature in the waveform, we could then use \acs{GW} observations to constrain the properties of \acs{DM}.

The small local densities of \acs{DM} might initially suggest that it has minimal impact on the evolution of compact binaries. However, the presence of a \acs{BH} that grows adiabatically in the center of a \acs{DM} configuration can result in the formation of a density cusp with a peak very close to the \acs{BH} horizon~\cite{Gondolo:1999ef, Ullio:2001fb, Sadeghian:2013laa}. The \acs{DM} density is expected to be significantly enhanced, possibly up to $10$ orders of magnitude at the center, and be scaling as $\rho_\text{DM} \sim r^{-\gamma}$, where $r$ is the radial distance to the \acs{BH} and $\gamma = 1.5-2.5$, depending on the \acs{DM} model considered. \acs{DM} annihilation, accretion, and other dynamical processes, like mergers, can deplete the overdensity and flat out the core of the \acs{DM} distribution~\cite{Merritt:2002vj, Bertone:2005hw}. The orbits of S-stars have already been used to place constraints on the properties of the \acs{DM} spike at the center of our own Milky Way~\cite{Lacroix:2018zmg, GRAVITY:2019tuf, GRAVITY:2021xju, Shen:2023kkm}. 

Despite its uncertainty, the large overdensity at galactic centers could make \acs{DM} accretion and dynamical friction relevant for binary inspirals. This mechanism has been suggested to explain the abnormally fast orbital decay of the two closest stellar-mass X-ray binary systems, which are inspiralling two orders of magnitude faster than expected from \acs{GW} emission~\cite{Chan:2022gqd}. As in accretion disks, most literature has applied results from Newtonian treaments~\cite{1943ApJ....97..255C} and applied it to \acs{EMRI}s to estimate its effect in the dephasing of compact binaries~\cite{PhysRevD.89.104059, Cardoso:2019rou}. Only recently have the first relativistic studies of dynamical friction in homogeneous mediums composed of an ultralight scalar field been conducted~\cite{Annulli:2020lyc, Vicente:2022ivh, Traykova:2021dua, Buehler:2022tmr, Traykova:2023qyv}.

When included in state-of-the-art waveform models~\cite{Chua:2020stf, Hughes:2021exa, Katz:2021yft}, these relativistic corrections can significantly impact the forecasts for the detectability and parameter estimation of an \acs{EMRI} evolving in a \acs{DM} environment during the \acs{LISA} mission lifetime~\cite{Speeney:2022ryg}. Another nuance to bear is that for binaries involving intermediate-mass \acs{BH}s, the energy dissipated via dynamical friction can become comparable to the gravitational binding energy of the \acs{DM} distribution. In this case, one needs to consider backreaction on the environment, which hinders the dephasing in the waveform~\cite{Kavanagh:2020cfn, Coogan:2021uqv}. 


Ultralight bosons can also extract rotational energy from rotating \acs{BH}s through a process known as superradiance and condensate into macroscopic clouds with a structure reminiscent of the hydrogen atom~\cite{Brito:2015oca}. These superradiant clouds are expected to create gaps in the mass-spin distribution of astrophysical \acs{BH}s~\cite{Brito:2014wla}. They also emit monochromatic \acs{GW}s which can appear as individual signals or stochastic backgrounds~\cite{Brito:2017wnc, Brito:2017zvb, Isi:2018pzk, Yuan:2021ebu}. \acs{LVK} has already placed constraints for masses around $\sim 10^{-13}\, \text{eV}$ based on null observations~\cite{Tsukada:2018mbp, Tsukada:2020lgt}. Nonetheless, when present in a binary, the superradiant clouds will interact with the companion of their \acs{BH} hosts, causing their deformation ~\cite{Cardoso:2020hca, Baumann:2018vus, Baumann:2019ztm, Baumann:2022pkl}. A proper understanding of the evolution of these structures in binary coalescences is necessary so that the constraints already imposed can be trusted.  

\subsection{Third wheels}

Finally, we have already mentioned that some \acs{GW}-events observed by \acs{LVK} should involve \acs{BH}s
which are remnants of previous coalescences~\cite{LIGOScientific:2020stg, Abbott:2020khf, Abbott:2020tfl, Abbott:2020uma}. These ``hierarchical mergers'' typically require the presence of a third body to induce coalescence~\cite{Liu:2020gif, Fragione:2020han, Martinez:2020lzt, Lu:2020gfh}. In fact, triple systems are expected in \acs{AGN}s~\cite{Bartos:2016dgn, 10.1093/mnras/stw2260, Chen:2018axp, Toubiana:2020drf} and other dense stellar environments~\cite{OLeary:2016ayz, 2016MNRAS.463.2109R, Portegies_Zwart_2000}, with around $90\%$ of low
mass binaries with periods shorter than three days belonging to some hierarchical structure~\cite{2006AA...450..681T, Pribulla:2006gk, Robson:2018svj}. Even at the Newtonian level, the presence of a third body is known to lead to interesting dynamical effects~\cite{poisson_will_2014}. It can accelerate the binary's center-of-mass (\acs{CM}), which causes a time-dependent Doppler shift~\cite{10.1093/mnras/stv172, Meiron:2016ipr, Randall:2018lnh, Wong:2019hsq, Han:2018hby} and relativistic beaming ~\cite{Torres-Orjuela:2018ejx, Torres-Orjuela:2020cly} in the waveform. It can also lead to secular changes in the orbital eccentricity and inclination through the Kozai-Lidov mechanism~\cite{1962AJ.....67..591K, doi:10.1146/annurev-astro-081915-023315, poisson_will_2014},  triggering periods of high eccentricity where \acs{GW} emission increases significantly~\cite{Hoang:2019kye, Randall:2019sab, Randall:2019znp, Deme:2020ewx}, causing periodic bursts of \acs{GW}s~\cite{PhysRevD.85.123005,Gupta:2019unn}. However, as with the previous environmental effects, most of these studies were restricted to the Newtonian regime and did not capture strong-field physics.

\section{Structure of the thesis}

The punchline of our discussion is that we lack the proper relativistic modeling of environmental effects in the evolution of binary systems necessary to fulfill the ambitious goals that \acs{GW} astronomy has set for the next decades. The Fundamental Physics Working Group of \acs{LISA} has recently emphasized this problem in their white paper~\cite{LISA:2022kgy} by dedicating an entire chapter to environmental effects (Chapter VII). The following \textit{burning questions and needed developments} were identified:

\begin{itemize}
	
\item  \textit{Can accretion, plasma effects or other stellar compact objects in the vicinity of an \acs{EMRI} induce observable changes in the \acs{GW} frequency evolution during the inspiral or ringdown that can spoil fundamental physics tests?}\\

\item \textit{Self-force calculations for generic \acs{BHB}s in vacuum or embedded in a background (e.g. \acs{DM} boson cloud) at second-order are necessary for proper modeling of \acs{EMRI}s, so that reliable waveforms are available to test for fundamental physics.}\\

\item \textit{A combination of the effects of bosonic \acs{DM} and modified gravity must be considered in order to be able to understand how more complex deviations from standard \acs{GR} can take place.}\\

\end{itemize}

In this Ph.D. thesis we will address some of these questions and point towards the necessary steps to further advance this research program. We end this introduction with an outline of how the thesis is organized. All my publications referenced below resulted from collaborations with my supervisor Vitor Cardoso. 

In Chapter~\ref{ch:theory} we present the theoretical framework on which most of our work is based, including the necessary equations of motion (\acs{EOM}) and numerical methods often employed. 

In Part~\ref{pt:tides} we focus on studying static tidal deformations to distributions of matter surrounding \acs{BH}s induced by external companions. In Chapter~\ref{ch:TLNs} we start by showing that the correction to waveforms due to tidal deformations of non-vacuum environments can be comparable to those signaling deviations from \acs{GR}. This can compromise tests on the nature of \acs{BH}s based on the measurement of tidal deformations, as an ultracompact object without a horizon would be indistinguishable from the presence of an accretion disk around a \acs{BH}~\cite{Cardoso:2019upw}. Chapter~\ref{ch:Cloud} follows from a collaboration with Taishi Ikeda~\cite{Cardoso:2020hca}, where we investigated the evolution of superradiant clouds when subject to static tidal fields from binary companions. We find they can deform the structure of the cloud or even disrupt it when too strong. This mechanism could be relevant for known \acs{BH} systems, such as Sagittarius A* at the center of our galaxy, or Cygnus X-$1$. 

Part~\ref{pt:light-ring} delves into the exquisite role that the \acs{LR} has on the physics of ultracompact objects, specifically its connection with the proper oscillation modes of a compact object. In Chapter~\ref{ch:LR}, based on Ref.~\cite{Cardoso:2021sip} co-authored with Arianna Foschi, we conclude that the appearance to distant observers of emitting sources of electromagnetic or gravitational waves being accreted by a \acs{BH} is controlled by the \acs{LR} properties and not particularly sensitive to near horizon details. Chapter~\ref{ch:Elephant} is a collaboration with Emanuele Berti, Mark Ho-Yeuk Cheung, Francesco Di Filippo, Paul Martens and Shinji Mukohyama~\cite{Berti:2022xfj}, where we revisited the problem of the stability of the quasinormal (\acs{QNM}) spectrum of \acs{BH}s, but from the point of view of time-domain observations. Our analysis reveals that changes in the amplitude of the fundamental mode in the prompt ringdown, which is the relevant portion of the signal for \acs{GW} observations, are parametrically small, even though formally the \acs{QNM} spectrum is unstable. In Chapter~\ref{ch:Greenhouse}, building upon Ref.~\cite{Cardoso:2022fbq}, we continue to explore the subtle differences between frequency and time-domain analysis and show the excitation of very sharp resonances in compact binaries may be hindered due to radiation-reaction effects which quickly move the system away from the resonant state. We end this part with Chapter~\ref{ch:tuningfork}, based on work with Gaurav Khanna~\cite{Cardoso:2021vjq}. In this chapter, we study \acs{GW} emission, in the strong-field regime, by a hierarchical triple system composed of a binary placed in the vicinity of a \acs{SMBH}. We observe interesting phenomena such as the resonant excitation of the \acs{QNM}s of the \acs{SMBH}, as in the resonant excitation of two tuning forks with matching frequencies. We also observe Doppler shifts, aberration, lensing, and strong amplitude modulations in the waveform. 

Finally, in the last part of the thesis, Part~\ref{pt:EMRIs}, we develop the first fully-relativistic framework to handle \acs{GW} emission in spherically-symmetric but otherwise generic spacetimes, including non-vacuum ones. This formalism was first laid out in Refs.~\cite{Cardoso:2021wlq, Cardoso:2022whc}, in collaboration with Andrea Maselli, Kyriakos Destounis and Rodrigo Panosso Macedo. It is based on applying \acs{BH} perturbation theory to spacetimes describing extended distributions of matter around \acs{BH}s, treating both matter and gravitational perturbations on an equal footing. This approach allows us to naturally account for accretion, gravitational drag, and halo feedback. We apply this newly developed framework to a relativistic solution of a non-rotating \acs{BH} immersed in a galactic \acs{DM} halo, as found in Ref.~\cite{Cardoso:2021wlq} using the \textit{Einstein Cluster} construction. We observe the manifestation of the spectral instability mentioned above for an astrophysical system. Moreover, our methods open up the possibility to infer galactic properties with \acs{EMRI}s. We conclude the thesis with remarks on our findings and discussions on future work. 

\chapter{Theory and Numerical Framework}\label{ch:theory}

In this chapter we give a brief presentation of the theoretical and numerical framework necessary for the rest of the thesis. 

\section{Theory}

\subsection{Action and Equations of Motion}

We consider the Einstein-Hilbert action in a $4$-dimensional spacetime~\cite{Wald:1984rg}
\beq \label{theory_action}
	S_\text{EH} = \int d^4x \sqrt{-g}\left[\frac{R- 2 \Lambda}{8 \pi} + \mathcal{L}_\text{M} \right]\,,
\eeq
where $R$ is the Ricci scalar, $g$ is the determinant of the metric $g_{\mu\nu}$, $\Lambda$ is the cosmological constant, and $\mathcal{L}_\text{M}$ is the Lagrangian density of matter fields.  

The variation of the Einstein-Hilbert action with respect to $g_{\mu\nu}$ leads to the Einstein's equations
\beq
	G_{\mu \nu}+\Lambda g_{\mu \nu}=8 \pi \,T_{\mu \nu} \label{Eins_EOM}\,,
\eeq
where~$G_{\mu \nu} = R_{\mu \nu} -\frac{1}{2} R\, g_{\mu \nu}$ is the Einstein tensor and~$T_{\mu \nu}$ is the energy-momentum tensor of matter
\beq
T^{\mu\nu} = - \frac{g^{\mu \alpha}g^{\nu \beta}}{\sqrt{-g}}\frac{\delta\left(\sqrt{-g}\,\mathcal{L}_\text{M}\right)}{\delta g^{\alpha \beta}} \, . \label{eq:EnergyTensorGeneral}
\eeq
The contracted Bianchi identities ($\nabla_\mu G^{\mu\nu}=0$) imply that $T_{\mu \nu}$ is divergenceless
\beq
	\nabla_\mu T^{\mu \nu}=0\, ,
\eeq
which dictates the equations of motion (\acs{EOM}) for matter.

A pedagogical example relevant for ultralight bosonic \acs{DM} is a complex scalar field $\Phi$ minimally coupled to gravity 
\beq
\mathcal{L}_\text{M} = \mathcal{L}_\Phi = - \nabla_\mu \Phi^* \nabla^\mu \Phi - \mathcal{U}_\Phi \left(\left| \Phi \right|^2 \right) - J_\Phi \left(\Phi^* + \Phi \right)\, ,
\eeq
where $\mathcal{U}_\Phi$ is the interaction potential of $\Phi$ (e.g. $\mathcal{U}_\Phi=\mu^2 \left| \Phi \right|^2 $, with $m_\Phi=\hbar\mu$ being the mass of the scalar field and $\mu$ the inverse Compton wavelength) and $J_\Phi$ is a putative current sourcing the scalar field. The corresponding energy-momentum tensor given by Eq.~\eqref{eq:EnergyTensorGeneral} is
\beq
T_\Phi^{\mu\nu} = \nabla^{(\mu} \Phi^* \nabla^{\nu)} \Phi - \frac{1}{2}g^{\mu\nu} \left( \nabla_\alpha \Phi^* \nabla^\alpha \Phi  +  \mathcal{U}_\Phi \left(\left| \Phi \right|^2 \right) \right) \, ,
\eeq
and its \acs{EOM} is the Klein-Gordon equation with some effective potential 
\beq
\Box \Phi = \frac{\delta \mathcal{U}_\Phi}{\delta \Phi^*} + J_\Phi \, .
\eeq

Another important case in the study of \acs{EMRI}s is when the matter Lagrangian represents a point particle of mass $m_p$ minimally coupled to gravity
\begin{equation} \label{matter_particles_Lagr}
\mathcal{L}_\text{M}=\mathcal{L}_p=- 2 m_p \int d\tau\, \frac{\delta^{(4)}\left(x^\alpha-x_p^{\alpha}\left(\tau\right)\right)}{\sqrt{-g}}\,,
\end{equation}
where $x_p\left(\tau\right)$ is the worldline of the particle parametrized by the proper time $\tau$. This Lagrangian implies the point particle follows geodesics of the background spacetime
\beq
\frac{d x^\alpha}{d\tau} \nabla_\alpha \left(\frac{dx^\mu}{d\tau}\right) = 0 \, , \label{eq:Geodesic}
\eeq
and has the energy-momentum tensor
\beq
	T_p^{\mu \nu} =  m_p \int d\tau \frac{\delta^{(4)}\left(x^\alpha-x_p^{\alpha}(\tau)\right)}{\sqrt{-g}} \,\frac{d{x}_p^{\mu}}{d\tau} \, \frac{d{x}_p^{\nu}}{d\tau}\, . \label{eq:PPTensor}
\eeq
%
%

\subsection{The Kerr spacetime}

We are interested in studying systems on timescales where the expansion of the Universe is negligible (e.g. the LISA mission lifetime). Hence, we can set $\Lambda = 0$. Additionally, in many scenarios we can treat the environment as a small perturbation to a particular vacuum \acs{BH} spacetime. In \acs{GR}, any stationary, axisymmetric, and asymptotically flat spacetime corresponding to a vacuum solution with an event horizon (i.e. a black hole solution) is a member of the $2$-parameter Kerr solution~\cite{PhysRevLett.26.331}. In Boyer-Lindquist coordinates $(t,r,\theta,\varphi)$ this solution is expressed as
\beq
ds^2_\text{Kerr}&=&-\left(1-\frac{2Mr}{\Sigma}\right)\,dt^2-\frac{4aMr}{\Sigma}\sin^2\theta \, dt\, d\varphi +\frac{\Sigma}{\Delta}\,dr^2 \nonumber \\
&+&\Sigma \, d\theta^2+\left(r^2+a^2+\frac{2Ma^2r\sin^2\theta}{\Sigma}\right)\sin^2\theta\, d\varphi^2 \, \label{eq:KerrLineElement} \, ,
\eeq
where $\Sigma = r^2 + a^2 \cos^2 \theta$ and $\Delta = r^2 + a^2 -2 M r$ \,. Far away from the $\acs{BH}$, $r \gg M, \, a$, one recovers spherical coordinates in Minkowski. The Kerr solution has a two-parameter group of isometries generated by the commuting Killing vector fields $\bm{k} = \partial/\partial t$ and $\bm{m} = \partial/\partial\varphi$. $\bm{k}$ is asymptotically timelike near infinity and is associated with the conserved mass $M$ of the \acs{BH}. $\bm{m}$ is asymptotically spacelike near infinity and generates rotations under the axis of symmetry of the \acs{BH}, leading to the conserved charge $J=a\,M$, where $J$ is the angular momentum of the \acs{BH}. Note that $0 \leq a/M < 1$, otherwise the \acs{BH} possesses a naked singularity.  

The Kerr solution has two horizons at the roots of $\Delta=0$, which are $r_{\pm} = M \pm \sqrt{M^2 - a^2}$. $r_+$ is the \acs{BH} event horizon while $r_-$ is a Cauchy horizon. Note also that the event horizon is a null hypersurface with normal $\bm{\xi} = \bm{k} + \Omega_\text{H} \bm{m}$, where
\beq
\Omega_\text{H} = \frac{a}{a^2 + r_+^2} \, , \label{eq:AngularBH}
\eeq
and one interprets $\Omega_\text{H}$ as the angular velocity of the \acs{BH}.

There is a region outside the \acs{BH} horizon, known as the \textit{ergoregion}, where $\bm{k}$ is spacelike and therefore no stationary observers exist. The ergoregion is limited by $r_+ < r < M + \sqrt{M^2 - a^2 \cos^2 \theta}$. In the ergoregion, the strong gravitational field causes a frame-dragging effect, which forces any observer to rotate along with the \acs{BH}.

It will also be useful to introduce the radial tortoise coordinate $r_*$
\beq
\frac{dr_*}{dr} &=& \frac {r^2 + a^2}{\Delta} \Rightarrow \nonumber \\
\Rightarrow r_* &=& r + \frac{r_+^2 + a^2}{r_+ - r_-}\log \left| \frac{r - r_+}{2M} \right| - \frac{r_-^2 + a^2}{r_+ - r_-}\log \left| \frac{r - r_-}{2M} \right| \, , \nonumber \label{eq:TortoiseKerr} \\
\eeq
which pushes the \acs{BH} horizon to $r_* \rightarrow -\infty$.

In the non-rotating limit ($a=0$), the Kerr metric reduces to Schwarzschild 
\beq
ds^2_\text{Schw}&=&-\left(1-\frac{2M}{r}\right)\,dt^2 + \frac{1}{1-\frac{2M}{r}}\,dr^2 + r^2d\Omega^2 \, \label{eq:SchwarzschildLineElement} \, , 
\eeq
with $d\Omega^2  = d\theta^2 + \sin^2\theta\, d\varphi^2$ the metric on the $2$-sphere. The event horizon goes to $r_+ = 2M$, there is no Cauchy horizon, and the tortoise coordinate becomes
\beq
r_* = r + 2M \log \left| \frac{r}{2M}-1 \right| \, .  \label{eq:TortoiseSchw}
\eeq

The Schwarzschild solution is a particular case of spherically-symmetric spacetimes, which in general can be described by the line element
\beq
ds^2_\text{Spherical}&=&-A\left(r\right)\,dt^2 + \frac{1}{B(r)}\,dr^2 + r^2d\Omega^2  \, \label{eq:SphericalLineElement} \, .
\eeq
The generalization of the radial tortoise coordinate for these spacetimes is
\beq
\frac{dr_*}{dr} = \frac{1}{\sqrt{A(r)\,B(r)}} \, .
\eeq
%

\subsection{Black-Hole Perturbations}\label{sec:BHPT}

\acs{EMRI}s can also be modeled as a point particle of mass $m_p$, representing the stellar-mass compact object, perturbing the \acs{SMBH} of mass $M$, where $m_p/M \ll 1$. To handle this, the natural framework is \acs{BH} perturbation theory~\cite{chandrasekhar1992mathematical, Pound:2021qin}, which describes the spacetime using a background metric $g_{\mu\nu}^{(0)}$, and adds a perturbation $h_{\mu\nu}$ such that the full metric is given by 
\beq
g_{\mu\nu}(x^\alpha) = g_{\mu\nu}^{(0)}(x^\alpha) + \epsilon \, h_{\mu\nu} (x^\alpha) + \mathcal{O}(\epsilon^2)\, , \label{eq:ExpansionH}
\eeq
where $\epsilon \ll 1$ is an expansion parameter. In the case of \acs{EMRI}s, this would be the mass ratio $q = m_p / M $. For accurate parameter estimation with LISA, this expansion will have to be carried until second order in the mass ratio $q$~\cite{Hinderer:2008dm, Pound:2019lzj, Spiers:2023cip}. 

\paragraph{Regge-Wheeler and Zerilli Equation\\}

We will often focus on spherically-symmetric backgrounds (e.g. Schwarschild), where any perturbation can be decomposed into irreducible representations of $SO(2)$. Then, gravitational perturbations can be expanded in ten spherical harmonics, which are the tensorial version of the standard spherical harmonics $Y^{\ell m} \left(\theta, \varphi\right)$ for scalars~\cite{NIST:DLMF}. The perturbations can be grouped into polar/electric/even type or axial/magnetic/odd type, depending on their behavior under parity transformations $\left(\theta, \varphi \right) \rightarrow \left(\pi - \theta, \pi - \varphi \right)$~\cite{Sago:2002fe, Martel:2005ir} 
\beq
\bm{h}^\text{axial}&=&\sum_{\ell,m}\frac{\sqrt{2\ell(\ell+1)}}{r} \bigg[i\,h_{1}^{\ell m }\bm{c}_{\ell m}
-h_{0}^{\ell m} \bm{c}^{0}_{\ell m} \nonumber + \frac{\sqrt{(\ell+2)(\ell-1)}}{2}h_{2}^{\ell m }\bm{d}_{\ell m} \bigg]\ , \\
\bm{h}^{\text{polar}}&=&\sum_{\ell,m} \bigg[A\,H^{\ell m }_0 \bm{a}^{0}_{\ell m}-i\sqrt{2}H^{\ell m}_1 \bm{a}^{1}_{\ell m} +\frac{1}{B}H_2^{\ell m }\bm{a}_{\ell m} +\sqrt{2}K^{\ell m } \bm{g}_{\ell m} \nonumber \\
&+& \frac{\sqrt{2\ell \left( \ell +1 \right)}}{r}\left(h^{(e) \ell m }_1 \bm{b}^1_{\ell m} - i h^{(e) \ell m }_0 \bm{b}^0_{\ell m} \right) \nonumber \\
&+& \left( \sqrt{\frac{(\ell +2 )(\ell +1) \ell (\ell -1 ) }{2} }\bm{f}_{\ell m} - \frac{\ell \left(\ell +1\right)}{\sqrt{2}} \bm{g}_{\ell m} \right) G^{\ell m} \bigg] \, , \nonumber \\ \label{eq:ExpansionMetric}
\eeq
where $\sum_{\ell,m}=\sum_{\ell=0}^{\infty}\sum_{m=-\ell}^\ell$. We are omitting the dependences on $(t,r,\theta,\varphi)$ to avoid cluttering, but the mode perturbations $h_1^{\ell m}, \, h_0^{\ell m}, \,...$ are only functions of $(t,r)$, while $\bm{a}^0_{\ell m}, \, \bm{a}_{\ell m}, \,...$, are the ten tensor spherical harmonics independent of $t$
\beq
\bm{a}^0_{\ell m}&=& 
\begin{pmatrix}
Y_{\ell m} & 0 & 0 & 0  \\
0 & 0 & 0 & 0 \\
0 & 0 & 0 & 0 \\ 
0 & 0 & 0 & 0 \\
\end{pmatrix}\, ,\\
\bm{a}^{1}_{\ell m}&=& \frac{i}{\sqrt{2}}
\begin{pmatrix}
0 & Y_{\ell m} & 0 & 0  \\
\text{Sym} & 0 & 0 & 0 \\
0 & 0 & 0 & 0 \\ 
0 & 0 & 0 & 0 \\
\end{pmatrix}\, ,\\
\bm{a}_{\ell m}&=&
\begin{pmatrix}
0 & 0 & 0 & 0  \\
0 & Y_{\ell m} & 0 & 0 \\
0 & 0 & 0 & 0 \\ 
0 & 0 & 0 & 0 \\
\end{pmatrix}\, ,\\
\bm{b}^{0}_{\ell m}&=&  \frac{i \, r}{\sqrt{2\ell \left(\ell+1\right)}}
\begin{pmatrix}
0 & 0 & \partial_\theta Y_{\ell m} & \partial_\varphi  Y_{\ell m}  \\
0 & 0 & 0 & 0 \\
\text{Sym} & 0 & 0 & 0 \\ 
\text{Sym} & 0 & 0 & 0 \\
\end{pmatrix}\, ,\\
\bm{b}_{\ell m}&=& \frac{r}{\sqrt{2\ell \left(\ell+1\right)}}
\begin{pmatrix}
0 & 0 & 0 & 0  \\
0 & 0 & \partial_\theta Y_{\ell m} & \partial_\varphi  Y_{\ell m} \\
0 & \text{Sym}  & 0 & 0 \\ 
0 & \text{Sym}  & 0 & 0 \\
\end{pmatrix}\, ,\\
\bm{c}^0_{\ell m}&=& \frac{r}{\sqrt{2\ell \left(\ell+1\right)}}
\begin{pmatrix}
0 & 0 & \frac{1}{\sin\theta}\,\partial_\varphi Y_{\ell m} & - \sin \theta \,\partial_\theta Y_{\ell m}  \\
0 & 0 & 0 & 0 \\
\text{Sym} & 0 & 0 & 0 \\ 
\text{Sym} & 0 & 0 & 0 \\
\end{pmatrix}\, ,\\
\bm{c}_{\ell m}&=& \frac{i \, r}{\sqrt{2\ell \left(\ell+1\right)}}
\begin{pmatrix}
0 & 0 & 0 & 0 \\
0 & 0 & \frac{1}{\sin\theta}\,\partial_\varphi Y_{\ell m} & - \sin \theta \,\partial_\theta Y_{\ell m}  \\
0 & \text{Sym}& 0 & 0 \\ 
0 & \text{Sym} & 0 & 0 \\
\end{pmatrix}\, ,\\
\bm{d}_{\ell m}&=&\frac{i r^2}{\sqrt{2\left(\ell+2\right)  \left(\ell+1\right) \ell  \left(\ell-1\right) }} 
\begin{pmatrix}
0 & 0 & 0 & 0 \\
0 & 0 & 0 & 0 \\
0 & 0 & - \frac{1}{\sin \theta}\, X_{\ell m} & \sin \theta\, W_{\ell m}\\ 
0 & 0 & \text{Sym} & \sin \theta \, X_{\ell m} \\
\end{pmatrix} \, , \nonumber \\
\eeq
\beq
\bm{f}_{\ell m}&=&\frac{r^2}{\sqrt{2\left(\ell+2\right)  \left(\ell+1\right) \ell  \left(\ell-1\right) }} 
\begin{pmatrix}
0 & 0 & 0 & 0  \\
0 & 0 & 0 & 0 \\
0 & 0 & W_{\ell m} & X_{\ell m} \\ 
0 & 0 & \text{Sym} & -\sin^2 \theta \, W_{\ell m} \\
\end{pmatrix}\, , \nonumber \\ \\
\bm{g}_{\ell m}&=&\frac{r^2}{\sqrt{2}} 
\begin{pmatrix}
0 & 0 & 0 & 0  \\
0 & 0 & 0 & 0 \\
0 & 0 & Y_{\ell m} & 0 \\ 
0 & 0 & 0 & \sin^2 \theta \,Y_{\ell m} \\
\end{pmatrix}\, .
\eeq
where `` $\text{Sym}$'' is the symmetric of that respective matrix entry and
\beq
X_{\ell m} &=& 2 \,\partial_\varphi \left( \partial_\theta -\cot \theta \right) Y_{\ell m} \, , \\
W_{\ell m} &=& \left(\partial^2_\theta - \cot \theta \, \partial_\theta - \frac{1}{\sin^2 \theta} \partial_\varphi^2 \right) Y_{\ell m} \, .
\eeq
Introducing the inner product $(\cdot,\cdot)$ on the two-sphere
\beq
\left(\bm{R}^{\ell' m'}, \bm{S}^{\ell m} \right) = \int_{S_2} d\Omega \, \left(R^{\ell' m'}_{\mu \nu}\right)^* \, S^{\ell m}_{\lambda \rho} \, \eta^{\mu\lambda} \,  \eta^{\nu \rho} \, ,
\eeq
where 
\beq
\eta_{\mu\nu} = \text{diag}\left(-1, 1, r^2, r^2 \sin^2 \theta \right) \, ,
\eeq
one can check that the harmonics defined above are orthonormal, i.e. $\left(\bm{R}^{\ell' m'},\, \bm{S}^{\ell m} \right) = \delta_{R\,S}\,\delta_{l'\,l}\,\delta_{m'\,m}$.

In a similar fashion, the energy-momentum tensor can be expanded as 
\beq
\bm{T}&=&\sum_{\ell,m}
\bigg[{\cal A}^{0}_{\ell m }\bm{a}^{0}_{\ell m}+{\cal A}^{1}_{\ell m}\bm{a}^{1}_{\ell m}
+{\cal A}_{\ell m }\bm{a}_{\ell m}+{\cal B}^{0}_{\ell m }\bm{b}^{0}_{\ell m}+{\cal B}_{\ell m }\bm{b}_{\ell m}\nonumber \\
&+&{\cal Q}^{0}_{\ell m }\bm{c}^{0}_{\ell m}+{\cal Q}_{\ell m }\bm{c}_{\ell m} +{\cal D}_{\ell m }\bm{d}_{\ell m}+{\cal G}_{\ell m}\bm{g}_{\ell m}+{\cal F}_{\ell m }\bm{f}_{\ell m}\bigg]\ .\label{harmonicexp} \nonumber \\
\eeq
where, for a given source, the expansion coefficients can be obtained by projecting the energy-momentum tensor on the respective spherical harmonic, e.g. ${\cal A}^{0}_{\ell m } = (\bm{a}^{0}_{\ell m}, \bm{T})$.

It is important to recall that \acs{GR} is invariant under diffeomorphisms. Infinitesimally
\beq
x_\mu \rightarrow x'_\mu = x_\mu  + \xi_\mu \Rightarrow
h_{\mu\nu} \rightarrow h'_{\mu\nu} = h_{\mu\nu} - 2\nabla_{(\mu} \xi_{\nu)} \, ,
\eeq
where $\bm{\xi}$ is the vector field generating the diffeomorphism. $\bm{\xi}$ can also be expanded in a set of polar and axial harmonics
\beq
\bm{\xi} &=& 
\sum_{\ell,m} \left( - \frac{\xi_t^{\ell m}}{A}  \, , \,  B \xi_r^{\ell m}  \, , \,  0 \, , \, 0 \right)Y_{\ell m} \nonumber \\
&+& \frac{\xi_{\Omega}^{\ell m}}{r^2 \sin\theta} \left( 0 \, , \, 0 \, , \, \sin \theta \partial_\theta Y_{\ell m} \, , \, \partial_\varphi Y_{\ell m} \right) \nonumber \\
&+& \frac{\xi_\text{ax}^{\ell m}}{\sqrt{2\ell \left(\ell + 1\right)}}  \left(0 \, , \, 0 \, , \, \frac{1}{r \sin \theta} \partial_\varphi Y_{\ell m} \, , \, - \frac{1}{r} \partial_\theta Y_{\ell m} \right) \, , 
\eeq
where $\xi_t^{\ell m}, \, \xi_r^{\ell m}, \, \xi_\Omega^{\ell m}, \, \xi_\text{ax}^{\ell m}$ are only functions of $t$ and $r$. The first two terms on the right-hand side are the polar ones and and the second is the single axial degree of freedom. Then
\beq
&&2\nabla_{(\mu} \xi_{\nu)} = \left( 2 \partial_t \xi_t  - A' B \xi_r \right) \bm{a}^{0} \nonumber \\
&-& i \sqrt{2} \left( \partial_r \xi_t  + \partial_t \xi_r  - \frac{A'}{A} \xi_t \right) \bm{a}^{1} \nonumber \\ 
&+& \left(2 \partial_r \xi_r + \frac{B'}{B} \xi_r \right) \bm{a} - i\frac{\sqrt{2\ell\left(\ell+1\right)}}{r}\left(\xi_t + \partial_t \xi_\Omega\right) \bm{b}^{0} \nonumber \\
&+& \frac{\sqrt{2\ell\left(\ell+1\right)}}{r}\left(\partial_r \xi_\Omega + \xi_r - \frac{2}{r}\xi_\Omega \right)\bm{b} + \partial_t \xi_\text{ax} \bm{c}^{0} \nonumber \\
&-& i \left( \partial_r \xi_\text{ax} - \frac{\xi_\text{ax}}{r}\right)\bm{c} + i \frac{\sqrt{\left(\ell +2\right)\left(\ell-1\right)}}{r}\xi_\text{ax} \bm{d} \nonumber \\ 
&+& \frac{\sqrt{2\left(\ell+2\right)  \left(\ell+1\right) \ell  \left(\ell-1\right) } }{r^2} \xi_\Omega \bm{f} \nonumber \\
&+& \frac{\sqrt{2}}{r^2} \left(2rB \xi_r - \ell \left( \ell +1 \right) \xi_\Omega \right)\bm{g}  \, , 
\eeq
where the prime denotes a derivative with respect to $r$, and from now on we omit $\left(\ell, \, m\right)$ indices unless necessary to avoid cluttering. Therefore, we can pick $\xi^\mu$ judiciously to eliminate four components of the metric perturbations, one in the axial sector and three in the polar one. A common choice is the \textit{Regge-Wheeler} gauge, where we set to zero terms involving angular derivatives of the highest order
\beq
h_2=h_0^{(e)}=h_1^{(e)}=G=0 \, . \label{eq:RWgauge}
\eeq

In the founding work of \acs{BH} perturbation theory, Regge and Wheeler employed this gauge to derive a decoupled wave equation for a master function $\Psi_\text{RW}$, which encodes all the dynamics of the axial sector around a Schwarzschild \acs{BH}~\cite{Regge:1957td}. Years later, Zerilli obtained the same for the polar sector~\cite{Zerilli:1970se, Zerilli:1970wzz}~\footnote{Zerilli's original papers contain numerous typos, namely in the definitions of the spherical harmonics. Ref.~\cite{Sago:2002fe} corrects them, and we follow it in our definitions.}. However, their derivation involves a lengthy manipulation of Einstein's equations, which we will revisit in Chapter~\ref{ch:GBH}, when we study \acs{EMRI}s in non-vacuum environments. For now, it is sufficient to state the final master wave equations for the axial (``RW'') and polar sectors (``Z'')~\cite{Berti:2009kk, Martel:2003jj}
\beq
\left(-\partial^2_t + \partial^2_{r_*} - V_\text{Z/RW} \right) \Psi_\text{Z/RW} = S_\text{Z/RW}\, . \label{eq:MasterZRW}
\eeq
$\Psi_\text{Z/RW}$ are the master functions, which are related to the metric perturbations above through
\beq
\Psi_\text{RW} &=& \frac{1}{r}\left(1 - \frac{2M}{r} \right)h_1 \, , \label{eq:MasterFuncRW} \\
\Psi_\text{Z}  &=& \frac{r}{\lambda +1}\left[K + \frac{1 - 2M/r}{\lambda + 3M/r}\left(H_2 - r \, \partial_r K \right) \right]  \, , \label{eq:MasterFuncZ}
\eeq
where $\lambda= \left(\ell-1\right)\left(\ell+2\right)/2$, while the effective potentials are~\cite{Berti:2009kk}
\beq
V_\text{RW} &=& \left(1-\frac{2M}{r}\right)\left[\frac{\ell \left(\ell + 1 \right) }{r^2} - \frac{6M}{r^3} \right] \, , \\
V_\text{Z} &=& \frac{2}{r^3} \left(1-\frac{2M}{r}\right) \frac{9M^3 + 3 \lambda^2 M \,r^2 + \lambda^2\left(1 + \lambda \right)r^3 + 9M^2 \lambda \, r}{(3M + \lambda r)^2}\, . \nonumber \\
\eeq 
Finally, $S_\text{Z/RW}$ are source terms which depend on the energy-momentum tensor. Assuming the source is localized, close to the \acs{BH} horizon and at large distances the master equations become
\beq
r \rightarrow 2M, \, \infty \Rightarrow \left(-\partial^2_t + \partial^2_{r_*} \right)  \Psi_\text{Z/RW} = 0 \, .
\eeq
Therefore, they admit two physical linearly independent solutions traveling at the speed of light. Physical boundary conditions are represented by ingoing waves at the horizon, $ A_\text{in} e^{-i\omega( t + r_*)}$, and the other outgoing waves at infinity,  $ A_\text{out} e^{-i \omega( t - r_*)}$.

$\Psi_\text{Z/RW}$ are gauge invariant quantities and control the two radiative degrees of freedom of \acs{GR} at large distances~\cite{Martel:2003jj, Martel:2005ir}
\beq
h_+ - i h_\times = \lim_{r\rightarrow \infty} \frac{1}{2r} \sum_{\ell = 2}^{\infty}\sum_{m = -\ell}^{\ell} \sqrt{\frac{(\ell + 2)!}{(\ell - 2)!}}\left(\Psi^{\ell m}_\text{Z} -2 i \int_{-\infty}^t \Psi_\text{RW}^{\ell m} \right)\, _{-2}Y^{\ell m} \ , \nonumber \
\eeq
with 
\beq
_{-2}Y^{\ell m}\left(\theta,\varphi\right) = \sqrt{ \frac{(\ell - 2)!}{(\ell + 2)!} }  \left(W^{\ell m}\left(\theta,\varphi\right) - \frac{i}{\sin \theta} X^{\ell m}\left(\theta,\varphi\right)  \right) \, , \nonumber
\eeq
the spin-$2$ weighted spherical harmonic. 

The total energy flux carried to infinity by \acs{GW}s can also be computed through these master functions~\cite{Martel:2003jj, Martel:2005ir} 
\beq
\dot{E}_\infty^{\ell m} &=& \frac{1}{64\pi} \frac{(\ell + 2)!}{(\ell - 2)!} \left[ \left| \dot{\Psi}^{\ell m}_Z \right|^2 + 4  \left| \Psi^{\ell m}_\text{RW} \right|^2  \right] \, , \nonumber \\
\dot{E}_\infty &=& \sum_{\ell = 2}^{\infty}\sum_{m = -\ell}^{\ell} \dot{E}_\infty^{\ell m} \, , \label{eq:FluxNonRot}
\eeq
where the overdot denotes differentiation with respect to time. 

The last two sums over the multipoles only started at $\ell = 2$. The monopole ($\ell = 0$) and dipole ($\ell = 1$) perturbations are pure gauge in \acs{GR} and do not contribute to the radiative degrees of freedom, though the perturbations may be non-zero in some spacetime regions. Additionally, we have that $\bm{b}^0_{\ell m}=\bm{b}_{\ell m}=\bm{c}^0_{\ell m}=\bm{c}_{\ell m}=0$ for $\ell = 0$, and $\bm{d}_{\ell m}=\bm{f}_{\ell m}=0$ for $\ell \leq 1$. So the Regge-Wheeler gauge~\eqref{eq:RWgauge} is not completely fixed for $\ell \leq 1$. 

This property is related to the fundamental nature of gravity, which according to \acs{GR} should be mediated by a massless spin-$2$ boson, the ``graviton''. As a result, \acs{GR} exhibits conservation of the energy-momentum tensor, analogous to vector current conservation in electromagnetism that is mediated by a massless spin-$1$ boson, the photon. Therefore, there is no dipolar radiation in \acs{GR}, similarly to the absence of monopole radiation in electromagnetism.

\paragraph{Teukolsky Equation\\}

Obtaining a master equation for linear perturbations in Kerr is more complicated since the loss of spherical symmetry means that separation in spherical harmonics is no longer possible. Instead, Teukolsky was able to decouple the equations governing any type of linear perturbations to Kerr using the Newman-Penrose formalism, in which tensors are projected onto a null tetrad $\{\bm{n},\bm{l}, \bm{m}, \bar{\bm{m}}\}$~\cite{Teukolsky:1973ha} (the overbar in $\bar{\bm{m}}$ denotes complex conjugation).The details of this formalism are not necessary to follow the thesis, and we refer the interested readers to Refs.~\cite{Teukolsky:1973ha, chandrasekhar1992mathematical, Mino:1997bx, Loutrel:2020wbw, Ripley:2020xby, Pound:2021qin} for more detailed descriptions. Here it will suffice to work with the final master equation describing any type of linear perturbations to Kerr, which in Boyer-Lindquist coordinates and using the Kinnersley tetrad
\beq
\bm{l} &=& \left( \frac{r^2 + a^2}{\Delta}, 1, 0 , \frac{a}{\Delta}\right) \, , \\
\bm{n} &=& \frac{1}{2\Sigma} \left(r^2+a^2,-\Delta, 0, a \right) \, , \\
\bm{m} &=& \frac{1}{\sqrt{2}\left(r+i a \cos \theta \right)}\left(i a \sin \theta , 0, 1, \frac{i}{ \sin \theta} \right) \, , 
\eeq
reads
\beq
&&\left[\frac{\left(r^2+a^2\right)^2}{\Delta}-a^2\sin^2\theta\right]\partial^2_t \Psi-\Delta^{-s}\partial_r\left(\Delta^{s+1}\partial_r \Psi\right) \nonumber \\
&+&\frac{4 a M r}{\Delta}\partial_\varphi\partial_t\Psi +2s\left[r+\frac{M\left(a^2-r^2\right)}{\Delta}+ia\cos\theta\right]\partial_t\Psi \nonumber \\
&-& \frac{1}{\sin\theta}\partial_\theta\left(\sin\theta\partial_\theta \Psi\right)-2s\left[\frac{a\left(r-M\right)}{\Delta}+\frac{i\cos\theta}{\sin^2\theta}\right]\partial_\varphi \Psi \nonumber \\
&-&\left[\frac{1}{\sin^2\theta}-\frac{a^2}{\Delta}\right]\partial^2_\varphi \Psi+\left(s^2\cot^2\theta-s\right)\Psi=4\pi\,\Sigma\, T \, .\label{eq:TeukolskyMaster}
\eeq
Here, $s$ determines the spin-weight of the master variable $\Psi$, which corresponds to $s=0$ for scalars, $s=\pm 1$ for electromagnetic perturbations (i.e. vectors), and $s=\pm 2$ for gravitational ones. $T$ encodes the source term and depends on the energy-momentum tensor. We particularize it below for a point-particle orbiting a Kerr \acs{BH}.
 	
For $s=0$ the Teukolsky equation reduces to the Klein-Gordon equation for a massless scalar field. For $s=\pm 2$, it describes perturbations to the Weyl tensor that govern the radiative degrees of freedom at the \acs{BH} horizon ($s=2$), and at infinity ($s=-2$). Since we are interested in studying \acs{GW} emission, we will mostly focus on the $s=-2$ case, for which the master variable is $\Psi = \rho^{-4}\psi_4$, with $\rho = -1/(r-ia \cos \theta)$. $\psi_4$ is then related to the \acs{GW} polarizations at large distances through 
\beq
\frac{1}{2}\left(\frac{\partial^2 h_+}{\partial t^2} - i \frac{\partial^2 h_\times}{\partial t^2} \right) = \lim_{r \rightarrow \infty} \psi_4 \, . \label{eq:RadiativeDegrees}
\eeq
The flux of energy carried by \acs{GW}s to infinity can also be computed through $\psi_4$ 
\beq
\dot{E}^\infty= \lim_{r\rightarrow \infty} \frac{r^2}{4\pi} \int_{S_2} d\Omega \int_{- \infty}^{t} dt' \Psi (t',r,\theta,\varphi) \, . \label{eq:FluxGW}
\eeq
%

\section{Numerical Framework}\label{sec:Numerics}

In general, analytic solutions to the master equations we discussed are not possible, and numerical methods must be used instead. In this thesis, we will often use a two-step Lax-Wendroff algorithm with second-order finite differences appropriate to solve ``wave-like'' partial differential equations in the time-domain~\cite{Krivan:1997hc, Pazos_valos_2005, Sundararajan:2007jg, Zenginoglu:2011zz}. In this section we provide an overview of the algorithm.

\subsection{The Lax-Wendroff algorithm}

Let us assume that we have manipulated our wave equation (e.g. the Teukolsky equation) so that in the homogeneous version it reads
\beq
\partial^2_\tau \Psi = \left[ \tilde{A}^{\tau\rho} \partial_\tau \partial_\rho  + \tilde{A}^{\rho\rho} \partial_\rho^2 + \tilde{A}^{\theta\theta} \partial_\theta^2 + \tilde{B}^\tau \partial_\tau + \tilde{B}^\rho \partial_\rho + \tilde{B}^\theta \partial_\theta + \tilde{C} \right] \Psi \, . \nonumber \\
\label{eq:GeneralWaveEq}
\eeq
Here, $\tau$ and $\rho$ are redefined time and radial variables, respectively. This coordinate $\rho$ should not be confused with the $\rho$ variable appearing in the definition of the master variable $\Psi$ in terms of $\psi_4$, and we adopt it for consistency with Ref.~\cite{Zenginoglu:2011zz}. The coefficients depend only on $\rho$ and $\theta$, which can be achieved even for rotating \acs{BH}s due to the axisymmetry of Kerr. In spherical symmetry, we could further eliminate the $\theta$ dependence by expanding in spherical harmonics, which would be absorbed in the $\tilde{C}$ coefficient. This procedure is easily generalizable for a system of two or more coupled wave equations, in which case the coefficients above would be promoted to matrices. 

Eq.~\eqref{eq:GeneralWaveEq} can be reduced to a system of first-order partial differential equations by defining the auxiliary variable $\Pi$
\beq
\Pi &=& \left( \partial_\tau + b \, \partial_\rho  \right) \Psi \, , \\
b &=& - \left( \tilde{A}^{\tau\rho} + \sqrt{ \left(\tilde{A}^{\tau\rho}\right)^2 + 4 \tilde{A}^{\rho\rho}  }   \right) / 2 \, . 
\eeq
We can then rewrite it as
\beq
\partial_\tau \bm{u} + \bm{M} \cdot \partial_\rho \bm{u} + \bm{L} \cdot \bm{u} + \bm{A} \cdot \bm{u}  = \bm{T} \, ,
\label{eq:VectorWaveEq}
\eeq
where we reintroduced the source term $\bm{T}$ and
\beq
\bm{u} = \left\{ \Psi_\text{R},  \Psi_\text{I},  \Pi_\text{R},  \Pi_\text{I} \right\} \, ,
\eeq
and the subscripts $\text{R}$ and $\text{I}$ refer to the real and imaginary part. The matrices $\bm{M}, \, \bm{A}\, , \bm{L}$ have the general structure
\beq
\bm{M}&=&
\begin{pmatrix}
b & 0 & 0 & 0 \\
0 & b & 0 & 0 \\
m_{31}& m_{32} & -b & 0 \\ 
-m_{32} & m_{31} & 0 & -b \\
\end{pmatrix}\, ,\\
\bm{A}&=&
\begin{pmatrix}
0 & 0 & -1 & 0 \\
0 & 0 &  0 & -1 \\
a_{31} & a_{32} & a_{33} & a_{34} \\ 
-a_{32} & a_{31} & -a_{34} & a_{33} \\
\end{pmatrix}\, ,\\
\bm{L}&=&
\begin{pmatrix}
0 & 0 & 0 & 0 \\
0 & 0 & 0 & 0 \\
l_{31} & 0 & 0 & 0\\ 
0 & l_{31} & 0 & 0 \\
\end{pmatrix} \, .
\eeq
Further below, we will make this coefficients explicit for the Teukolsky equation. The matrix $\bm{L}$ contains the angular derivatives in $\theta$ (in spherical symmetry $\bm{L}=0$). This decomposition is particularly useful for inferring the hyperbolicity of the system by computing the eigenvalues/eigenvectors of $\bm{M}$. Since $\bm{L}$ contains second-order derivatives, hyperbolicity is not guaranteed even if $\bm{M}$ has a complete set of linearly independent eigenvectors with real eigenvalues. However, this method was numerically well-behaved and convergent in all the cases we studied.

We can finally construct a time-explicit evolution scheme based on the two-step, second-order Lax-Wendroff finite-difference method. First, we rewrite Eq.~\eqref{eq:VectorWaveEq} in the form of an advection equation
\beq 
\left(\partial_\tau + \bm{D}\,\partial_\rho\right)\bm{u}=\bm{S}\, , 
\eeq
with
\beq 
\bm{D}&=&\text{diag}\left(b,b,-b,-b\right) \, , \\
\bm{S}&=&-\left(\bm{M}-\bm{D}\right)\cdot \partial_{\rho}\bm{u}-\bm{L}\cdot \bm{u} -\bm{A}\cdot \bm{u} + \bm{T} \, .
\eeq

We discretize this equation on a uniform two-dimensional grid, with grid steps $\delta \rho$ and $\delta \theta$. Each iteration of the numerical integration has two steps. In the first step, we compute an intermediate solution between the main grid points
\beq
\bm{u}^{n+1/2}_{i+1/2}&=&\frac{1}{2}\left(\bm{u}^n_{i+1}+ \bm{u}^n_{i}\right)- \frac{\delta \tau}{2}\left[\frac{1}{\delta \rho}\bm{D}^n_{i+1/2}\left(\bm{u}^n_{i+1}-\bm{u}^n_i\right) - \bm{S}^n_{i+1/2} \right] \, . \nonumber \\
\eeq
We omit angular angular indexes to avoid cluttering and $\delta \tau$ is the time step. The fields are centered in the angular direction and angular derivatives are approximated by a centered second-order difference stencil 
\beq
\partial_\theta \bm{u}_j &=& \frac{\bm{u}_{j+1} - \bm{u}_{j-1} }{2 \, \delta \theta} + \mathcal{O}\left( \delta \theta^2 \right)\, , \\
\partial_\theta^2 \bm{u}_j &=& \frac{\bm{u}_{j+1} - 2\bm{u}_{j} + \bm{u}_{j-1} }{\delta \theta^2}  + \mathcal{O}\left( \delta \theta^2 \right) \, ,
\eeq
while the radial derivatives are approximated using centered second-order differences on the values $i$ and $i+1$. The algebraic terms in $\bm{D}^n_{i+1/2}$ and $\bm{S}^n_{i+1/2}$ are given by an average between the values at $i$ and $i+1$. This intermediate solution is then used to update the solution at the next time step
\beq
\bm{u}^{n+1}_{i}&=&\bm{u}^n_{i} - \delta \tau \left[\frac{1}{\delta \rho}\bm{D}^{n+1/2}_{i}\left(\bm{u}^{n+1/2}_{i+1/2}-\bm{u}^{n+1/2}_{i-1/2}\right) - \bm{S}^{n+1/2}_{i} \right] \, ,\nonumber\\
\eeq
while now the centered radial differences and averages are taken on the values $\bm{u}^{n+1/2}_{i+1/2}$ and $\bm{u}^{n+1/2}_{i-1/2}$. The only missing step is to impose appropriate boundary conditions for $\rho$ and $\theta$. We will illustrate this with a concrete example below. 

\subsection{A practical case: the Teukolsky equation}\label{sec:TeukNum}

Let us now apply this method to the Teukolsky equation~\eqref{eq:TeukolskyMaster}. 

\paragraph{Coordinate transformations.\\} 

Boyer-Lindquist coordinates suffer from several pathologies that need to be cured for numerical schemes. The radial coordinate becomes singular at the \acs{BH} horizon and there is a frame-dragging effect along $\varphi$ due to the \acs{BH} rotation. Furthermore, the asymptotic behavior of $\Psi$ at both the \acs{BH} horizon and infinity needs to be considered to prevent a numerical blow-up of the solution. The general behavior for spin-weight $s$ is different for ingoing and outgoing waves 
\beq
\lim_{r\rightarrow +\infty} \left| \Psi\right | &\sim&  
\begin{cases}
1/r^{2s+1} \quad \, \, \text{outgoing} \\
1/r \qquad \quad\text{ingoing} \\
\end{cases}\, , \label{eq:AsymptoticBehaviorOuter} \\
\lim_{r\rightarrow r_+} \left| \Psi\right | &\sim & \,  
\begin{cases}
1 \qquad \quad \,\text{outgoing} \\
\Delta^{-s}  \qquad \text{ingoing}\\
\end{cases}\, .
\label{eq:AsymptoticBehaviorInner}
\eeq

To deal with the radial singularity at the \acs{BH} horizon we use the radial tortoise coordinate $r_*$ introduced in Eq.~\eqref{eq:TortoiseKerr}. As mentioned above, $r_*$ pushes the \acs{BH} horizon to $- \infty$, which in the numerical domain is put at some finite distance where artificial ingoing boundary conditions are imposed. 

To handle the angular twisting in the ergoregion, we introduce a modified azimuthal coordinate
\beq
d\tilde{\varphi} &=& d\varphi + \frac{a}{\Delta}dr \Rightarrow \nonumber \\
\Rightarrow  \tilde{\varphi} &=& \varphi + \frac{a}{r_+ - r_-} \log \left|\frac{r-r_+}{r-r_-} \right| \, .
\eeq

Finally, we rescale the master variable according to the respective asymptotic behavior at large distances~\eqref{eq:AsymptoticBehaviorOuter}. The axisymmetry of the Kerr background also allows us to separate the azimuthal dependence in $\tilde{\varphi}$ with the mode number $m$
\beq
\Psi\left(t,r,\theta,\varphi\right) = e^{i m \tilde{\varphi}}r^{\left(-2s+1\right)}\psi\left(t,r,\theta\right) \, . 
\eeq

Applying these transformations to the Teukolsky equation~\eqref{eq:TeukolskyMaster} we arrive at
\beq
&&\left[\frac{\left(r^2+a^2\right)^2}{\Delta}-a^2\sin^2\theta\right]\partial^2_t \psi-\frac{\left(r^2+a^2\right)^2}{\Delta}\partial^2_{r^*} \psi \nonumber \\
&-&\partial^2_\theta\psi-\cot\theta\,\partial_\theta\psi-\frac{2}{\Delta}\bigg[M \, s \left(r^2-a^2\right)-rs\Delta \nonumber \\
&-&ia\left(s\Delta\cos\theta+2Mmr\right)\bigg]\partial_t\psi-\frac{1}{r\Delta}\bigg[2iamr\left(r^2+a^2\right) \nonumber \\
&-&2rs\left(r^2+a^2\right)\left(M-r\right) -\left(4\left(r^2+a^2\right)s+2a^2\right)\Delta  \bigg]\partial_{r^*}\psi \nonumber \\
&-&\frac{1}{r^2\Delta}\bigg[2\left(1+s\right)\left(1+2s\right)\Delta^2-r\Delta\big[r\left(s\cot\theta+m\csc\theta\right)^2  \nonumber \\
&-&r\left(s-2\left(1+s\right)\left(1+2s\right)\right)-2M\left(1+s\right)\left(1+2s\right)\big] \nonumber \\
&-&2iamr\left[2rs\left(M-r\right)+\left(1+2s\right)\Delta \right]\bigg]\psi=4\pi \, \Sigma\, r^{2s+1} e^{-im\tilde{\varphi}} T  \, \nonumber . \\
\label{eq:TeukolskyAdapted}
\eeq
Multiplying it by $\Delta / \sigma^2$, where
\beq
\sigma^2 = \left(r^2+a^2\right)^2  - \Delta \, a^2 \sin^2\theta \, ,
\eeq
we finally bring it to the form in Eq.~\eqref{eq:GeneralWaveEq} with the following coefficients
\beq
&& b = \frac{r^2+a^2}{\sigma} \nonumber \quad ,\\
&& m_{31} = -2\frac{rs\left(M-r\right)\left(r^2+a^2\right)+\left(2\left(r^2+a^2\right)s+a^2\right)\Delta}{r\sigma^2} \nonumber \\
&&\qquad -b2s\frac{M\left(a^2-r^2\right)+r\Delta}{\sigma^2} + b \partial_{r^*}b \nonumber \quad , \\
&& m_{32} = 2am\frac{r^2+a^2}{\sigma^2}+b2a\frac{s\Delta \cos \theta+2 M m r}{\sigma^2} \quad , \nonumber \\
&&a_{31} =\frac{\Delta}{r^2 \sigma^2}\big[r^2\left(s\cot\theta+m\csc\theta\right)^2-2Mr\left(1+s\right)\left(1+2s\right) \nonumber  \\
&&\qquad +\,r^2\left(4s^2+5s+2\right)-2\left(1+s\right)\left(1+2s\right)\Delta \big] \quad , \nonumber \\
&& a_{32} = -2am\frac{2rs\left(M-r\right)+\left(1+2s\right)\Delta}{r \sigma^2}  \quad , \nonumber \\
&&a_{33} = 2s\frac{M\left(a^2-r^2\right)+r\Delta}{\sigma^2} \quad , \quad  a_{34} = - 2a\frac{2mMr+s\Delta\cos\theta}{\sigma^2} \quad , \nonumber \\
&&l_{31} = - \frac{\Delta}{\sigma^2}\partial_\theta^2 -  \frac{\Delta}{\sigma^2} \cot \theta \,\partial_\theta \quad .
\eeq

\paragraph{Boundary conditions\\} 

To evolve the system described above, we need to impose appropriate boundary conditions for $\psi$. The physical solution corresponds to having ingoing waves at the \acs{BH} horizon and outgoing waves at infinity. At the inner boundary, we exploit the asymptotic behavior of ingoing waves $\psi \sim \Delta^{-s}$ as in Eq.~\eqref{eq:AsymptoticBehaviorInner}. For $s=-2$, which describes the gravitational perturbations of interest, we can then set $\psi=\Pi=0$. Imposing outgoing boundary conditions at the outer boundary is more complicated and often one obtains spurious reflections there. One solution is to set the outer boundary far enough away so that it does not affect the interior domain in the maximal Cauchy development of our initial data. For example, if we evolve the system for $t=10^{3} M$ and extract the fields at $r_*^\text{ext}=500M$, then the outer boundary should be placed further than $r_*^\text{out} = 10^3M$ to prevent any signal from being reflected back and affect the field values at the extraction radius. The same strategy can be used at the inner boundary when the asymptotic behavior of $\psi$ is less trivial, as in the scalar case where it asymptotes to a constant. 

Finally, we use the symmetries of spheroidal harmonics, which are the generalization of spherical harmonics to axisymmetric backgrounds, to impose boundary conditions on the rotation axis~\cite{NIST:DLMF} 
\beq
\psi \Big|_{\theta = 0,\pi} = 0 &\quad& , \quad m\, \text{odd} \\
\partial_\theta \psi \Big|_{\theta = 0,\pi} = 0 &\quad& , \quad m  \, \text{even} \, .
\eeq

This numerical method has been used extensively in the literature for over two decades. Initially, it was used to study the late-time polynomial tail decay of perturbations in Kerr~\cite{Krivan:1997hc, Pazos_valos_2005, Burko:2010zj, Zenginoglu:2012us}, which require high precision and stable codes over long computational times. The method exhibited second-order convergence and a Courant condition $\delta t \leq \left(\delta r_*\, ,\,  5 \delta \theta \right)$. More recently, it has been applied to the study of \acs{EMRI}s~\cite{Sundararajan:2008zm, Sundararajan:2010sr, Zenginoglu:2011zz,Barausse:2011kb}, as we will do. Typical choices for the grid discretization are $\delta r_*/M = 0.05$, $\delta \theta = \pi / 64$, and $\delta t = \delta r_* /2$. 

\paragraph{Point-Particle Source Term\\}

We will now explain how to incorporate point-particle source terms in the Teukolsky equation, which can model the secondary object in an \acs{EMRI}. The source term $T$ appearing in Eq.~\eqref{eq:TeukolskyAdapted} is~\cite{Teukolsky:1973ha, Sundararajan:2008zm}
\beq
T &=& 2 \rho^{-4} T_4\, , \\
T_4 &=& \left( \tilde{\Delta} + 3\gamma - \bar{\gamma}+ 4\mu + \bar{\mu} \right) \left(\tilde{\Delta} + 2\gamma - 2 \bar{\gamma} + \bar{\mu} \right)T_{\bar{m}\bar{m}} \nonumber \\ 
&-&\left(\tilde{\Delta} + 3\gamma - \bar{\gamma} + 4\mu + \bar{\mu}\right) \left( \bar{\delta} - 2 \bar{\tau} + 2 \alpha \right)T_{n\bar{m}} \nonumber \\
&+& \left(\bar{\delta} - \bar{\tau} + \bar{\beta} + 3 \alpha + 4\pi \right)\left(\bar{\delta} - \bar{\tau} + 2\bar{\beta} + 2\alpha \right)T_{nn} \nonumber \\
&-& \left(\bar{\delta} - \bar{\tau} + \bar{\beta} + 4\pi \right)\left(\tilde{\Delta} + 2 \gamma + 2 \bar{\mu} \right)T_{n\bar{m}} \, ,
\eeq
where we alert again that now this $\rho = -1(r-i a \cos \theta)$ and not the general $\rho$ radial coordinate we used in the description of the Lax-Wendroff method. Also $T_{n\bar{m}} = n^\mu \bar{m}^\nu T_{\mu\nu} $, $T_{nn}= n^\mu n^\nu T_{\mu\nu} $, $T_{\bar{m}\bar{m}} = \bar{m}^\mu \bar{m}^\nu T_{\mu\nu}$. $\tilde{\Delta}$ and $\bar{\delta}$ are the differential operators 
\beq
\tilde{\Delta} &=& n^\mu \partial_\mu \nonumber \\
			   &=&\frac{\rho^2 \left(r^2+a^2\right)}{2}\frac{d}{dt} - \frac{\rho^2 \Delta}{2}\frac{d}{dr} + \frac{a i m \rho^2}{2} \, , \\
\bar{\delta} &=& \bar{m}^\mu \partial_\mu \nonumber \\
			 &=& - \frac{i a \sin \theta \rho^2 \left(r+ i a \cos \theta \right)}{\sqrt{2}}\frac{d}{dt} \nonumber \\
			 &+& \frac{\left(r + i a \cos \theta \right)\rho^2}{\sqrt{2}}\frac{d}{d\theta} + \frac{m \rho^2 \left( r + i a \cos \theta \right)}{\sqrt{2} \sin \theta} \, , 
\eeq
and the other coefficients are~\cite{Teukolsky:1973ha}
\beq
\beta &=& \frac{1}{r+i a \sin \theta}\frac{\cot \theta}{2\sqrt{2}}\, , \\
\pi &=& i a \frac{1}{(r-i a \sin \theta)^2} \frac{\sin \theta}{\sqrt{2}}\, , \\
\mu &=&  -\rho^2 \frac{1}{r- i a \sin \theta} \frac{\Delta}{2}\, , \\
\gamma &=& \mu + \rho^2 \frac{r-M}{2}\, , \\ 
\tau &=& -i a \rho^2 \frac{\sin \theta}{\sqrt{2}}\, , \\
\alpha &=& \pi - \bar{\beta}\, .
\eeq
The energy-momentum tensor of the point-particle can be integrated in $t$ and rewritten as
\beq
T_p^{\mu \nu} &=& \frac{m_p }{\Sigma \sin\theta} \frac{dt_p}{d\tau}\frac{dx^\mu}{dt}\frac{dx^\nu}{dt} \delta\left(r-r_p(t)\right)\delta\left(\theta - \theta_p(t) \right)\delta\left(\varphi - \varphi_p(t) \right) \, . \nonumber \\
\eeq
Axial symmetry also allows for the mode separation of the energy momentum tensor
\beq
T_{\mu\nu} = \sum_{m=0}^\infty T^m_{\mu\nu}e^{im \varphi} \, .
\eeq

Finally, the Dirac delta distributions need to be represented in the numerical grid. For the azimuthal decomposition, we use the representation in modes
\beq
\delta \left[\varphi - \varphi_p (t) \right] = \frac{1}{2\pi} \sum_{m=0}^\infty e^{im (\varphi-\varphi(t))} \, .
\eeq

For the radial and $\theta$ directions, we approximate the Dirac delta by a narrow Gaussian distribution 
\beq
\delta \left( r - r_p \left(t\right) \right)& = &\frac{\delta \left[ r^* - r^*_p \left(t\right) \right]}{\left| dr/dr_* \right|} \nonumber \\
&=& \frac{\left| dr^*/dr \right|}{\sqrt{2\pi}\lambda_{r_*}}\exp\left[ - \left(r^* - r_p^* \left(t\right)^2 \right)/2\lambda_{r_*}^2 \right]\ ,  \\
\delta \left( \theta - \theta_p \left(t\right) \right)& = & \frac{1}{ \sqrt{2\pi} \lambda_\theta} \exp\left[ - \left(\theta - \theta_p \left(t\right) \right)^2/2\lambda_{\theta}^2 \right] \, , 
\eeq
where $\lambda_{r_*}$ and $\lambda_\theta$ are varied to ensure convergence of the numerical results as they approach $0$. Numerical experience shows $\lambda_{r_*} \approx 4 dr_*$ and $\lambda_\theta \approx 4 d\theta$ yield the best results~\cite{Lopez-Aleman:2003sib}. Although there are more refined numerical representations of the Dirac delta in the literature~\cite{Sundararajan:2007jg}, the one we follow is more practical and versatile for modeling different orbital motion.

When doing simulations with point-particles, we generate them at $t=0$ and prescribe as initial data for the field $\psi(t=0)=\partial_t \psi(t=0) =0$. This leads to a burst of initial junk radiation which needs to dissipate before the ``physical'' solution is observed. The junk radiation can, however, be reduced by multiplying the source by a starting ``window'' function that varies between $0$ and $1$ on some timescale $T$, such as $\left(1 - \exp(-t/T) \right)^4$.

\subsection{Hyperboloidal layers}\label{sec:Hyperboloidal}

One drawback of the method described above is that the outer boundary needs to be pushed to very large radius to avoid spurious reflections in the numerical solution. Additionally, \acs{GW}s are only defined in a gauge-invariant manner at null infinity. Therefore, to study radiation numerically, one often needs to go to very large values of the radial coordinate $r_*$ to ensure that extraction is performed in a region where the fields already behave like outgoing waves.

These obstacles have motivated the study of how to include null infinity in the computational domain. A successful strategy is hyperboloidal foliations. These are surfaces everywhere spacelike, that still approach null infinity~\cite{Zenginoglu:2007jw, Zenginoglu:2011zz, PanossoMacedo:2019npm}. We will follow the method of scri-fixing gauges~\cite{Zenginoglu:2007jw}, in which null infinity is fixed at some spatial coordinate independent of the time coordinate. This can even coincide with the numerical outer boundary, eliminating the need for imposing boundary conditions. 

The first step in this construction is to compactify the radial coordinate
\beq
r_* = \frac{\rho}{\Omega (\rho)} \, , \label{eq:Compact}
\eeq
where $\Omega$ is a conformal factor whose zero set $S$ corresponds to infinity in $r_*$ and $\rho$ is the new radial coordinate. It obeys to $\Omega(S) = 0$ and $d\Omega(S)/d\rho \neq 0$.

The second step is to introduce a new time coordinate $\tau$ that preserves the timelike Killing vector field, i.e. $\partial_\tau = \partial_t$. This is achieved by a transformation $\tau = t - h(r, \theta, \varphi)$, where $h$ is called the \textit{height} function. Since we are interested in studying the emission of \acs{GW}s, it is useful to pick $h$ so that it depends only on $r_*$
\beq
\tau = t - h(r_*) \, . \label{eq:heightfunc}
\eeq
In this way, surfaces of constant $\tau$ are hyperboloidal.

One advantage of this method is that the strong-field region where the motion of the source takes place can be kept unaltered. At some radius, a truncated hyperboloidal layer is introduced, which must be matched in a sufficiently smooth way. The matching procedure we adopt follows that of Ref.~\cite{Zenginoglu:2011zz}
\beq
\Omega = 1 - \left( \frac{ \rho -  R_\text{layer} }{S - R_\text{layer}}  \right)^4 \Theta \left(  \rho - R_\text{layer} \right) \, , 
\eeq
where $\Theta$ is the Heaviside function and $S$ is the location of the outer boundary in the numerical domain. $\Omega = 1$ for $\rho < R_\text{layer}$, so $\rho = r_*$ in this region. 

To choose the height function we require that outgoing null waves have the same representation in the strong field region and the exterior layer, which means $t-r=\tau-\rho$. Eq.~\eqref{eq:heightfunc} then implies
\beq
h = t - \tau = r_* - \rho(r_*) = \frac{\rho}{\Omega(\rho)} - \rho \, . \label{eq:heightfuncOUR}
\eeq
For $\rho < R_\text{layer}$, $\Omega = 1$ and therefore $h=0$, so that in interior region $\tau = t$.

How does this affect the Teukolsky equation? Suppose we have it written in the form 
\beq
\left[ A^{tt} \partial_t^2 + A^{t r_*} \partial_t \partial_{r_*}  + A^{r_*r_*} \partial_{r_*}^2 + A^{\theta\theta} \partial_\theta^2 + B^t \partial_t + B^{r_*} \partial_{r_*} + B^\theta \partial_\theta + C \right] \Psi = 0\, . \nonumber \\\eeq
The coordinate transformations in Eq.~\eqref{eq:Compact} and~\eqref{eq:heightfunc} change the derivative operators
\beq
\partial_t = \partial \tau \quad , \quad \partial_{r_*} = -H \,\partial_\tau + \left(1-H \right)\partial_\rho \, ,
\eeq
where $H$ is the boost function
\beq
H = \frac{dh}{dr_*} \, , 
\eeq
which for our choice of the height function in Eq.~\eqref{eq:heightfuncOUR} is 
\beq
H = 1 - \frac{d\rho}{dr_*} = 1 - \frac{\Omega^2}{\Omega-\rho \frac{d\Omega}{d\rho}} \, .
\eeq
The transformed Teukolsky equation now has the form
\beq
\left[A^{\tau \tau} \partial^2_\tau + A^{\tau\rho} \partial_\tau \partial_\rho  + A^{\rho\rho} \partial_\rho^2 + A^{\theta\theta} \partial_\theta^2 + B^\tau \partial_\tau + B^\rho \partial_\rho + B^\theta \partial_\theta + C \right] \Psi = 0\, , \nonumber \\ \label{eq:WaveEqAlmost}
\eeq
with the new coefficients 
\beq
A^{\tau \tau } &=& A^{t t} - H A^{t r_*} +  H^2 A^{r_* r_*} \, , \\
A^{\tau \tau } &=& \left(1-H \right) \left(A^{t r_*} - 2H A^{r_* r_*} \right) \, , \\
A^{\rho\rho} &=& \left(1-H\right)^2 A^{r_* r_*} \, , \\
B^\tau &=& B^t - H B^{r_*} - \frac{dH}{d\rho} \left(1-H\right) A^{r_* r_*}  \, , \\
B^\rho &=& \left(1-H\right)\left(B^{r_*} - \frac{dH}{d\rho} A^{r_*r_*}\right) \, .
\eeq
Note that Eq.~\eqref{eq:WaveEqAlmost} can be put in the form of the original wave equation~\eqref{eq:GeneralWaveEq} by dividing everything by $-A^{\tau\tau}$. 

This set of hyperboloidal coordinates preserves the regularity of the Teukolksy equation, and all the coefficients appearing in it are finite at the outer boundary $\Omega(S) = 0$~\cite{Zenginoglu:2011zz}. Moreover, it ensures that the equation does not admit ingoing solutions at the outer boundary~\cite{Zenginoglu:2011zz}.

\cleardoublepage
\part{Tides}\label{pt:tides}
\chapter{Deformability of black holes immersed in matter}\label{ch:TLNs}

Tidal interactions are responsible for many astrophysical phenomena that have caught our attention since the dawn of Newton’s theory of gravitation~\cite{poisson_will_2014}. An obvious example is ocean tides, caused by differences in the gravitational field produced by the Moon at different Earth locations. Their tidal interaction also explains why the Earth is losing angular momentum to the Moon, resulting in longer days. Tidal effects play a crucial role in close binary systems, as demonstrated by the spectacular tidal disruption events of stars that orbit \textit{too} close to \acs{SMBH}s~\cite{vanVelzen:2016jsk, Holoien:2019zry}.

In the first part of this thesis, we study two problems where the matter surrounding a \acs{BH} gets tidally deformed by a companion in a binary system and examine their implications for \acs{GW} astronomy. 

\section{A Brief History of Tidal Love Numbers}

The tidal distortion of a compact object by an external gravitational field is quantified, at a linear level, through its tidal Love numbers (\acs{TLN}s)~\cite{poisson_will_2014}. They are the gravitational analogue of the electric susceptibility. The \acs{TLN}s depend only on the dynamics of the gravitational field, i.e. the underlying theory of gravity and the internal structure of the deformed body. They appear in the orbital equation of motions of a binary system at leading Newtonian order~\cite{Mora:2003wt, Vines:2010ca}, and introduce corrections in the gravitational waveform at fifth post-Newtonian order~\cite{Flanagan:2007ix, Hinderer:2016eia}. The prospects of using \acs{GW}s measurements to understand the structure of more compact objects have motivated the development of a relativistic theory of \acs{TLN}s\cite{Binnington:2009bb, Damour:2009vw, Hinderer:2007mb}.

Initially, works on \acs{TLN}s focused on neutron stars and provided access to the equation of state above the currently understood nuclear densities~\cite{Baiotti:2019sew, LIGOScientific:2018cki}. More recently, tidal deformations have been proposed as a good candidate to test strong-field gravity, the \acs{BH} paradigm and to search for new exotic, compact objects (\acs{ECO}s)~\cite{Cardoso:2017cfl,Maselli:2017cmm,Maselli:2018fay,Pani:2019cyc, Datta:2021hvm}. A crucial aspect of this is the fact that \acs{TLN}s of \acs{BH}s vanish in \acs{GR}~\cite{Binnington:2009bb, Damour:2009vw, Gurlebeck:2015xpa, Landry:2015zfa}, even when the \acs{BH} is rotating~\cite{Pani:2015hfa, PhysRevD.103.084021, PhysRevLett.126.131102, Chia:2020yla}. This property has been geometrically linked to hidden near-horizon enhanced symmetries~\cite{Charalambous:2021kcz, BenAchour:2022uqo, Hui:2022vbh}. Therefore, a measurement of a nonvanishing \acs{TLN} is evidence for new physics: either the object is not a \acs{BH}, or \acs{GR} is not the most accurate description of gravity. 

Consider the first possibility. Quantum corrections at the horizon scale or exotic matter could form horizonless \acs{ECO}s~\cite{Cardoso:2019rvt}. As the \acs{BH} limit is approached,
\beq
\mathcal{C} = M/R \rightarrow 1/2 \, ,
\eeq
where $M$ and $R$ are, respectively, the mass and radius of the \acs{ECO}, its
\acs{TLN}s generically converge to the \acs{BH} limit (zero), but, for many models, logarithmically~\cite{Cardoso:2019rvt}.
\acs{LVK} and the Einstein Telescope can only set constraints on low compact \acs{ECO}s, but \acs{LISA} would probe tidal deformability almost up to the \acs{BH} limit~\cite{Maselli:2017cmm,Maselli:2018fay,Pani:2019cyc,Cardoso:2019rvt}.
Nonzero \acs{TLN}s may also signal corrections to \acs{GR}. Extra degrees of freedom create extra tidal fields for which a theory of \acs{TLN}s is still poorly formulated~\cite{Bernard:2019yfz}, but in some of the extensions studied in the literature, \acs{BH}s in modified gravity theories can have nonzero \acs{TLN}s~\cite{Cardoso:2018ptl,Cardoso:2019rvt} (e.g. Chern-Simons gravity~\cite{Alexander:2009tp}).

There is, however, a third unexplored option that could be responsible for the (apparent) nonvanishing of \acs{TLN}s of a \acs{BH}: the presence of external matter. 
Any astrophysically plausible self-gravitating object will be surrounded by some matter, which could contribute with small but nonzero effective \acs{TLN}s. In the rest of this chapter, we will quantify this contribution and conclude whether the matter surrounding a binary coalescence limits our ability to test \acs{GR} with \acs{TLN}s.
%

\section{Newtonian Shell}\label{sec:NewShell}

As a proxy for the relativistic case, let us start by studying the tidal deformability of a spherical shell of matter in Newtonian gravity. Consider an object formed by a perfect fluid (isotropic with no shear stresses and viscosity) with matter density $\rho$, pressure $p$ and velocity $u^j$, that obeys the Poisson-Euler equations~\cite{poisson_will_2014}
\beq
\partial_i \partial^i \Phi &=&- 4\pi G\, \rho	\, ,\label{eq:PoissonEq} \\
\rho \frac{du_j}{dt} &=& \rho \partial_j \Phi-\partial_j p \, ,\label{eq:Euler}
\eeq
where $\Phi$ is the Newtonian gravitational potential and we temporarily recover Newton's gravitational constant $G=6.67\times 10^{-11} \, \text{N}\cdot \text{m}^2/\text{kg}^2$. This system is complemented with a mass continuity equation
\beq
\frac{\partial \rho}{\partial t}+\partial^i \left(\rho\,u_i \right)=0 \, .	\label{eq:Masscontinuity}
\eeq
The isotropy inherent to a perfect fluid implies that, in equilibrium, it is spherically symmetric and we can use spherical coordinates centered at the body's center of mass. 

Next, we introduce an external tidal field, $V$, which perturbs the equilibrium configuration of the body. We assume the regime of static tides, 
meaning that the time variations of the tidal perturber are small compared to the dynamical timescale of the system, and thus tides are independent of time. In this regime, the condition of hydrostatic equilibrium becomes 
\beq
\partial_j p=\rho \partial_j \Phi \, .	\label{eq:HydroEq}
\eeq
In order to exploit the spherical symmetry of the system, it is useful to define the mass function $m(r)$
\beq
\frac{dm(r)}{dr} = 4\pi r^2 \rho \, ,
\eeq
and rearrange Eq.~\eqref{eq:HydroEq} as 
\beq
\frac{dp}{dr}=-\rho \frac{Gm}{r^2} \, .
\eeq
For the condition of hydrostatic equilibrium to hold, the tidal field must be sufficiently far away from the central body. In fact, we will assume that it is located in vacuum and satisfies Laplace's equation
\beq
\partial_i \partial^i V=0 \,. \label{eq:Laplace}
\eeq
which admits as solution
\beq
V=\sum_{\ell, m}\frac{4\pi}{2\ell+1}d_{\ell m}r^{\ell}Y^{\ell m} ,
\eeq
where $d_{\ell m}$ are called the \textit{tidal moments}.

At this point, we introduce fluid perturbations. We follow a surface of constant density, $\rho_0$, which in the unperturbed configuration is at radius $r_0$. Then, we need to consider perturbations in the mass density, $\delta \rho$, and in the radius of the surface, $\delta r$. There are two possible ways to approach this. The Eulerian/macroscopic framework compares quantities at the same position in space, while the Lagrangian/microscopic framework describes changes in the same fluid element as it is perturbed. We will not delve into how to handle the differences between them and refer the reader to Ref.~\cite{poisson_will_2014} for a more detailed treatment. 

If we follow a spherical surface of density matter $\rho$ in the microscopic description, the following macroscopic statements are true
\beq
\delta \rho &=& -\rho' \delta r \, , \label{eq:DensityFrac}\\
\delta p &=& -p' \delta r   \, , \label{eq:PressureFrac}
\eeq
where primes denote derivatives with respect to $r$.

The fluid perturbations change the body's gravitational potential so that a perturbed Poisson equation holds
\beq
\partial_i \partial^i \delta \Phi = -4\pi G \delta\, \rho \, .\label{eq:PerturbedPoisson}
\eeq
Outside the body, where $\delta \rho $ is zero, the solution to this equation is 
\beq
\Phi^\text{out}_{lm}=\frac{4\pi G}{2\ell+1}\frac{I_{\ell m}}{r^{\ell+1}} \, ,\label{eq:ExternalPotentialNewton}
\eeq
where $I_{\ell m}$ are the body's multipole moments. The \acs{TLN}s are defined as the ratio
\beq
k_\ell = \frac{1}{2}\left(\frac{c^2}{G\,M}\right)^{2\ell+1}\frac{G\,I_{\ell m}}{d_{\ell m}} \, ,
\eeq
where we also recovered the speed of light $c$ and $M$ is the total mass of the object
\beq
M = \lim_{r\rightarrow \infty} m\left(r\right) \, .	\label{eq:Mass}
\eeq
To actually compute the \acs{TLN}s explicitly, we need to solve Eq.~\eqref{eq:PerturbedPoisson} inside the body and then match the internal and external potential perturbations at the body's surface. 

To solve the internal problem, we start by decomposing every perturbation in spherical harmonics~\cite{NIST:DLMF}
\beq
\delta r &=& \sum_{\ell, m} r f_{\ell m}\left(r\right)Y_{\ell m}\left(\theta,\varphi\right) \, , \label{eq:RadiusExpansion} \\
\delta X&=& \sum_{\ell, m}\delta X_{\ell m}\left(r\right)Y_{\ell m}\left(\theta,\varphi\right) \,, \label{eq:PotentialExpansion}
\eeq
with $X=\rho, p, \Phi$ or $V$. Inserting in Eq.~\eqref{eq:PerturbedPoisson} 
\beq
r^2 \delta \Phi_{\ell m}''+2r \delta \Phi_{\ell m}'-\ell\left(\ell+1\right)\delta \Phi_{\ell m}=-4\pi G r^2 \delta \rho_{\ell m} \,.	\label{eq:PoissonPert}
\eeq
Euler's equation~\eqref{eq:Euler} expanded to first order gives
\beq
\delta p'_{\ell m}&=&-\frac{G m}{r^2}\delta \rho_{\ell m}+\rho\left(\delta \Phi'_{\ell m}+ V'_{\ell m}\right) \, , \\
\delta p_{\ell m}&=&\rho\left(\delta \Phi_{\ell m}+ V_{\ell m}\right) \, .
\eeq
Differentiating the second equation and inserting it in the first one, and using Eq.~\eqref{eq:DensityFrac} and \eqref{eq:PressureFrac}, we arrive at
\beq
\frac{Gm}{r}f_{\ell m}=\delta \Phi_{\ell m}+ V_{\ell m} \, .	\label{eq:RelationModes}
\eeq

Finally, we match this expression with the external one \eqref{eq:ExternalPotentialNewton} at the body's surface by demanding that the gravitational potential has to be smooth. In practice, we impose continuity of $\delta \Phi$ and its first derivative, arriving at
\beq
k_\ell=\left(\frac{c^2\,R}{G\,M}\right)^{2\ell+1}\frac{\ell+1-\eta_\ell \left(R\right)}{2\left(\ell+\eta_\ell\left(R\right)\right)} \, , 	\label{eq:TLNRad}
\eeq	
where $\eta_\ell$ is called the Radau's function
\beq
\eta_\ell\left(r\right)= r \frac{ f_{\ell m}'\left(r\right)}{f_{\ell m}\left(r\right)} \, .
\eeq	
We find that the fractional deformation modes $f_{\ell m}$ completely determine the structure of the tidally deformed body. To compute them, we transform the differential equation for $\Phi_{\ell m}$ \eqref{eq:PoissonPert} into one for $f_{\ell m}$ by making use of Eq.~\eqref{eq:RelationModes}
\beq
r^2f_{\ell m}''+6\mathcal{D}\left(r\right)\left(rf'_{\ell m}+f_{\ell m}\right)-\ell\left(\ell+1\right)f_{\ell m}=0\, ,	\label{eq:Clairaut}
\eeq
where
\beq
\mathcal{D}\left(r\right)=\frac{4\pi\rho\left(r\right)r^3}{3\,m\left(r\right)} \, .
\eeq
$\mathcal{D}\left(r\right)$ contains the information about the internal structure of the deformed body, namely it depends on its equation of state. 

We will now solve this problem for a spherical shell model given by Vogt and Letelier~\cite{Vogt:2010ad}. This model is represented by the gravitational potential and matter density
\beq 
	\Phi\left(r\right) &=& -\frac{G M}{\left(r^n+r_0^n\right)^{1/n}}	\, , \label{eq:ShellPotential}\\
	\rho\left(r\right) &=& \frac{M\left(n+1\right)b^nr^{n-2}}{4\pi\left(r^n+r_0^n\right)^{2+1/n}} \, , \label{eq:ShellDensity}
\eeq 
where $r_0$ is a parameter with units of length, $M$ is the total mass of the shell \eqref{eq:Mass} and $n>0$. For $n>2$, $\rho$ vanishes at $r=0$ and the mass distribution indeed represents a shell. As $n$ increases, the shell becomes thinner and localized around $r=r_0$. In the limit $n\rightarrow \infty$ this model describes an infinitesimal thin shell located at $r=r_0$.

The formalism developed to compute the \acs{TLN}s relies on making a match at the surface of the compact object. However, this shell does not possess a hard surface. A possible solution to this problem occurs if the matter density is sufficiently localized so that the matching is well defined in the limit $R\rightarrow \infty$. This occurs in boson stars, whose tidal deformations in Newtonian gravity and \acs{GR} were studied in Refs.~\cite{Mendes:2016vdr,Cardoso:2017cfl}. 

The solution of Eq.~\eqref{eq:Clairaut} for this model which is regular at $r=0$ is
\beq
f_{\ell m}\left(y\right)&=&c_1 \, y^{-d}\, _2\widetilde{F}^1\left(a,b,1+\frac{c}{n};-y^n\right) \, ,	\label{eq:SolutionNewtonShell} \\
	y&=&\frac{r}{r_0} \, , \\
	a&=&-1-\frac{\ell}{n}+\frac{c-3}{2n} \, ,\\
	b&=&-1+\frac{\ell}{n}+\frac{c-1}{2n} \, , \\
	c&=&\sqrt{-7+4n\left(n-1\right)+4\ell\left(\ell+1\right)} \, ,\\
	d&=&n+\frac{1-c}{2}  \, ,
\eeq	
where $_2\widetilde{F}^1\left(a,b,c;x\right)$ are the regularized hypergeometric functions~\cite{NIST:DLMF}. The \acs{TLN}s of the shell can then be computed by plugging this solution in Eq.~\eqref{eq:TLNRad} and taking the limit $R\rightarrow \infty$. We conclude that the solution only converges for $n>2\ell+1$. For smaller values of $n$, the equality \eqref{eq:RelationModes} is not respected in the $R \rightarrow \infty$ limit. When the problem is well-posed, we find 
\beq
		k_\ell &=& -\frac{\left(1+n\right)\left(1-2n+c\right)}{2\,n^2\left(b+1\right)}\frac{\Gamma\left(a-b\right)}{\Gamma\left(b-a\right)}\nonumber \\
		&\times& \frac{\Gamma\left(b\right)}{\Gamma\left(a+1\right)}\frac{\Gamma\left(2+\frac{1-2\ell}{n}+b\right)}{\Gamma\left(2+\frac{3+2\ell}{n}+a\right)}\left(\frac{c^2\, r_0}{G\,M}\right)^{2\ell+1} \label{eq:TLNNewtonShell} \nonumber \, .
\eeq
We conclude that the \acs{TLN}s of this Newtonian shell are of order $\mathcal{O}\left(\frac{c^2\, r_0}{G\,M}\right)^{2\ell+1}$ and bounded below in the thin shell limit by
\beq
\lim_{n\rightarrow \infty} k_\ell	= \frac{\ell+2}{2\left(\ell-1\right)} \left(\frac{c^2\,r_0}{G\,M}\right)^{2\ell+1}\, .	\label{eq:ThinShellNewtonian}
\eeq
For the quadrupole mode $\ell=2$, which typically dominates \acs{GW} emission, we obtain the scaling $k_2\propto r_0^{5}$.

\section{Tidal deformability in General Relativity} \label{sec:FormalismTLN}

We now move to the theory of tidally deformed objects in \acs{GR}. The starting setup is the same, an isolated, self-gravitating compact object perturbed by an external tidal field. As before, we wish to describe the gravitational field of the two bodies in terms of their multipole moments, but now using the geometric point of view of \acs{GR}. We follow Thorne's approach, which holds for stationary, asymptotically flat spacetimes and requires adopting asymptotically Cartesian and mass-centered coordinates at the isolated object~\cite{Thorne:1980ru, Damour:2009vw, Binnington:2009bb}. 

First, we decompose the external tidal field as
\beq
	\mathcal{E}_{a_1...a_\ell}&=& \left[\left(\ell-2\right)!\right]^{-1}\left<C_{0a_1;a_3...a_\ell}\right> \, , \\
	\mathcal{B}_{a_1...a_\ell}&=& \left[\frac{2}{3}\left(\ell+1\right)\left(\ell-2\right)!\right]^{-1}\left<\epsilon_{a_1bc}C^{bc}_{a_20;a_3...a_\ell}\right> \, ,
\eeq
where $C_{abcd}$ is the Weyl tensor, $\epsilon_{abc}$ is the permutation tensor and the angular brackets denote symmetrization and trace removal. $\mathcal{E}_{a_1...a_l}$ ($\mathcal{B}_{a_1...a_l}$) are the polar (axial) moments, and since we will only study spherical symmetric configurations, they can be expanded in spherical harmonics.

To describe the deformation induced by the tidal field on the equilibrium configuration of the compact object, we use the framework of linear \acs{BH} perturbation theory as introduced in Sec.~\ref{sec:BHPT}. Since we have a spherically symmetric background, we can separate perturbations in the polar and axial sectors and adopt the Regge-Wheeler gauge~\eqref{eq:RWgauge}. The tidal fields and induced multipole moments are then extracted from the asymptotic behavior of the full metric
\beq
	g_{tt}=-1+\frac{2M}{r}&+&\sum_{\ell\geq 2}\Bigg( \, \frac{2}{r^{\ell+1}}\left[\sqrt{\frac{4\pi}{2\ell+1}}M_\ell\,Y^{\ell 0}+\left(\ell'<\ell \, \text{pole}\right)\right]\nonumber \\
	&-&\frac{2}{\ell \left(\ell-1\right)}r^\ell\left[\mathcal{E}_\ell Y^{\ell 0} +\left(\ell'<\ell \, \text{pole}\right) \right]   \Bigg) \,,\\ \label{eq:PolarMetricExpansion}
	g_{t\varphi}=\frac{2J}{r}\sin^2\theta&+&\sum_{\ell\geq 2}\Bigg( \, \frac{2}{r^{\ell}}\left[\sqrt{\frac{4\pi}{2\ell+1}}\frac{S_\ell}{\ell}\,\sin \theta \partial_\theta Y^{\ell0} + \left(\ell'<\ell \, \text{pole}\right)\right]\nonumber \\
	&+&\frac{2r^{\ell+1}}{3\ell\left(\ell-1\right)}\left[\mathcal{B}_\ell \sin \theta \partial_\theta Y^{\ell0} +\left(\ell'<\ell \, \text{pole}\right) \right]   \Bigg) \, ,\label{eq:AxialMetricExpansion}
\eeq
where $M_\ell$ are the mass multipole moments, $S_\ell$ are the current multipole moments, and $\mathcal{E}_\ell$ and $\mathcal{B}_\ell$ are, respectively, the amplitudes of the polar and axial components of the external field with harmonic number $\ell$, where spherical symmetry was used to fix $m=0$.

Finally, we define the polar and axial \acs{TLN}s, respectively, as the dimensionless ratios \cite{Cardoso:2017cfl}
\beq
	k_\ell^E&=&-\frac{1}{2}\frac{\ell\left(\ell-1\right)}{M^{2\ell+1}}\sqrt{\frac{4\pi}{2\ell+1}}\frac{M_\ell}{\mathcal{E}_{\ell0}} \, , \label{eq:PolarTLNs} \\
	k_\ell^B&=&-\frac{3}{2}\frac{\ell\left(\ell-1\right)}{\left(\ell+1\right)M^{2\ell+1}}\sqrt{\frac{4\pi}{2\ell+1}}\frac{S_\ell}{\mathcal{B}_{\ell0}}  \,  , \label{eq:AxialTLNs}
\eeq
where $M$ is the mass of the deformed object. Note that the axial \acs{TLN}s have no Newtonian analogous. Also, most references~\cite{Damour:2009vw, Binnington:2009bb, Hinderer:2007mb} normalize the \acs{TLN}s in powers of the object radius $R$ instead of $M$, because they study bodies with a hard surface (e.g. neutron stars). Here, we instead adopt the convention of Ref.~\cite{Cardoso:2017cfl} since the radius of distributions of matter surrounding \acs{BH}s is in general ill-defined, as occurs for some \acs{ECO}s. The two definitions are related by
\beq
k^\ell_{\text{ours}}=\left(\frac{R}{M}\right)^{2\ell+1}k^\ell_{\text{standard}}\,.
\label{eq:TLNdiff}
\eeq
%

\section{Black holes surrounded by matter\label{sec:DirtyBH}}

\subsection{Black holes with short hair}\label{sec:ShortHair}

We consider two models for \acs{BH}s surrounded by matter. The first one is a static, spherically symmetric spacetime containing an anisotropic fluid surrounding a \acs{BH}, which satisfies both the weak and strong energy condition~\cite{Brown:1997jv}. Its line element is of the form in Eq.~\eqref{eq:SphericalLineElement} with
\beq
A\left(r\right)&=&B\left(r\right)= 1 - \frac{2M}{r} - \frac{Q^{2k}}{r^{2k}} \,,\\
\rho &=& \frac{Q^{2k}\left( 2k-1 \right)}{8 \pi r^{2k+2}} \quad , \quad P = k\rho \label{eq:shorthairmatter}\, ,
\eeq
where $\rho$ and $P$ are, respectively, the matter density and the pressure on the isotropic $\theta$-$\varphi$ surfaces, while $Q$ is a constant. The energy-momentum tensor of the fluid is
\beq
T_{\mu\nu}=\rho \left( u_\mu u_\nu - n_\mu n_\nu \right) + P \sigma_{\mu\nu} \, , \label{eq:StressTensorFluid}
\eeq
where $u^\mu$ is the fluid's $4$-velocity, while $n_\nu$ and $\sigma_{\mu\nu}$ are, respectively, the unit normal and metric of the isotropic $2$-spheres
\beq
	u^\mu&=&\left(-\frac{1}{\sqrt{A\left(r\right)}},0,0,0\right) \, , \\
	n_\nu&=&\left(0,\frac{1}{\sqrt{B\left(r\right)}},0,0\right) \, , \\
	\sigma_{\mu\nu}&=&\text{diag}\left(0,0,r^2,r^2\sin^2\theta\right)\, .
\eeq
For $k=1$, this class of \acs{BH}s yields the Reissner-Nordstr\"{o}m solution~\cite{Brown:1997jv}. For $k>1$, the parameter $Q$ corresponds to a \textit{matter-hair}~\cite{Brown:1997jv,Barausse:2014tra}, which can be arbitrarily short by taking $k$ to be arbitrarily large.

As in the Newtonian analysis, to determine the \acs{TLN}s of this configuration, we need to complement the gravitational perturbations with the ones from matter, where any equilibrium background quantity $X=X_0$ gets perturbed by the external tide, $X\to X_0+\delta X(t,r,\theta,\varphi)$.
Again, we consider static tides, so all the perturbations introduced are independent of the coordinate time $t$. This immediately fixes $u^\mu$ and $n^\mu$ by imposing the normalizations $u^2=-1$ and $n^2=1$, and that $u^\mu$ remains proportional to the timelike killing vector field $\partial/\partial t$
\beq
	\delta u^\mu&=&\sum_{\ell, m}\left(\frac{1}{2\sqrt{A}}H_0^{\ell m}Y^{\ell m},0,0,0\right) \, , \\
	\delta n_\mu&=&\sum_{\ell, m}\left(0,\frac{1}{2\sqrt{B}}H_2^{\ell m}Y^{\ell m},0,0\right) \, .
\eeq
For $\sigma_{\mu\nu}$, we allow one more degree of freedom that respects the background spherical symmetry
\be
\delta \sigma_{\mu\nu}=\sum_{\ell, m}\text{diag}\left(0,0,r^2 \, K_2^{\ell m}(r) \, Y^{\ell m},r^2\sin^2\theta\, K_2^{\ell m}(r)\,Y^{\ell m}\right)	\,.
\ee	
Finally, we perturb
\beq
\rho&=&\rho_0+\sum_{\ell, m}\delta \rho^{\ell m}\left(r\right)\,Y^{\ell\,m} \, , \\
P&=&P_0+\sum_{\ell, m}\delta P^{\ell m}\left(r\right)\,Y^{\ell m}\,.
\eeq
%
\subsubsection{Axial perturbations}
The axial sector of stationary gravitational perturbations is entirely decoupled from matter perturbations~\cite{Damour:2009vw,Binnington:2009bb, 1968ApJ...152..673T}. Consequently, the $t\varphi$-component of Einstein's equations gives a decoupled ordinary differential equation for $h_0$
\beq
&&r^2\left(1-\frac{2M}{r} + \frac{Q^{2k}}{r^{2k}}\right)h_0'' = \nonumber\\
%
&=&\left(\ell\left(\ell+1\right)-\frac{4M}{r}+2k \left(1+2k\right) \frac{Q^{2k}}{r^{2k}}\right)h_0 \, . \nonumber \\
\label{eq:Aniaxialmaster}
\eeq

We now follow a similar approach to that in Ref.~\cite{Cardoso:2017cfl} and treat the matter-hair perturbatively. We expand the metric perturbations in powers of the adimensionalized coupling $Q^{2k}/M^{2k}$,
\beq
h_{\mu\nu}=h_{\mu\nu}^{\left( 0 \right)}+ \frac{Q^{2k}}{M^{2k}}\, h_{\mu\nu}^{\left( 2 \right)} \, , \label{eq:expansionmatterhair} \, 
\eeq
where $h_{\mu\nu}^{\left( 0 \right)}$ is the vacuum \acs{GR} solution. For $\ell=2$, the $0$th-order axial perturbation regular at the horizon is
\beq
h_0^{\left( 0 \right)}=\frac{\mathcal{B}_2}{3}r^3\left(1-\frac{2M}{r}\right)\, .
\eeq
Expanding Eq.~\eqref{eq:Aniaxialmaster} to order $\mathcal{O}\left(\epsilon\right)$ we find
%
\beq
&&\left(\frac{d^2}{dr^2}+2\frac{2M-3r}{r^2\left(r-2M\right)}\right)h_0^{\left( 2 \right)} =  \frac{2\mathcal{B}_2}{3}\left(\frac{M}{r} \right)^{2k} \times\nonumber\\
&\times&\frac{r\left(\left(2k^2+k-3\right)r-\left(2k^2+k-1\right)2M\right)}{r-2M}\, .\label{eq:Axialperturbation}
\eeq
This equation admits a solution in closed form in terms of the homogeneous solution and a hypergeometric function.
From it, we read the \acs{TLN}s
\beq
k_2^B=\frac{1}{5}\frac{2^{5-2k}\left(2k-1\right)}{2k^2-9k+10}\frac{Q^{2k}}{M^{2k}}\,,\quad k>2\,.\label{eq:AxialTLNs_fluid}
\eeq
For $k=1$ we find $k_2^B=0$ which agrees with the literature for the charged \acs{BH} solution~\cite{Cardoso:2017cfl}. For $k=2$, we find
a new dominant logarithmic term $\log(r)/r^2$, for which we lack a physical interpretation.
We can express the above in terms of the mass $\delta M\sim \left(Q^{2k}/M^{2k}\right)M$ contained in the fluid: $k_2^B \sim \delta M/M$.

\subsubsection{Polar perturbations}
In the polar sector, matter perturbations are no longer decoupled from gravitational ones. The $tr$- and $\theta$$\theta$-component of Einstein's equations, respectively give
\beq
H_1=0 \,,\quad H_2=H_0\, .
\eeq
The $\theta$-component of the energy-momentum tensor conservation fixes the pressure perturbation to be
\beq
\delta P=2k\left(2k-1\right)\frac{Q^{2k}}{r^{2k}} \frac{K-K_2}{16\pi r^2}\, .
\eeq

The $tt$, $rr$, and the $\theta\theta$ component of Einstein's equations provide expressions for $K_0''$, $ K_0'$ and $K_0$ in terms of $H_0''$, $H_0'$, $H_0$. Substituting these in the $tr$-component of Einstein's equations gives the following decoupled ordinary differential equation for $H_0$
\beq
&&r^2\left(1-\frac{2M}{r}+\frac{Q^{2k}}{r^{2k}} \right)^2 H_0''\nonumber \\
&+& 2r\left(1-\frac{2M}{r}+\frac{Q^{2k}}{r^{2k}} \right)\left(1-\frac{M}{r}+\left(1-k\right)\frac{Q^{2k}}{r^{2k}} \right) H_0' \nonumber \\
&=&\Bigg[\ell\left(\ell+1\right)+\frac{4M^2}{r^2}-2\ell\left(\ell+1\right)\frac{M}{r} \nonumber \\
&+& \left(\ell\left(\ell+1\right)+2k\left(1-\frac{6M}{r}\right) \nonumber-4k^2\left(1-\frac{2M}{r}\right) \right)\frac{Q^{2k}}{r^{2k}}\nonumber + 2k \frac{Q^{4k}}{r^{4k}}\Bigg] H_0\, . \nonumber \\
\label{eq:PolarFluid}
\eeq

Following the same approach as in the axial case, we treat the matter-hair as a perturbation to \acs{GR} using the expansion in Eq.~\eqref{eq:expansionmatterhair}. For $\ell=2$, the polar perturbation regular at the horizon is
\beq
H_0^{\left(0\right)}=-\mathcal{E}_2\,r^2\left(1-\frac{2M}{r}\right)\, , 
\eeq
and Eq.~\eqref{eq:PolarFluid} can be written as
%
\beq
\left(\frac{d^2}{dr^2}+\frac{\left(r-M\right)}{r\left(r-2M\right)}\frac{d}{dr}-\frac{2\left(2M^2-6Mr+3r^2\right)}{r^2\left(r-2M\right)}\right)H_0^{\left(2\right)}=\mathcal{S}_P^{\left(2\right)}\,,\nonumber
\eeq
with
%
\beq
\mathcal{S}_P^{\left(2\right)}&=&2\frac{M^{2k}}{r^{2k}}\mathcal{E}_2\frac{c_1-c_2 r+\left(3+k\left(2k-3\right)\right)r^2}{\left(r-2M\right)^2}\,,\label{eq:PolarFluidPert} \\
c_1&=&2\left(3+4k\left(k-2\right)\right)M^2 \, , \\
c_2&=&2\left(4+k\left(4k-7\right)\right)M \, .
\eeq

Even though this differential equation admits a solution in closed form, it is simpler to work in terms of Green's functions. The two linearly independent solutions to the homogeneous equation are
\beq
	\Psi_-&=&\frac{3A_1}{M^2}r^2\left(1-\frac{2M}{r}\right) \, , \\
	\Psi_+&=&\frac{A_2}{M^2r\left(r-2M\right)}\bigg(\left(r-M\right)\left(3r^2-6Mr-2M^2\right)M \nonumber\\
	&+&3r^2\left(r-2M\right)^2\text{arctanh}\left(1-\frac{M}{r}\right)\bigg) \, ,
\eeq
with $A_1$ and $A_2$ constants. The Wronskian is
\beq
W\left(r\right)=\Psi_+'\left(r\right)\Psi_-\left(r\right)-\Psi_+\left(r\right)\Psi_-'\left(r\right)=\frac{24MA_1A_2}{r\left(2M-r\right)}\, .
\eeq
$\Psi_-\left(r\right)$ is regular at the horizon and $\Psi_+\left(r\right)$ at infinity. Imposing the correct physical boundary conditions, we find the solution to the inhomogeneous problem directly
\beq
H_0^{\left(2\right)}\left(r\right)&=&\Psi_+\left(r\right)\int_{2M}^{r}dr'\, \frac{\mathcal{S}_P^{\left(2\right)}\left(r'\right)\Psi_-\left(r'\right)}{W\left(r'\right)} \nonumber \\
&+&\Psi_-\left(r\right)\int_{r}^{\infty}dr'\, \frac{\mathcal{S}_P^{\left(2\right)}\left(r'\right)\Psi_+\left(r'\right)}{W\left(r'\right)} \,.\label{eq:GreenH0}
\eeq

For $k > 2$ the first integral converges as $r\rightarrow \infty$, and we find that the second one does not contribute to the induced mass quadrupole moment. 
Again, the \acs{TLN}s vanish for $k=1$ as expected, but in general
\beq
k_2^{E}=\frac{1}{5}\frac{2^{5-2k}\left(2k-1\right)}{2k^2-9k+10}\frac{Q^{2k}}{M^{2k}}=k_2^B \,,\quad k>2\,. \label{eq:ShellTLNNew}
\eeq
Remarkably, the polar \acs{TLN}s are the same as the axial ones. This feature was already present in the \acs{TLN}s of \acs{ECO}s in the \acs{BH} limit~\cite{Cardoso:2017cfl}~\footnote{We are grateful to Lam Hui for highlighting this property.}.
A similar procedure can be used to obtain the octupolar $\ell=3$ or higher \acs{TLN}s.

\subsection{Matter away from the horizon: Thin shells} \label{sec:ThinShell}

While the previous results are interesting, astrophysical \acs{BH}s should have surrounding matter distributions localized away from the horizon. It is challenging to construct stationary solutions describing astrophysically realistic \acs{BH} spacetimes. As a surrogate for those setups, we will pack all the interstellar material in a (infinitesimally) thin shell surrounding a Schwarzschild \acs{BH}. The dynamics of thin shells are a vastly explored subject, both in \acs{GR}~\cite{Israel:1966rt,Brady:1991np,Poisson:1995sv,Lobo:2005zu,Dias:2010uh,LeMaitre:2019xez} and in modified theories of gravity; we refer the interested reader to Ref.~\cite{Poisson:2009pwt} for a pedagogical introduction to the subject. As physical systems, thin shells are nothing more than very crude approximations. However, their mathematical description is much simpler than more realistic distributions of matter and they often present the key features of these. While there are many studies regarding the stability of thin shells, little has been made in studying the explicit form of gravitational perturbations in spacetimes containing them~\cite{Pani:2009ss,Leung:1999iq,Leung:1999rh,Barausse:2014tra,Uchikata:2016qku,McManus:2020lgm}. 

Let us then consider the tidal deformation of a distribution of matter whose metric is again given by the general spherically symmetric line element in Eq.~\eqref{eq:SphericalLineElement} with
\beq
\begin{cases}
	A\left(r\right)=\bar{\alpha}\left(1-\frac{2M}{r}\right) \, , \, B\left(r\right)=\frac{1}{\bar{\alpha}}A\left(r\right) \, , \,  r<r_0\\
	A\left(r\right)=\left(1-\frac{2M_0}{r}\right) \, \, \, , \, B\left(r\right)=A\left(r\right) \, \, , \, r>r_0
\end{cases}	\, , 
\eeq
where $r_0$ is the radius at which the shell is located, $\bar{\alpha}=\frac{1-2M_0/r_0}{1-2M/r_0}$, $M$ is the \acs{BH} horizon mass, $M_0$ the ADM mass and for future reference we define the shell energy
\beq
	\delta M = M_0-M \, . 
\eeq
%

\subsubsection{Unperturbed solution}

We start by analyzing the unperturbed configuration. The wordline of matter elements of the shell is parametrized by
\beq
	x_{\pm}^\mu=x_{\pm}^\mu\left(y^a\right) \, ,
\eeq
where $y^a$ are the intrinsic coordinate functions of the shell, the subscript $+$ or $-$ refers to, respectively, the coordinate chart used outside and inside the shell, and momentarily latin indices denote objects defined on the $3$D hypersurface of the shell. We choose the intrinsic coordinates of the shell to be
\beq
	y^{a}= \left(T,\Theta,\Phi \right)\, ,
\eeq
and the unperturbed shell is located at
\beq
	x_{+}^\mu&=& \left(T,r_0,\Theta,\Phi \right)\, , \\
	x_{-}^\mu&=& \left(A_T\,T,r_0,\Theta,\Phi \right)\, .
\eeq
The constant $A_T$ reflects a possible time-rescaling so that the proper time of the shell is the same for both the exterior and interior coordinate chart. These two regions have to be matched according to the Darmois-Israel junction conditions, which relate the discontinuities on the metric functions with the matter properties of the thin shell \cite{Israel:1966rt}. The first of these imposes that the induced metric, $\gamma_{ab}$, on the $3$D hypersurface defined by the shell is continuous
\beq
	\left[\left[\gamma_{ab}\right]\right]=0\, ,
	\label{eq:DarmoiIsraelFirst}
\eeq
where $\left[\left[...\right]\right]$ denotes a jump on a quantity across the shell
\beq
	\left[\left[E\right]\right] = E\left(r_{0_+}\right) - E\left(r_{0_-}\right) \, .
\eeq

The induced metric can be computed through
\beq
	\gamma_{ab} = g_{\mu\nu}\,e_a^\mu \, e_b^\nu \, ,
\eeq
where $e_a^\mu$ are a set of three linearly independent tangent vectors to the shell 
\beq
	e^{\mu}_a=\frac{\partial x^\mu}{\partial y^a} \, .
\eeq

The second junction condition determines the stress-energy tensor of the shell, $S_{ab}$, in terms of the jump of the extrinsic curvature $K_{ab}$
\beq
	S_{ab}& = &- \frac{1}{8\pi}\left( \left[\left[K_{ab}\right]\right]-\gamma_{ab}\left[\left[K\right]\right] \right) \, ,
	\label{eq:DarmoiIsraelSecond} \\
	K_{ab}&=& e^\mu_a \, e^\nu_b \, \nabla_\mu n_\nu \, , \\
	K&=&\gamma_{ab}K^{ab} \, ,
\eeq
where $n^\mu$ is the unit normal to the thin shell
\beq
n_\mu\,e^\mu_a=0 \,,\qquad  n^\mu n_\mu=1 \, .\label{eq:normalThinShell}
\eeq

The first junction equation \eqref{eq:DarmoiIsraelFirst} yields
\beq
A_T^2=\frac{A_+\left(r_0\right)}{A_-\left(r_0\right)} \, ,
\eeq
which for our model gives $A_T=1$. Since the configuration is stationary, we can always rescale time such that this is verified and we assume it hereafter. From the second junction condition \eqref{eq:DarmoiIsraelSecond} we obtain
\beq
	S_{TT}&=&-\frac{1}{4\pi r_0}\left[\left[A\sqrt{B}\right]\right] \, , 
	\label{eq:StressTensorUnperturbed1}\\
	S_{\Theta\Theta}&=&\frac{1}{8\pi}\left[\left[\sqrt{B}\right]\right] +\frac{r_0}{16\pi}\left[\left[ \frac{A'}{A} \sqrt{B}\right]\right] 	\, .
	\label{eq:StressTensorUnperturbed2}
\eeq

If we consider the thin shell to be composed of a perfect fluid, its stress-energy tensor is simply
\beq
	S_{ab}=\left(\sigma+ p \right)u_a u_b + p \, \gamma_{ab} \, ,
\eeq	
where $\sigma$ is the surface energy density, $p$ the surface tension and $u^a$ is the fluid's velocity (normalized as $u^a u_a = -1$). For the unperturbed configuration, the latter is given by
\beq
	u^a=\left(\frac{1}{\sqrt{A\left(r_0\right)}},0,0\right) \, .
\eeq

Using Eqs.~\eqref{eq:StressTensorUnperturbed1}-\eqref{eq:StressTensorUnperturbed2} the surface energy density and pressure are determined by
\beq
	\sigma = -\frac{1}{4\pi r_0}\left[\left[\sqrt{B}\right]\right] \, , \\
	\sigma + 2 p =\frac{1}{8\pi}\left[\left[\frac{A'}{A}\sqrt{B}\right]\right]  \, ,
\eeq
which agrees with previous results on thin shell dynamics \cite{Uchikata:2016qku,Pani:2009ss,Brady:1991np,Visser:2003ge}.

\subsubsection{Perturbed configuration}
To compute the \acs{TLN}s of this object, we need to derive the junction conditions for the stationary, axisymmetric perturbed configuration when the external tidal field is introduced. First, we perturb the shell radius by
\beq
	\delta r_\pm =\sum_{\ell,m} \delta r_\pm^{\ell m}Y^{\ell m}\left(\Theta,\Phi \right) \, .
\eeq
The junction condition \eqref{eq:DarmoiIsraelFirst} evaluated at first order yields
\beq
	\left[\left[h_0\right]\right]&=&0 \, , \label{eq:junctionh0}\\
	\left[\left[H_0\right]\right]&=&\left[\left[\frac{\delta r \, A'}{A} \right]\right]\, , \label{eq:junctionH0}\\
	\frac{2}{r_0}\left[\left[\delta r \right]\right]&=& -\left[\left[ K \right]\right]\, , \label{eq:junctionK}
\eeq
where from now on, we omit the harmonic indexes $\ell,\,m$ in the junction conditions to avoid cluttering. 

To apply the second junction condition, we need to consider perturbations to the surface energy density, $\delta \sigma$, and the surface tension, $\delta p$. These are scalars and therefore can be expanded as
\beq
(\delta \sigma,\delta p)=\sum_{\ell, m} (\delta \sigma^{\ell m},\delta p^{\ell m})Y^{\ell m}\left(\Theta,\Phi \right) \,.
\eeq

Finally, we need to perturb the fluid velocity and the unit normal to the shell. The former is determined by imposing the correct normalization and the stationarity condition as in the previous sections
\beq
	\delta u^a=\sum_{\ell,m}\frac{1}{2\sqrt{A}}\left(H_0^{\ell m}-\frac{A'}{A}\delta r^{\ell m} ,0,0\right)Y^{\ell m}  \, ,
\eeq
while the latter is computed using \eqref{eq:normalThinShell}
\beq
	\delta n_{\mu_{\pm}}=\sum_{\ell,m}\frac{1}{\sqrt{B}}\left(0,\frac{1}{2}H_2^{\ell m}\,Y^{\ell m},-\delta r_{\pm}^{\ell m}\, \partial_\theta Y^{\ell m} ,0\right)  \, .
\eeq

The second junction condition \eqref{eq:DarmoiIsraelSecond} gives
\beq
	&&\left[\left[h_{1}\sqrt{B}\right]\right]=0 \, , \\
	&&\frac{1}{2}\left[\left[h_0'\sqrt{B}\right]\right]-\frac{2}{r_0}\left[\left[\sqrt{B}\right]\right]h_0 
	-\frac{1}{2}\left[\left[\frac{A'}{A}\sqrt{B}\right]\right]h_0= 8\pi\sigma\, h_0 \,. \label{eq:junctionh0'} \nonumber \\
\eeq
While the first of these agrees with previous results \cite{Uchikata:2016qku,Pani:2009ss}, as far as we are aware, the second equation above has not been presented in this form anywhere.

The polar sector couples to matter perturbations and we find more complicated junction conditions
\beq
&&\left[\left[H_{1} \sqrt{B} \right]\right]=\left[\left[\frac{\delta r}{\sqrt{B}}\right]\right]=0 \, , \\
&&\frac{2}{r_0^2}\left[\left[\delta r \sqrt{B}\right]\right] + \frac{2}{r_0}\left[\left[H_0\sqrt{B}\right]\right]+\frac{1}{r_0}\left[\left[H_2\sqrt{B} \right]\right] \nonumber \\
&-&\left[\left[K'\sqrt{B}\right]\right]-\frac{1}{r_0}\left[\left[\frac{\delta r \, B'}{\sqrt{B}}\right]\right]-\frac{2}{r_0}\left[\left[\delta r\frac{ A'}{A}\sqrt{B}\right]\right]  \nonumber \\
&=&8\pi\, \delta \sigma+ 8\pi\,\sigma\left(\frac{A'}{A}\delta r-H_0\right) \, ,\\
&&\frac{1}{2\,r_0^2}\left[\left[\delta r\sqrt{B}\right]\right]-\frac{1}{2 r_0}\left[\left[H_2\sqrt{B}\right]\right] +\frac{2}{r_0}\left[\left[K\sqrt{B}\right]\right] \nonumber \\
&&-\frac{1}{4}\left[\left[H_2\frac{A'}{A}\sqrt{B}\right]\right]+\frac{1}{2}\left[\left[K\frac{A'}{A}\sqrt{B}\right]\right]+\frac{1}{2}\left[\left[A'\sqrt{B}\right]\right] \nonumber \\
&&-\frac{1}{2}\left[\left[H_0'\sqrt{B}\right]\right] + \frac{1}{2r_0}\left[\left[\frac{\delta r\, B'}{\sqrt{B}}\right]\right] +\frac{1}{r_0}\left[\left[\delta r\frac{ A'}{A}\sqrt{B}\right]\right] \nonumber \\
&&+\frac{1}{2}\left[\left[\delta r \frac{A'}{A}\sqrt{B}\right]\right]' = 8\pi\, \delta p + 8\pi\, p\left(K+2\frac{\delta r}{r_0} \right) \, .
\eeq
We have to complement this with an equation of state 
\beq
\delta p=v_s^2 \, \delta \sigma \, , \qquad v_s^2=\left(\frac{dp}{d\sigma}\right)\Bigr\vert_{\sigma_0} \, .
\eeq
For ordinary matter, $v_s$ is the sound of speed of the fluid and ranges between $0<v_s^2<1$. Again, the first two of the above conditions agree with previous results \cite{Uchikata:2016qku} while we could not find the last two written in this manner anywhere.


\subsubsection{Axial TLNs}
The exterior spacetime has the form of a Schwarzschild metric, so using known results \cite{Cardoso:2017cfl,Binnington:2009bb,Pani:2009ss}
\beq
h_1^{\text{ext}}&=&0 \, , \\
h_0^{\text{ext}}&=&A_1 r^2 _2F^1\left(1-\ell,\ell+2;4;\frac{r}{2M_0}\right) \nonumber\\
&+& A_2 G_{2,0}^{2,2} \left( \frac{r}{2M_0}  \bigg | \begin{matrix}1-\ell & \ell+2 \\
                                        		-1 & 2 
                          						\end{matrix} \right) \, ,
\eeq
where $G_{2,0}^{2,2}$ is the Meijer function~\cite{NIST:DLMF}. The first term of $h_0^{\text{ext}}$ corresponds to the external tidal field and the second to the object's response.

For the interior region, the final equation for $h_0^{\text{int}}$ is similar to that in the exterior, with $M_0$ replaced by $M$
\beq
\left(h_0^{\text{int}}\right)''=\frac{4M-\ell\left(\ell+1\right)r}{r^2\left(2M-r\right)}h_0^{\text{int}} \, .
\eeq
Consequently, the solution is of the form above substituting $M$ by $M_0$. Imposing regularity of $h_0$ at the \acs{BH} horizon means the term with Meijer function has to vanish
\beq
h_0^{\text{int}}&=&A_3 r^2 _2F^1\left(1-\ell,\ell+2;4;\frac{r}{2M}\right) \, ,
\eeq
and $h_1^{\text{int}}=0$.

Now, we can impose the junction conditions derived previously. For $\ell=2$, the general large-distance behavior of $h_0$ is given by a complicated expression. In the limit where the shell is far away
\beq
k_2^B=\frac{\delta M}{5M_0}\frac{r_0^4}{M_0^4} \,,\quad r_0\rightarrow \infty \,. \label{eq:ShellAxialFarAway}
\eeq
Notice that when the shell disappears, $\delta M\rightarrow 0$, $k_2^B\to 0$. This agrees with the vanishing of the \acs{TLN}s of a \acs{BH}~\cite{Binnington:2009bb,Cardoso:2017cfl}. The \acs{TLN} is proportional to the mass in the shell, as we had found for the ``short-hair'' solution. However, the presence of a length scale $r_0$ now implies that the \acs{TLN}s are very sensitive to the location of the matter. In fact, the $r_0^4/M^4$ dependence is expected on general dimensional grounds and from comparison with the \acs{TLN}s of extended configurations, such as boson stars.

In the \acs{BH} limit, when $M_0\rightarrow M$ and $r\rightarrow 2M$
\beq
k_2^B \to \frac{8}{5}\frac{\delta M}{M}\left(\frac{r_0}{M}-2\right) \, ,\label{eq:ShellAxialBHLimit}
\eeq
which is also compatible with the result for an isolated \acs{BH}.

It is also interesting to see the system's behavior when we start without a \acs{BH}, i.e. $M= 0$. In this case, one finds the exact result 
\beq
k_2^B&=&\frac{8\,\xi}{10\,\mathcal{C}\left(3-3\mathcal{C}-2\mathcal{C}^2+2\mathcal{C}^3\sqrt{\frac{1}{\xi}}\right)+15\xi\,\log\xi}\,,\nonumber \\
\xi &=& 1-\frac{2M_0}{r_0} \,,\qquad	\mathcal{C} = \frac{M_0}{r_0}\,.\nonumber
\eeq
This result seems to be at odds with the claims of Ref.~\cite{Cardoso:2019rvt} that the general scaling of the \acs{TLN}s of an \acs{ECO} in the \acs{BH} limit is $k\sim 1/\log\xi$ (see their discussion around Eq.~$(95)$). The proof presented there relies on imposing Robin-type boundary conditions, $a\Psi+b\Psi'=c$, on the Zerilli function $\Psi$, at the surface of the compact object, where $a$, $b$ and $c$ depend on the background spacetime. However, the true scaling goes as $k\propto 1/\left(b+\log\xi\right)$, so if in the \acs{BH} limit $b$ is diverging faster than the logarithm, the claim does not hold. Notice that the factor $b$ is related with the term containing information about the derivatives of the perturbations \textit{at} the boundary. For a thin shell, the perturbations will not be differentiable at such boundary. Therefore, it is not clear how we can rephrase the boundary conditions imposed in Eqs.~\eqref{eq:junctionh0} and \eqref{eq:junctionh0'}, which relate quantities on both sides of the boundary but which are not well defined \textit{at} it, in terms of Robin-type boundary conditions for which the result of Ref.~\cite{Cardoso:2019rvt} applies.

\subsubsection{Polar perturbations}\label{sec:PolarShell}
For the polar sector, the behavior of the perturbations inside and outside the shell is similar to the axial case. They are~\cite{Cardoso:2017cfl,Binnington:2009bb,Pani:2009ss}
\beq
	H_0^{\text{ext}}&=&A_1 P^2_\ell\left(r/M_0-1\right)+A_2Q^2_\ell\left(r/M-1\right) \, , \\
	H_0^{\text{int}}&=&A_3 P^2_\ell\left(r/M-1\right) \, , \\
	H_1^{\text{int}}&=&H_1^{\text{ext}}=0 \, ,
\eeq
where regularity of $H_0^{\text{int}}$ at the \acs{BH} horizon fixes one of the constants. $K$ is determined by the field equations
\beq
K&=&\frac{\left(4M_i^2+2\left(\ell^2+\ell-4\right)M_i r-\left(\ell^2+\ell-2\right)r^2\right)H_i}{\left(\ell^2+\ell-2\right)\left(2M_i-r\right)r}  \nonumber\\
&+&\frac{2M_i\left(2M_i-r\right)H'_i}{\left(\ell^2+\ell-2\right)\left(2M_i-r\right)r} \, ,
\eeq
where $i$ labels the interior or exterior solution which correspond, respectively, to $M$ or $M_0$.

We can impose the junction conditions and obtain the polar \acs{TLN}s. For $\ell=2$, the large distance behavior of $H_0$ is again given by a complicated expression. However, in the large shell radius pressureless limit ($v_s=0$), the polar \acs{TLN} is simply
\beq
k_2^E=\left(1+\frac{M_0}{M}\right)\frac{\,\delta M}{2\,M}\frac{r_0^5}{M_0^5}  \quad , \quad r_0 \,  \to\infty \, ,\label{eq:ShellPolarBigR}
\eeq
in such a way that $k_2^E$ vanishes when $\delta M\rightarrow 0$, as it should.
We note an important dependence on the speed of sound $v_s$. As consequence, $k_2^E$ is positive for small $v_s$ (in the Newtonian limit), but can become negative at large values of $v_s$. Negative \acs{TLN}s have been found in other models involving infinitesimal thin shells~\cite{Uchikata:2015yma, Uchikata:2016qku} and extended configurations of \acs{ECO}s~\cite{Cardoso:2017cfl}. They are usually interpreted as leading to a prolation of the deformed compact object instead of a more intuitive oblate shape. We find the same scaling $k_2^E \sim r_0^5$ as in the Newtonian analysis~\eqref{eq:ShellTLNNew}.

In the \acs{BH} limit, $M_0\to M$ and $r_0\to 2M$, we find
\beq
k_2^E \to \frac{8\left(3-8v_s^2\right)}{5}\frac{\delta M}{M}\left(\frac{r_0}{M}-2\right) \, ,\label{eq:ShellPolarBHLimit}
\eeq
which has a similar dependence as the axial case \eqref{eq:ShellAxialBHLimit}. Although the numerical coefficients do not exactly match, as occurred for the short hair and \acs{ECO}s \cite{Cardoso:2017cfl}, we can attribute this difference to the lack of specification of the equation of state. There is perfect agreement between the $\ell=2$ axial \acs{TLN} \eqref{eq:ShellAxialBHLimit} and the corresponding polar one \eqref{eq:ShellPolarBHLimit} when $v_s^2=0.25$, which is in the allowed range for $v_s$. Also when $v_s^2 > 3/8$, the $\ell=2$ polar \acs{TLN} becomes negative.

If we start without a \acs{BH}, i.e. $M= 0$, and analyze now the \acs{BH} limit $r_0\rightarrow 2M_0$ we obtain 
\beq
k_2^E \to \frac{8}{5\left(9+\sqrt{\frac{2}{\xi}}+4v_s^2+3\log\xi\right)}\,.
\eeq
%

\section{Implications for tests of fundamental physics}

We showed that the leading tidal deformability of a thin shell of matter surrounding a \acs{BH} scales with 
the shell radius as $k_2^E \propto r_0^5$. Let us then extrapolate it for more generic matter distributions. We would be led to conclude that \acs{TLN}s diverge when matter is located sufficiently far away, as $r_0\rightarrow \infty$, which would have a massive impact on \acs{GW} signals. This sounds physically unreasonable and has not been observed. 

Consider a binary system composed of two objects of masses $M_1$ and $M_2$ (total mass $M_\text{tot}=M_1+M_2$) at a Newtonian level. To simplify, consider that both bodies only develop a non-negligible mass quadrupole moment through tidal interactions. The \acs{EOM} for the relative position between the objects, $r^j=r^j_1-r^j_2$, to linear order in the quadrupole moments is \cite{poisson_will_2014, Vines:2010ca}
\beq
\frac{d^2 r^j}{dt^2}&=&-\frac{M_\text{tot}}{r^2}\left(1+\frac{9}{r^5}\left(\lambda_{1}\frac{M_2}{M_1}+\lambda_{2}\frac{M_1}{M_2}\right)\right)n^j\, , \label{eq:EoMBinary}
\eeq
where
\beq
\lambda_i= \frac{2}{3}k_{2_i} M_i^5 \, ,\quad  r=\left|r^j\right| \, ,\quad	n^j=\frac{r^j}{r} \,,
\eeq
being $k_{2_i}$ the $\ell=2$ polar \acs{TLN} of each object.

Simplifying even further, take only object ``$1$'' to be immersed in matter, the other being ``isolated.'' This fixes $k_{2_2}=0$  \cite{Binnington:2009bb,Cardoso:2017cfl}. Then, inserting our results for the $r_0 \rightarrow \infty$ limit of the $\ell=2$ polar \acs{TLN}s of a \acs{BH} surrounded by a thin shell \eqref{eq:ShellPolarBigR} in the \acs{EOM} \eqref{eq:EoMBinary}, we expect a dependence as
\beq
	\frac{d^2 r^j }{dt^2} \sim \frac{\delta M}{M_1}\frac{r_0^5}{r^5}\frac{M_2}{M_1}n^j \, ,
	\label{eq:ShellEffect}
\eeq
where we have used that in realistic astrophysical scenarios $\delta M \ll M_0$.

We assumed that tidal interactions are weak and can be treated perturbatively. The external tidal field is caused by a body in a region far away from the deformed one, which fixes $r_0/r \ll 1$. However, this condition might not be sufficient. From Eq.~\eqref{eq:EoMBinary} and the results for the asymptotic behavior of  $k_2^E$ in the limit $r_0\rightarrow \infty$ \eqref{eq:ShellPolarBigR}, to treat the tidal terms as perturbations we can only consider matter in a region around the compact objects such that
\beq 
\frac{r_0}{r} \ll \text{min}\left(1\, , \,\left(\frac{M_1}{\delta M}\frac{M_1}{M_2}\right)^{1/5}\right) \,.
\eeq
Although this does not fix $r_0$ to an unambiguous value, it justifies why the divergence of the \acs{TLN}s with $r_0$ is not problematic. 

\subsection{Binaries in astrophysical settings} \label{sec:Astro}
Let us now consider a realistic astrophysical system in which the environment may have a measurable impact. As discussed in the previous section, the leading order effect of tidal interactions in the dynamics of a binary comes from the polar $\ell=2$ \acs{TLN}. We also concluded that to use our results for the \acs{TLN}s of a thin shell, we had to consider a lengthscale $r_0$ for the environment smaller than the typical separation $r$ between the binary objects. 

\acs{LISA} will observe in the frequency band $\left[10^{-4},1\right]\, \text{Hz}$. For a circular binary, the relation between the orbital separation $r$ and the \acs{GW} frequency $f_\text{GW}$ is
\beq
r \sim \left(\frac{G M_\text{tot}}{\left(\pi f_\text{GW}\ \right)^2}\right)^{1/3}\, .
\eeq
Consequently, the lower bound of the \acs{LISA} frequency band corresponds to binaries separated by $r \sim 10^{6} \left(M_\text{tot}/M_{\odot} \right)^{1/3} \, \text{km}$ . 

To obtain the properties of the environmental matter, we can use the steady-state model of a Shakura-Sunyaev thin accretion disk~\cite{Shakura:1972te, Abramowicz:2011xu, Barausse:2014tra}. This is an axisymmetric, vertically thin disk, i.e. $H<r$ being $H$ the height of the disk. Following Ref.~\cite{Barausse:2014tra}, we parametrize the mass accretion rate with the mass Eddington ratio $f_{\text{Edd}}$, which for thin disks varies between $10^{-2} \lesssim f_{\text{Edd}} \lesssim 0.2 $. The surface density of the thin disk $\Sigma_ {\text{disk}}$ and the disk height $H$ can be written as

%
\beq
	\frac{\Sigma_ {\text{disk}}}{10^9}&\approx& \frac{f_{\text{Edd}}^{7/10}}{\tilde{r}^{3/4}}\left(1-\sqrt{\frac{\tilde{r}_{\text{in}}}{\tilde{r}}} \right)^{7/10}\left(\frac{0.1}{\alpha} \right)^{4/5}\left(\frac{M}{10^6 M_{\odot}} \right)^{1/5} \text{kg}\cdot \text{m}^{-2} \, , \nonumber \\ \\
	\frac{10^3 H}{GM/c^2}&\approx &  f_{\text{Edd}}^{3/20}\left(1-\sqrt{\frac{\tilde{r}_\text{in}}{\tilde{r}}} \right)^{3/20}\left(\frac{0.1}{\alpha}\right)^{1/10}\left(\frac{10^6 M_{\odot}}{M}\right)^{1/10} \tilde{r}^{9/8} \, , \nonumber \\
\eeq
where $M$ is the mass of the accreting object, $\tilde{r}=r/\left(GM/c^2\right)$, $\alpha \sim 0.01\, -\, 0.1$ is the viscosity parameter and $\tilde{r}_\text{in}\sim 6$ is the radius of the inner edge of the disk. The total mass of the disk is then
\beq
\delta M &\approx& 2\pi \int_{r_\text{in}}^{r_\text{out}}  \Sigma_ {\text{disk}}  r\,dr \,  ,
\eeq
where $r_\text{out}$ is the radius of the disk's outer edge. 

\subsection{On the minimum measurable TLN}
%
\begin{figure}[t]
\centering
\includegraphics[width=0.9\linewidth]{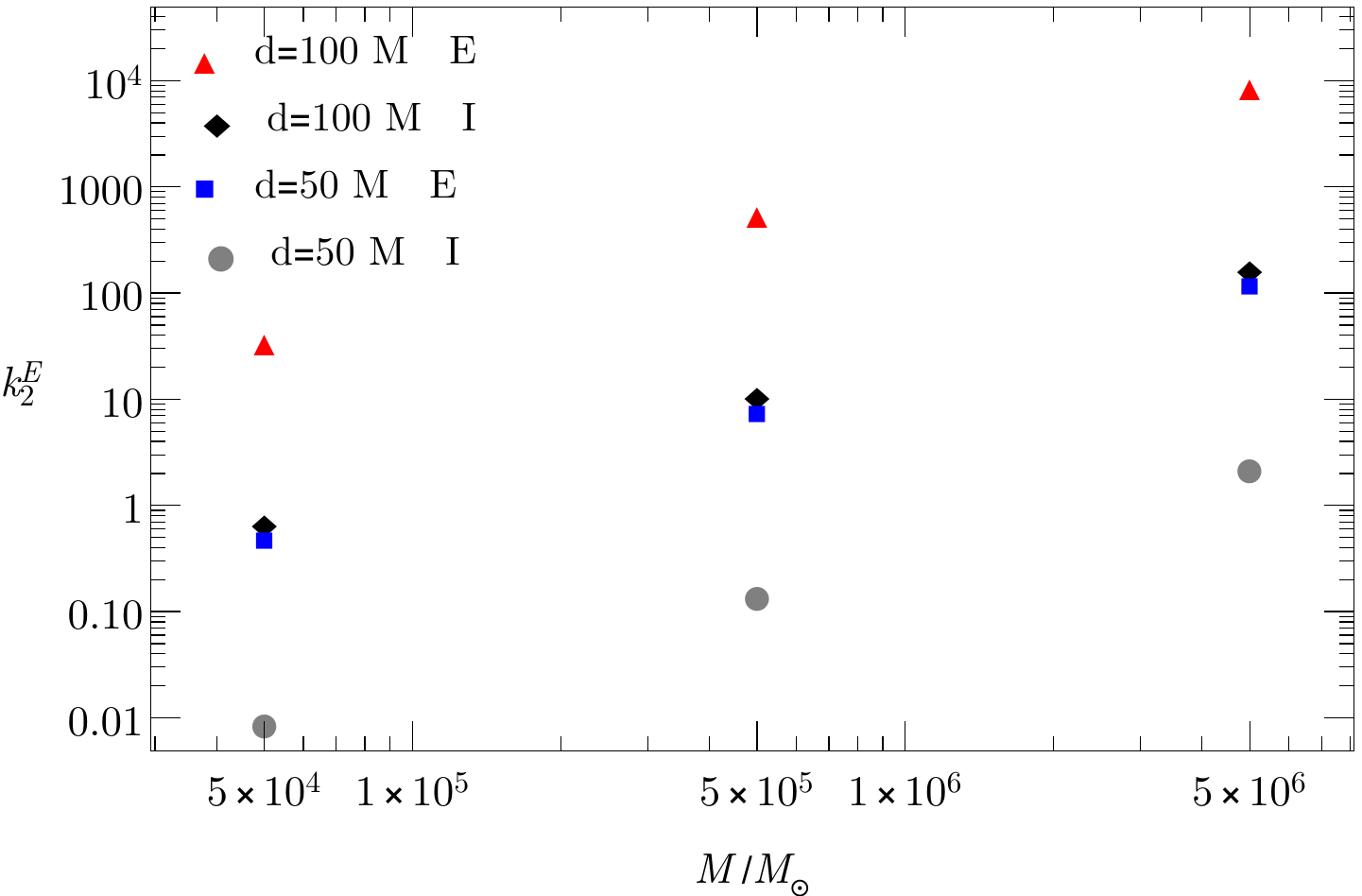}
\caption{The quadrupolar tidal Love number $k_2^E$ for a central \acs{BH} of mass $M$ surrounded by a Shakura-Sunayev thin accretion disk in a circular binary and a companion of the same mass. We present two different binary distances $d$ for $M/M_\odot=5\times 10^4$, $M/M_\odot=5\times 10^5$ and  $M/M_\odot=5\times 10^6$. For each $d$ we present estimations in the most efficient accretion scenario (labeled by $E$), $f_{\text{Edd}}=0.2$ and $\alpha=0.01$, and in the most inefficient one (labeled by $I$)  $f_{\text{Edd}}=0.01$ and $\alpha=0.1$.}
\label{fig:LoveAccretion}
\end{figure}

We can finally compute the dominant \acs{TLN}, $k_2^E$, of a ``dirty'' \acs{BH}. Fig.~\ref{fig:LoveAccretion} shows $k_2^E$ for representative values of the distance $d=(50\,, 100)M$ and for two different accretion scenarios, an efficient (``E'') with $f_{\text{Edd}}=0.2,\,\alpha=0.01$, and an inefficient (``I'') with $f_{\text{Edd}}=0.01,\,\alpha=0.1$.

The lesson to be learned from Fig.~\ref{fig:LoveAccretion} is that massive objects are typically surrounded by enough
matter that they are perceived as having \acs{TLN}s of order $\gtrsim 1$. Thus, extreme care and account of environmental effects must be considered when inferring the properties and nature of ultracompact objects from a measurement of \acs{TLN}s~\cite{Cardoso:2017cfl,Maselli:2017cmm,Maselli:2018fay}. This is especially important for \acs{EMRI}s, where the long time in band would ideally allow for extremely precise constraints on \acs{TLN}s~\cite{Pani:2019cyc}. More recent studies for other types of matter distribution have confirmed our conclusions~\cite{DeLuca:2021ite, DeLuca:2022xlz, Torres:2022fyf, Katagiri:2023yzm}.

%
%


\chapter{Tidal effects in superradiant clouds}\label{ch:Cloud}

As discussed in the Introduction, ultralight bosonic fields are predicted in various extensions of the Standard Model~\cite{PhysRevLett.38.1440, PhysRevD.81.123530, Brito:2015oca, Hui:2021tkt} and have been proposed as a component of \acs{DM}~\cite{Hui:2021tkt}. Through superradiance~\cite{Brito:2015oca}, they can extract rotational energy from spinning \acs{BH}s and grow into macroscopic clouds. This mechanism only requires the bosonic field to be minimally coupled to gravity. However, for it to be efficient in astrophysical timescales, the \acs{BH} radius needs to be of the order of the Compton wavelength $G/(c^2 \mu)$ of the field, where $\mu = G m_\text{B}	 / (c \hbar)$ and $m_\text{B}$ is the mass of boson. Restoring geometric units, we therefore require $M\mu \sim 1$. Considering \acs{BH}s in the Universe appear across ten orders of magnitude, superradiance allows to constrain the existence of new bosonic fields by the same range~\cite{Arvanitaki:2010sy, Brito:2014wla, Brito:2015oca}. 

As also mentioned before, the existence of superradiant clouds would lead to observable signatures, such as peculiar holes in the mass-spin plane of \acs{BH}s~\cite{Arvanitaki:2010sy,Brito:2014wla}, monochromatic emission of \acs{GW}s~\cite{Arvanitaki:2016qwi,Brito:2014wla}, and a significant stochastic background of \acs{GW}s~\cite{Brito:2017zvb,Brito:2017wnc}. 
They can also leave dynamical imprints through Lindblad and co-rotation resonances~\cite{Ferreira:2017pth,Boskovic:2018rub}, or through floating or sinking orbits~\cite{Cardoso:2011xi,Zhang:2018kib,Zhang:2019eid,Baumann:2019ztm}.
However, there are a few factors that could alter, in a significant way, the formation of boson clouds around \acs{BH}s. For example, in the presence of couplings with standard model fields the cloud growth can be suppressed, while stimulating bursts of light~\cite{Ikeda:2019fvj,Boskovic:2018lkj}. 

In this chapter, we will focus on the effects that a companion object, like a \acs{BH}, has on the structure of the boson cloud. Previous works have looked into this problem from the analytical standpoint, restricting the analysis to Newtonian dynamics and non-relativistic fields~\cite{Arvanitaki:2014wva,Zhang:2018kib,Zhang:2019eid,Baumann:2018vus,Berti:2019wnn,Baumann:2019ztm}. At specific orbital frequencies, the motion of the binary can induce resonant transitions between growing and decaying modes, that enhance the cloud's depletion or transfer energy and angular momentum to the companion~\cite{Cardoso:2012zn}. This would leave distinctive imprints in the \acs{GW} signal emitted by the binary, both in the monochromatic signal from the cloud or as modifications in the \acs{GW} waveform of the binary, due to finite-size effects like variations on the spin-induced quadrupole or the \acs{TLN}s~\cite{Baumann:2018vus}. 

\section{Setup}

%
\begin{figure}[t]
\centering
\includegraphics[width=0.9\linewidth]{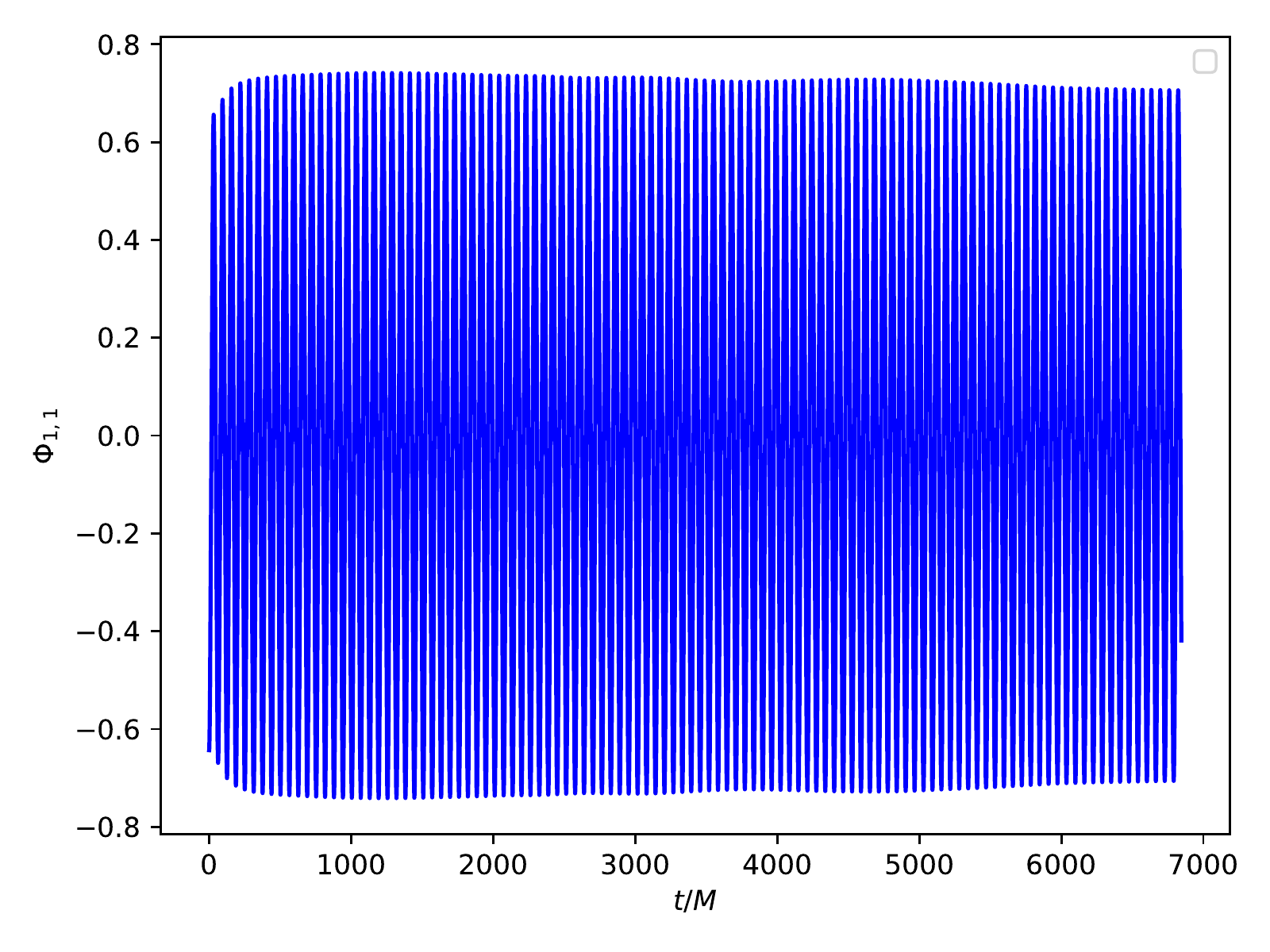}
\caption{A dipolar scalar cloud around a Schwarzschild \acs{BH}. This figure shows the time evolution of initial conditions~\eqref{Eq.axion cloud initial data}
for a dipole with gravitational coupling $M\mu=0.1$ around a non-spinning \acs{BH}, and in absence of a companion ($\epsilon=0$).
The field is extracted at $r=60M$. For the timescales of interest for our problem, the amplitude of the field varies by only a few percent and is therefore a good description of a quasi-stationary state.
\label{fig:a0_Mc0_mu01}}
\end{figure}
\begin{figure}[t]
\centering
\begin{tabular}{cc}
\includegraphics[width=5.5cm]{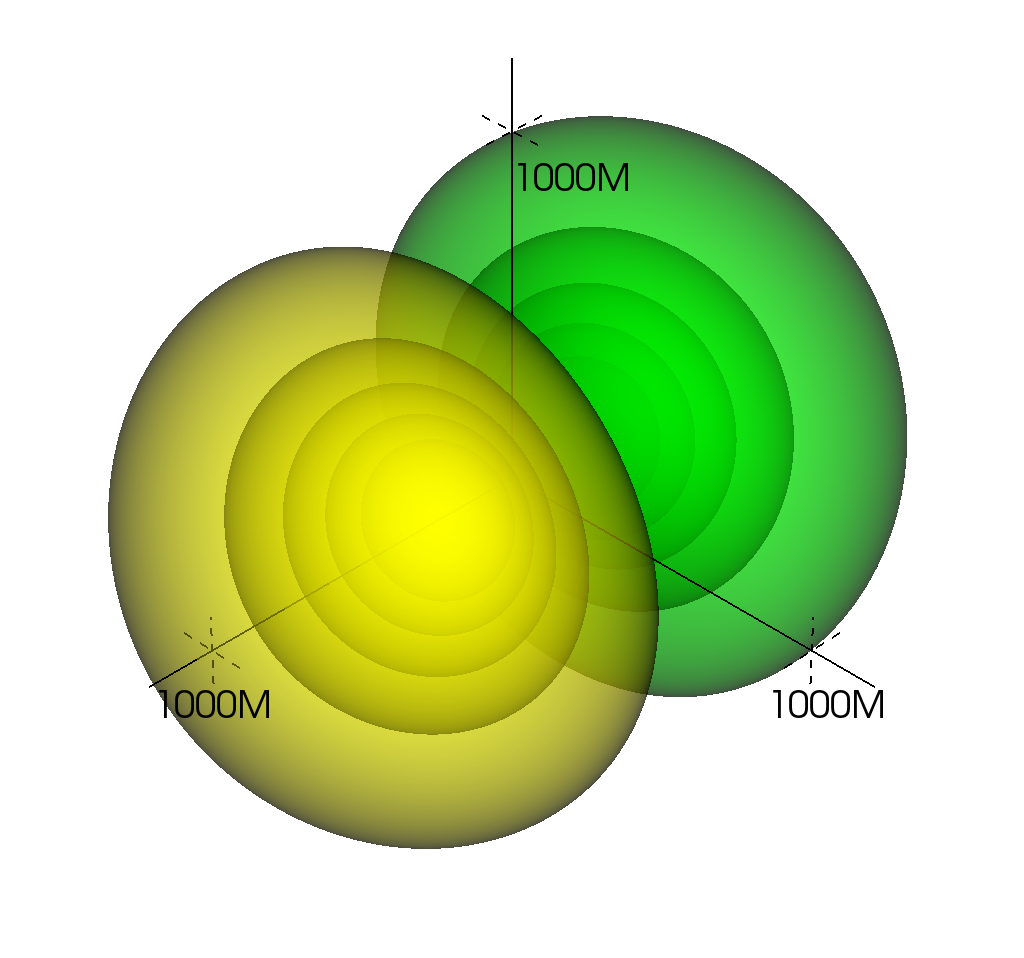}&
\includegraphics[width=5.5cm]{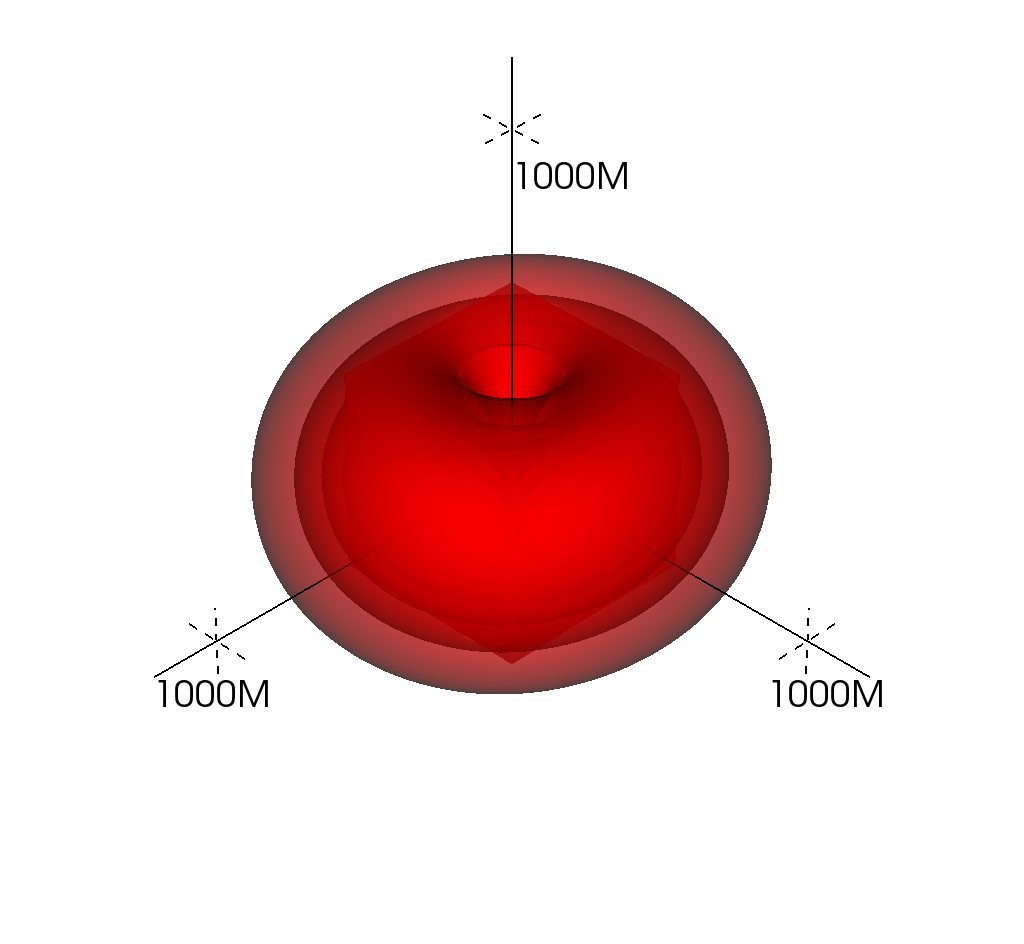}
\end{tabular}
\caption{Field (left) and energy density (right) distribution along the equatorial plane for the same initial data as Fig.~\ref{fig:a0_Mc0_mu01}. The field is dipolar, as expected, whereas the energy density at the equator is almost -- but not exactly -- symmetric along the rotation axis. The length scale of these images is of order $1000M$. 
\label{fig:a0_Mc0_mu01_snapshot}
}
\end{figure}
\begin{figure}[t]
\centering
\includegraphics[width=0.9\linewidth]{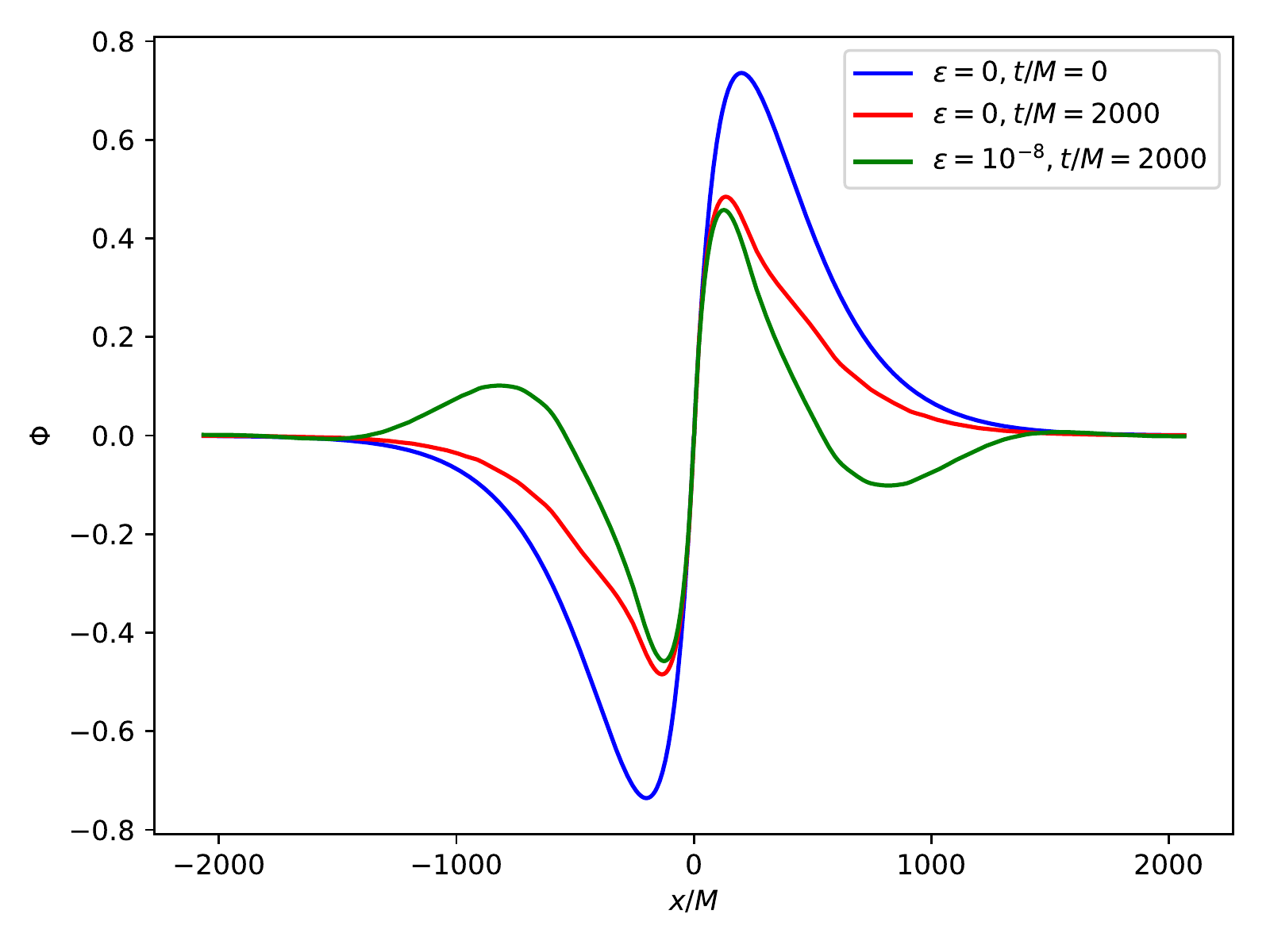}
\caption{Dependence of the field $\Phi$ along the $x-$ axis at different instants for a coupling $M\mu=0.1$. In the absence of a companion ($\epsilon=0$) and despite a slight change in the profile, the field has no nodes. It has a local extremum at $\sim r=200M$
as predicted by a small $M\mu$ expansion for the fundamental mode.  
When a weak tidal field is turned on ($\epsilon=10^{-8}$) the field develops a different radial profile with one node, pointing to a significant component of overtones. Our results indicate a sizeable excitation of the second excited state $n=4$, which has extrema at $r/M\sim 170, 850$ (see Appendix~\ref{app:Cloud_PT}).
\label{fig:overtones}}
\end{figure}
\begin{table}[t]
\centering 
\begin{tabular}{c|c|c|c}
    ($n \, \ell \, m$)   & $t_{\text{trans}}/M$    & $\frac{c_{n\ell m}}{c_{211}}$  ($\frac{\phi_{n\ell m}}{\phi_{211}}$) &  $\frac{c_n^{\text{Num}}}{c_2^{\text{Num}}}$ \\
    \hline
    $3 1 1 $  	   	   &    $ 1888 $    	      & 	$1.03	(0.85)$	&	$0.221$\\
    $4 1 1 $  	   	   &   $  458	$          & 	$0.236	(0.13)$	&	$0.094$\\
    $5 1 1$   	   	   &    $ 173	$          & 	$0.113	(0.046)$	&	$0.058$\\
\end{tabular}
\caption{Timescales $t_{\text{trans}}$ and relative amplitudes predicted by time-independent perturbation theory (see Appendix~\ref{app:Cloud_PT}) and those obtained from numerical simulations data at $t=1000M$, for the most relevant $1^{st}$ order transitions from the initial dipolar state ($\ell=m=1$) for a gravitational coupling of $M\mu=0.1$. 
The second column shows the timescale to transition from the initial to the $(n\ell m)$ state, as obtained from time-dependent perturbation theory. The third column shows the relative amplitude of the overtones with respect to the fundamental mode, predicated by time-independent perturbation theory (in parenthesis is the corresponding ratio of the field components at $r=60M$). The last column shows this same quantity but for our numerical results. They agree with perturbation theory within a factor of two, except for the $\ell=m=1, n=3$ mode. In this case, the timescale needed for excitation is larger than the instant at which the coefficients were extracted.} 
\label{tab:overtone_excitation}
\end{table}
\begin{figure}[t]
\centering
\begin{tabular}{cc}
\includegraphics[width=5.9cm]{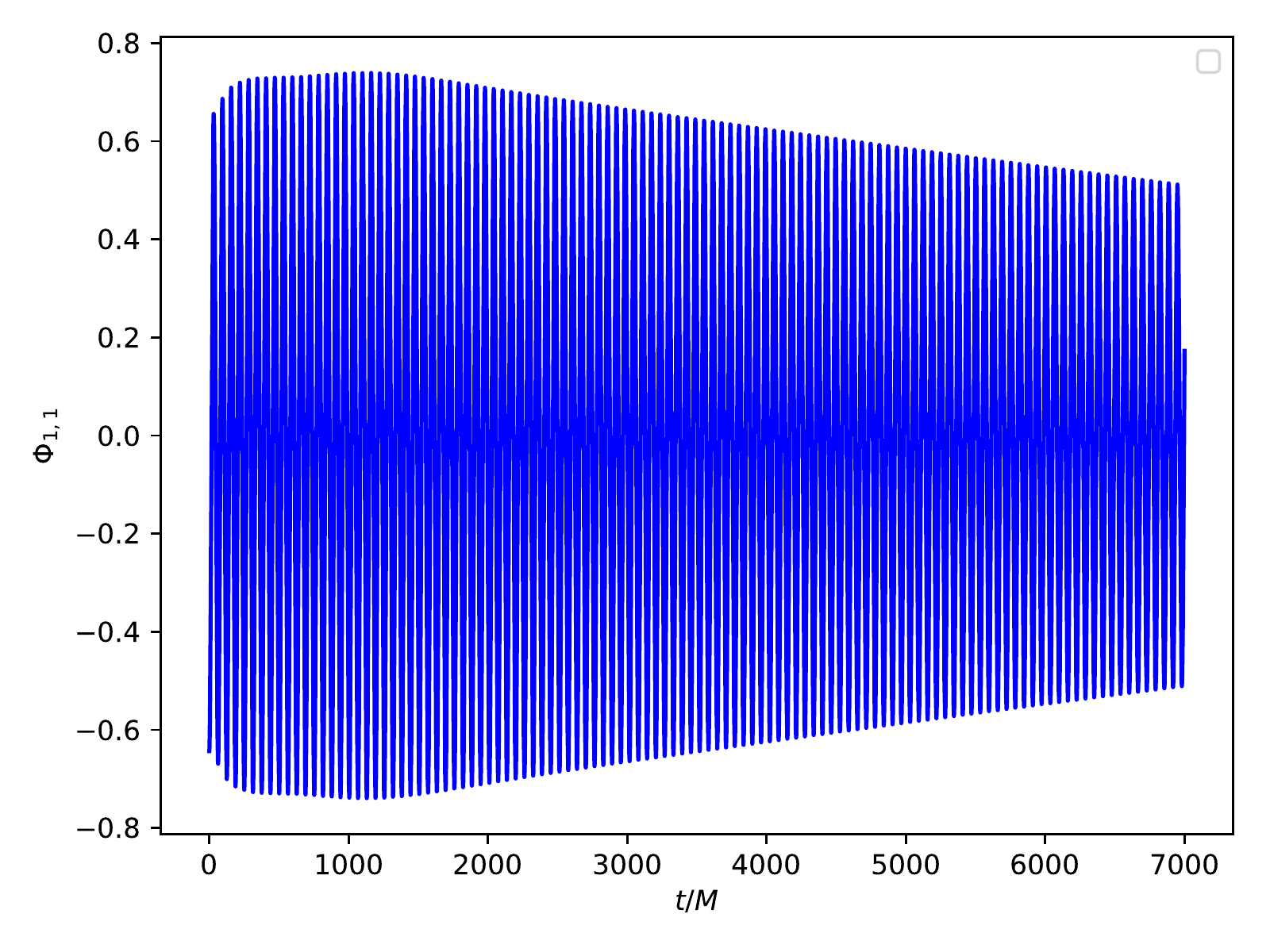} 
\includegraphics[width=5.9cm]{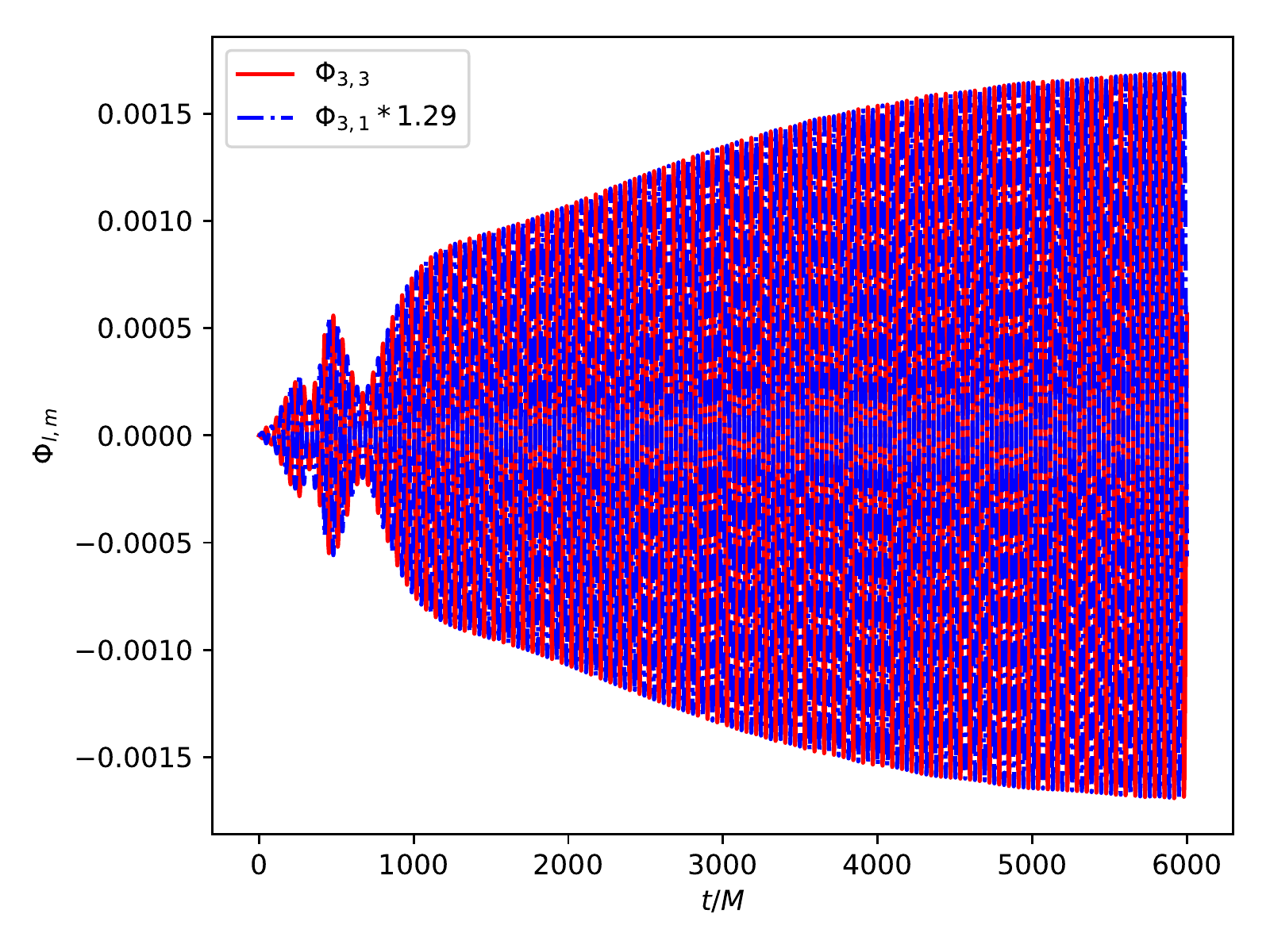}
\end{tabular}
\caption{Dipolar ($\ell=m=1$, left) and octupolar ($\ell=3,\, m=1,3$, right) component of the scalar cloud when in the presence of a weak tidal field $\epsilon=10^{-8}$, for the same initial conditions as in Fig.~\ref{fig:a0_Mc0_mu01} (non-spinning \acs{BH} and gravitational coupling $M\mu=0.1$). The $\ell=3, m=1$ mode amplitude relative to the $\ell=m=3$ was rescaled by the perturbation theory prediction ($\sqrt{5/3}\sim 1.29$). The agreement is very good throughout the evolution.
\label{fig:a0_Mc001_mu01}}
\end{figure}
\begin{figure}[t]
\centering
\includegraphics[width=0.9\linewidth]{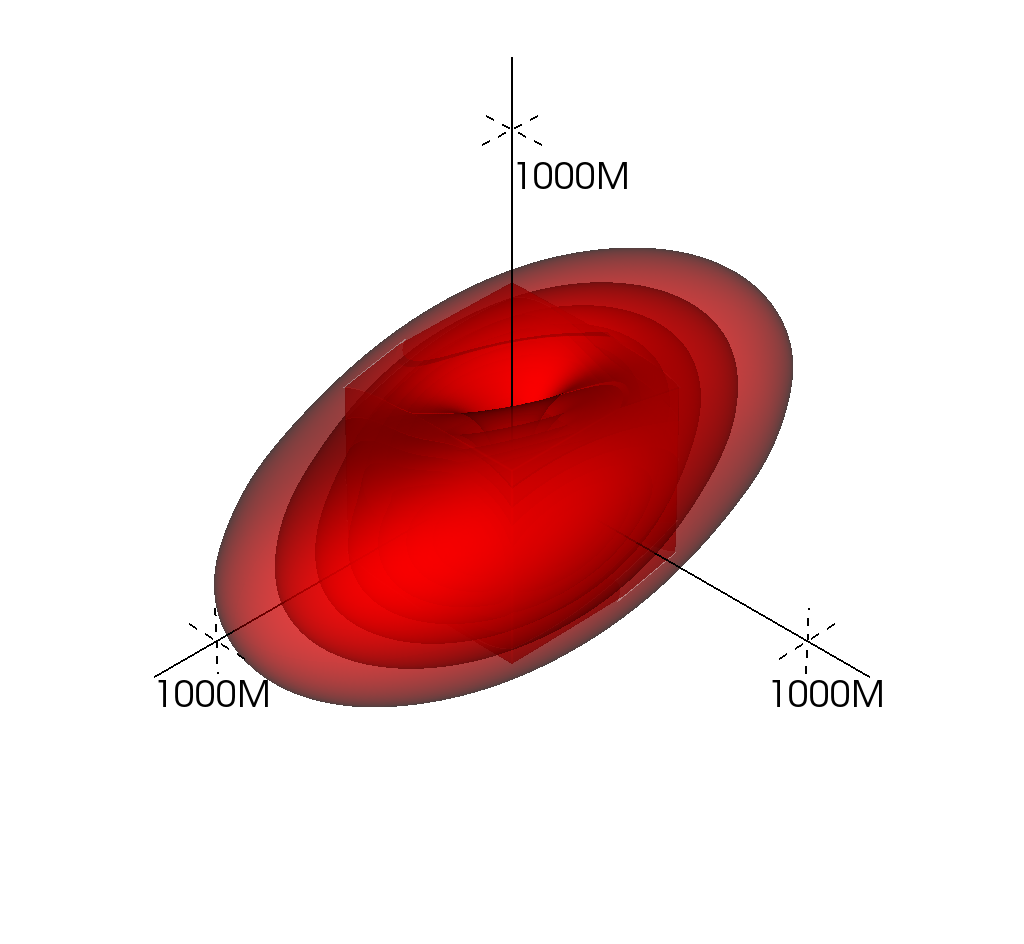}
\caption{Snapshot of a tidally deformed scalar cloud with $\epsilon=10^{-8}$ for the strength of the tidal field and $M\mu=0.1$ for the gravitational coupling. In the absence of a companion, the energy density is almost spherical and remains so for thousands of dynamical timescales. Instead, here we observe that after $7000M$ the system settles to a new stationary configuration. 
\label{fig:a0_Mc001_mu01_snapshot}}
\end{figure}

Our starting point is that of a Kerr \acs{BH} surrounded by a superradiant cloud being perturbed by a distant companion at a fixed location. We also assume the test-field approximation in which the backreaction of the cloud in the spacetime geometry is negligible. This is a good approximation because the timescales we will probe
are much shorter than any superradiant-growth timescales~\cite{Brito:2014wla} ($\tau^s_\text{growth} \sim (M\mu)^{-9} M$ for scalar fields~\cite{Detweiler:1980uk}). The unperturbed background is therefore assumed to be described by a Kerr \acs{BH} (line element in Eq.~\eqref{eq:KerrLineElement}).

The tidal field is created by a companion of mass $M_c$, at a distance $R$, and located at $\theta=\theta_c,\,\varphi=\varphi_c$ in the \acs{BH} sky. The companion induces a change $\delta ds_\text{tidal}^2$ in the geometry, so our full spacetime is
\beq
ds^2=ds^2_\text{Kerr}+\delta ds_\text{tidal}^2\,.
\eeq
For the tidal perturbation, we consider the non-spinning approximation and import the results from the previous chapter. Taking into account only the dominant quadrupole term~\cite{Cardoso:2017cfl,Cardoso:2019upw,Taylor:2008xy}
\beq
\delta ds_\text{tidal}^2&=&\sum_{m=-2}^{2} r^2{\cal E}_{2m}Y_{2m}(\theta,\phi)(A^2dt^2+dr^2+(r^2-2M^2)d\Omega^2)\nonumber\\
%
%
{\cal E}_{2m}&=&\frac{8\pi \epsilon}{5 M^2}\,Y^*_{2m}(\theta_c,\phi_c)\,,\label{eq:MetricPerturbation}
\eeq
where $A(r)=1-2M/r$ and we neglect subdominant magnetic-type contributions and higher multipoles. We introduce a dimensionless tidal parameter
\beq
\epsilon=\frac{M_{c}M^{2}}{R^3}\,,
\label{eq:TidalParam}
\eeq
which measures the strength of the tidal moment.

The tidal field we are considering is not accurate close to the central \acs{BH}, where spin effects will change the tidal potential. Yet, in the parameter space we will explore, most of the cloud is localized far away from the horizon, where these effects are minimal and should not affect our qualitative conclusions. As in the previous chapter, we will focus exclusively on static tides, which are independent of coordinate time $t$. 
We stress that we are using coordinates adapted to the \acs{BH}: the companion position should in general be time-dependent, but we focus exclusively on slowly moving companions (or equivalently, large separations $R$).

The superradiant cloud is described by a massive, minimally coupled scalar field $\Phi$ evolving on the above fixed geometry
\beq
\Box \Phi = \mu ^2 \Phi \, .
\eeq

We evolve this equation numerically using a $3+1$ decomposition in Cartesian Kerr-Schild coordinates $(t,x,y,z)$. We refer the reader to Ref.~\cite{Witek:2012tr} for more details on the numerical implementation.~\footnote{The numerical implementation of this work was performed by Taishi Ikeda.}

We will be interested in extracting from our numerical simulation the multipolar components of the scalar $\Phi$
\beq
\Phi_{\ell,m}(t,r)=\int_{S_2} d\Omega \, \Phi(t,r,\theta,\varphi) Y_{\ell,m}(\theta,\varphi)\,.
\eeq

As initial data for our numerical evolutions we will use a fundamental superradiant dipolar mode adequate to describing quasi-stationary states around a \acs{BH}~\cite{Yoshino:2013ofa,Brito:2014wla}
\beq
\label{Eq.axion cloud initial data}
\Phi(t,r,\theta,\varphi)=A_{0}\,rM\mu^{2}e^{-rM\mu^{2}/2}\cos(\varphi -\omega_\text{R}t)\sin\theta\,.
\eeq
$A_{0}$ is an amplitude related to the mass in the axion cloud, and $\omega_\text{R}\sim \mu$ is the bound-state frequency.

The spacetime of a real astrophysical binary is asymptotically flat. However, because we are using only an approximation to the full problem, where the companion is supposed to be far away, the geometry~\eqref{eq:MetricPerturbation} is no longer asymptotically flat. To avoid unphysical behavior at large distances, we force the geometry to be asymptotically flat by replacing the far region with
%
%
%
\beq
ds^2=ds^2_\text{ Kerr}+\left(1-\mathcal{W}\right)\delta ds^2_\text{ tidal}\,,
\eeq
where $\mathcal{W}=\mathcal{W}(\tilde{r})$ is a following piecewise function
\beq
\mathcal{W}(\tilde{r})=\left\{
\begin{array}{ll}
1&(\tilde{r}>1)\\
\mathcal{W}_{5}&(0<\tilde{r}<1)\\
0&(\tilde{r}<0).\\
\end{array}
\right.
\eeq
Here, $\tilde{r}=(r-r_\text{ th})/w$ and $\mathcal{W}_{5}(\tilde{r})$ is chosen to match smoothly with the required asymptotic behavior, so we choose a 5th-order polynomial satisfying $\mathcal{W}_{5}(1)=1,\mathcal{W}_{5}(0)=\mathcal{W}_{5}'(0)=\mathcal{W}_{5}''(0)=\mathcal{W}_{5}'(1)=\mathcal{W}_{5}''(1)=0$.
The transition region has a width $w=500M$ and is located at $r_\text{th}/M\simeq \sqrt{0.9\times 5/(8\pi \epsilon)}$.
These parameters were chosen to ensure that the bosonic cloud sits entirely in a region described by Eq.~\eqref{eq:MetricPerturbation}.
Accuracy requirements, finite size of the numerical grid and computational power all contribute to limit the timescales that one is able to access numerically. Here, we evolve these systems for timescales $\sim 7000M$.

Although we have results for general \acs{BH} spin, we focus mostly on states around a \textit{non-spinning} \acs{BH}.
These states are not superradiant in origin and arise due to the fine-tuned initial data. However, they are extremely long-lived (the decay timescale is of the order of the superradiant growth timescale if the \acs{BH} was spinning), as we show below, with a lifetime that far exceeds that of all the tidally-induced transitions studied here. Thus, \acs{BH} spin is important to generate the scalar clouds but has little impact on \textit{some} of the physics of tides. In addition, the tidal field in Eq.~\eqref{eq:MetricPerturbation} is adapted to a non-spinning \acs{BH}. Our numerical results show only a very mild dependence on \acs{BH} spin. With the exception of Ref.~\cite{Berti:2019wnn}, all previous results on tidal effects in superradiant clouds focus on the small $M\mu$ coupling parameter, consider a flat background on which the superradiant states evolve, and have only used linearized analysis for small tidal fields. Our framework can go beyond all these limitations.

\section{Weak tides: transitions to new stationary states}

\begin{figure}[t]
\centering
\begin{tabular}{cc}
\includegraphics[width=5.9cm]{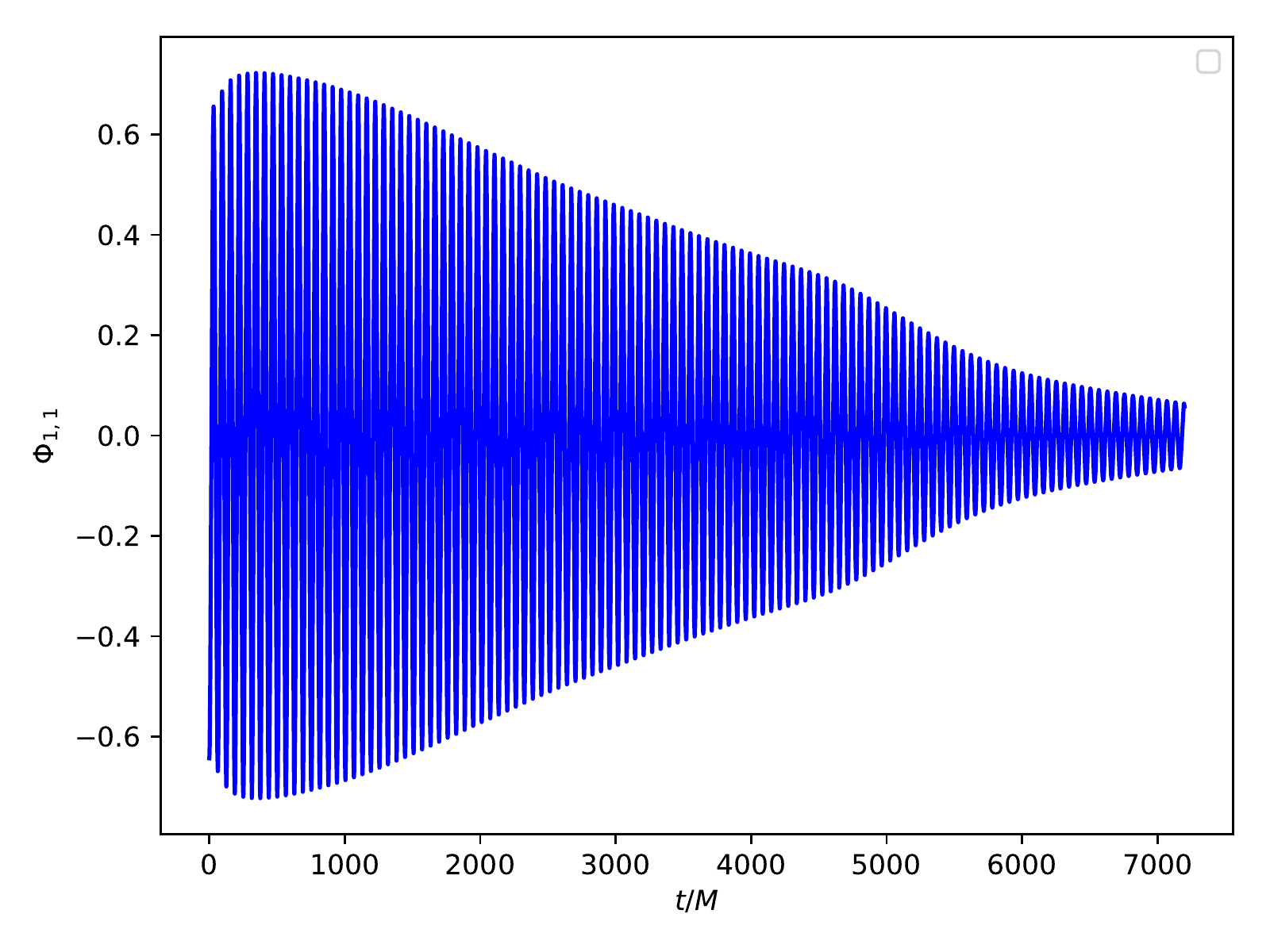}
\includegraphics[width=5.9cm]{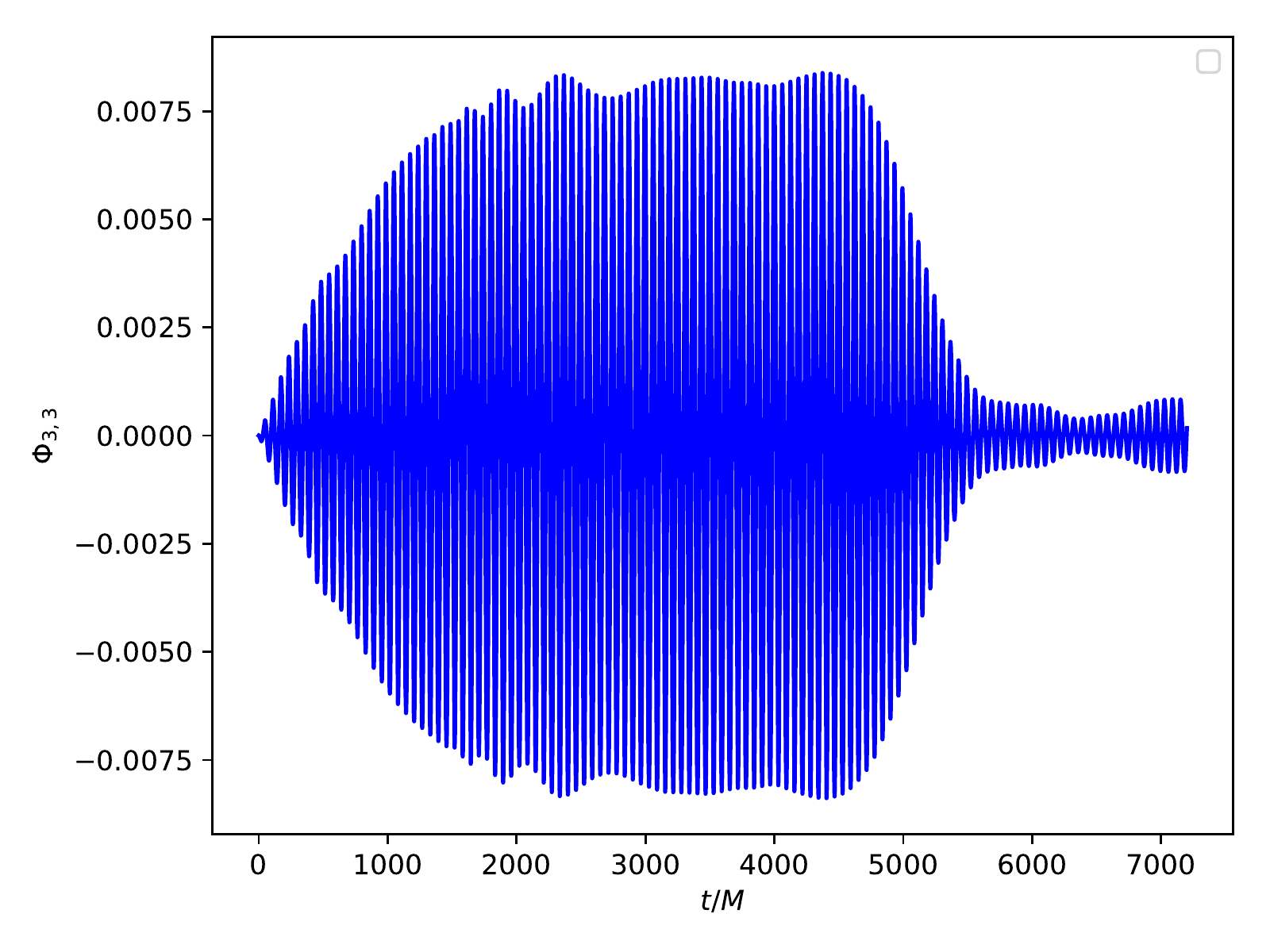}
\end{tabular}
\caption{Cloud being tidally disrupted as it loses energy to asymptotically far-away distances from the central \acs{BH}. The field cascades to smaller angular scales, and thus transitions to higher multipoles. This figure shows the time evolution of the dipolar ($\ell=m=2$) and octupolar ($\ell=m=3$) components of the scalar field for a gravitational coupling $M\mu=0.1$, and a companion with $\epsilon=10^{-7}$. The extraction radius is $r=60M$.
\label{fig:a0_Mc01_mu01}}
\end{figure}
\begin{figure}[t]
\centering
\includegraphics[width=0.9\linewidth]{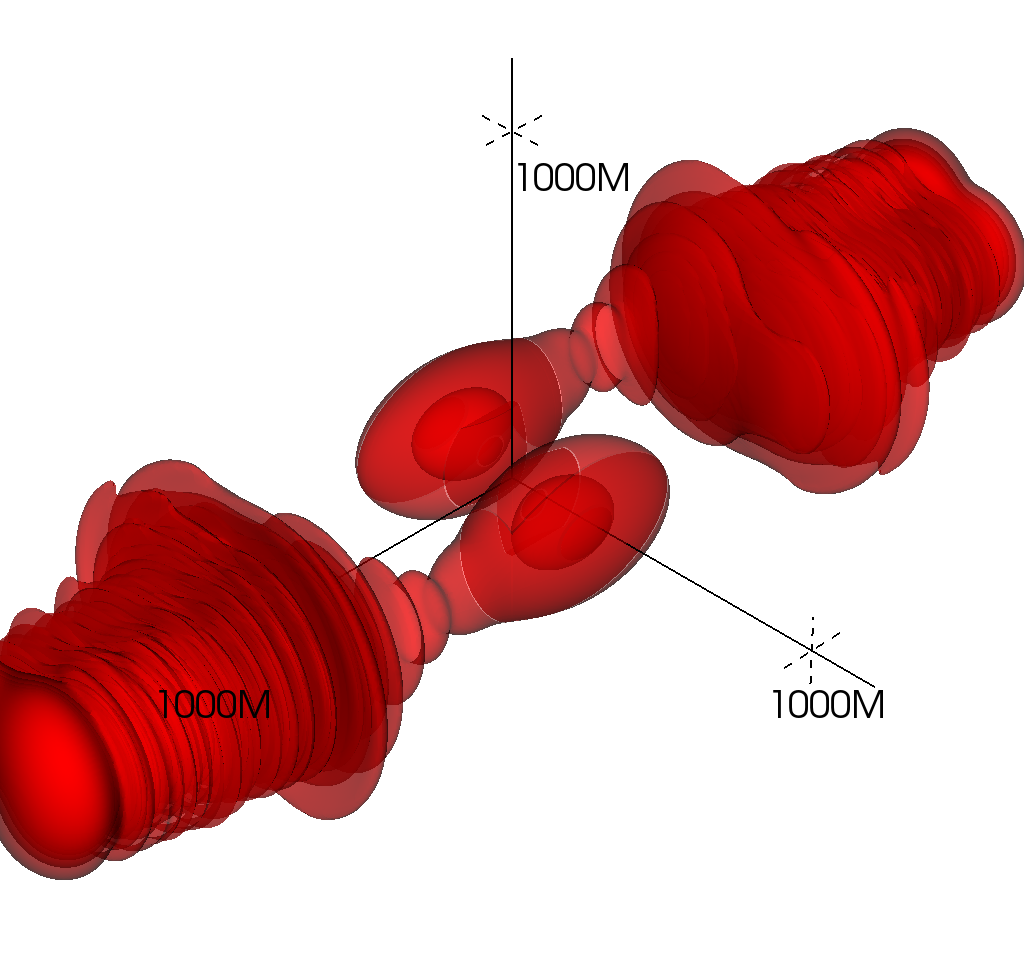}
\caption{Snapshot of a tidally disrupting cloud. The snapshot depicts the energy density along the equator of a scalar cloud which was set initially around a non-spinning \acs{BH}. 
In the absence of a companion mass, the energy density is almost spherical and remains so for thousands of dynamical timescales.
Here, the simulation starts with one symmetric initial scalar energy distribution, but in the presence of a star for which $\epsilon=10^{-7}$. The gravitational coupling is $M\mu=0.1$.
The snapshot is taken after $7000M$ and is leading to the disruption of the cloud. 
\label{fig:a0_Mc01_mu01_snapshot}}
\end{figure}

To show that our initial dipolar state is a (quasi-)stationary state in the timescales of interest of our problem, we start by evolving the initial data described above~\eqref{Eq.axion cloud initial data} around an isolated, non-spinning \acs{BH} ($\epsilon=0$). Its non-vanishing multipolar component, the dipole $\ell=m=1$, is shown in Fig.~\ref{fig:a0_Mc0_mu01}. The amplitude of the field varies by a few percent over a time interval of $\sim 7000M$, which corresponds to $\sim 100$ scalar field periods of oscillation ($\sim 2\pi /\mu$). The scalar field and energy density along an equatorial slice are shown in Fig.~\ref{fig:a0_Mc0_mu01_snapshot} at $t=7000M$. The density is almost (but not exactly) symmetric along this slice. 

We now turn on a tidal field with $\epsilon=10^{-8}$ produced by a companion star on the $x$-axis. We consider this to be a weak tide since no nonperturbative feature is seen on timescales of $\sim 6000M$. The first feature we observe is the transition from the fundamental dipolar state to higher excited overtones, in the same spirit as previous analytical studies~\cite{Baumann:2018vus,Berti:2019wnn}. The $n$-th overtone is localized at a radius $r_\text{ Bohr}\sim n^2/(M\mu^2)$, which means their excitation leads to an expansion of the cloud. Fig.~\ref{fig:overtones} shows the $x-$ dependence of the field initially and at $t=2000M$. If the cloud stayed on the fundamental dipolar mode it should not have any node along the radial direction, as it happens for $t=0M$. Instead, we observe one node at $t=2000M$, meaning the cloud is transitioning to excited states. This profile also includes the octupolar $\ell=3$ component, though this is two orders of magnitude smaller than the dipolar term. 

In Appendix~\ref{app:Cloud_PT} we outline how to study the transition between bound states of the cloud using standard perturbation theory in Quantum Mechanics. In the small coupling limit, $M\mu \ll 1$, the description of the cloud state is analogous to that of the hydrogen atom, with $\alpha=M \mu$ playing the role of a gravitational fine-structure constant. The properties of the radial distribution of the cloud can then be studied using the typical radial ``hydrogenic'' functions $R_{n\ell}\left(r\right)$. From them, we know the $n=3$ state has extrema at $r/M=175, 1024$, which are not apparent in Fig.~\ref{fig:overtones} (in our convention, states are labeled by an integer $n = \ell + 1,\,  \ell + 2, \,...$). Instead, the second excited state $n=4$ is known to have extrema at $r/M=170, 875, 2155$, which agrees with those of the numerical results (the last point is challenging to confirm, as the grid size and spurious reflections affect a proper evaluation of eigenfunctions at large distances). 

Moreover, since in this limit the eigenstates of the cloud are orthonormal, we can extract the amplitude $c_n$ of a specific overtone from the numerical data by a simple projection
\beq
c_n=\int_0^\infty dr \, r^2 \, R_{n1}^*\left(r\right)\Phi\left(r\right) \, , 
\label{eq:AmpOvertone}
\eeq	
where $\Phi\left(r\right)$ corresponds to the numerical data at a given radial direction (e.g. $\theta=\pi/2$ and $\varphi=0$). We are implicitly assuming that our data is only composed of $\ell=1$ modes, which again, as we will see below, is a reasonable approximation. Since our grid size is limited, we can only capture a few nodes along the radial direction and for later times our data at large radius will be contaminated by spurious reflections from the outer boundary. For this reason we restrict this analysis to times $t\lesssim 1000M$. 

Our results are shown in Table~\ref{tab:overtone_excitation}. They are within a factor two from the estimates of perturbation theory, which is an excellent agreement considering: $\bm{1}$. perturbation theory uses a small $M\mu$ expansion which is already inaccurate at $M\mu\gtrsim 0.1$ (e.g. Fig.~$1$ in Ref.~\cite{Berti:2019wnn} shows how factors of two can easily arise from such an approximation); $\bm{2}$. there can be intermediate transitions between states which complicate the analysis.
Nevertheless, perturbation theory justifies why the first excited state, the $n=3$ mode, is not yet dominant, since the timescale for its excitation is larger than higher overtones. Our numerical results are consistent with transition occurring on the timescales predicted from the table for the $n=3, 4, 5$ overtones.

Another important feature of our simulations are transitions to octupolar and higher multipoles, as illustrated in Fig.~\ref{fig:a0_Mc001_mu01}. There we show the time evolution of the dipolar $\ell=m=1$ and octupolar $\ell=m=3$ mode. It is apparent that the magnitude of the dipolar mode is now decreasing, and that a fraction of this energy is going into higher multipoles, specifically the octupolar $\ell=3, \, m=1,3$. Such migration changes the spatial distribution of energy density, as depicted in Fig.~\ref{fig:a0_Mc001_mu01_snapshot}.

Again, our results are consistent with the prediction from perturbation theory, in particular that the amplitude of the $\ell=3$ mode scales with the external tide $\epsilon$~\cite{Baumann:2018vus,Berti:2019wnn}. One of the cleanest indications of the validity of the perturbative framework is the excitation of the $\ell=3,m=1$ mode. Perturbation theory predicts that the relative amplitude of the $\ell=m=3$ mode is $\sqrt{5/3}\sim 1.29$ larger than that of the $\ell=3, m=1$ mode, and this depends exclusively on an angular matrix with no radial dependence. Our results show a relative amplitude across all times and extraction radii consistent with such prediction, as shown in the figure.

\section{Strong tides: tidal disruption of clouds}

We now move to study stronger tidal fields. If too strong, we expect the scalar configuration to be tidally disrupted. A star of mass $M_*$, radius $R_*$, in the presence of a companion of mass $M_c$ at distance $R$ is on the verge of disruption when the tidal force surpasses its own self-gravity, known as the \textit{Roche} limit. Up to numerical factors of order one this is
\beq
M_*/R_*^2=2M_cR_*/R^3\,\label{Roche}\, .
\eeq
When the mass in the scalar cloud is a fraction of that of the \acs{BH}, $M_*\approx M$ and its radius is of the order of $R_*\gtrsim 5/(M\mu^2)$ (see Appendix~\ref{app:Cloud_PT}). This corresponds to a critical tide
\beq
\epsilon_{\text{crit}}\approx\frac{(M\mu)^6}{250}\,.\label{tide_scaling}
\eeq

The simulations presented in Figs~\ref{fig:a0_Mc01_mu01}--\ref{fig:a0_Mc01_mu01_snapshot} corroborate this behavior. For large masses of the companion, the initial dipolar mode quickly transfers energy to the octupole, which then drains into higher and higher multipoles. This signals a transfer of energy to smaller and smaller angular scales, as the cloud is disrupted and loses mass to far-away distances, as illustrated in Fig.~\ref{fig:a0_Mc01_mu01_snapshot}.

Extracting precise values for the critical tide from our simulations is complicated because: $\bm{1.}$ these are extended scalar configurations, and to understand whether there is mass being lost to large distances requires large numerical grids; $\bm{2.}$  a seemingly stable cloud on the simulation timescale can eventually be disrupted when evolved for longer times. Typically, our numerical simulations last for $\sim 6000M$. With this in mind, we estimate $\epsilon_\text{crit}\sim  2\times 10^{-7}$ for $M\mu=0.2$, for which disruption is clear after $4000M$. This critical tide agrees remarkably well with our crude estimate of Eq.~\eqref{tide_scaling}. On the other hand, for $M\mu=0.1$ we see disruption on the simulated timescales only for $\epsilon\gtrsim 2\times 10^{-8}$, a factor of four bigger than Eq.~\eqref{tide_scaling}. Disruption may occur for smaller tides, but on timescales that we are currently unable to probe. Disruption can also be stimulated by transitions to overtones which would ``puff up'' the cloud and increase its size to a few times the estimate $\sim 1/(M\mu^2)$. Consequently, the critical tide would be reduced.

To summarize, our results are consistent with the behavior of Eq.~\eqref{tide_scaling}, though longer evolutions are needed to understand better the correct prefactor in the critical tide $\epsilon_\text{crit}$.

\section{Astrophysical systems}\label{sec:Astro_Cloud}

We will now apply our results to some known astrophysical systems. Examples of \acs{BH}s with companions are the Cygnus X-$1$ system and Sagittarius A* at the center of our galaxy. Cygnus X-$1$ is a binary system composed of a \acs{BH} of mass $M_{\text{BH}}\sim 15 M_\odot$ and a companion with $M_c\sim 20 M_\odot$ at a distance $R\sim 0.2 \,{\text{AU}} \sim 3\times 10^{10}  \,{\text{ m}}$~\cite{Orosz_2011}. Plugging these in Eq.~\eqref{eq:TidalParam} we find $\epsilon \sim 5\times 10^{-19}$. Then, for it to sit at the critical tide, $M\mu\sim 2\times 10^{-3}$. The growth timescale $\tau$ of scalar clouds via superradiance is of order $\tau^s_\text{growth} \sim (M\mu)^{-9} M \sim  \left(M\mu/(2\times 10^{-3})\right)^{-9}\left(M/10^5 M_\odot\right) 10^{16}\, \text{years}$~\cite{Brito:2015oca}, which for Cygnus X-$1$ is too large to be meaningful compared to the age of the universe ($\sim10^{10}$ years). However, for vector fields the growth timescale is smaller, $\tau^v_\text{growth} \sim (M\mu)^{-7} M$~\cite{Pani:2012vp,Witek:2012tr,Cardoso:2018tly}. If our results extrapolate to vector fields, this mechanism becomes astrophysically relevant for this system. Note also that the tide is small enough that it should not affect any of the constraints derived from the non-observation of \acs{GW}s emitted by the cloud~\cite{Yoshino:2014wwa,Sun:2019mqb}.

Moving to Sagittarius A*, there is a supermassive \acs{BH} at the center of our galaxy with mass $\sim 4\times 10^6 M_{\odot}$, The closest known companion is the S$2$ star~\cite{Abuter:2018drb,Naoz:2019sjx}, with mass $M_{\text{S}2} \sim 20 M_{\odot}$ and a pericenter distance of $\sim 1400M_{\text{BH}} \sim 120\, \text{AU}$~\cite{GRAVITY:2018ofz}. Then $\epsilon\sim 2\times 10^{-15}$, which corresponds to a critical coupling for disruption of $M\mu \sim 9\times 10^{-3} \sim  \left(M\mu/ 10^{-2}\right)^{-9}\left(M/10^5 M_\odot\right) 10^{10}\, \text{years}$. This means tidal disruption may occur in astrophysically relevant timescales, which would affect the estimates of \acs{GW} emission using just the dipolar mode. However, note that our approximations always require that the companion sits outside the cloud ($R>R_*$, or the approximation in Eq.~\eqref{eq:MetricPerturbation} would break down). Using Eq.~\eqref{Roche}, disruption together with such a condition always requires $M_c>M/2$, which is not the case here.

At the verge of tidal disruption, the binary system composed of the central \acs{BH} and the companion is emitting \acs{GW}s which carry energy at a rate approximately given by the quadrupole formula
\beq
\dot{E}_{\text{binary}}= \frac{32}{5}\frac{M_c^2M^3}{R^5}\,,
\eeq
where we assume the companion to be much lighter than the \acs{BH}. On the other hand, the \acs{GW} flux emitted by the cloud-\acs{BH} system scales as~\cite{Yoshino:2013ofa,Brito:2014wla,Brito:2017zvb} 
\beq
\dot{E}_{\text{cloud}}\sim \frac{1}{50}\left(\frac{M_S}{M}\right)^2\left(M\mu\right)^{14} \,,
\eeq
with $M_S$ being the mass of the scalar cloud. Thus, \acs{GW} emission by the binary dominates the signal when
\beq
\frac{M_c}{M_S}\gtrsim \left(\frac{M_S}{M}\right)^5 \left(\frac{5}{2}M\mu\right)^{12}\,,
\eeq
Therefore, in the context of \acs{GW} emission and detection, disruption will not affect our ability to probe the system:
if it was visible via monochromatic emission by the cloud before disruption, it would be seen after disruption as a binary.

Note that tidal disruption of the cloud is a relevant possibility for these systems, since
the cloud is generically not depleted due to mode mixing by the time the system reaches the Roche radius.
In fact, for cloud depletion due to mode mixing to be effective, the system needs to be in a resonant epoch for a long time~\cite{Baumann:2018vus}. This requires a particular combination of the mass ratio and gravitational coupling $M\mu$, which can only be realized in a small region in the possible parameter space (see Figs.~$7$ and $8$ in Ref.~\cite{Baumann:2018vus}). 

\section{Discussion}\label{sec:Discussion_Cloud}

In this chapter, we performed the first numerical study of the impact of a possible companion star or \acs{BH} on the development of such superradiant cloud. We observe transitions to higher overtones and to higher multipoles, which \textit{stretch} and \textit{deform} the cloud. Weak tidal fields slightly deform the cloud, affecting \acs{GW} emission by the system. These changes have not been computed. For tidal fields larger than the threshold of Eq.~\eqref{tide_scaling}, the companion simply breaks the cloud apart. This could potentially occur for \acs{BH} systems such as the one at the center of our galaxy or the Cygnus X-$1$ binary system.

Our results generalize to a number of situations. Although we have discussed only tides acting along the equator, we have performed evolutions for polar tides along the $z-$axis, and found the same phenomenology. This includes overtone excitation and transitions between multipoles and tidal disruption, even if quantitatively different.
Our setup is that of a real scalar field, but the results generalize to complex scalars. 

The phenomena we studied are in similar spirit to the \textit{cloud ionization} discussed recently in Refs.~\cite{Baumann:2021fkf, Baumann:2022pkl, Tomaselli:2023ysb}. In this case, the cloud is depleted by the transition from bounded to unbounded states induced by the orbital motion of the binary. The necessary energy for this transition comes from gravitational binding energy of the binary, so ionization accelerates the inspiral and acts as an effective dynamical friction.


\cleardoublepage
\part{Light-ring and the ringdown}\label{pt:light-ring}
\chapter{One ring to rule them all}\label{ch:LR}

In the previous chapters, we discussed various physical processes involving \acs{BH}s, where we assumed a stationary ``background'', where ``matter probes'' and evolves. These setups are, therefore, particularly apt at providing detailed information on the geometry and underlying theory of gravity. For instance, Earth's gravitational multipole moments can be determined in this way by studying the motion of orbiting satellites~\cite{tapley2004gravity,drinkwater2003goce,ciufolini2012overview}. In astrophysics, accretion flows around \acs{SMBH}s can also be well described as flows on a fixed Kerr geometry. The matter density outside the \acs{BH} is so low that its backreaction can be safely neglected for practical purposes~\cite{Cardoso:2016ryw}. 

Thus, a fixed Kerr geometry suffices to understand and study the physics associated with observations by the Event Horizon Telescope~\cite{Akiyama:2019cqa} or GRAVITY~\cite{Abuter:2020dou}. The appearance of \acs{BH}s is determined by photons reaching far-away observers~\cite{1972ApJ...173L.137C,Luminet:1979nyg,Falcke:1999pj,Cardoso:2019dte,Gralla:2020srx,Cunha:2018acu}. It is therefore no surprise that the separatrix between photons escaping to infinity and those eventually plunging into the \acs{BH} horizon plays an important role in \acs{BH} imaging. 
Photons sent in from large distances with a decreasing impact parameter will be deflected with a larger angle, probing stronger-gravity regions before being scattered to far-away observers. Below a critical impact parameter, they fall onto the \acs{BH}. At the critical impact parameter, the photon circles the \acs{BH} an infinite number of times. These trajectories asymptote to a closed, unstable, circular orbit, known as the light-ring (\acs{LR}). For non-rotating \acs{BH}s, it is located at radius $r_\text{LR}=3M$.

The \acs{LR} is therefore associated with the amount of information that one can gather related to the \acs{BH} geometry, and this does not restrict just to electromagnetic waves. If we give a little ``kick'' to a \acs{BH}, it will relax to stationarity as it vibrates according to some proper frequencies - its \textit{quasinormal modes} (\acs{QNM}s). They are quasinormal because there is a loss of energy to both infinity and the \acs{BH} horizon, leading to an imaginary part in their frequencies. Formally, the \acs{QNM} modes are the eigenfunctions of the Teukolsy equation~\eqref{eq:TeukolskyMaster} (or the Zerilli/Regge-Wheeler for non-rotating \acs{BH}s) with ingoing behavior at the \acs{BH} horizon and outgoing at far-away distances. The Zerilli and Regge-Wheeler equation are actually \textit{isospectral}, i.e. they have the same \acs{QNM} frequencies~\cite{Chandrasekhar:1975zza}. 

In a binary coalescence, the final \acs{BH} remnant approaches stationarity as it \textit{ringdowns} and emits \acs{GW}s that can be described by a superposition of exponentially damped sinusoidals~\cite{Vishveshwara:1970zz, 1971ApJ...170L.105P, Leaver:1985ax, Leaver:1986gd, Kokkotas:1999bd, Berti:2009kk}. Each multipole $(\ell, m)$ is described by a sum over possible overtones (labeled by $n$)
\beq
\psi_{\ell m}(t) = \sum_n A_{\ell m n}\,e^{-i\left[\omega_{\ell m n}\left(t-t_\text{start} \right)+\phi_{\ell m n} \right]} \, , \label{eq:Ringdown}
\eeq
where $\omega_{\ell m n}$ are the \acs{QNM} frequencies, which depend only on the mass $M$ and spin $a$ of the remnant. On the other hand, $t_\text{start}$ is an arbitrary starting time and $A_{\ell m n}$ and $\phi_{\ell m n}$ are, respectively, an amplitude and phase that depend on the source of the perturbation, i.e. the full history of the merger. By extracting the frequencies $\omega_{\ell m n}$ from the waveform we can then estimate the mass and spin of the remnant, and if multiple overtones are detected, perform tests of the no-hair theorem. This program is dubbed \textit{\acs{BH} spectroscopy}~\cite{Detweiler:1980gk,Dreyer:2003bv,Berti:2005ys,LIGO:2016lio}. Its importance was accurately foretold by Detweiler in 1980~\cite{1980ApJ...239..292D, Baibhav:2023clw} 
\begin{displayquote}
\textit{After the advent of \acs{GW} astronomy, the observation of [the
\acs{BH}’s] resonant frequencies might finally provide direct evidence of \acs{BH}s with the same certainty as, say, the {21} cm
line identifies interstellar hydrogen.}
\end{displayquote}

So far, the \acs{LVK} has only confirmed the presence of the fundamental $\ell = m = 2$ mode in its ringdown catalog~\cite{LIGOScientific:2020tif,LIGOScientific:2021sio}. These observations have provided constraints on \acs{BH} charge~\cite{Carullo:2021oxn}, the Bekenstein-Hound bound on the \acs{BH} information emission rate~\cite{Carullo:2021yxh}, and modifications to \acs{GR}~\cite{Maselli:2019mjd, Carullo:2021dui}. Independent analyses have suggested the inclusion of overtones in the ringdown model improves agreement with numerical relativity simulations up to (or even before) the peak amplitude of radiation~\cite{Giesler:2019uxc, Ota:2019bzl, Bhagwat:2019dtm, Dhani:2020nik}. Subsequent works claimed the presence of one overtone in GW$150914$~\cite{Isi:2019aib, Isi:2020tac, Isi:2022mhy, Ma:2023vvr, Finch:2022ynt}. However, including these overtones in the early ringdown appears to be an overfit of the signal which lacks physical significance~\cite{Buonanno:2006ui, London:2014cma, Cotesta:2022pci, Baibhav:2023clw}. More recently, two independent studies found nonlinearities in the early ringdown of numerical relativity waveforms~\cite{Cheung:2022rbm, Mitman:2022qdl}~\footnote{We recommend the interested reader the introduction of Ref.~\cite{Baibhav:2023clw} for a more detailed discussion on this topic}.

Returning to our discussion, what is the connection between the \acs{QNM}s and the \acs{LR} after all? It turns out the \textit{ringdown} can be interpreted in terms of high-frequency waves trapped in unstable orbits at the \acs{LR} that slowly leak out to infinity. The real part of the \acs{QNM}s, $\omega_\text{QNM} = \omega_\text{R} + i \omega_\text{I}$, is determined by the angular velocity at the \acs{LR}, while the imaginary part is related to the instability timescale of the orbit, i.e. how fast particles can escape from the \acs{LR}~\cite{1971ApJ...170L.105P, Ferrari:1984zz,Cardoso:2008bp}. Then, the presence of a non-trivial environment can change the structure of geodesics, including the \acs{LR}, but also interact and excite the waves that are trapped there. In this second part of the thesis, we will explore the latter possibility and draw consequences for the observation of compact objects.

\section{Light-rings: the key to compact objects}\label{sec:LRKey}

Let us again consider generic spherically symmetric spacetimes described by the line element in Eq.~\eqref{eq:SphericalLineElement}. Null geodesics are described by Eq.~\eqref{eq:Geodesic} together with the condition they are null, i.e. $g_{\mu\nu}\frac{dx^\mu}{d\lambda}\frac{dx^\nu}{d\lambda}=0$, where $\lambda$ parametrizes the geodesic. Spherically symmetry allows to restrict movement to the $\theta = \pi/2$ plane, while the other \acs{EOM} are~\cite{Wald:1984rg, chandrasekhar1992mathematical, Cardoso:2008bp}
\beq
\frac{dt}{d\lambda} &=& \frac{E}{A(r)} \, , \\
\frac{d\varphi}{d\lambda} &=& \frac{L}{r^2} \, , \\
\frac{dr}{d\lambda} &=& E \sqrt{1- A(r)\frac{b^2}{r^2}} = \sqrt{V_r(r)} \, , \\
b &=& \frac{L}{E} \, . \label{eq:EOMNullParticle}
\eeq
where $E, \, L$ are , respectively, the ``energy'' and ``angular momentum'' constants of motion, and $b$ is the \textit{impact parameter}. Circular orbits are defined by the condition 
\beq
\frac{dr}{d\lambda} = \frac{d^2r}{d\lambda^2} = 0 \Rightarrow V_r = \frac{dV_r }{dr} = 0 \, , \label{eq:CircularOrbit} 
\eeq
which have as implicit solution
\beq
r_\text{LR} &=& 2\frac{A(r_\text{LR})}{A'(r_\text{LR})} \, , \\
b^2_c &=& \frac{r_\text{LR}^2}{A(r_\text{LR})} \, , \\
\Omega_\text{LR} &=& \frac{d\varphi}{dt} = \sqrt{\frac{A'(r_\text{LR})}{2r}} \frac{\sqrt{A(r_\text{LR})}}{r_\text{LR}} \, .
\eeq
For Schwarzschild they are
\beq
r_\text{LR} &=& 3M \, , \\
b_c &=& 3\sqrt{3}M\, , \label{eq:Criticalb}\\
\Omega_\text{LR} &=& \frac{1}{3\sqrt{3}M}\, \label{eq:LRFreq}.
\eeq
Since we are considering spherically symmetric spacetimes, the \acs{LR} defines more broadly a photonsphere, where high-frequency waves can be trapped, whether they are photons or \acs{GW}s. It is an unstable trapping since any small perturbation $r=r_\text{LR} + \delta$ grows exponentially. Expanding the potential close to the \acs{LR} one finds
\beq
\left(\frac{d \delta}{d\lambda}\right)^2 = V_r(r_\text{LR}) +  V'_r(r_\text{LR}) (r-r_\text{LR}) + \frac{V''_r(r_\text{LR})}{2}(r-r_\text{LR}) + ... 
\eeq
By the definition of circular orbit~\eqref{eq:CircularOrbit} the first two terms vanish and one is left with
\beq
\left(\frac{d \delta}{d\lambda}\right)^2=\frac{\delta^2}{2}V''_r\left(r_\text{LR} \right) \, . 
\eeq
Rewriting it as
\beq
\frac{d\delta/dt}{\delta} = \sqrt{\frac{V''_r\left(r_\text{LR} \right) }{2 (dt/d\lambda)^2} } \, , 
\eeq
we arrive at the solution
\beq
\delta &\sim& \delta_0 \, e^{\lambda_L \, t}  \, , \\
\lambda_L &=& \sqrt{\frac{V''_r\left(r_\text{LR} \right) }{2 (dt/d\lambda)^2} } = \frac{A\left(r_\text{LR} \right)}{E}\sqrt{\frac{V''_r\left(r_\text{LR} \right)  }{2 } } \,.
\eeq
$\lambda_L$ is known as the Lyupanov exponent, and for Schwarzschild $\lambda_L = \Omega_\text{LR}$ . In other words, a null ray slightly displaced off the \acs{LR} will orbit on a timescale $t\sim \log \delta /\lambda_L$, during which the null particle does a number of orbits
\beq
N\sim \frac{\Omega_\text{LR} t}{2\pi}=-\frac{\log{\delta}}{2\pi} \,,
\eeq
close to the \acs{LR}.

Because of the above trapping properties, \acs{LR}s play a crucial role in our understanding of \acs{BH}s. They are, for many purposes, the inner surface probed by high-frequency observations. Ref.~\cite{Cardoso:2008bp} formalized this relation by showing that in this high-frequency limit the \acs{QNM} frequencies of spherically symmetric, asymptotically flat spacetimes are 
\beq
\omega_\text{QNM} = \Omega_\text{LR} \ell - i \left(n + \frac{1}{2}\right)\lambda_L \, ,  \label{eq:EikonalQNM}
\eeq
where $n=0, \,1, \, ...,$ labels the overtone. This correspondence is only valid in the eikonal limit ($\ell \gg 1$) but gives excellent predictions even for low values of $\ell$~\cite{PhysRevD.35.3632} (relative differences below $5~\%$ already for $\ell=4$). 

In this chapter, we will show that the \acs{LR} also dictates the late-time behavior of the luminosity of sources plunging into \acs{BH}s. Such events appear to occur periodically in the vicinities of Sagittarius A*~\cite{Baubock:2020dgq,Abuter:2020fpy}. Similar events were reported in the past for Cygnus X-$1$. In particular, dying pulses from \acs{BH} accretion were discussed in the context of Cygnus X-$1$ years ago~\cite{2001PASP..113..974D,2011arXiv1104.3164D}.
Emitters falling onto \acs{BH}s may also radiate in the \acs{GW} window. These could be, for example, a hierarchical triple system where the \acs{CM} of a small binary is inspiralling onto a \acs{SMBH}~\cite{Cardoso:2021vjq}.

Previous studies on the dynamical appearance of bright sources follow a number of approximations and are restricted to spherically symmetric gravitational collapse~\cite{Novikov:1965sik, 1965SvA.....8..868P, 1968ApJ...151..659A}. Here, we extend those and investigate how a pointlike source that emits \acs{GW}s or electromagnetic waves fades out as it is accreted by a \acs{BH}.

\section{How do bright objects fade out?}

\subsection{An outward-pointing beam}
%
\begin{figure}[t]
\centering
\includegraphics[width=0.9\linewidth]{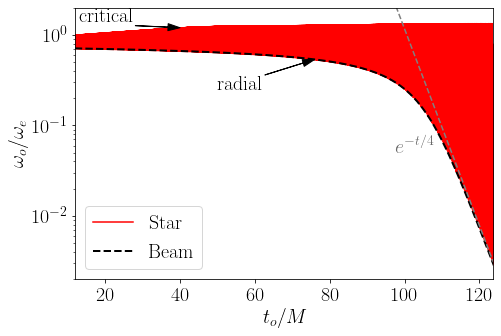}
\caption{
Redshift of two different sources as they plunge radially into a Schwarzschild \acs{BH}, emitting null particles (e.g. photons, gravitons) of fixed proper frequency $\omega_e$.
The source, located in the equatorial plane at $\theta=\pi/2,\,\varphi=0$ begins from rest at infinity, but (for numerical purposes) starts emitting only when it crosses $r=30.65M$. 
\textbf{ Beam:} this source emits only radially outwards. The observer is located at $r_o=100M$, $\theta=\pi/2$,
$\phi=0$, and receives particles whose frequency/energy decreases with time. At late times, the frequency $\omega_o$ measured by the observer decays exponentially as $\omega_o\sim \omega_e e^{-t/(4M)}$, according to our analytic prediction~\eqref{redshift_laser}. 
\textbf{ Isotropic star:} the second source is a pointlike ``star'' emitting isotropically in its local rest frame. At a fixed instant, far-away observers distributed along the sphere at $r_o=100M$ receive a wide range of redshifts. The lower part of the curve is due to radially propagating null particles, whereas the top part of the curve is due to particles with a near critical impact parameter $b_c\approx 3\sqrt{3}M$ that linger close to the \acs{LR}, which can be blueshifted~\cite{Cardoso:2019dte}.
}
\label{fig:Infall_geometricoptics} 
\end{figure}
\begin{figure}[t]
\centering
\includegraphics[width=0.9\linewidth]{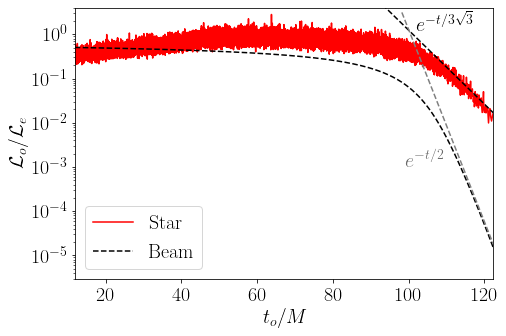}
\caption{
Normalized luminosity ($\mathcal{L} = dE/dt$) of the two different sources discussed in Fig.~\ref{fig:Infall_geometricoptics}. The observed luminosity of the radial beam scales as ${\cal L}\sim e^{-t/(2M)}$ at late times, again in agreement with our prediction. The luminosity of the isotropic star was calculated by ``binning'' null particles in packets of $20$, to avoid large scatters. At late times, the luminosity is dominated by those particles lingering on the \acs{LR}, hence ${\cal L}\sim e^{-t/(3\sqrt{3} M)}$. 
}
\label{fig:Infall_geometricoptics2} 
\end{figure}
\begin{figure}[t]
\centering
\includegraphics[width=0.9\linewidth]{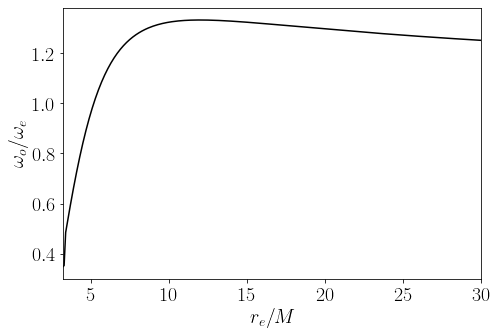}
\caption{The blueshift distribution of photons with near-critical impact parameter, emitted from an object freely-falling onto a \acs{BH}. The blueshift is maximum, $\omega_o=4\omega_e/3$,
at $r_0=12M$, and is unit at $r_e=3\sqrt{3}M$ (see also Ref.\cite{1982Ap&SS..88..307C}).
}
\label{fig:Redshift_absolute} 
\end{figure}

We start by focusing on the geometric optics regime, where one considers that high-frequency waves follow null geodesics on a fixed background geometry, independently of their nature.

Let us imagine a laser pointer shooting a ``beam'' of light outwards as it falls radially onto a Schwarzschild \acs{BH} from rest at infinity. The $4$-velocity of this source and its radial position in terms of the proper time is
\beq
v^\mu_e&=&\frac{dx_e^\mu}{d\tau_e}=\left(	 \frac{1}{ 1- 2M/r_e },-\sqrt{\frac{2M}{r_e}},0,0\right )\,, \\
\label{radial_plunge}
r_e&=&2M\left(-\frac{3\tau_e}{4M}\right)^{2/3}\,,\label{re_tau}
\eeq
and the coordinate time at the source position is given implicitly by
\beq
\frac{dt_e}{d\tau_e} = \frac{1}{1-2M/r_e} \, .
\eeq

Now, consider a photon with $4$-momentum $k^\mu$ such that its proper frequency is $\omega_e = - v_e^\mu k_u$. This photon will intersect the world-line of an observer with $4$-velocity $v_o$ which measures a frequency $\omega_o = - v_o^\mu k_\mu$. For a static observer at large distances and for radial null geodesics followed by the photon
\beq
v_o^\mu &=& \left(1, 0,0,0 \right) \, , \\
k^\mu &=& E\left(\frac{1}{1-2M/r},1,0,0\right) \,,
\eeq
which implies
\beq
\omega_o=\omega_e \left(1-\sqrt{\frac{2M}{r_e}} \right)\,.\label{redshift}
\eeq

We now want to compute the redshift as seen by a distant observer as a function of the coordinate time $t$. We need to take into account that the null particle 
is being emitted by a source that is getting closer to the horizon and which also needs time to reach the observer.
An outward-directed photon obeys
\beq
\frac{dt_\text{travel}}{dr}=\frac{1}{1-2M/r}\,.
\eeq
We can integrate this to find the arrival time of the null particle as measured by a far-away observer
\beq
t_o=t_e+(r_o-r_e)+2M\log\left(\frac{r_o-2M}{r_e-2M} \right)\,.\label{to_te}
\eeq
%

We can obtain the behavior when the source is close to the \acs{BH} horizon by solving
\beq
\frac{dr_e}{dt_e}=-\sqrt{\frac{2M}{r_e}}\left(1-\frac{2M}{r_e}\right)\,,\label{dre_dte}\,  
\eeq
when $r_e \sim 2M$. We find $t_e \sim -2M \log \left(r_e - 2M\right)$, and plugging these in Eq.~\eqref{to_te} we find $r_e-2M\propto e^{-t_o/(4M)}$. Therefore, at late times 

\beq
\frac{\omega_o}{\omega_e}\sim e^{-t_o/(4M)}\,.\label{redshift_laser}
\eeq

The total luminosity $dE_o/dt_o$ can be calculated similarly. At late times $dE_o/dt_o\sim e^{-t_o/(2M)}$.

Figure~\ref{fig:Infall_geometricoptics}-\ref{fig:Infall_geometricoptics2} shows the numerical solution of this problem (black dashed line). An emitter starts falling at $r_i=30.65M$ and sources $20000$ null particles, one every (proper) time interval $\delta \tau_e=4\times 10^{-3}M$. These particles are collected by an observer at $r_o=100M$. 
Our numerical results show that at late times the frequency as measured by far-away observers decreases exponentially as described by Eq.~\eqref{redshift_laser}. 
Note that $\omega_o$ is always \textit{redshifted}. The same applies to the luminosity, shown in Fig.~\ref{fig:Infall_geometricoptics2}. 

\subsection{An isotropically-emitting star}

%
\begin{figure}[t]
\centering
\includegraphics[width=0.9\linewidth]{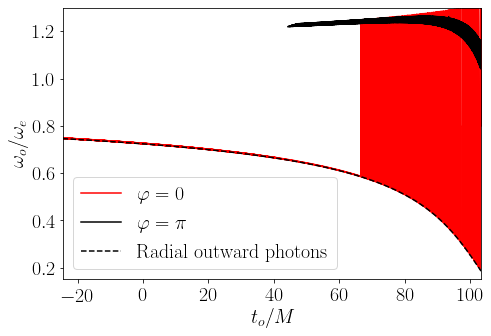}
\caption{The redshift distribution of light emitted by an infalling (isotropic) star as measured by observers at $r=100M$ on the infalling axis. For $\varphi=0$ the star is between the \acs{BH} and the observer, whereas observers at $\varphi=\pi$ only see the star due to gravitational lensing, as the \acs{BH} sits between them and the star. Note the delay with which the $\varphi=\pi$ observer receives the first signal, with respect to $\varphi=0$. Note also that the signal is mostly Doppler blueshifted for $\varphi=\pi$, as the observer sees light emitted from an approaching source.
Some of the details of this figure, in particular the graininess and isolated points, are due to insufficient number of null particles being sent from the star.
}
\label{fig:Star_Angle_Redshift} 
\end{figure}

Most astrophysical sources of radiation are not ``beams''. Even if they can be studied within the geometric optics limit, they should be emitting in different directions. The easiest generalization one can then make is to consider an isotropic pointlike source. Now, there are two new physical structures: the \acs{BH} horizon, which captures null particles, and the \acs{LR}, which traps them in unstable geodesics. 

We consider a similar situation as before, where a body is falling radially from rest at infinity onto a \acs{BH}. However, now it constitutes a luminous ``hot spot'' which emits radiation isotropically \textit{in its rest frame}, where it has total luminosity ${\cal L}_e$. To compute the total luminosity ${\cal L}_o$ measured by a stationary far-away observer as a function of time, we need to follow each particle emitted. 

Spherical symmetry allows us to focus on the emission on the equatorial plane without loss of generality. The conserved quantities $E$ and $L$ characterizing each emitted particle should be related to observables in the local freely falling frame, where they have energy $\omega_e$ and are emitted at an angle $\alpha$ with respect to the radial direction. In Appendix~\ref{app:LR}, we outline how to obtain the following relations 
\beq
\omega_o&=&\omega_e\left(1+\sqrt{\frac{2M}{r}}\cos\alpha\right)\,,\label{eq:redshift_star}\\
b&=&r_e\frac{\sin\alpha}{1+\sqrt{\frac{2M}{r}}\cos\alpha}\,.\label{eq_b_alpha}
\eeq
We can now study the infall of an isotropic star by shooting null particles uniformly distributed in $\alpha$ and collecting them at some fixed radius $r_o$. Particles with impact parameter smaller than the critical impact parameter $b_c$~\eqref{eq:Criticalb}
%
%
will fall into the \acs{BH} and are not considered in our calculation. 

Equations \eqref{eq:redshift_star}-\eqref{eq_b_alpha} can be solved for the redshift of particles with critical impact parameter
\beq
\frac{\omega_o}{\omega_e}=\frac{r_e^3+\sqrt{2M}\sqrt{r_e^5-b_c^2r_e^2(r_e-2M)}}{2Mb_c^2+r_e^3}\,.
\eeq
This relation is shown in Fig.~\ref{fig:Redshift_absolute} as the star falls. For most of the fall, radiation with near-critical impact parameter is blueshifted for values of around $1.2-1.3$ (in particular, it is larger than $1.2$ for $6.7M<r_e<49M$). The blueshift peaks at $r_e=12M$ and crosses unit at $r_e=b$
\beq
\frac{\omega_o}{\omega_e}\biggr\rvert_\text{ max}&=&\frac{4}{3}\,,\quad r=12M\,,\\
\frac{\omega_o}{\omega_e}&=&1\,,\quad r=3\sqrt{3}M\,,
\eeq
in agreement with previous results~\cite{1982Ap&SS..88..307C}.

We repeated the same numerical analysis as in the collimated beam but now distributing $1600$ particles uniformly in the emission angle $\alpha$. The results are illustrated in red in Figs.~\ref{fig:Infall_geometricoptics}-\ref{fig:Infall_geometricoptics2}. 

The first difference with respect to the ``beam'' is that now radiation reaches far-away observers (distributed along the whole sky) with different redshifts, depending on their propagation history. As can be seen in Fig.~\ref{fig:Infall_geometricoptics}, the most redshift occurs for photons emitted radially outwards (the lower part of the red region that is limited by the dashed-black line corresponding to the collimated ``beam''). On the other hand, some null rays can also be \textit{blueshifted} ($\omega_o/\omega_e > 1$). These occur due to the extreme bending of rays close to or at the \acs{LR}, where rays with near-critical impact parameter can make a U-turn. This example is similar to having a moving source and a mirror, as studied in Ref.~\cite{Cardoso:2019dte}, where a similar blueshift was observed. Note that as the critical impact parameter is approached, the rays linger longer and closer to the \acs{LR}, and consequently take more time to reach the observer. 

Now, we focus on a fixed angular position to understand what a particular observer measures. We selected among all the outgoing photons those that reach the observer with $\cos\varphi>0.99$ (which we label ``$\varphi=0$'') and those with $\cos\varphi<-0.99$ (which we label ``$\varphi=\pi$''). The former sees the \acs{BH} behind the source, while the latter sees the opposite. The corresponding redshift distributions are shown in Fig.~\ref{fig:Star_Angle_Redshift} for observers at $r_o=100M$. At early times, ``$\varphi=0$'' observers see only radially-outward particles, with maximum redshift. At late times, 
particles with near-critical impact parameter circle the \acs{BH} and return to reach this observer, with maximum blueshift. Recall that the source starts infalling at $r_i\sim 30M$. The first blueshifted particles should then arrive at a time $\Delta t_1\sim T_\text{ LR}/2+60M\sim 76M$ after the first outwards particles reach the observer, with $T_\text{ LR}/2$ being the time it takes to circle the \acs{LR} and come back in the opposite direction. On the other hand, an observer on the opposite side of the \acs{BH} would see the first null particles to be always blueshifted, since the observer sees an approaching source, a time $\Delta t_2 \sim 60 M$ after the first signal arrives at the $\varphi=0$ observer. These estimates do not take into account Shapiro time delay, but the estimate $\Delta t_1-\Delta t_2\sim T_\text{ LR}/2\sim 16M$ should be more reliable. All these features are apparent in Fig.~\ref{fig:Star_Angle_Redshift}.

The total luminosity is shown in Fig.~\ref{fig:Infall_geometricoptics2} and follows the same trend. Note that due to the finite number of ``photons''
that we used in our numerical study, the total luminosity is not smooth. The jagged features carry no physical information and are purely a
result of the numerical method used to estimate the luminosity. We opted to ``bin'' $20$ particles at a time, and we have explicitly checked that larger binnings produce smoother luminosity functions, as it should. For realistic sources the true curve is single-valued and smooth. 
At late times our results are consistent with a decay controlled by \acs{LR}
${\cal L}\sim e^{-t_o/(3\sqrt{3}M)}$ (recall the \acs{LR} frequency in Eq.~\eqref{eq:LRFreq}, which as we see below is generic for similar sources.

\subsection{An isotropic body emitting scalar waves} \label{sec:Scalars}

%
\begin{figure}[htb]
\centering
\includegraphics[width=0.9\linewidth]{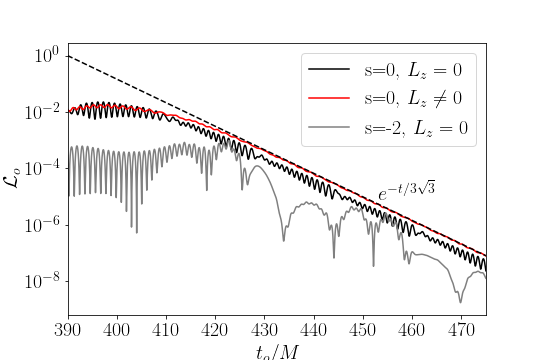}
\caption{
Total luminosity in scalar waves ($s=0$) and \acs{GW}s ($s=-2$) from a source plunging into a Schwarzschild \acs{BH}, and emitting at fixed proper frequency $M \omega_e = 2.5$. The source is located on the equatorial plane at $\theta_e = \pi/2$ and starts from rest at $r_e(t=0)=35M$. We consider both a radial plunge ($L_e=0$)  and one with finite angular momentum ($L_e=3.0M$). For both processes, the luminosity follows the exponential decay at late times dictated by the \acs{LR}, ${\cal L}_o\sim e^{-t/(3\sqrt{3}M)}$, in accordance with the geometric optics limit for the isotropic star in Fig.~\ref{fig:Infall_geometricoptics2}. The different features between scalar and \acs{GW}s are due to the source structure in both scenarios. The low frequency oscillations in the \acs{GW} spectrum come from the plunge of the \acs{CM} of the binary system. The high-frequency content of the signal for both scalar and \acs{GW}s is dominated by frequencies around $M\omega_o \sim 3.0$, blueshifted with respect to $\omega_e$ by a factor $\sim 1.2$, which is consistent with Fig.~\ref{fig:Redshift_absolute}.
}
\label{fig:GWs} 
\end{figure}
\begin{figure}[t]
\centering
\includegraphics[width=0.9\linewidth]{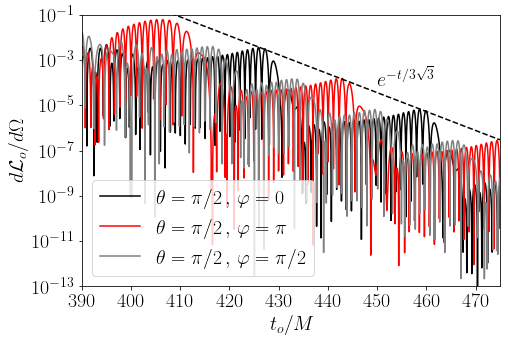}
%
\caption{Energy flux for a scalar source plunging radially into a \acs{BH} (from $r_e(t=0)=35M$ as in Fig.~\ref{fig:GWs}), extracted at specific angular positions on the equator. All signals exhibit the same global exponential decay dictated by the \acs{LR} as seen in Figs.~\ref{fig:Infall_geometricoptics2} and Fig.~\ref{fig:GWs}. Stationary observers now see a periodic structure, whose period may differ for different observers (notice that at $\varphi=\pi/2$ the period is half that at $\varphi=0,\pi$). Once again, the high frequency content of the spectrum corresponds to waves with $M\omega_o \sim 3.0$, in accordance with the blueshift predictions of Fig.~\ref{fig:Redshift_absolute}.
}
\label{fig:Comparison_Angular_Flux} 
\end{figure}

With the geometric-optics limit in control, we generalize the problem by solving for the full dynamics of wave propagation. We first consider a toy-model where a source with \textit{scalar} charge is emitting \textit{scalar} waves. This problem can be modeled by the Klein-Gordon equation for a massless scalar field, which is the $s=0$ case of the Teukolsky equation~\eqref{eq:TeukolskyMaster}. For the source term, we take the trace $T$ of the stress-energy tensor of a pointlike body of mass $m_p$~\eqref{eq:PPTensor} vibrating at constant proper frequency $\omega_e$ (and therefore emitting spherical waves in its rest frame)~\cite{Nambu:2015aea,Nambu:2019sqn}. This body couples to the scalar field through a scalar charge $q$
\beq
\square \Psi = q\, T \, \sin \left(\omega_e \tau_e(t) \right)\,,\label{eq:KG} 
\eeq
which we can set to $1$ in our analysis without loss of generality. Again, spherical symmetry allows us to restrict the motion of the point-particle to the equator. The emitter can have non-zero angular momentum $L_e$, so its \acs{EOM} are
\beq
\frac{dr_e}{dt} &=& -\sqrt{E_e^2 - \left(1-\frac{2M}{r_e}\right)\left(1+ \frac{L_e^2}{r_e^2}\right)}  \, , \label{eq:rCM} \\ 
\frac{d\varphi_e}{dt} &=& \frac{L_e}{E_e}\frac{1-2M/r_e}{r_e^2} \, .
\eeq
We solved this problem using the time-domain code described in Sec.~\ref{sec:Numerics}.

In Fig.~\ref{fig:GWs} we show the total luminosity for this system for a monochromatic source with $M\omega_e=2.5$, with and without angular momentum. This flux of energy is computed through 
\beq
\dot{E}^\infty = \mathcal{L}_o &=& \lim_{r\rightarrow \infty} \int_{S_2} d\Omega \sqrt{-g} T^{tr}_\Phi = \nonumber \\
&=& \lim_{r\rightarrow \infty} \int_{S_2} d\Omega\, r^2 \sin \theta \, \partial_t \Phi \, \partial_r \Phi \, . \label{eq:ScFlux}
\eeq
Even though the source is now emitting radiation whose wavelength is comparable to the \acs{BH} size, the late time behavior is still described by the exponential decay, $\mathcal{L}_o \propto e^{-t_o/(3\sqrt{3}M)}$, independently of whether the body falls with non-zero angular momentum or not.

The luminosity per solid angle at different angular positions is presented in Fig.~\ref{fig:Comparison_Angular_Flux}. The global \acs{LR} decay is the same but we notice the presence of additional structure. In particular, there are periodic oscillations whose period may differ for different observers. Their frequency is a multiple of half of the frequency of the \acs{LR} $M\omega_{\text{LR}}\approx 0.192$ (corresponding to a period $T_{\text{LR}}\approx 32.6M $). Each of these \acs{LR} pulsations is succeeded by a sharp, fast transition, lasting for $\sim5M$, a behavior and timescale that we do not fully understand.

As we might have anticipated, the spectral content is dominated by blueshifted radiation emitted in the past with a near-critical angle, which is absorbed by the \acs{LR} and re-emitted later. Referring to Fig.~\ref{fig:Redshift_absolute}, such radiation is blueshifted to $\omega_o\sim 1.2-1.3 \omega_e$, in this case corresponding to $M\omega_o \sim 3.0-3.1$ during most of the infall. 

\subsection{An infalling binary}\label{sec:GWs}
%
\begin{figure}[t]
\centering
\includegraphics[width=0.9\linewidth]{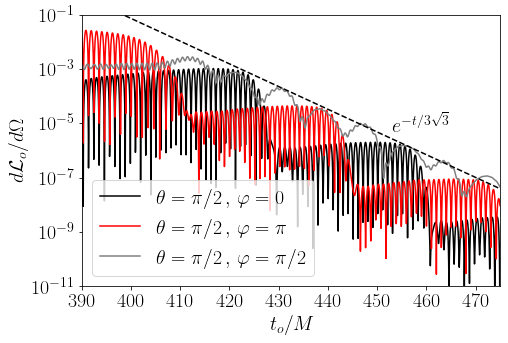}
\caption{Same as Fig.~\ref{fig:Comparison_Angular_Flux}, but for a source emitting high-frequency \acs{GW}s. The source is a binary, and is plunging radially onto a massive \acs{BH}, while emitting \acs{GW}s of proper frequency $M \omega_e = 2.5$. The frequency of the signal measured by far away observers is blueshifted to $M\omega_o\sim 3$.}
\label{fig:Comparison_Angular_Flux_Grav} 
\end{figure}
\begin{figure}[t]
\centering
\begin{tabular}{cc}
\includegraphics[width=5.5cm]{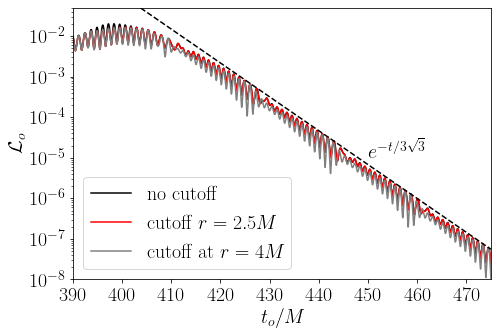}
\includegraphics[width=5.5cm]{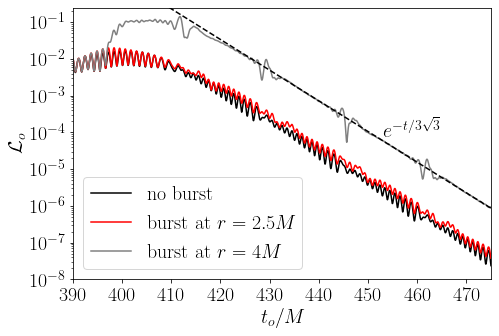}
\end{tabular}
\caption{\textbf{Left panel:} Luminosity in scalar waves for the system studied in Section~\ref{sec:Scalars}. Now, the source is turned off below a certain radius (which we selected to be either $r=2.5M$ or $r=4M$).
When the source is turned off inside or close to the \acs{LR}, the flux is nearly unchanged, as it is controlled by waves emitted in the past and lingering close to the photonsphere.
\textbf{Right panel:} Luminosity for the scalar system studied in Section~\ref{sec:Scalars} but whose source is suddenly increased by a factor of $10$ at the same radii as in the left panel. In flat spacetimes, this would correspond to a luminosity $100$ times higher.
However, since the process takes place close to the \acs{LR}, the luminosity is very weakly affected and has the same global exponential decay. As expected, when the increase in amplitude occurs
deep inside the light ring, the increase in the luminosity is less significant.
}
\label{fig:Comparison_Scalars} 
\end{figure}
\begin{figure}[t]
\centering
\includegraphics[width=0.9\linewidth]{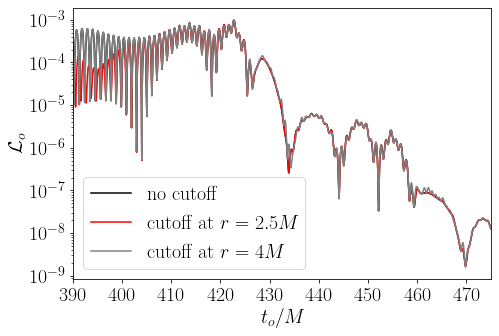}\\
\caption{Luminosity in \acs{GW}s from the system described in Section~\ref{sec:GWs}, for which the binary is shut off below a certain radius, signaling for example a sudden merger of the binary. In line with the findings for scalar waves in Fig.~\ref{fig:Comparison_Scalars}, near-horizon details are irrelevant for the appearance of these objects, and it is the \acs{LR} that controls the late time signal.
}
\label{fig:Comparison_GWs} 
\end{figure}
The final system we will study is a possible astrophysical realization of the previous toy model. We consider a \acs{BHB} whose \acs{CM} is falling onto a massive \acs{BH}. We consider that the binary is composed of two pointlike masses, which can be treated within linear perturbation theory on the background spacetime of the massive \acs{BH}. This system constitutes a hierarchical triple system and it emits \acs{GW}s which can be studied using the Teukolsky equation, where the source term is characterized by two pointlike particles. 

To simplify the analysis, we assume the binary has a very eccentric orbit while its \acs{CM} is radially plunging into the central \acs{BH}
\beq
r^\pm &=& r_\text{CM}(t) \, ,\qquad \theta^\pm = \theta_\text{CM}(t) \, ,  \\
\varphi^\pm &=& \varphi_\text{CM} + \epsilon \sin(\omega_e \tau_e)\, ,
\eeq 
where $\pm$ refers to the two bodies composing the binary and $\epsilon=\epsilon(r_\text{CM})$ defines the axis of the very eccentric ellipse followed by the binary
\beq
\epsilon= \left(1-\frac{2M}{r_\text{CM}} \right) \frac{\delta r }{r_\text{CM}} \, ,
\eeq
where $\delta r$ is the proper length of the binary axis around its \acs{CM}. For the examples we will discuss, we fix $\delta r = 0.1M$, but our qualitative conclusions are independent of this value. The two point-particles enter as sources of the Teukolsky equation for $s=-2$ as described in Sec.~\ref{sec:TeukNum}.

The flux of energy carried by \acs{GW}s to infinity (cf. Eq.~\eqref{eq:FluxGW}) is shown in Fig.~\ref{fig:Comparison_Angular_Flux_Grav} for different angular locations. Even though it is impossible to decouple the contribution of the motion of the \acs{CM} from that of the binary itself, we see again the late time exponential decay~$d{\cal L}_o/d\Omega \sim e^{-t/(3\sqrt{3}M)}$. 

The peculiar nature of gravity is manifest in the low-frequency components in Fig.~\ref{fig:Comparison_Angular_Flux_Grav}. They are contributions to the flux due to the motion of the \acs{CM}, which modulates the high-frequency components of the signal emitted by the small binary. As a consequence, for certain directions, such as $\theta=\pi/2$, $\varphi = 0, \pi$, the high-frequency content dominates the spectrum, whereas for others the signal is controlled by the lower frequencies from the plunge of the \acs{CM}.

\section{Testing Horizons}

Our results strengthen the point of view that \acs{LR}s control how dynamical processes
look to outside observers, while horizons play a secondary role on this. We can test this further by studying how our results change when the near-horizon physics of the compact object is altered, or when we change the properties of the source close to or within the \acs{LR}. We have tested this with two different processes:
\begin{itemize}
\item First, we turn off the source after it reaches a selected radius (e.g. $r=4M,\, 2.5M$ ). This could represent a merging binary before its \acs{CM} plunges on the \acs{BH}. Our results are summarized in Figs.~\ref{fig:Comparison_Scalars}-\ref{fig:Comparison_GWs}. For both types of waves, the spectrum is mostly independent of the cutoff radius if this is located close or inside the \acs{LR}. The late-time decay is still given by the exponential law we observed previously, which reinforces the interpretation that it really describes waves trapped close to the \acs{LR} (which accumulated during the infall), and are slowly leaking out to infinity (and the horizon). This result also indicates that two different compact objects with similar \acs{LR} structures may be hard to distinguish based on observations of matter surrounding it, whether they possess a horizon or not.

\item In the second case the source becomes suddenly brighter, increasing its proper luminosity after it falls within some radius. In Fig.~\ref{fig:Comparison_Scalars} we show results when the proper luminosity is increased by a factor of $100$. As before, the late-time behavior decay is unaltered and the change in the luminosity measured by far-away observers is small when the burst occurs inside the \acs{LR}. 
\end{itemize}

The punchline is \textit{What happens inside the light ring stays inside the light ring}, or more seriously, near horizon details are mostly irrelevant for how matter accreted onto a \acs{BH} appears to distant observers.

\section{Discussion}\label{sec:LR}

In this chapter, we showcased how the properties of \acs{LR}s control the appearance and late-time dynamics of \acs{BH}s and other compact objects~\cite{1965SvA.....8..868P,1968ApJ...151..659A,1972ApJ...173L.137C,Luminet:1979nyg,Falcke:1999pj,Cardoso:2008bp,Yang:2012he,Cardoso:2016rao,Cardoso:2016oxy,Cunha:2018acu,Cardoso:2019rvt,Cardoso:2019dte,Gralla:2020srx,Yang:2021zqy}. The \acs{LR} can be seen as a region where high frequency waves are trapped on timescales~$\sim 3\sqrt{3}M$ or more~\cite{Cardoso:2019rvt}. A \acs{GW} or bright electromagnetic source falling onto a \acs{BH} will ``heat up'' this cavity as it falls. As the source enters the photonsphere, the transfer of energy from the source to the \acs{LR} is maximum. From then onwards, the source gets progressively redshifted away, and the energy leaking from the \acs{LR} dominates emission. Thus, observers see a late-time appearance of infalling stars dominated by the \acs{LR} cooling down process: the signal has a spectral content dominated by frequencies slightly blueshifted with respect to the proper frequency of the source, and a luminosity dying off as ${\cal L}\sim e^{-t/(3\sqrt{3}\,M)}$.

This behavior is best understood within a null-particle approach. Literature on waves around \acs{BH}s usually discusses, instead, a mode analysis where the late-time behavior is dominated by \acs{QNM} ringdown and power-law tails from the backscattering of radiation from curvature~\cite{Leaver:1986gd,Berti:2009kk}. \acs{GW}s are emitted by coherent motion of sources, and usually excite only a few modes. For high-frequency sources, however, a large number of multipoles are excited. The quasinormal frequencies at large mode number $\ell$, are described by Eq.~\eqref{eq:EikonalQNM}~\cite{Berti:2009kk}, and the ringdown amplitude is given by Eq.~\eqref{eq:Ringdown}.
%
%
If we plug the asymptotic expression above in this sum over all the multipoles, we obtain a ringdown stage with a global modulation given by $\Phi \propto e ^{-\Omega_{\text{LR}}\, t/2 }$, which in the Schwarzschild case corresponds exactly to the decay in luminosity we observed ($\mathcal{L} \propto |\Phi|^2$) . In other words, both results (geometric optics and wave propagation) are compatible. 
Finally, late-time polynomial tails are extremely challenging to observe in the presence of these sources, as their amplitude is expected to be many orders of magnitude below the ringdown signal~\cite{Harms:2014dqa}. Consequently, they should only appear at later timescales than the ones we probed and for this reason are not expected to be astrophysically relevant.

The decay timescale is controlled by the \acs{LR}, whose properties depend on the \acs{BH} spin. We studied only non-spinning \acs{BH}s, but geometric-optics approximation can be used to predict that rapidly spinning \acs{BH}s will
show a much larger relaxation timescale, and a breaking of degeneracy with respect to different angular directions~\cite{Cardoso:2008bp,Yang:2012he}. This raises the interesting possibility of determining the \acs{BH} spin from the ratio of amplitudes of different redshifts.

Dying pulses from \acs{BH} accretion were discussed in the context of Cygnus X-$1$, years ago~\cite{2001PASP..113..974D,2011arXiv1104.3164D}.
These works assume that light from such pulses mimics the motion of the source, which as we discussed is not correct.
It is challenging to explain such observations through \acs{LR} properties, since timescales seem to be off by almost an order of magnitude. Nevertheless, these observations show how \acs{LR} relaxation could show up in observations with enough precision. This is relevant for \acs{BH} imaging~\cite{1642263, 2019ApJ...875L...1E, Gralla:2020srx, Johnson:2019ljv}, in particular for the next generation Event Horizon Telescope~\cite{Galison:2023qlm, Johnson:2023ynn}. This collaboration plans to add $10$ new observation points to the current Event Horizon Telescope, which will increase the precision of current images and open the possibility of producing movies monitoring the evolution of matter being accreted by \acs{BH}s, which could be compared with our general results for the luminosity decay.


\chapter{The Elephant the Flea}\label{ch:Elephant}

In the previous chapter, we discussed \acs{BH} spectroscopy, the research program that aims to measure the characteristic frequencies of the late-time ringdown of a binary coalescence, where the waveform is described by a superposition of exponentially damped sinusoidals, and compare them with the theoretically predicted \acs{QNM} frequencies of \acs{BH}s in \acs{GR}. However, for this comparison to make sense, the \acs{QNM} spectrum should be stable, meaning that the \acs{QNM} frequencies should not be very sensitive to possible perturbations in the spacetime. Otherwise, small environmental perturbations could produce large deviations in the \acs{QNM} spectrum and hide hypothetical signatures of new physics. 

It turns out this is exactly the case! Recent computations of the pseudospectrum~\footnote{The pseudospectrum of an operator are the level sets of numbers that are ``close" to the eingenvalues of the operator, where the meaning of close can be put more formally. This notion is particularly useful to study non-self-adjoint operators, as those arising in \acs{BH} spacetimes, where the eingenvectors are not complete and orthogonal~\cite{davies_2007}.} in \acs{GR} confirm that the \acs{QNM} spectrum is unstable under small perturbations~\cite{Jaramillo:2020tuu, Jaramillo:2021tmt, Destounis:2021lum} (see also~\cite{PhysRevD.53.4397} for a preliminary suggestion of this result). For example, the fundamental \acs{QNM} of Schwarzschild, which dominates the ringdown, can have order $\mathcal{O}(1)$ corrections if a ``tiny'' perturbation is added to the gravitational potential (see Fig.~$1$ and~$2$ in Ref.~\cite{Cheung:2021bol}).

However, the instability has been demonstrated in the frequency domain. This only describes the late-time behavior of a binary coalescence. Although results in the frequency domain should translate to the time domain by a Fourier transformation, in practice, this could be challenging to achieve in an experiment, as one would need to observe the \acs{GW} signal for very long times and with precision above what is achievable. 

This is what happens for horizonless exotic (ultra)compact objects (\acs{ECO})~\cite{Cardoso:2019rvt}. To recall, these are \acs{BH} mimickers where the \acs{BH} horizon is substituted by reflecting surfaces, and the exterior geometry and dynamics are usually left unchanged. This leads to a \acs{QNM} spectrum very distinct from that of a \acs{BH}. In particular, \acs{ECO}s develop cavity modes that are trapped between the \acs{LR} and their surface/interior, which decay much slower than the fundamental \acs{QNM} of Schwarzschild. However, by causality, the time-domain response of a \acs{BH} and an \acs{ECO} is the same for the time necessary for a perturbation to travel to the surface of the \acs{ECO}, which should be close to the \acs{BH} horizon limit, and then be reflected back~\cite{Cardoso:2016rao, Mark:2017dnq, Hui:2019aox}. The prompt ringdown is instead controlled by the \acs{LR}~\eqref{eq:EikonalQNM}, in line with our results from the previous chapter.

In this chapter, we are thus interested in understanding how the \acs{QNM} spectrum instability manifests itself in \acs{GW} observations, and if it jeopardizes the \acs{BH} spectroscopy program.

\section{Perturbations of the potential}
%
\begin{figure}[t]
    \centering
    \includegraphics[width=0.9\linewidth]{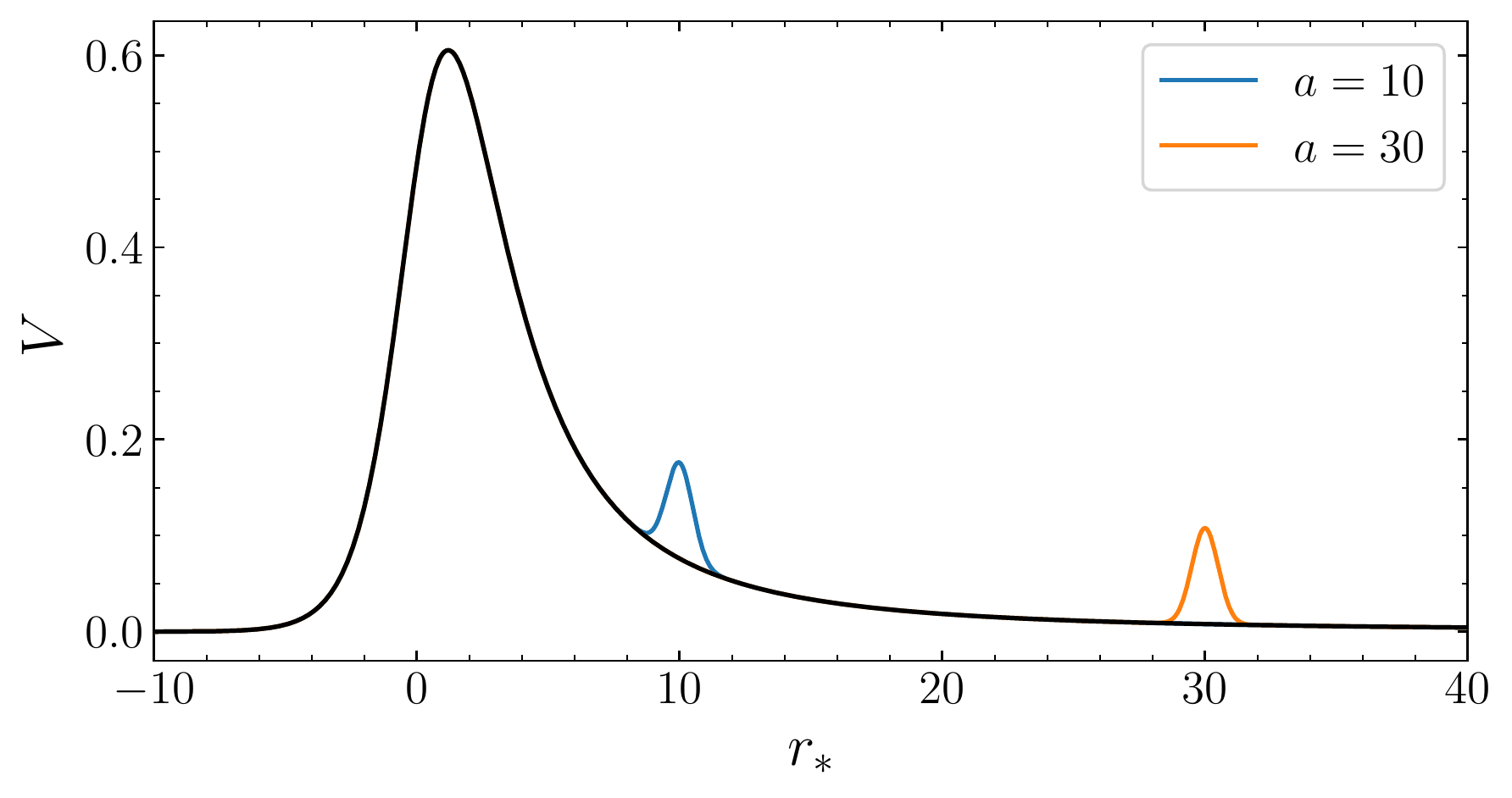}
    \caption{The unperturbed ($\epsilon=0$) and perturbed potentials used in this study. The Regge-Wheeler potential with $\ell = 2$ is shown in black, while two perturbative bumps $\epsilon V_\text{bump}$, with $\epsilon = 0.1$, are shown in blue ($a = 10$) and orange ($a=30$). The unperturbed potential has a peak close to the \acs{LR}.
  }%
    \label{fig:Potential}
\end{figure}
In order to do that, we will study a modified version of the equations that govern (linear) gravitational perturbations around nonrotating \acs{BH}s in \acs{GR}, which were introduced in Sec.~\ref{sec:BHPT}~\cite{Regge:1957td, Zerilli:1970wzz}. There, we saw they have the general homogeneous form 
\beq
- \frac{\partial^2 \Psi}{\partial t^2} + \frac{\partial^2 \Psi}{\partial r_*^2}  - V \, \Psi = 0\,,
\label{eq:MasterEq}
\eeq
where $\Psi$ is the complex ``master" function. $V=V(r)$ is the perturbed effective gravitational potential
\beq
V = V_0 +  \epsilon V_\text{bump} \, .
\label{eq:Potential}
\eeq
$V_0$ denotes the unperturbed potential of a Schwarzschild \acs{BH}, and $\epsilon V_\text{bump} $, with $\epsilon \ll 1$, represents a small perturbation or ``bump'' to it. These perturbations can arise in different physical scenarios, such as modifications to \acs{GR}~\cite{Cardoso:2019mqo,McManus:2019ulj}, environmental matter~\cite{Leung:1997was,Leung:1999iq,Barausse:2014tra, Chung:2021roh}, and non-linearities~\cite{Gleiser:1998rw,Campanelli:1998jv,Zlochower:2003yh,Nakano:2007cj,Ioka:2007ak,Okuzumi:2008ej,Pazos:2010xf,Sberna:2021eui, Cheung:2022rbm}. For concreteness, we will mostly focus on the Regge-Wheeler case with $\ell = 2$. Also, in this chapter, we will exceptionally work in units where $M=1/2$, to be consistent with the work of Ref.~\cite{Cheung:2021bol} on this problem.

Following Ref.~\cite{Cheung:2021bol}, we assume that the bump is localized around some radius $r_* = a$ and that it goes to zero faster than $V_0$ as $r_* \rightarrow \infty$. We will study a Gaussian bump of the form
\beq
V_\text{bump} = \exp\left[-\frac{(r_*-a)^2}{2\sigma^2}\right] \, .\label{bump_width}
\eeq
The specific form of the bump in Eq.~\eqref{bump_width} is chosen as an illustrative example. However, we have considered other perturbations, such as: $\bm{1.}$ a bump which is concave rather than convex, corresponding to negative $\epsilon$; $\bm{2.}$ a bump which decays as $r^{-3}$ at large distances and which is exactly zero for small $r$, mimicking a perturbation due to a thin shell of matter. 
The qualitative behavior of the time-domain signal is the same in all these different cases, so we will not report all the results here. In fact, the only relevant aspect of the perturbation is that it introduces a second small peak. When this is not the case, e.g., when $a$ is very small, the instability is not present~\cite{Cheung:2021bol}. Note also that the isospectrality of the Zerilli and Regge-Wheeler potentials is broken even if we use the {\em same} perturbative bump in both equations~\cite{Cardoso:2019mqo,McManus:2019ulj}, though in realistic astrophysical scenarios the odd- and even-parity perturbative ``bumps'' are not expected to be the same.

To solve the perturbed Regge-Wheeler equation, we again use the numerical framework introduced in Sec.~\eqref{sec:Numerics}. We prescribe as initial data for the perturbation, a Gaussian pulse of the form
\beq
\Psi \Big|_{t=0} = 0 \quad , \quad 
\frac{\partial \Psi}{\partial t} \Big|_{t=0} = e^{-(r_* - 5)^2/2} \, , \label{eq:id_gaussian}
\eeq
but our qualitative conclusions are independent of this choice.

We also want to extract the spectrum at late times, when the time-domain signal is decaying as a linear combination of exponentially damped sinusoids. Their frequencies and damping times can be extracted by fitting the waveform with the $N$-mode template~\footnote{The fitting procedure and frequency-domain analysis was conducted by Mark Ho-Yeuk Cheung.}
\beq
\Psi(t) &=& \text{Re} \sum_{n=0}^{N-1} A_n e^{-i (\omega_n t - \phi_n)} \label{eq:modes}\\
&=& \sum_{n=0}^{N-1} A_n e^{\omega_{n I} t} \cos(\omega_{n R} t - \phi_n)\, ,
\label{eq:template}
\eeq
where the index $n$ labels the different modes we find by fitting, and it does not necessarily coincide with the overtone number. Each mode is characterized by four parameters: an amplitude $A_n$, a phase $\phi_n$, and the real and imaginary parts of the \acs{QNM} frequency $\omega_n = \omega_{nR} + i\omega_{nI}$.
We will find that several \acs{QNM}s could have similar decay times, and hence comparable amplitudes. In this situation, a good fit of the waveform requires a relatively large number of modes $N$. The largest number of modes we will look for is $N = 8$, corresponding to $8 \times 4 = 32$ fitting parameters. 

The eigenfrequencies $\omega_n$ of Eq.~\eqref{eq:modes} can also be computed directly from Eq.~\eqref{eq:MasterEq} with a Laplace transform
\beq\label{eq:MasterFD}
\frac{\partial^2 \Psi}{\partial r_*^2} +\left(\omega^2 - V\right)\Psi =0\,.
\eeq
The \acs{QNM} frequencies $\omega_n$ correspond to the poles of the Green's function of Eq.~\eqref{eq:MasterFD} with boundary conditions of ingoing waves at $r_* = -\infty$ (the \acs{BH} horizon), and outgoing waves at $r_* = +\infty$ (far-away distances).
The frequencies can be found by a shooting method~\cite{Chandrasekhar:1975zza}: starting from each of the two boundaries and numerically integrate inwards or outwards iteratively searching for the values of $\omega$ for which the two solutions match smoothly in an intermediate region.

We have performed this direct-integration analysis using a modification of the \textit{Mathematica} notebook used in Ref.~\cite{Molina:2010fb} and available online~\cite{RDwebsites}.

\section{Stability of the waveform}
%
\begin{figure}[t]
	\centering
		\includegraphics[width=12.0cm]{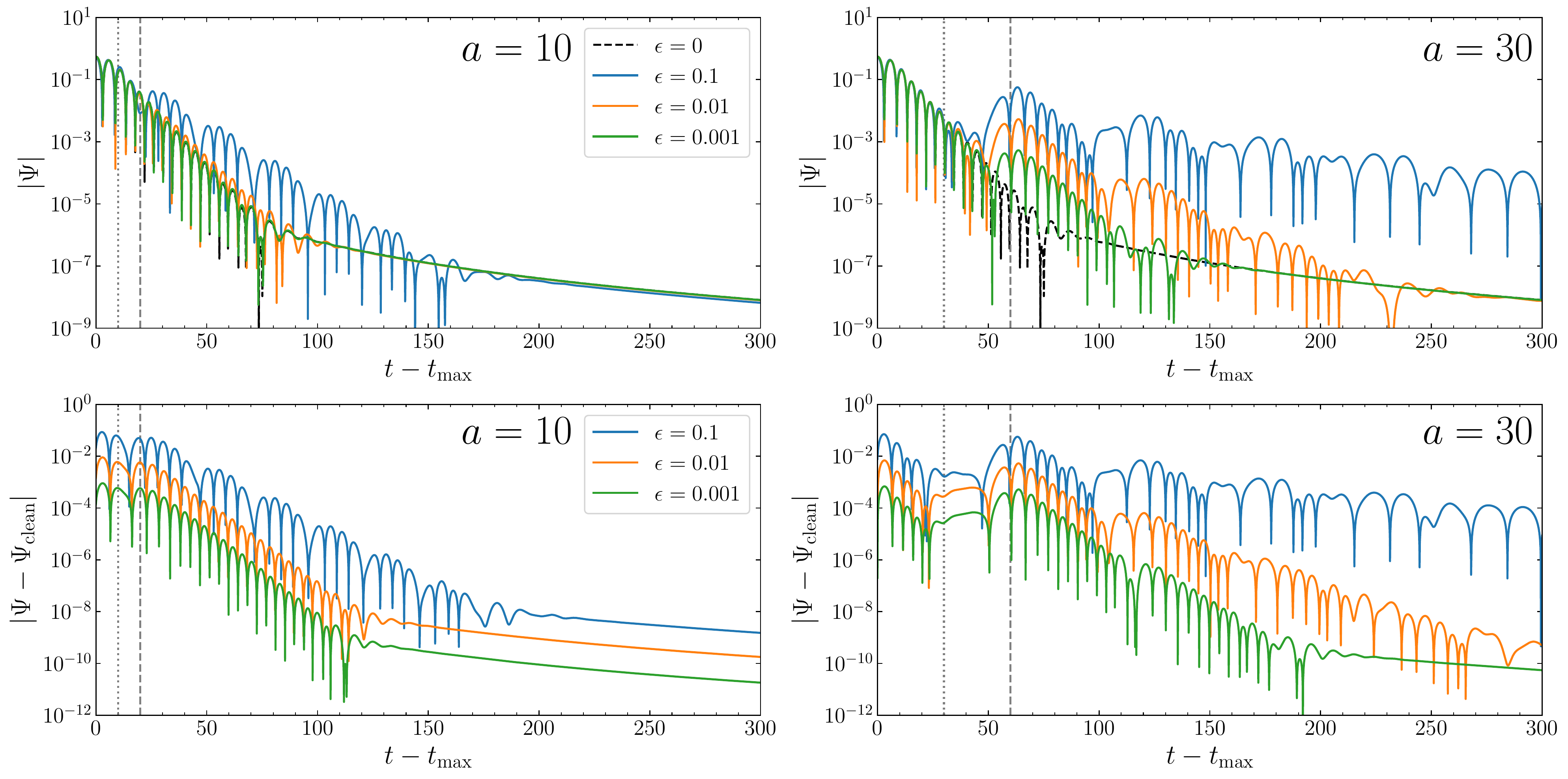} 
	\caption{\textit{Top panels}: absolute value of the waveform arising from the scattering of the Gaussian pulse of Eq.~\eqref{eq:id_gaussian} for ``bumps'' with different amplitudes $\epsilon$, located at two selected distances $a$ from the main peak. The bump width $\sigma$ in Eq.~\eqref{bump_width} is fixed at $\sigma = 0.5$. 
    ``Echoes'' are apparent when the bumps are located at large distances ($a = 30$). The dotted and dashed vertical gray lines correspond to $t - t_\text{max} = a$ and $2a$, and they illustrate how the delay between echoes is related to the size of the ``cavity'' between the two maxima in the perturbed potential.
    \textit{Bottom panels}: absolute value of the difference between the waveforms shown in the top panels and the unperturbed clean waveform without a bump ($\epsilon = 0$).}
	\label{fig:WaveformsEps}
\end{figure}
\begin{figure}[htb]
	\centering
		\includegraphics[width=0.9\linewidth]{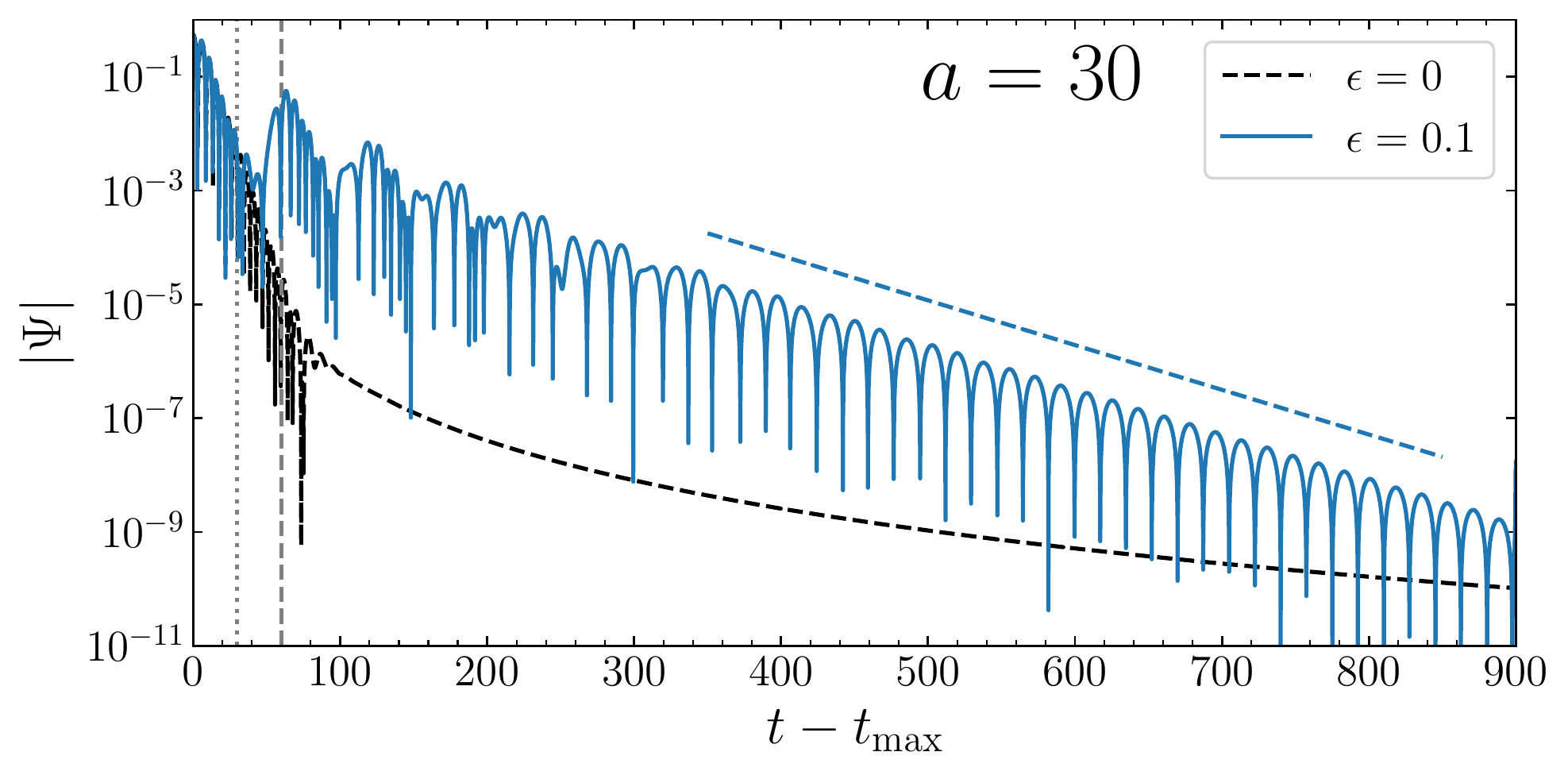} 
	\caption{Waveform for $a=30$ and $\epsilon = 0.1$ over long time windows, where the late-time behavior is dominated by the new fundamental mode \acs{QNM}. The dashed blue line represents its expected decay, which corresponds to the bottom blue cross with smallest $\left| \omega_{I}\right|$ in the top panel of Fig.~\ref{fig:FitCompFrequenciesa30}.}
	\label{fig:LongWaveform}
\end{figure}

We will say that ``destabilization'' occurs when a quantity that characterizes the \acs{BH}'s' response (e.g. the \acs{QNM} frequency or the waveform amplitude) changes by an amount much larger than the magnitude of the perturbation. We will now show that while the fundamental \acs{QNM} mode of Schwarzschild is spectrally unstable, in accordance to Ref.~\cite{Cheung:2021bol}, the change in the time-domain waveform amplitude is of the same order as the size of the perturbation. In this sense, the time-domain waveform itself is stable under perturbations. Furthermore, the {\em observed} \acs{QNM} frequency is also stable because the {\em early part} of the waveform is not significantly affected by the perturbation. 

In Fig.~\ref{fig:WaveformsEps}, we show the waveform resulting from the scattering of Gaussian pulses for different locations $a$ and magnitudes $\epsilon$ of the bump. $t_\text{max}$ is the time at which $\left| \Psi \right|$ reaches its maximum value, and we only show the signal afterward. In the ``clean'' Schwarzschild case ($\epsilon = 0$ in black), we observe the typical exponentially decaying ringdown followed by an expected power-law tail coming from the backscattering of radiation at large distances~\cite{PhysRevD.5.2419, Leaver:1986gd}. As expected, the larger the $\epsilon$, the larger the difference in the waveform with respect to Schwarzschild, even at early times. For times $t-t_\text{max} \lesssim a$, the absolute difference of the waveform with respect to the unperturbed potential scales linearly with $\epsilon$. The \textit{prompt} ringdown signal close to the waveform peak is affected by environmental disturbances but not destabilized in the sense defined above. This is the portion of the ringdown with the most interest to \acs{GW} astronomy, meaning such differences are not expected to be observable with the \acs{SNR}s achievable by today's interferometers~\cite{Barausse:2014tra,Cardoso:2019rvt}.   

After this initial regime, for $t_\text{max} \gtrsim 2a$, we start to observe \textit{echoe}s of the original ringdown due to the leakage of waves being reflected between the peak of the potential and the bump. At late times, the bump can drastically modify the signal's frequency content. At this stage, the waveform is well-described by a superposition of the long-lived \acs{QNM}s of the perturbed potential, as illustrated in Fig.~\ref{fig:LongWaveform}. For smaller $\epsilon$, the late time power-law tail can hide the difference in the late-time behavior with respect to the unperturbed waveform. 

In general, when we add a large bump in the potential, the \acs{QNM}s have a longer damping time, and hence they survive longer before ``diving below the tail''. 
When the bump is located far from the original potential peak, we first observe lower amplitude echoes of the original pulse, which eventually give way to a different ringdown signal. For large $a$, there is a clear separation of timescales between the ringdown pulse produced at the \acs{LR} and the light travel time characterizing the ``cavity'' between the \acs{LR} and the bump. Thus, we have a pulse bouncing back and forth within the cavity and gradually losing its high-frequency component, which tunnels out more easily. This produces a sequence of echoes repeating at a characteristic frequency defined by the cavity size and damped on a timescale defined by the transmission coefficient of the small peak, as shown in Fig.~\ref{fig:WaveformsEps} (see also Refs.~\cite{Cardoso:2016rao,Cardoso:2016oxy,Cardoso:2017cqb,Cardoso:2019rvt} for similar behavior when the bump is arbitrarily close to the horizon). These two scales determine the \acs{QNM} spectrum of the bumpy potential, which can be nonperturbatively different from the $\epsilon=0$ case. A simple rule of thumb for the echoes to be visible is that the prompt ringdown lifetime $\sim 9\sqrt{3}M$ (allowing for three e-folding times) should be smaller than the travel time within the cavity $\sim 2a$~\cite{Cardoso:2017cqb,Cardoso:2019rvt}, and therefore we should require $a\gtrsim 4$ (in units where $M=1/2$). Note that, however, the amplitude of the echoes and the induced \acs{QNM} ringing is proportional to $\epsilon$.


The prompt ringdown is excited mainly at the peak of the potential, which broadly coincides with location of the \acs{LR}. If the bump is placed close to the peak, it will change its shape and consequently the frequency content of the prompt ringdown. When the bump is placed farther, the spectrum still changes because the \acs{QNM}s are sensitive to the entire potential. However, the wave train excited at the \acs{LR} should be similar to the one in the case of the unperturbed potential. As it meets the bump, part of it will be reflected and the other will tunnel out to far-away distances. The reflection coefficient is $\mathcal{O}(\epsilon)$ for any frequency, and consequently the change in the prompt ringdown also scales with this factor. 

In the prompt ringdown, the relative difference in the waveform $\left|\Psi - \Psi_\text{clean} \right| / \left| \Psi \right|$ scales linearly with $\epsilon$, while in the echo-dominated regime the difference is larger. The reason for this is the following: the original ringdown signal decays as $e^{-\omega_I t}$. Each reflection of the waves in the cavity reduces their order of magnitude by $\sim\epsilon$. On the other hand, each back-and-forth bounces inside the cavity occurs on timescales of $t_\text{bounce}\sim 2a$. This means the amplitude of the \textit{n}-th echo will be larger than the ringdown by a factor of $\left(\epsilon / e^{-\omega_I t_\text{bounce}}\right)^n$, or $\epsilon^n e^{n2a \left|\omega_I\right|}$. When the power-law tail starts dominating, the modification returns to order $\epsilon$ because the tail can be formally seen as a ``direct'' zero-frequency signal.

\section{Extraction of QNM frequencies}

%
\begin{figure}[t]
\centering
	\begin{tabular}{c}
		\includegraphics[width=0.9\linewidth]{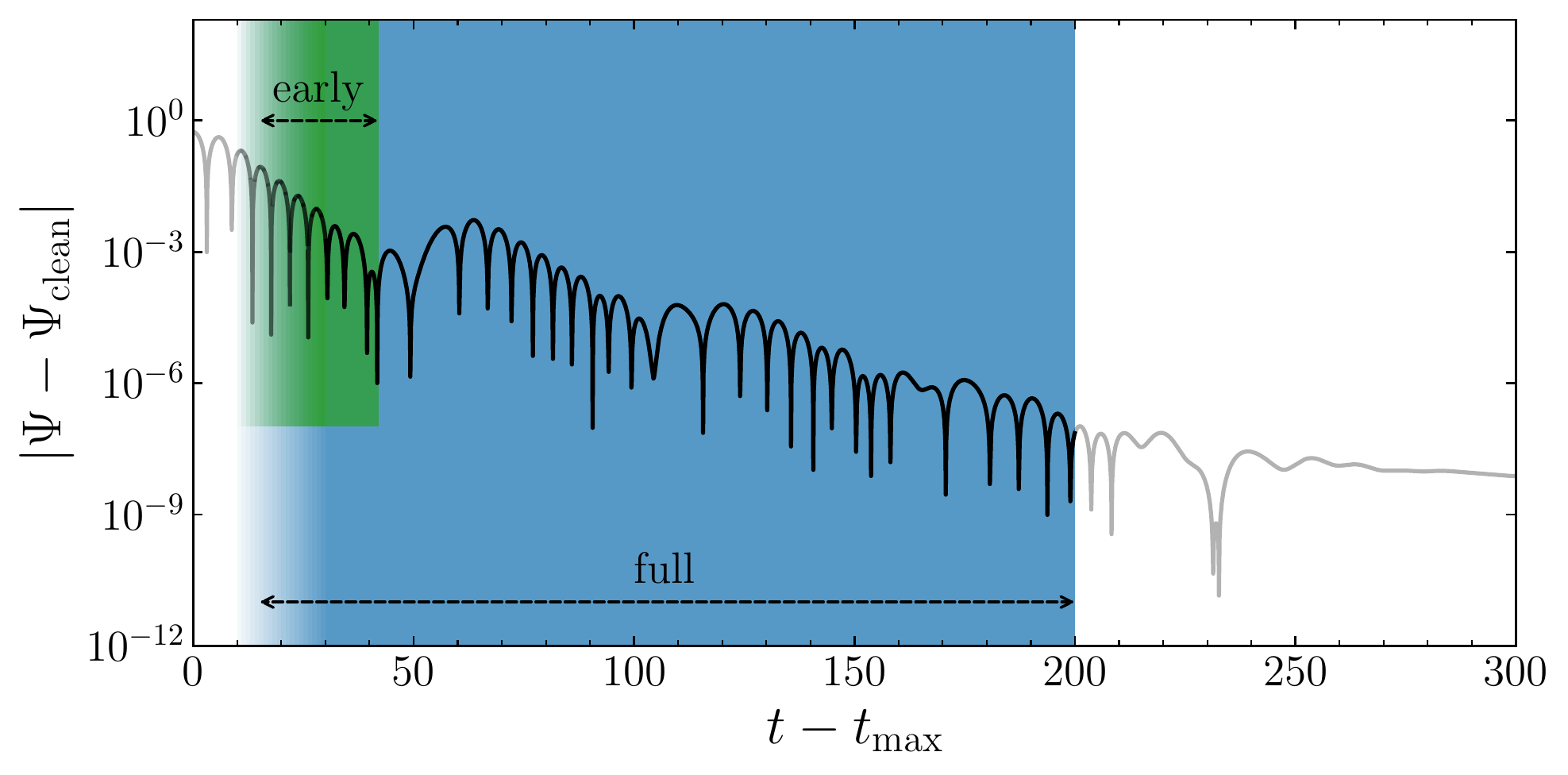} 
	\end{tabular}
	\caption{The portion of the waveform used for the damped-sinusoid fitting in the two different regimes of interest. For completeness, this waveform corresponds to the case $a=30, \, \epsilon =0.01$, but the same procedure applies to other examples.}
	\label{fig:FitRange}
\end{figure}
\begin{figure}[t]
\centering
	\begin{tabular}{c}
		\includegraphics[width=0.9\linewidth]{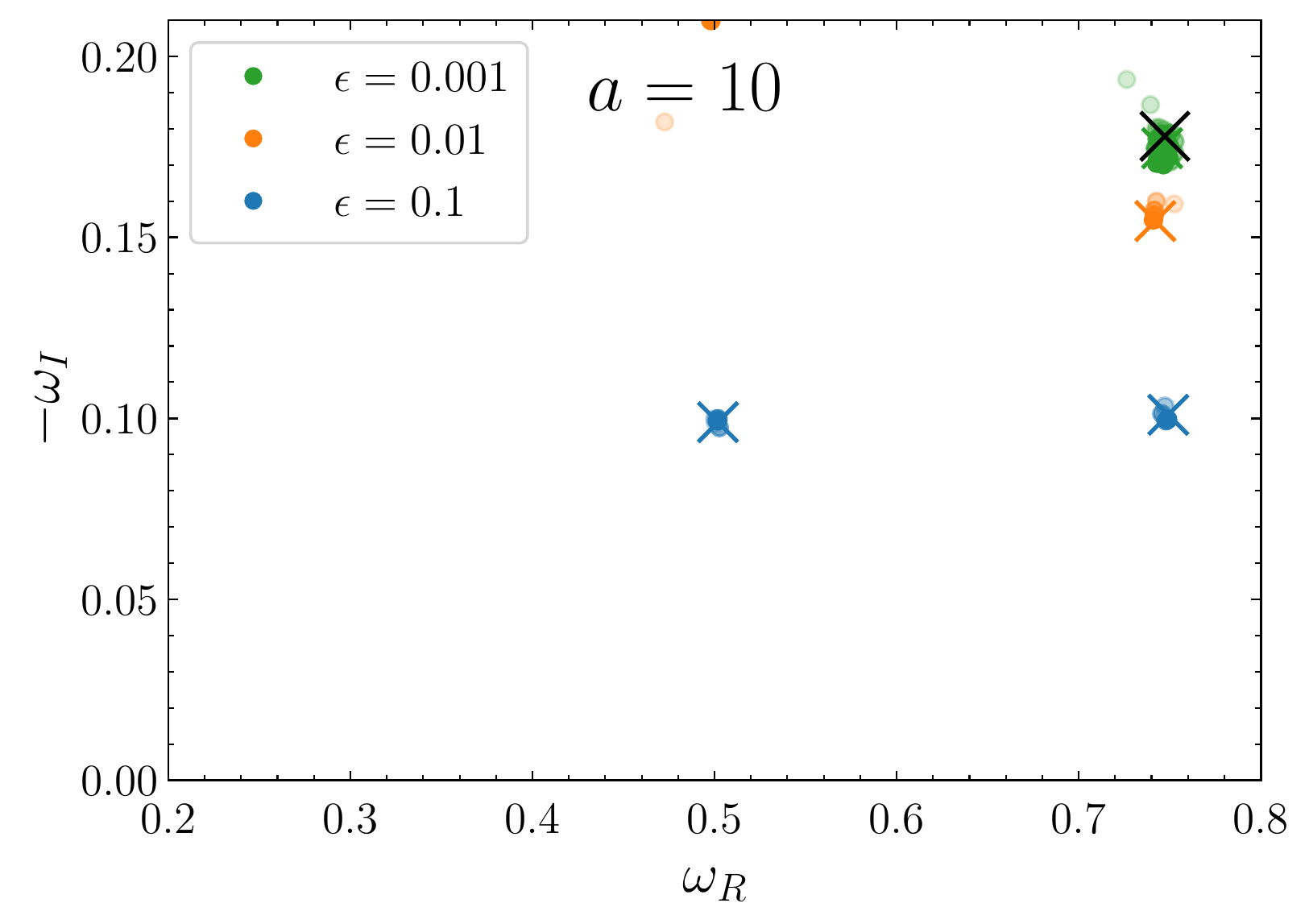} 
	\end{tabular}
	\caption{Comparison between the \acs{QNM} frequencies computed with the shooting method in the frequency domain (crosses) and those extracted by fitting the full time domain waveform (dots). The black cross is the unperturbed clean fundamental \acs{QNM} of Schwarzschild, with $M\omega=0.374 - 0.089 i$. The starting times used are $t-t_\text{max} = 10, \, 15, \, 20, \, 25, \, 30M$. 
	}
	\label{fig:FitCompFrequenciesa10}
\end{figure}
\begin{figure}[ht]
\centering
		\includegraphics[width=0.775\linewidth]{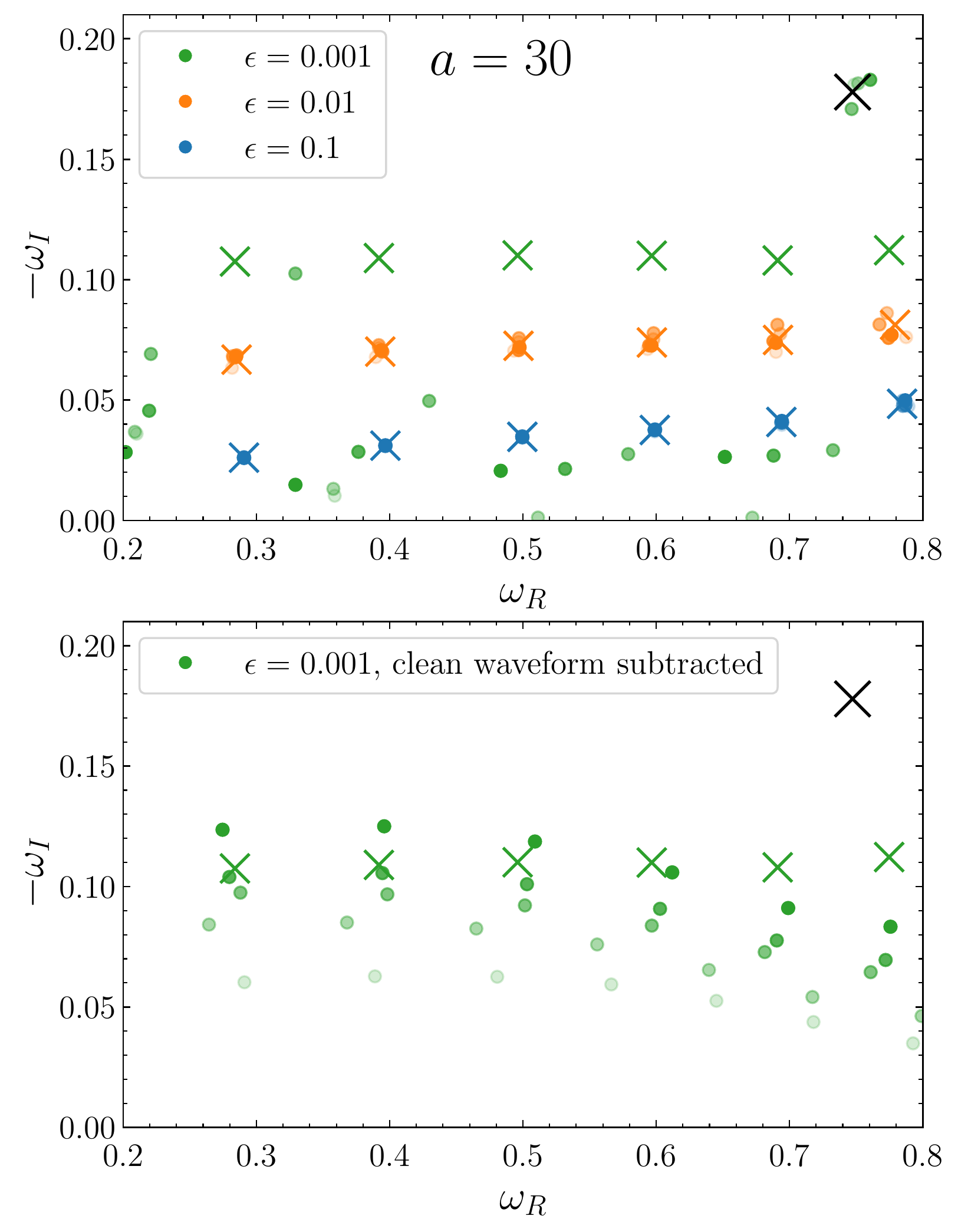}
	\caption{
	\textit{Top panel}: same as in Fig.~\ref{fig:FitCompFrequenciesa10} but with $a=30$. For $\epsilon = 0.001$, the full time-domain fits can only confidently detect a mode close to the fundamental mode of Schwarzschild, because the power-law tail dominates the signal before it transitions to the new set of \acs{QNM}s. \textit{Bottom panel}: The same analysis as in the \textit{top panel} for $\epsilon = 0.001$ but subtracting the unperturbed clean waveform to the waveform before performing the fit. The frequencies extracted do not converge exactly to the \acs{QNM}s, but the structure is more similar, in particular for the real part.}
	\label{fig:FitCompFrequenciesa30}
\end{figure}

We now move to the core problem of \acs{BH} spectroscopy, which is the comparison between the \acs{QNM} frequency theoretically predicted and those extracted by fitting the waveform with damped sinusoids. 

Regarding the fitting of time-domain waveforms, we wish to answer two separate questions:
\begin{enumerate}
\item How the spectral instability affects the prompt ringdown radiation emitted in binary coalescences, which is the louder portion of the signal and consequently more easily detectable by \acs{GW} interferometers.

\item If the full waveform, in particular the late-time portion, is well described by the destabilized \acs{QNM} spectrum, including the long-lived cavity trapped modes.
\end{enumerate}

In Fig.~\ref{fig:FitRange} we highlight the portions of the signal used in the investigation of both problems. In all cases, we discard times $t-t_\text{max}\lesssim 5$, where there is contamination from the direct signal coming from the initial data. For the full signal (shaded in blue), we also discard the portion of the waveform dominated by the power-law tail. For the prompt ringdown analysis (shaded in green), we only consider the portion of the signal before the appearance of the first echo ($t-t_\text{max} \lesssim a$). In both cases the starting time of the fit is varied to ensure convergence of the frequencies obtained.

\subsection{The full time-domain signal}

We start by addressing the second problem, where the full signal is used (shaded in blue in Fig.~\ref{fig:FitRange}).

In Fig.~\ref{fig:FitCompFrequenciesa10}, we present results for a bump located close to the peak of the unperturbed potential ($a=10$) and in Fig.~\ref{fig:FitCompFrequenciesa30} another case where the bump is farther away ($a=30$). The frequencies recovered from the fitting procedure using different starting times are shown as dots with different shades, where the darker the dot, the later the starting time. The crosses correspond to the \acs{QNM} frequencies computed with the shooting method in the frequency domain, with the black cross representing the fundamental \acs{QNM} of the unperturbed potential ($\epsilon = 0$). Finally, different colors refer to different values of $\epsilon$ used. 

The fitted frequencies for the bump closer to the potential peak are in very good agreement with the ones predicted by the frequency-domain computations for all values of $\epsilon$ presented. The minor discrepancies can be attributed to numerical error and contamination by the initial data and the power-law tail. 

For the further bump, the destabilization of the spectrum is more noticeable. Since in this case the new \acs{QNM}s are longer lived, after one echo the waveform transitions to a combination of new \acs{QNM}s from the cavity in the effective potential until their amplitude becomes so small that they are masked by the late-time power-law tail. For $\epsilon = 0.1$ it can even transition to a clean exponential decay controlled by the new fundamental mode, as illustrated in Fig.~\ref{fig:LongWaveform}. Because of this, we are able to extract multiple slowly decaying modes from the fit as long as $\epsilon \gtrsim 0.01$.

The time-domain fit is more difficult when $\epsilon \lesssim 0.001$. Our fits can only confidently detect a mode close to the clean fundamental mode of the unperturbed potential, because the \acs{QNM}s have a very short decay time for such small $\epsilon$, and the waveform does not have time to transition to the ``new'' trapped \acs{QNM} spectrum before decaying below the power-law tail. Therefore, the fitting algorithm can only pick up the clean mode, which is excited promptly at the \acs{LR} and therefore observable at early times. To remove the contribution of the tail, in the bottom panel of Fig.~\ref{fig:FitCompFrequenciesa30} we subtract the clean ($\epsilon=0$) waveform from the signal and repeat the fit using the green curve in the bottom-right panel of Fig.~\ref{fig:WaveformsEps}, which as we can see contains more \acs{QNM} oscillation periods that were previously hidden below the late-time power-law tail. The fitted modes do not converge as in the cases with $\epsilon \geq 0.01$, but their general structure is now in good agreement with the trapped \acs{QNM} spectrum computed in the frequency domain.

\subsection{The prompt ringdown}
%
\begin{figure}[t]
\centering
	\begin{tabular}{c}
		\includegraphics[width=0.95\linewidth]{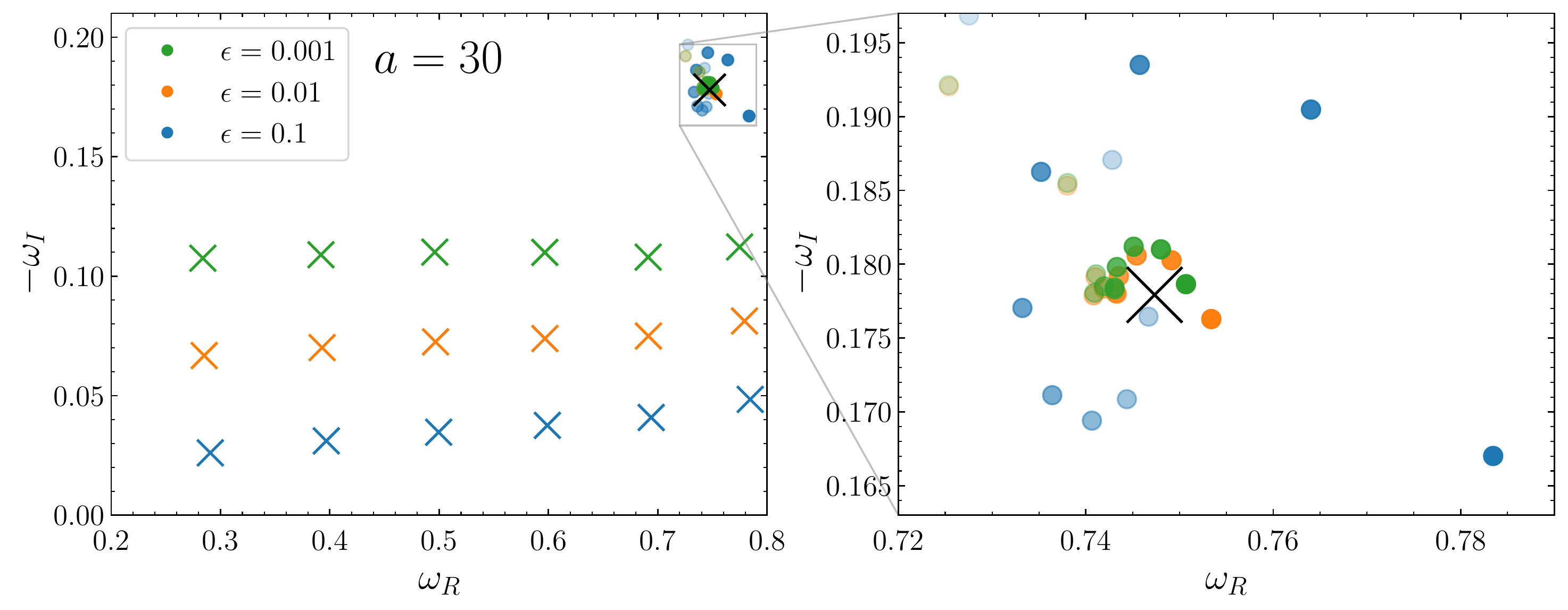} 
	\end{tabular}
	\caption{Same as Fig.~\ref{fig:FitCompFrequenciesa30}, but we only fit the first train of the initial ringdown without echoes. The starting times are $t - t_\text{max} = 10, \,11,\, 12,\, \dots,\, 20M$. All the dots obtained using the fitting method now cluster around the clean fundamental mode of the unperturbed potential. The zoom-in in the right panel shows that a perturbation of order $\epsilon$ can induce systematic errors (approximately of order $\epsilon$) in the measurement of the fundamental mode's frequency and damping time.}
	\label{fig:FitCompPrompt}
{}	
\end{figure}

As mentioned before, in a real detector we do not have access to the full signal due to noise. Also, for astrophysical systems, the perturbation bump should be rather small. The presence of matter typically introduces corrections on the potential of amplitude $\epsilon V_\text{bump} \sim \rho$, where $\rho$ is the matter density~\cite{Kokkotas:1999bd,Cardoso:2021wlq,Cheung:2021bol}. In units of the \acs{BH} mass
\beq
\rho M^2 = 1.6 \times 10^{-18} \frac{\rho}{\rho_\text{water}} \frac{M^2}{M^2_\odot}\, , 
\eeq
so $\epsilon$ will be very small for most realistic scenarios. Consequently, with current \acs{SNR}s we should only have access to the prompt ringdown.

Considering this, in Fig.~\ref{fig:FitCompPrompt} we repeated the analysis of the previous section but restricting it to \textit{prompt} ringdown, i.e. the portion of the signal before the appearance of the first echo ($t-t_\text{max} \lesssim a$, shaded in green in Fig.~\ref{fig:FitRange}). We find it is well fitted by a \textit{single} mode, whose frequency appears to converge to the frequency of the unperturbed fundamental \acs{QNM}, instead of the rich \acs{QNM} spectrum recovered using the full signal. We then conclude that as the waveform itself, the \textit{observed} \acs{QNM} frequency associated with the prompt ringdown is not destabilized.

\section{Discussion}\label{sec:Discussion_Elephant}

In this chapter, we studied how the \acs{QNM} spectral instability manifests in time-domain waveforms. Formally, our analysis should be equivalent to frequency-domain results if we could observe the full \acs{GW} signal, which in practice is not achievable due to noise. We concluded that it is necessary to include the late-time portion of the signal in the fitting procedure to recover the ``correct'' destabilized \acs{QNM} spectrum predicted in the frequency domain. Even that might not be sufficient if the power-law tail dominates over the amplitude of the perturbations at late times. Thus, all calculations of \acs{QNM} frequencies using modified
potentials or modified boundary conditions should be complemented by time-domain studies to verify that these modifications affect the prompt ringdown. In the problem we studied, if we only analyze this early part of the signal, which is the portion relevant for \acs{GW} astronomy, we instead detect the fundamental \acs{QNM} with corrections of the same order as those introduced in the potential. Therefore,  even though the \acs{QNM} spectrum of \acs{BH}s is unstable, the \acs{BH} spectroscopy program is not compromised. Nonetheless, an immediate question that arises is what is the needed SNR to observe structure beyond the prompt ringdown. In principle, this can be computed using specific astrophysical models, such as the one studied in Chapter~\ref{ch:GBH} for a \acs{BH} immersed in a galactic \acs{DM} halo. We leave this exploration for future work.

In addition to that, it is important to note that we restricted the analysis to the spectral instability of the \textit{fundamental} \acs{QNM}. Overtone destabilization tends to be more dramatic for the type of short-wavelength perturbations used in our study~\cite{Jaramillo:2021tmt}, and these modes affect the signal at early times. Yet, the extraction of overtones from waveforms is highly nontrivial, even at the linear level, and is a subject of current research and debate~\cite{Isi:2021iql, Cotesta:2022pci, Baibhav:2023clw}. Any future investigation of the impact of overtone spectral instability in \acs{GW} signals needs to be complemented by a better understanding of how to fit them accurately.


\chapter{Resonances in black hole mimickers}\label{ch:Greenhouse}

In our previous discussions, we have been assuming that very compact dark objects are \acs{BH}s described by \acs{GR}. However, we have already seen how the absence a horizon can dramatically change the dynamics of compact objects. Horizonless exotic compact objects (\acs{ECO}s) have a \acs{QNM} spectrum very different from that of \acs{BH}s. Even though the prompt ringdown should be the same, these \acs{BH} mimickers exhibit late-time echoes in the ringdown similar to the ones we observed in the previous chapter~\cite{Cardoso:2016rao,Cardoso:2016oxy,LIGOScientific:2020tif,Cardoso:2019rvt, Oshita:2018fqu, Oshita:2020dox, Oshita:2020abc}. Echoes of \acs{ECO}s have already been searched in \acs{GW} data with conflicting conclusions~\cite{Abedi:2020sgg, Nielsen:2018lkf, Uchikata:2019frs, Abedi:2016hgu, Abedi:2018npz, LIGOScientific:2020tif,LIGOScientific:2021sio}. In Chapter~\ref{ch:TLNs} we also mentioned that \acs{ECO}s have nonzero \acs{TLN}s, while \acs{BH}s in \acs{GR} have vanishing \acs{TLN}s. Finally, since they lack a horizon, \acs{ECO}s absorb radiation very differently from \acs{BH}s, which could impact the inspiral of a binary~\cite{Maselli:2017cmm,Datta:2019epe}. 

Additionally, it has been observed that massive bodies on stable orbits around \acs{ECO}s could resonantly excite the small-frequency, long-lived \acs{QNM}s that characterize them~\cite{Cardoso:2019nis,Maggio:2021uge,Fransen:2020prl,Fang:2021iyf,Sago:2021iku, Heidmann:2023ojf}. This cannot occur in \acs{BH}s because their proper modes are localized close to the \acs{LR}, and their frequency is always greater than that of stable orbits. 

However, previous analyses of detectability of resonance-crossing in \acs{BH} mimickers were conducted in the frequency domain~\cite{Cardoso:2019nis,Maggio:2021uge,Fransen:2020prl,Fang:2021iyf,Sago:2021iku}, assuming that the field is stationary and superposing an adiabatic evolution to evolve the binary, as driven by \acs{GW} emission. However, as we have seen in the previous chapter, the frequency and time domain results only coincide when the physical process occurs for an ``infinite'' time. Here, infinite refers to a time much longer than the relevant timescales of the problem. For the resonant excitation of \acs{ECO}s in a binary, the time it takes for the resonance to develop compete with the inspiral timescale. The resonance does not have time to grow if the latter is much shorter than the former .

In this chapter, we will complement the previous studies on inspirals around \acs{BH} mimickers but with a time domain analysis using our numerical framework.

\section{Toy Model: A Constant-Density Star}

For the background spacetime of our \acs{BH} mimicker we use constant-density stars. They are spherically symmetric and, therefore, their interior is described by the line element in Eq.~\eqref{eq:SphericalLineElement}. The metric functions~\cite{Shapiro:1983du} are
\beq
A&=&\left(\frac{3}{2}\left(1-\frac{2M}{R}\right)^{1/2}-\frac{1}{2}\left(1-\frac{2Mr^2}{R^3}\right)^{1/2}\right)^2\,,\\
B&=&1-\frac{8\pi\rho}{3}r^2\,,
\eeq
where $R$ is the star's radius, $M$ is its mass, and $\rho=3M/(4\pi R^3)$ is its density. Outside the star, Birkhoff's theorem asserts that the geometry is Schwarzschild, described by the line element in Eq.~\eqref{eq:SchwarzschildLineElement}.

The geometry above only describes ``realistic" stars when $R>9M/4$. Otherwise, the pressure diverges somewhere inside the star. Above some compactness, the geometry admits two \acs{LR}s at the roots of $2A=rA'$~\cite{Cardoso:2008bp}~\footnote{If needed revisit the discussion on Sec.~\ref{sec:LRKey} where we discussed null circular orbits}. When $R<3M$, they are located at
\beq
r_\text{LR}^+&=& 3M\,,\\
r_\text{LR}^-&=&\frac{R\sqrt{4R^2-9MR}}{\sqrt{9MR-18M^2}}\,,
\eeq
where $r_\text{LR}^+$ coincides with the unstable \acs{LR} in Schwarzschild, and the second solution corresponds to a stable \acs{LR} located inside the star.

%
%

We can also compute the transit time between the unstable \acs{LR} and the center of the star, which dictates the period of trapped oscillations, and thus of the ensuing echoes in the waveform~\cite{Cardoso:2016rao,Cardoso:2016oxy,Cardoso:2019rvt}. It turns out that this time, $T_\text{echo}$, is also approximately
$T_\text{echo}\approx 2\pi/\Omega^+_\text{ LR}$, although this may be a fortuitous aspect of very compact constant density stars~\cite{Pani:2018flj}.

For definiteness, we will primarily focus on a configuration with $R=2.26 M$ and compare it with less compact geometries. This choice is close to the maximum possible compactness for this equation of state (the so-called Buchdahl limit), and the spacetime has two photonspheres, sufficiently compact to mimic some aspects of \acs{BH}s.

We will perturb the constant density star with the toy model already used in Chapter~\ref{sec:Scalars}, the massless scalar field being sourced by the trace of a point-particle coupled to the scalar field.

\section{The build-up time of black hole mimickers\label{sec:build-up}}

\subsection{A scattering approach\label{subsec:greenhouse}}

%
\begin{figure}[t]
\centering
\includegraphics[width=0.925\columnwidth]{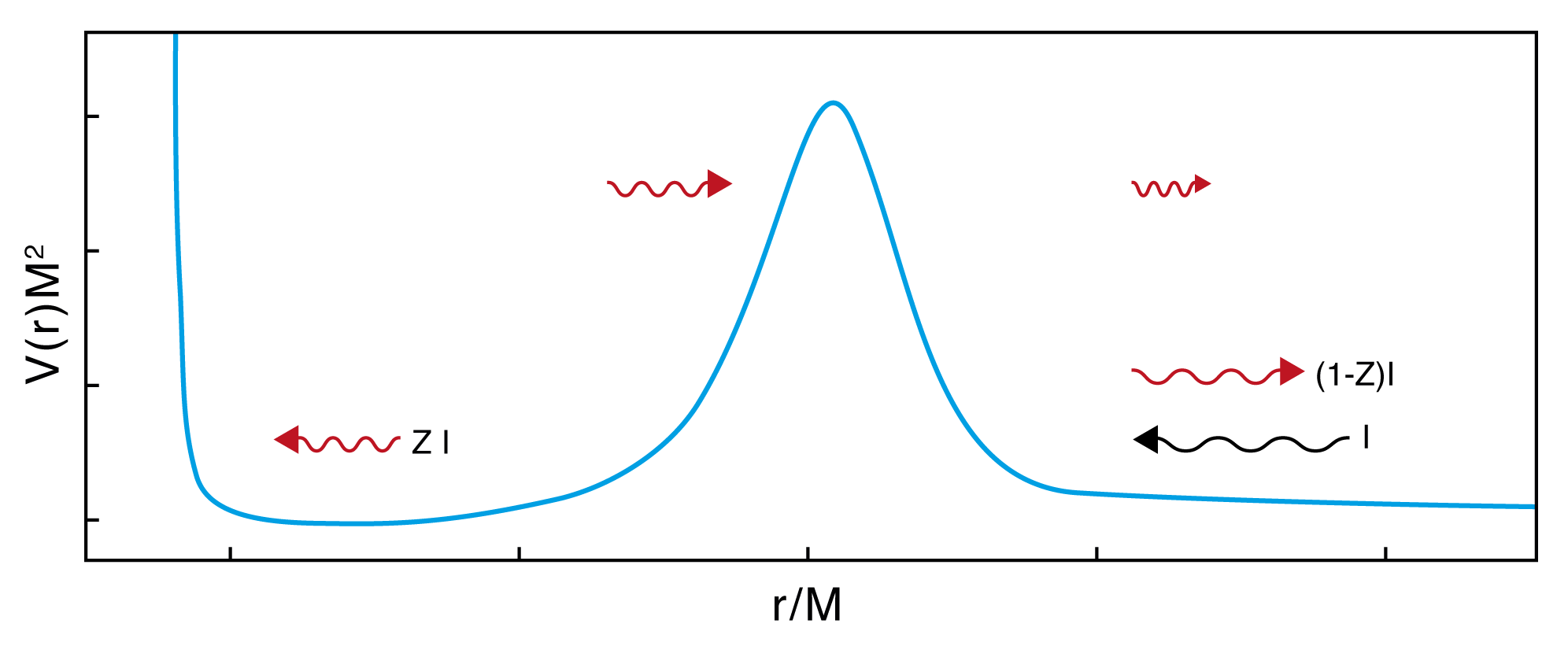}
%
\caption{Effective potential governing massless fields on a horizonless ultracompact spacetime. The peak of the potential is close to the location of the unstable \acs{LR}, and the potential in its vicinities is indistinguishable from that of a \acs{BH} spacetime (compare with the black curve in Fig.~\ref{fig:Potential}). The centrifugal barrier in the object's interior produces an effective cavity in spacetime, from which waves can slowly tunnel out. As discussed below, a cavity illuminated from the exterior ``heats-up'', akin to a greenhouse effect.}
\label{fig:potential} 
\end{figure}

An object sufficiently compact as to develop photonspheres is expected to behave as a cavity~\cite{Keir:2014oka,Cardoso:2014sna,Cardoso:2016rao,Cardoso:2016oxy,Cardoso:2017cqb,Cardoso:2019rvt}, in the sense that radiation is trapped in its interior, bouncing back and forth between its center (or surface) and the unstable photonsphere. Massless fields are subject to an effective potential with two ``barriers'', as illustrated in Fig.~\ref{fig:potential}, in contrast with \acs{BH}s where the potential still has the peak near the \acs{LR} but asymptotes to zero in both boundaries. 

Now let us consider the following \textit{gedankenexperiment}: we bombard a cavity with a constant flux of radiation ${\cal I}$ from a spin-$s$ wave carrying angular mode $\ell$. This flux may correspond to radiation emitted directed toward the central object by the secondary body in orbit. When the wave meets the barrier, a small fraction $Z$ tunnels in, and another part is reflected back. Conservation of energy implies that the reflected flux is $(1-Z){\cal I}$, where the absorption coefficient $Z$ is frequency-dependent ($Z=Z(\omega)$).

The transmitted part will then meet the interior barrier and be reflected. For simplicity, consider this interior is a reflecting mirror, so reflection is total (there is no absorption). After a roundtrip time within the cavity, $T_\text{echo}$, radiation will now be impinging the outside barrier from within. Again, a fraction $Z$ of this incident radiation tunnels out, corresponding to $Z^2{\cal I}$ of the initial one. This outgoing re-transmitted fraction of radiation will add up to the flux of outgoing radiation that is directly emitted to far-away distances. 

The reflection/outward transmission keeps happening inside the cavity. After $N$ reflections (or a time interval $NT_\text{echo}$), the outgoing flux of radiation is
\beq
{\cal F}_{NT_\text{echo}}=(1-Z){\cal I}+Z^2 {\cal I}\sum_{j=0}^N(1-Z)^j ={\cal I}-{\cal I}Z(1-Z)^{N+1}\,.            
\eeq
This tell us that the flux at large distances should be increasing in time steps of $T_\text{echo}$ and relaxing to the final state on a timescale 
\beq
\tau_\text{relax}=\frac{T_\text{echo}}{Z}\,.\label{scattering_prediction}
\eeq
The final state is that of an outgoing flux ${\cal I}$, as it should be since the object is not absorbing.

The absorption factors can be computed analytically in the low-frequency limit~\cite{Starobinski:1973,Starobinski2:1973,Brito:2015oca} from solving the Teukolsky equation~\eqref{eq:TeukolskyMaster} 
\begin{eqnarray}
Z_{s\ell m} &=& C\left[\frac{(\ell-s)!(\ell+s)!}{(\ell!)^2}\right]^2 \,, \label{sigma}\\
C &=&4\left(2M\omega\right)^{2\ell+2}\left[\frac{(\ell!)^2}{(2\ell)!(2\ell+1)!!}\right]^2\prod_{k=1}^\ell \left[1+\frac{16M^2\omega^2}{k^2}\right]\,. \nn\\ \label{sigma0}
\end{eqnarray}
For example, for small frequencies where we can ignore the terms involving $16M^2\omega^2/k^2$ 
\beq
Z_{010}=\frac{16M^4\omega^4}{9}\,,\, Z_{020}=\frac{64M^6\omega^6}{2025}\,,\, Z_{220}=\frac{256M^6\omega^6}{225}\,.\nonumber
\eeq
For binary systems, $M \omega \sim 10^{-2}-10^{-1}$, so we conclude that the build-up time can be very large. This puts in question the assumption of stationarity for the evolution of astrophysical binaries. As a side note, this calculation is very similar to how the greenhouse effect for planet Earth is estimated in a naïve approach.

The timescale in Eq.~\eqref{scattering_prediction} is the time the system needs to ``settle''. We will see below that this also corresponds to the resonant timescale, which is implied through the \acs{QNM}s.

\subsection{Resonances and forced oscillators\label{subsec:forcedoscillator}}

Before diving into our results, it is pedagogical to recall the results of a simple forced system with resonances, the driven harmonic oscillator~\cite{georgi1993physics}
\beq
\frac{d^2\Psi}{dt^2}+\Gamma\frac{d\Psi}{dt}+\omega_0^2\Psi=F_0\cos\omega t\,,
\eeq
with $F_0$ being a generic force per unit mass, $\omega_0$ the natural frequency of the system and $\Gamma$ a dissipation coefficient. The solution that starts off at $\Psi (t=0)=\partial_t \Psi (t=0)=0$ is 
\beq
\Psi(t)&=&F_0\frac{(\omega_0^2-\omega^2)}{(\omega_0^2-\omega^2)^2+\Gamma^2\omega^2}\left(\cos\omega t-e^{-\Gamma t/2}\cos\omega_\Gamma t\right)\nonumber\\
&+&F_0\frac{\Gamma \omega}{(\omega_0^2-\omega^2)^2+\Gamma^2\omega^2}\left(\sin\omega t-e^{-\Gamma t/2}\sin\omega_\Gamma t\right)\,,
\eeq
where $\omega_\Gamma=\sqrt{\omega_0^2-\Gamma^2/4}$. When $\Gamma t\ll 1$ and for small damping $\Gamma\ll \omega_0$, $\Psi$ grows on a timescale
\beq
\tau_\text{DHO}\approx \frac{2\pi}{\omega-\omega_0}\,.
\eeq
This is valid for short timescales and off the resonance. On resonance, i.e. $\omega=\omega_0$, the field attains a maximum on a timescale of $\tau\sim 1/\Gamma$. $\Gamma$ is intrinsic to the resonating system and corresponds roughly to $\omega_\text{I}$, so as we suggested above, for compact horizonless objects one should identify $1/\Gamma$ with the relaxation timescale in Eq.~\eqref{scattering_prediction}. 

\section{Numerical Results}

\subsection{A point particle orbiting a compact object}

%
\begin{table}[ht!]
\centering
\begin{tabular}{c c c} \hline\hline
\multirow{2}{*}{$M\omega_\text{QNM}$}
& \multicolumn{2}{c}{$r_p/M$}  \\ 
\cline{2-3}
& $a=0M$ & $a=0.9M$ \\
\hline \hline
$0.0881 - i1.197\times 10^{-7}$& $5.051$ & $4.780$ \\
$0.1259 - i2.687\times 10^{-6}$& $3.981$ & $3.674$ \\
$0.1633 - i2.470\times 10^{-5}$& $3.347$ & $3.011$ \\
\hline\hline
\end{tabular}
\caption{The lowest $\ell=1$ scalar quasinormal frequencies of a uniform-density relativistic star with $R=2.26M$. We also show the corresponding orbital radius at which the mode would be excited, calculated by equating the orbital frequency $\Omega$ in Eq.~\eqref{eq:AngularFreq} to the real part of the \acs{QNM} frequency and solving for $r_p$.
The value of $a$ corresponds to the used in the expression for the orbital frequency $\Omega$~\eqref{eq:AngularFreq}. For less compact stars, resonant frequencies are impossible to excite with matter on circular orbits outside the object. For example, for $R=6M$ the lowest dipolar \acs{QNM} frequency is $M\omega=0.262189 - i\,0.204880$. 
}
\label{tab:QNMs}
\end{table}

We now place a pointlike particle of mass $m_p$ in a circular orbit around a constant-density star,
\beq
r_p(t)=\text{const} \, ,\quad \theta_p(t) =\frac{\pi}{2} \, ,\quad \varphi_p(t) = \Omega\, t \,.
\eeq
The Schwarzschild geometry admits stable timelike circular geodesics for radius larger than the \acs{ISCO} at $r_\text{ISCO}=6M$~\cite{chandrasekhar1992mathematical} with $\Omega_\text{ISCO}=\sqrt{M/r_\text{ISCO}^3} \sim 0.068M $. While they can excite some proper modes of very compact constant-density stars (reference values are shown in Table~\ref{tab:QNMs}), the timescales of these resonances are too large to be probed by our numerical setup in a reasonable time frame. The only possibility would be to consider unstable circular geodesics, which have larger frequencies and can excite modes that grow on smaller timescales. However, since we eventually want to understand the impact of energy loss on the orbit, unstable motion is not the best option to study. 

To circumvent this, we consider non-geodesic motion. To keep the analysis simple and satisfy the requirement that it excites resonant modes, we consider the orbital motion to be equivalent to that around a Kerr \acs{BH} with mass $M$ and spin $a$. While this is not geodesic motion, prescribing it allows us to numerically investigate resonances and resonance-crossing scenarios in feasible timescales with acceptable accuracy. The actual nature of the motion is not relevant  for the excitation of the resonances. We therefore take~\cite{Wald:1984rg} 
\beq
\Omega &=& \frac{\sqrt{M}}{r_p^{3/2} + a\sqrt{M}} \, , \label{eq:AngularFreq} \\
\eeq
where $0\leq a/M \leq 1$ should be seen as a free ``knob'' (which, were the central object a Kerr \acs{BH}, would be the \acs{BH} spin). The energy  $E$ and angular momentum $L$ of these orbits is
\beq
\frac{E}{m_p}=\frac{r_p^{3/2} - 2M r_p^{1/2}+ a \sqrt{M}}{r_p^{3/4}\sqrt{r_p^{3/2} - 3M r_p^{1/2}   + 2 a \sqrt{M} }  } \, , \label{eq:Energy}\\
\frac{L}{m_p} = \frac{\sqrt{M}\left(r_p^2 - 2 a \sqrt{M}\,r_p^{1/2} + a^2 \right)}{r_p^{3/4}\sqrt{r_p^{3/2} - 3M r_p^{1/2}   + 2 a \sqrt{M} }  } \label{eq:Lz}.
\eeq

We will also show in Appendix~\ref{app:Greenhouse} that the timescale associated with the excitation of the resonance is independent of this choice, being completely controlled by the frequency of the circular orbit. Our imposition of this artificial motion is purely pragmatic, as this is a simple way to make circular orbits have higher frequency without changing the geometry of the central object. 

To solve this problem in the time-domain, we again used the numerical framework presented in Sec.~\ref{sec:Numerics}. The only difference is that we also solve the Klein-Gordon equation inside the star and impose regular boundary conditions for the scalar field at the center of the star. To compare both results, we also solved the problem in the frequency domain using standard Green function techniques~\cite{Davis:1971gg,Mino:1997bx,Cardoso:2002ay,Berti:2010ce}. Similar techniques have been employed in the past for the problem of particle scattering by the constant density stars~\cite{Kokkotas:1995av,Andrade:1999mj,Ferrari:2000sr,Tominaga:2000cs}, where a transient excitation of \acs{QNM}s can also be observed.

\begin{figure}[ht!]
\centering
\begin{tabular}{c}
\includegraphics[width=0.5\columnwidth]{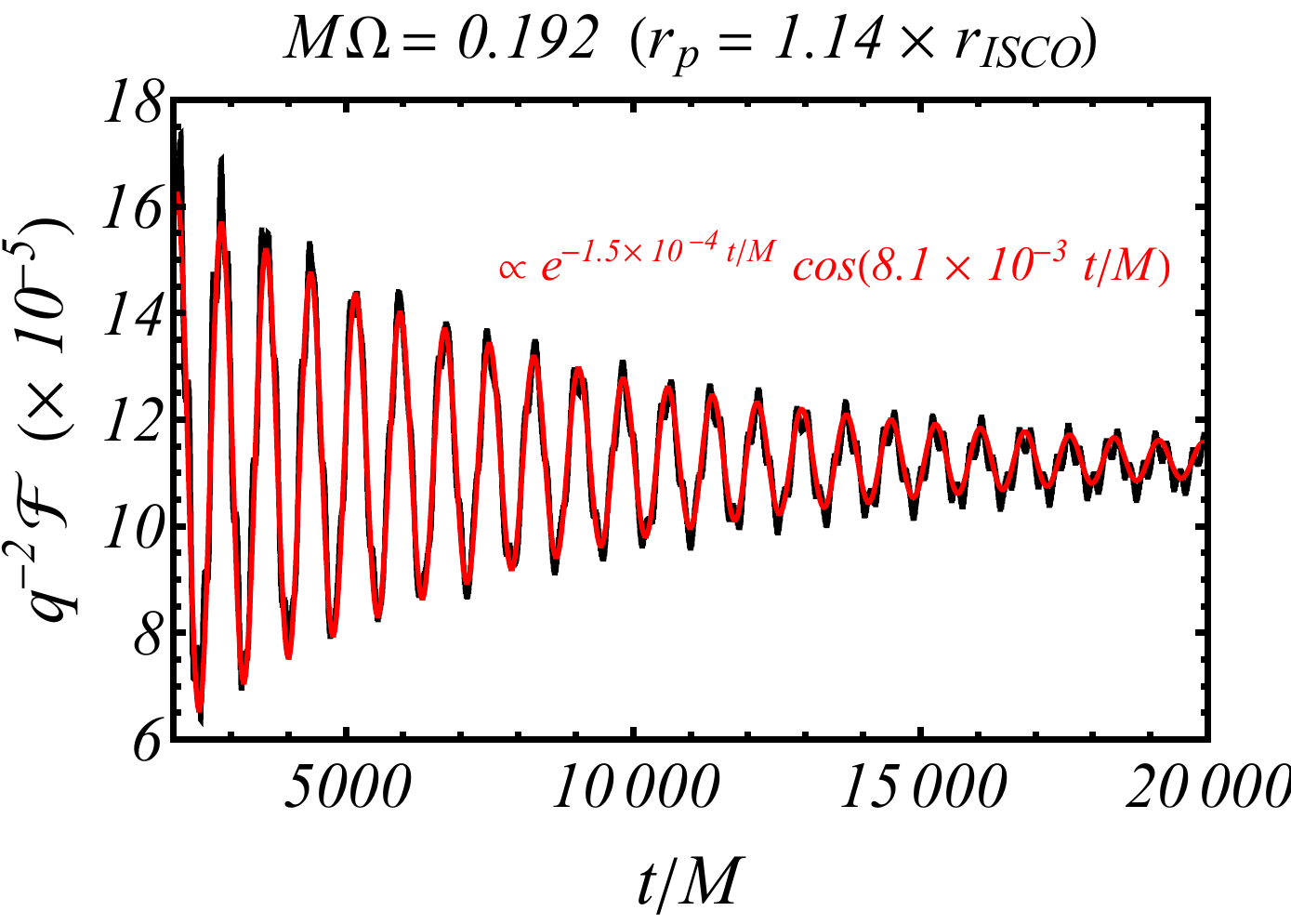}
\includegraphics[width=0.5\columnwidth]{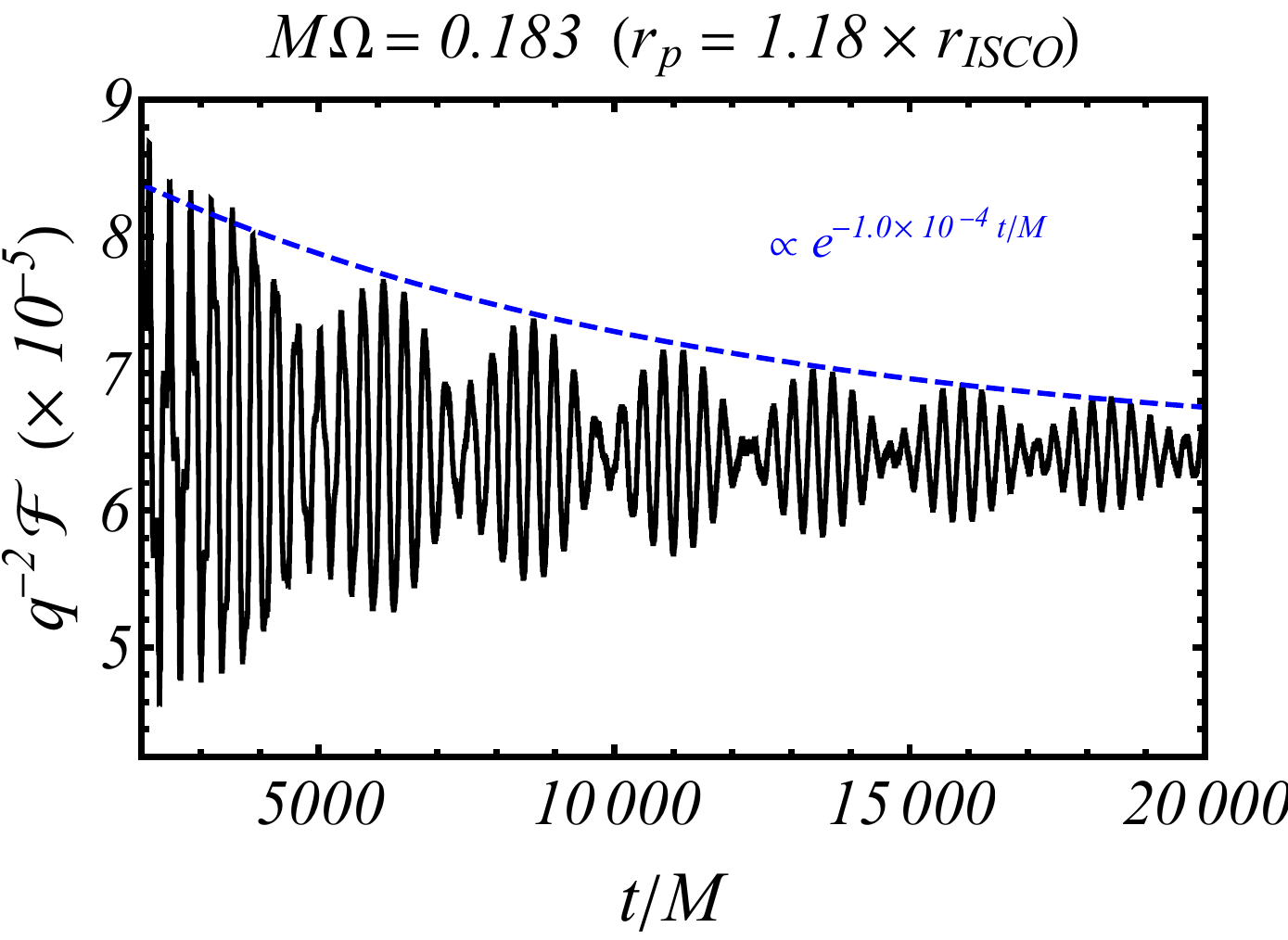} \\
\includegraphics[width=0.5\columnwidth]{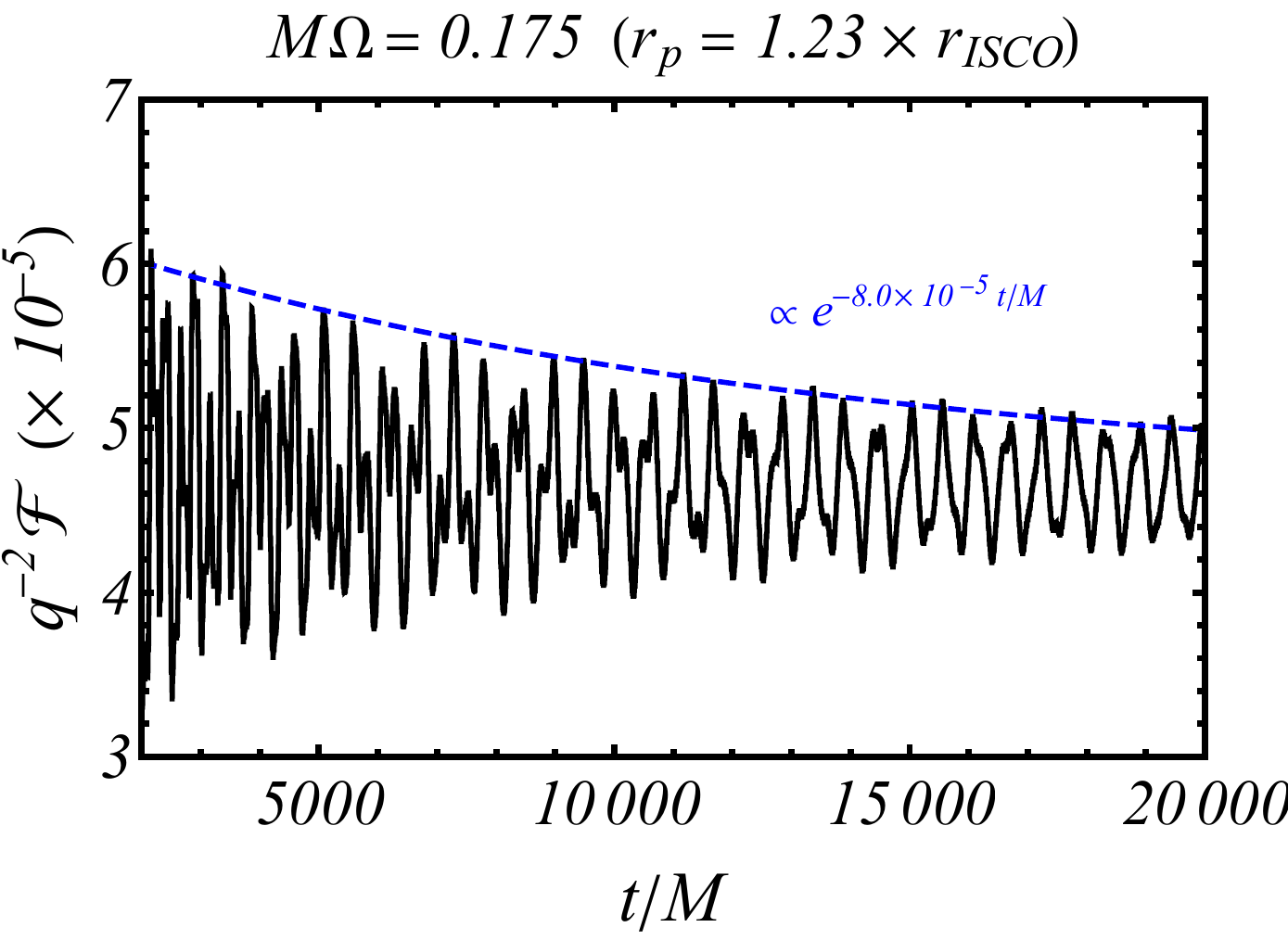} 
\includegraphics[width=0.5\columnwidth]{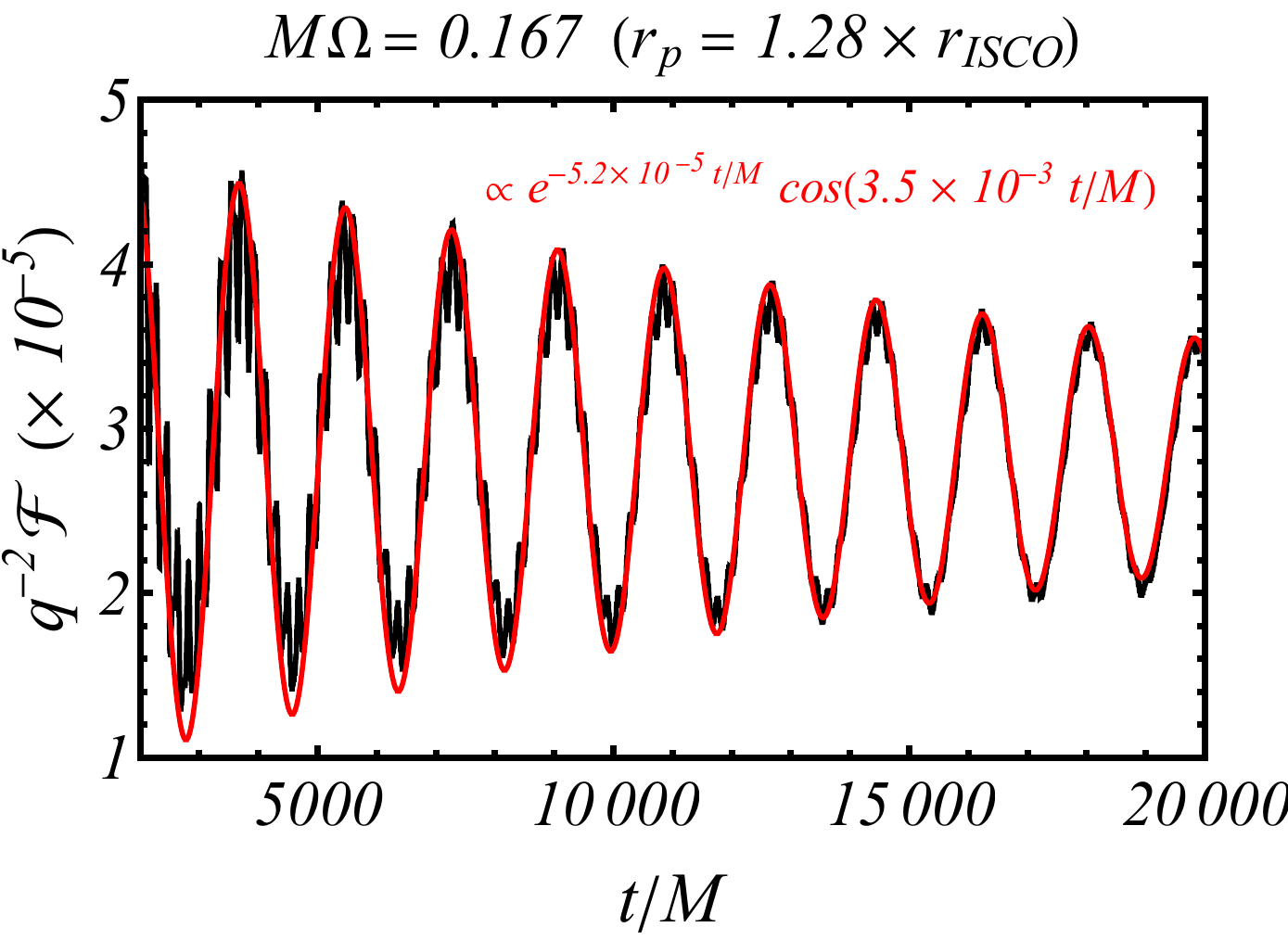} \\
\includegraphics[width=0.5\columnwidth]{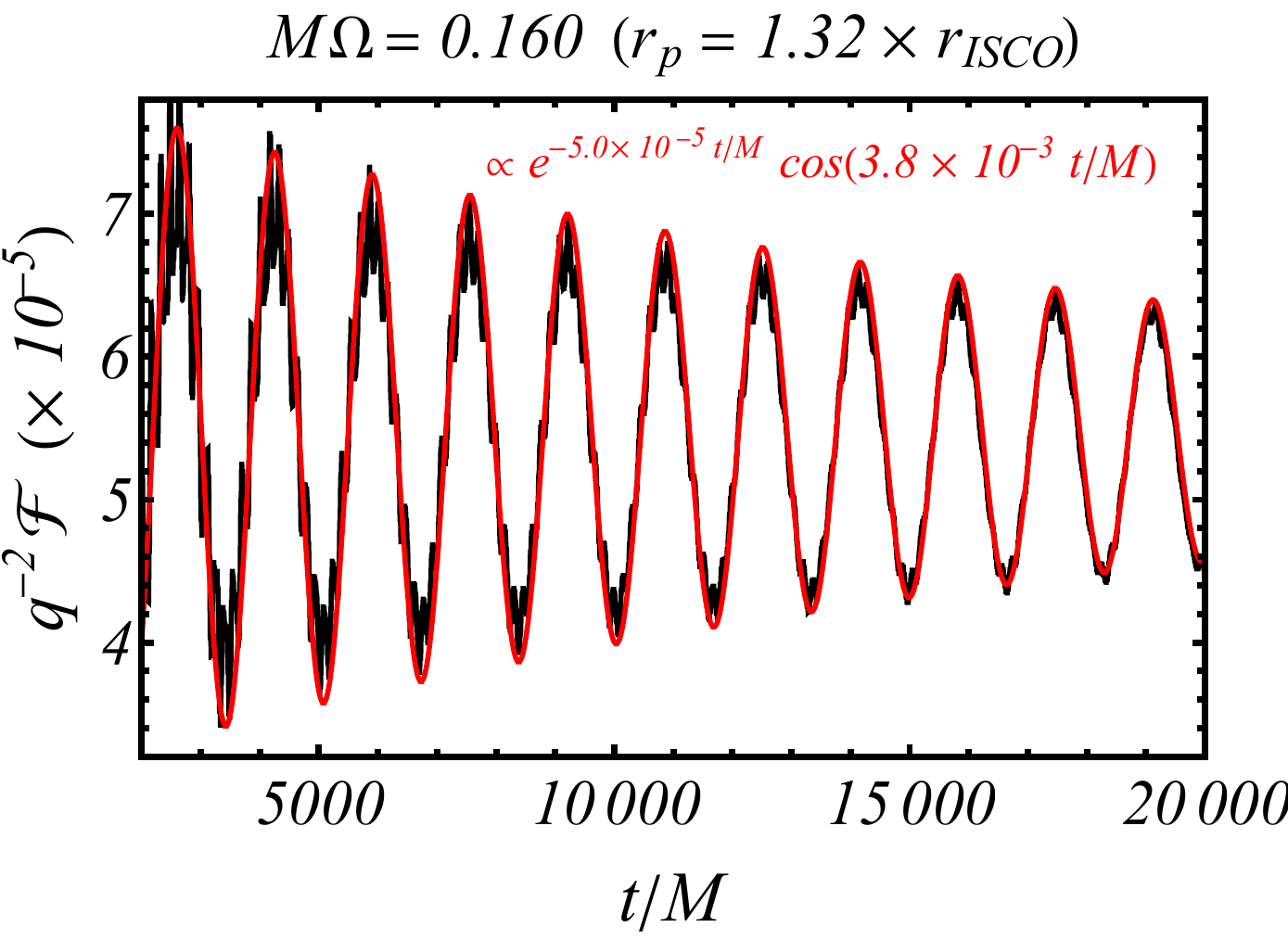} 
\includegraphics[width=0.5\columnwidth]{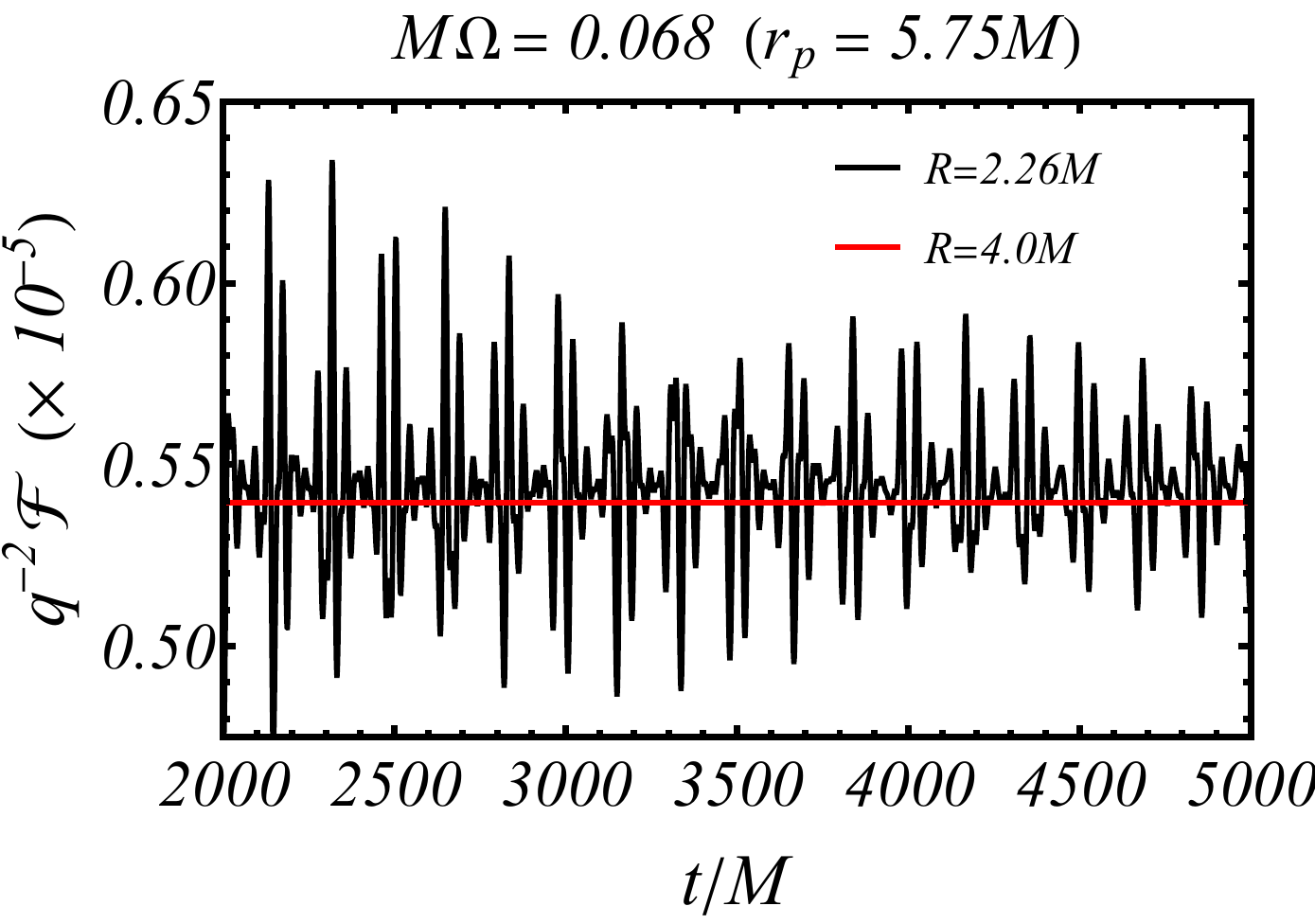} 
\end{tabular}
\caption{Evolution of the scalar energy flux $\mathcal{F}$ (Eq.~\eqref{eq:ScFlux}) emitted by a point-particle of mass $m_p$, made to orbit a constant-density star of mass $M$ on a circular orbit of constant radius $r_p$ (the orbit is not allowed to evolve). The flux is normalized by the mass ratio $q=m_p/M$. The results refer to the dipolar mode ($\ell=1$), but results are similar for higher multipoles. 
Except for the right bottom panel, the star has radius $R=2.26M$, and the frequency $\omega$ corresponds to the angular frequency $\Omega$ of the circular orbit, with $a=0.9M$ in Eq.~\eqref{eq:AngularFreq} ($r_\text{ISCO}\approx2.321M$). At late times, the flux asymptotes to a constant that agrees with the value computed in the frequency domain. The relaxation time is large for stars with photonspheres, but very short for less compact stars, where the system quickly becomes stationary, as seen in the right-bottom panel. 
}
\label{fig:l1} 
\end{figure}
%

\subsection{The build-up time}\label{sec:build-up_numerical}

%
\begin{figure}[t]
\centering
\begin{tabular}{c}
\includegraphics[width=0.9\columnwidth]{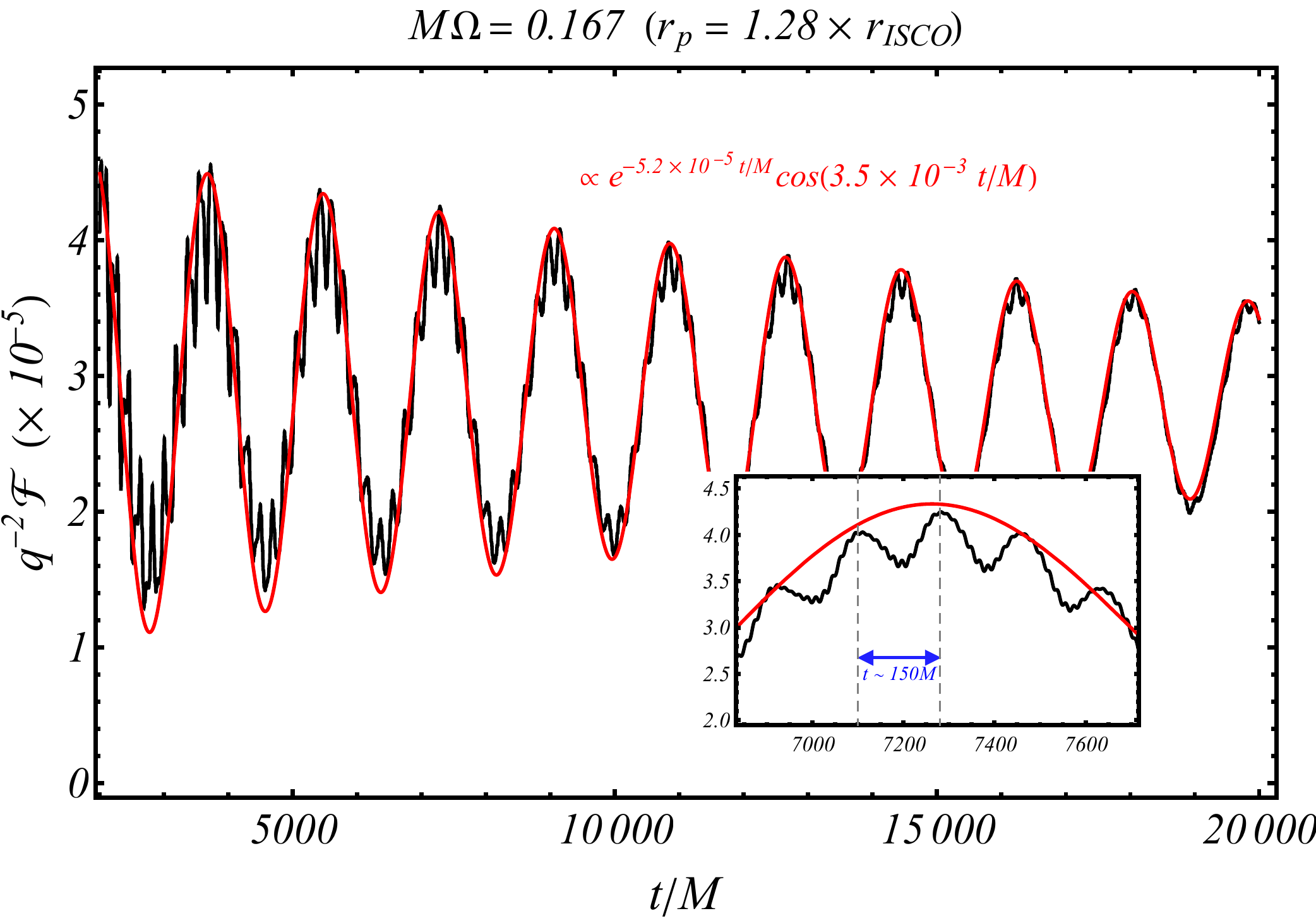} 
\end{tabular}
\caption{Scalar energy flux emitted by a point particle in circular orbit at $r_p=1.28\, r_\text{ISCO}$, with angular frequency $\Omega$ given by Eq.~\eqref{eq:AngularFreq} with $a=0.9M$, around a constant density star of radius $R=2.26M$ (the orbit is not evolving, the particle remains at fixed $r_p$). There are three different timescales in the signal: a high-frequency component corresponding to the ``direct signal'' with an orbital period $T_\text{Orb}/2= \pi / \Omega \sim 19M $ (the $1/2$ factor appears since we are showing fluxes); the traveling time $T_\text{echo} \sim 150M$ of waves inside the cavity potential; a lower frequency ``envelope'' corresponding to the excitation of the \acs{QNM} of the constant density star with frequency $M\omega_\text{QNM}=0.16333 - i 2.470 \times 10^{-5}$. This leads to a beating whose frequency is given by the semi-difference between the orbital and the \acs{QNM} frequency $\tau_\text{beating} \sim 2\pi/(\Omega - \omega_\text{QNM}) \sim 1800M $.}
\label{fig:Inset} 
\end{figure}
\begin{figure}[t]
\centering
\includegraphics[width=0.9\columnwidth]{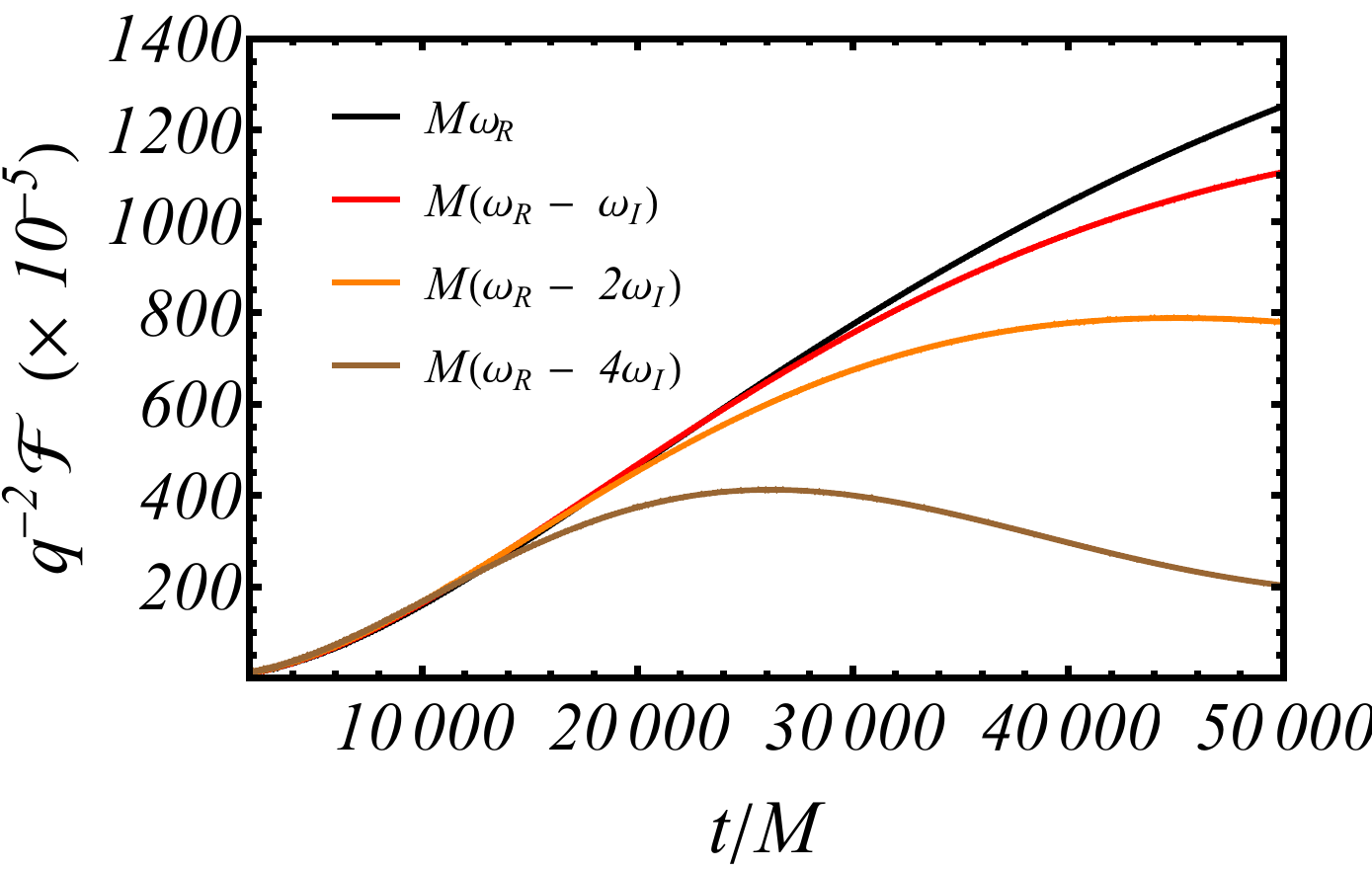} 
%
\caption{Resonant excitation of the dipolar \acs{QNM} of a constant-density star of radius $R=2.26M$ with frequency $M\Omega=M\omega_\text{R}=0.1633$ (cf. Table~\ref{tab:QNMs}), corresponding to a point particle at $r_p=3.011M$. A small deviation of this radius resulting in a frequency shift of $\delta \Omega \gtrsim 2 \omega_\text{I}$ can significantly hinder the excitation of the resonance. This agrees with standard results for the driven-harmonic oscillator, where the frequency bandwidth of the resonance peak is $\delta \Omega\sim \omega_\text{I}$.}. 
\label{fig:Resonance} 
\end{figure}

We consider $a=0.9M$ in Eq.~\eqref{eq:AngularFreq} because it is one of the smallest values for the spin that allow us to probe a fast-growing resonance while keeping the circular motion stable. Our time-domain numerical results are summarized in Figs.~\ref{fig:l1}-\ref{fig:Inset}. 

For the less compact spacetimes, which do not have trapping regions (in this context without \acs{LR}s), the initial data relaxes on a few dynamical timescales to a final stationary result, which coincides with that obtained via a frequency-domain approach. This behavior is apparent for the dipolar mode of an $R=4M$ uniform-density star in Fig.~\ref{fig:l1} (results are similar for other modes).

By contrast, for spacetimes sufficiently compact as to have photonspheres, the approach to stationarity is a long process. As explained above, the photonsphere is responsible for a potential barrier, through which waves need to tunnel and ``build-up'' until a stationary state is reached. The very first stages of this process are -- in accordance with the analysis of Section~\ref{subsec:greenhouse}-- a slow growth of the outgoing flux in steps of $T_\text{echo}$, the light travel time inside the photonsphere (see also the inset of Fig.~\ref{fig:Inset}, where the steps are clear). The relaxation timescale is also in good agreement with our \textit{greenhouse} estimate made in Section~\ref{subsec:greenhouse}. For $R=2.26M$ and $r_p=1.14 \, r_\text{ISCO}$ ($M\Omega= 0.193$  as set by Eq.~\eqref{eq:AngularFreq}) our results indicate a relaxation timescale $\tau_\text{relax} \sim 6500M,\, 4500M$ for $\ell=1,\,2$, whereas Eq.~\eqref{scattering_prediction} would indicate $\tau_\text{relax} \sim 4000M, \, 3100M$, respectively. The relaxation time increases when the circular orbit radius increases, again in line with our prediction~\eqref{scattering_prediction}. 

Our results also show finer details, in particular beatings and finer structure at small timescales, apparent in Fig.~\ref{fig:l1}. A zoom-in for $M\Omega=0.167M$ ($r_p=1.28r_\text{ISCO}$) is shown in Fig.~\ref{fig:Inset} for the dipolar mode. These features can be understood with the three different scales of the problem: $\bm{1}$. the orbital timescale, $T_\text{Orb}/2= \pi / \Omega \sim 19M$ shows up as the smallest timescale in the problem and is clear in the inset of Fig.~\ref{fig:Inset} (the $1/2$ factor appears since we are discussing fluxes); $\bm{2}$.  the orbital frequency $M\Omega=0.1668$ is close to the resonant \acs{QNM} frequency $M\omega_{R}=0.1633$ (see Table~\ref{tab:QNMs}). By our parallelism with the driven harmonic oscillator, we then anticipate a beating mode of frequency $\Omega-\omega_{R}$, i.e. a beating period $\tau_\text{beating}\sim 1800M$, in good agreement with our numerics; $\bm{3}$  the travel time of waves inside the cavity. This is clear in Fig.~\ref{fig:Inset} where we see steps of $T_\text{echo}\sim 150M$ for the build-up of the field in the cavity.

To excite resonances, we need to tune the orbital frequency closer to the resonant \acs{QNM}. Our results are shown in Fig.~\ref{fig:Resonance}, for the dipolar mode. The flux reaches amplitudes which are two orders of magnitude larger, but large timescales of order $\sim \mathcal{O}\left[\text{min} \left( \frac{1}{\omega_\text{I}}, \frac{2\pi}{\Omega-\omega_\text{R}}\right) \right]$ are required for this build-up. The frequency needs to be very fine-tuned in order to properly excite the resonance, since as expected from the driven-harmonic oscillator, the bandwidth of the resonance peak is $\delta \omega \sim \omega_\text{I}$. Hence, when $\omega_\text{I}$ is very small, as it happens for the proper modes of horizonless ultracompact objects, the region of the parameter space where the resonance can be triggered is very limited and the resonance takes a long time to grow. These two conditions can jeopardize the ability to excite a resonance in a binary coalescence effectively .

One could question the generality of our results considering the artificial motion we took for the point particle. However, in Appendix~\ref{app:Greenhouse} we repeat the analysis for $a=0M$, which makes motion geodesic. By placing the particle at radii that yields the same orbital angular frequency as the ones presented in Fig.~\ref{fig:l1}, we observe that the timescales involved are exactly the same for every single case, with only the relative magnitude between the fluxes changing. Note that in order to excite the \acs{QNM} with $M\omega_\text{QNM}=0.16333 - i2.470\times 10^{-5}$ with a circular geodesic, the point particle would have to be placed at $r_p=3.347M$.

\section{Consequences for gravitational-wave physics}

We have shown that very compact objects can effectively absorb \acs{GW}s for a significant time due to the spacetime geometry that traps waves within the photonsphere. This trapping process occurs on timescales of the order of Eq.	~\eqref{scattering_prediction}, after which radiation is re-emitted. The physics of these objects must take into account such delay, which has not been considered with due care in the literature~\cite{Cardoso:2019nis,Maggio:2021uge,Fransen:2020prl,Fang:2021iyf,Sago:2021iku}.

First, we consider the dynamics away from resonances. When they happen on short timescales, such as the final stages of a coalescence, then
the cavity has no time to ``fuel up'' and absorbs most of the impinging radiation. In this regime, horizonless compact objects behave similarly to \acs{BH}s, with equivalent absorption properties, and possibly indistinguishable from them. 

The second effect concerns the crossing of resonances, a generic effect not particular to compact horizonless objects. We show that frequency-domain adiabatic evolutions do not capture the entire physics and must be complemented by additional constraints when time evolutions are prohibitive.

\subsection{Adiabatic evolution of orbits and energy balance}\label{sec:AdiaEv}

To study \acs{GW}-driven inspirals, we consider adiabatic evolutions, where the point particle is always on a circular orbit with some associated energy $E$ and angular momentum $L$. We place the particle at some initial radius $r_0$, and determine its initial energy and angular momentum as dictated by the \acs{EOM}. Then, we need to evaluate the backreaction on these due to energy emission (and angular momentum). As argued, the flux needs to include the energy loss to infinity, but it should also include the energy piling up within the cavity.
However, considering the effects of the cavity is a challenging problem that we will not address here. We will only consider the energy radiated away to infinity, but we insist that the cavity may play an important role. For circular orbits, the angular momentum net balance is completely determined by the energy balance, so we only need to solve
\beq
\frac{dE}{dt}=-\mathcal{F} \, , 
\eeq
with the appropriate initial energy and use this to evolve
\beq
\frac{dr}{dt}=-\mathcal{F} \left(\frac{dr}{dE} \right) \, , 
\eeq
again with the appropriate initial conditions. Having the updated value for $r_p$, we can compute the angular frequency $\Omega$ again.
%
%

This procedure can be applied both for the time and the frequency domain. However, the flux computed in the frequency domain implicitly assumes stationarity, i.e. that the oscillations around the average flux vary out to zero much faster than the timescale on which the particle inspirals. For the systems we are discussing, this implies the cavity has had time to fuel-up. For the time domain instead, the energy balance is done at every instant and therefore can account for the inhomogeneities in the flux as the star is relaxing or the cavity is fueling-up.

\subsection{Off resonance}

As a binary coalesces, its orbital frequency changes. For objects on quasi-circular orbit millions of years prior to the merger, a ``stationary state'' (to be read as where the frequency-domain calculation yields the same result as the time-domain) is reached. However, in the late stages of the inspiral, the frequency varies rapidly, hence not allowing the compact object to ``fuel up''. This happens whenever the frequency changes by $\Delta \omega \gtrsim \omega_\text{R}$ and the corresponding inspiral time is small enough that it does not allow for relaxation. 

Let us start the inspiral at some radius $r_p(t=0)=r_i$. Then, for quasi-circular orbits, and including only the leading terms in \acs{GW} reaction~\cite{Peters:1964zz} 
\beq
r_p(t)&=& (r_i-4\beta t)^{1/4}\,,\\
\beta&=&\frac{64}{5}M^2m_p\,.
\eeq
The time taken to inspiral from $r_0$ to $r_p(t)$ can also be written in terms of the initial and final GW frequency $\omega_i,\,\omega_f$ as~\cite{Peters:1964zz}
\beq
t_\text{inspiral}=\frac{2^{2/3}M^{4/3}\left(1-(\omega_i/\omega_f)^{8/3}\right)}{\beta \omega_i^{8/3}}\,,
\eeq
and therefore 
\beq
\frac{t_\text{inspiral}}{\tau_\text{relax}}\sim 10^{-2} \frac{100M}{T_\text{echo}}\frac{10^{-5}M}{m_p}\left(\frac{M\omega}{0.06}\right)^{10/3}\,.
\eeq
Considering the typical values for \acs{EMRI}s ($m_p/M \lesssim 10^{-5}$, $M\omega \sim 10^{-2}-10^{-1}$), the result above implies that cavity effects should be taken into account in the evolution of the inspiral. Time or frequency domain analysis should include the temporary pile-up of energy in the cavity in the computation of radiation-reaction effects as the inspiral progresses. 

Our results also show how the \acs{BH} limit is approached naturally when $T_\text{echo} \rightarrow \infty$ in the previous construction.
In this limit, the central object is a perfect absorber during the entire inspiral. We forecast that properly handling the cavity problem in radiation-reaction should recover the \acs{BH} result continuously.

Previous works suggested that the cavity would only be important for the evolution of the binary when the traveling time inside it is comparable (or larger) than the radiation-reaction timescale \cite{Maselli:2017cmm, Maggio:2021uge}. However, as discussed above, energy can be trapped by being reflected back and forth in the cavity until it saturates. This process corresponds to multiple travel times, as dictated by Eq.~\eqref{scattering_prediction}. In general, this timescale can be much bigger than the travel time inside the cavity, making the latter more relevant for larger mass-ratios than previously considered.

\subsection{Crossing resonances}

The above results strongly suggest that in order to excite a resonance, the system needs to spend at least a time $\sim 1/\omega_\text{I}$ in a frequency band $\delta \omega \sim \omega_\text{I}$~around the resonance at $\omega_\text{R}$. Rigorous estimates for simple linear differential equations were obtained in Refs.~\cite{Fowler:1921,Nayfeh:book}. We can work out the consequences for \acs{GW} science: the time $\delta t_\text{cross}$ that the system takes to cross the resonance is~\cite{Cardoso:2019nis}
\beq
\delta t_\text{cross} \sim \omega_\text{I}/\left(d\Omega/dt \right) \, ,
\eeq
with
\beq
\frac{d\Omega}{dt} = \left(\frac{d\Omega}{dr}\right)\left(\frac{dr}{dE}\right)\mathcal{F} \,.
\eeq
Then, for the resonance to be effectively excited 
\beq
&&\omega_\text{I}\delta t_\text{cross} \gtrsim 1 \nonumber\\
&\Leftrightarrow& q \lesssim q_\text{max} = \frac{(M \omega_\text{I})^2}{q^{-2} \mathcal{F}}\left(\frac{1}{m_p} \frac{dE}{dr}\right)\bigg/\left(M \frac{d\Omega}{dr} \right)  \, . \label{eq:MassCondition} 
\eeq 
In this estimate, orbital quantities on the right-hand side are meant to be evaluated at the radius where the resonance is excited, and the flux is to be taken \textit{outside} the resonance since this is the actual energy flux emitted by the binary while the resonance grows. 

\section{Discussion}\label{sec:Discussion_Greenhouse}

%
\begin{figure}[t]
\centering
\begin{tabular}{c}
\includegraphics[width=0.9\columnwidth]{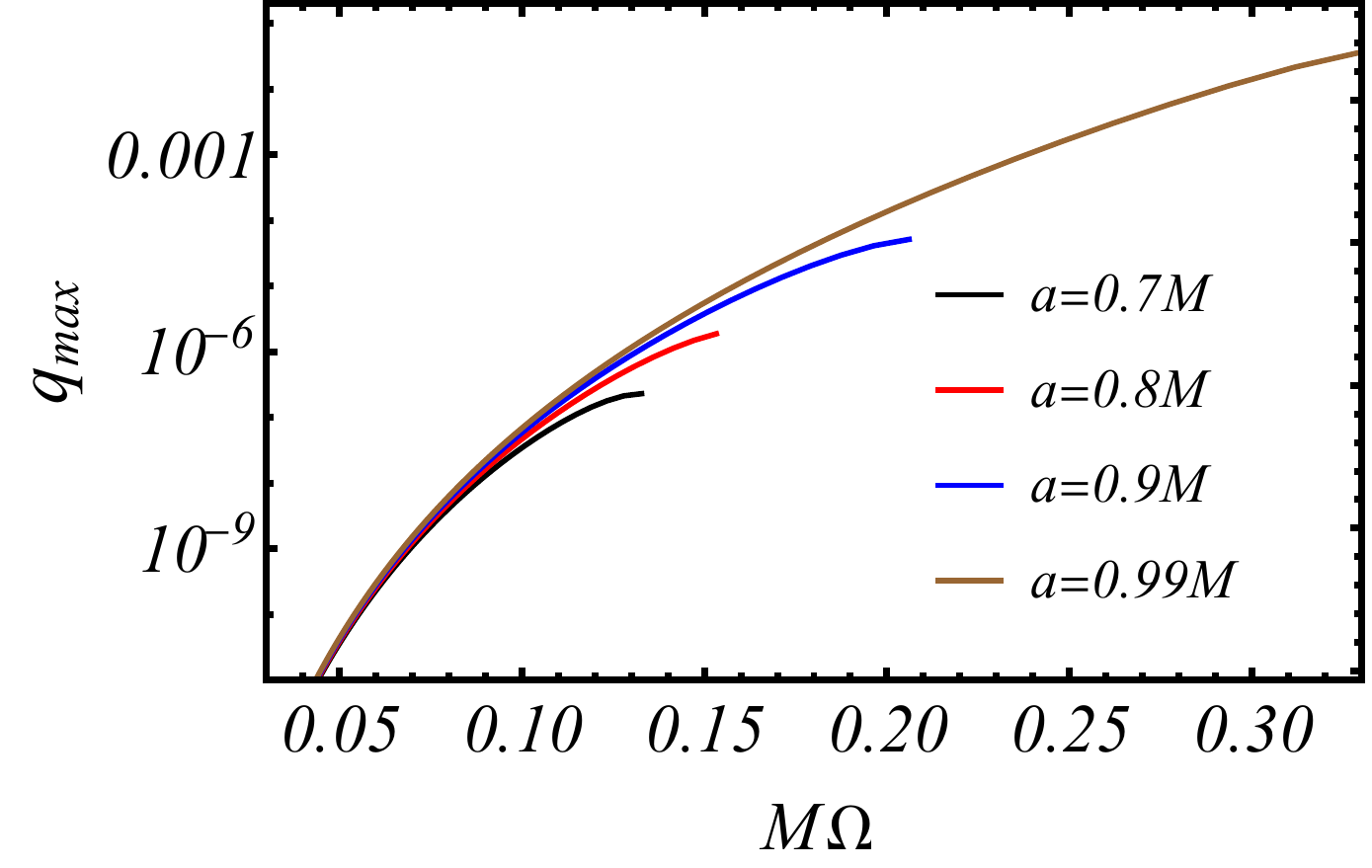}
\end{tabular}
\caption{Largest mass ratio $q_\text{max}$ predicted by the estimate~\eqref{eq:MassCondition} that would allow the resonant excitation, by \acs{GW}s, of a \acs{QNM} of an \acs{ECO} with frequency $\omega_\text{R} = 2\Omega$. We considered that $\omega_\text{I} \sim \omega_\text{R}^{2l+3}$ \cite{Cardoso:2019nis,Maggio:2021uge} and only took into account the quadrupolar mode in the energy flux, making this estimate conservative, since $\mathcal{F}$ will increase if higher multipoles are considered. We show orbital frequencies $\Omega$ corresponding to radius of the particle from $r_p=10M$ up to almost the \acs{ISCO} for each spin $a$. For mass ratios larger than these limits, the particle crosses the resonance too quickly for it to grow effectively.}
\label{fig:qmax}
\end{figure}

In this chapter, we have demonstrated that spacetimes with \acs{LR}s behave as cavity resonators. They have large ``build-up'' times, determined by the transmission amplitudes at the \acs{LR}, and are prone to resonances. To properly evolve binaries composed by \acs{ECO}s, it is necessary to take into account the energy piling up within the photonsphere. For systems evolving rapidly under radiation reaction, these objects act as effective absorbers and may mimic \acs{BH}s. A proper modeling of this process and the full evolution of an \acs{EMRI} is therefore an open problem. Additionally, the secondary object can excite resonances of these objects, but for the resonances to fully develop, the binary has to evolve slower than the time needed for the resonance to grow. 

We focused on a simple toy-model of a scalar field around a constant-density star and considered an artificial motion for the point-particle that is not dictated by its equations of motion. Nonetheless, all our analytical estimates for the relaxation timescale \eqref{scattering_prediction}, resonance growth time, and the upper limit on the mass ratio that excites a resonance in an inspiralling binary \eqref{eq:MassCondition} are model indepedent, and agree with the numerical results for the particular system that we studied. Therefore, our conclusions should be applicable to any astrophysical system. 

Refs.~\cite{Maggio:2021uge, Sago:2021iku} studied \acs{EMRI}s around spinning horizonless compact objects (see also \cite{Cardoso:2019nis,Fransen:2020prl} for the non-spinning case). As in our toy-model, the low-frequency \acs{QNM}s of the spinning \acs{ECO} can be resonantly excited during the inspiral, leading to non-negligible effects in the waveform that must be considered for the detection and parameter estimation of these sources. However, their work in the frequency domain ignores the growth time of the resonance, implicitly assuming stationarity at all instants. This approximation is only correct when the mass ratio of the system obeys the condition of Eq.~\eqref{eq:MassCondition}.

In Fig.~\ref{fig:qmax}, we apply this estimate to the type of systems studied in Refs.~\cite{Maggio:2021uge, Sago:2021iku}. Typically, the resonance width for these exotic compact objects is $\delta \omega \sim \omega_\text{I} \sim \omega_\text{R}^{2\ell+3}$ \cite{Cardoso:2019nis, Maggio:2021uge}. In a binary system, the frequency of the emitted \acs{GW}s is determined by the orbital frequency and, for circular orbits, corresponds to $\omega_\text{GW} = 2\Omega$. We can then compute, for every orbital frequency $\Omega$ (or radial location of the particle), how light the point particle must be to resonantly excite an \acs{ECO} with a \acs{QNM} of frequency $\omega_\text{R} = 2\Omega$, $\omega_\text{I} \sim (2\Omega)^{2\ell+3}$. For the off-resonance flux $\mathcal{F}$, we used the same values as in Kerr since the relative difference with respect to a horizonless ultracompact object should be small (though non-negligible when accumulated over many orbits). We only took into consideration the quadrupolar mode $\ell=2$, and higher multipoles will typically increase $\mathcal{F}$, therefore placing even more stringent limits on the mass ratio. We conclude that for the reference value of $q=3 \times 10^{-5}$ used in most results presented in Ref.~\cite{Maggio:2021uge}, the particle should only be able to excite resonances in \acs{ECO}s with spins $a>0.9M$, and on a limited region of the parameter space where it is very close to the central object.   

Our conclusions should apply to other systems where resonances are excited, such as massive scalar theories \cite{Arvanitaki:2009fg, RevModPhys.84.671,deRham:2014zqa}. In these theories, matter orbiting a Kerr \acs{BH} can resonantly excite superradiant modes, which might lead to so-called \textit{floating orbits} \cite{Press:1972zz, Brito:2015oca}, where the energy absorbed by the horizon is positive and counterbalances the loss of energy to infinity \cite{Cardoso:2011xi,Yunes:2011aa}. As a consequence, the inspiral freezes and the radiated energy is solely provided by the rotational energy of the \acs{BH}. These resonances occur for $\omega^2_\text{res} = \mu_s^2 - \mu_s^2 \left(M \mu_s /(\ell+1+n) \right)^2$, where $\mu=m_s/\hbar$ is the reduced mass of the scalar field, and have typical widths of $\delta \omega \sim \omega_\text{I} \propto \mu_s^{4\ell+5}$ \cite{Cardoso:2011xi, Detweiler:1980uk}. These are even more narrow than the \acs{QNM}s of \acs{ECO}s we have been discussing. Generically, the off-resonance energy flux is dominated by \acs{GW}s, which means that for the same orbital frequency, the mass ratios needed to properly excite superradiant resonances of massive scalars would be even smaller than the ones in Fig.~\ref{fig:qmax}. 

Additional dissipation mechanisms could undermine the fueling-up of the cavity and the excitation of resonances. However, \acs{GW}s are known to interact very weakly with matter, with effects only being relevant at the Hubble timescale \cite{1971ApJ...165..165E, 1985ApJ...292..330P, Kocsis:2008aa, Loeb:2020lwa}. Hence, any additional channel of dissipation should be subdominant with respect to the emission of waves to infinity and the trapping of energy by the central object on the timescales of interest for these systems. We cannot rule out, however, that extremely stiff equations of state giving rise to large viscosities and large sound speeds strongly suppress resonances in compact objects. Even in such case, our results still apply to other systems, e.g. the resonances of massive boson fields around spinning \acs{BH}s discussed in the Introduction of Chapter~\ref{ch:Cloud}~\cite{Baumann:2019ztm, Baumann:2022pkl}. 

Our conclusions have obvious implications to \acs{GW} astronomy, and highlight the necessity of a better understanding of \acs{GW} emission in less conventional systems that are not typical \acs{BHB}s in vacuum \acs{GR}. The proper modeling of ``transient'' resonances which do not have time to fully develop in binary inspirals has already been tackled in Refs.~\cite{Bonga:2019ycj, Yang:2019iqa, Gupta:2022fbe, Speri:2021psr}, in the context of tidal resonances induced by a third companion. The steps laid there could be adapted for \acs{EMRI}s involving \acs{ECO}s. 



\chapter{Gravitational tuning forks}\label{ch:tuningfork}

At the end of the Introduction, we discussed how hierarchical triple systems are common in a variety of astrophysical scenarios, such as globular clusters, \acs{AGN}s, and other dense stellar environments\cite{Zevin:2018kzq,Martinez:2020lzt, Bartos:2016dgn, 10.1093/mnras/stw2260, Chen:2018axp, Toubiana:2020drf, OLeary:2016ayz, 2016MNRAS.463.2109R, Portegies_Zwart_2000}. ``'Hierarchical'' here refers to the distinct length scales between the orbit of a binary and the one of its \acs{CM} around the third body. Recalling, around $90\%$ of low mass binaries with periods shorter than $3$ days are expected to belong to some hierarchical structure~\cite{2006AA...450..681T, Pribulla:2006gk, Robson:2018svj}.

This has motivated recent studies on the dynamics and \acs{GW} emission in hierarchical triple systems.
Kozai-Lidov resonances, in particular, have attracted some attention~\cite{1962AJ.....67..591K, doi:10.1146/annurev-astro-081915-023315, poisson_will_2014}. As we had already mentioned, these describe secular changes in the binary eccentricity and inclination with respect to the orbit described by its \acs{CM} around the third object. This mechanism triggers periods of high eccentricity, where \acs{GW} emission increases significantly, potentially inducing coalescence in eccentric orbits detectable by \acs{LISA}~\cite{Hoang:2019kye, Randall:2019sab, Randall:2019znp, Deme:2020ewx}, that then enter in the frequency band of ground-based detectors still at high eccentricities~\cite{Antonini:2012ad,Antonini_2016, Hoang_2018, Zevin:2018kzq}. A direct integration of the \acs{EOM} 
confirms these systems have unique \acs{GW} signatures~\cite{PhysRevD.85.123005,Gupta:2019unn}, which may be detected indirectly via radio observations of binary pulsars~\cite{Suzuki:2020zbg}. There are also attempts at modeling the effects of a third body directly into the waveform. These include Doppler shifts~\cite{10.1093/mnras/stv172, Meiron:2016ipr, Randall:2018lnh, Wong:2019hsq, Han:2018hby}, relativistic beaming effects~\cite{Torres-Orjuela:2018ejx, Torres-Orjuela:2020cly}, gravitational lensing~\cite{Ezquiaga:2020dao, Ezquiaga:2020gdt} and other dynamical effects~\cite{Yu:2020dlm, Bonga:2019ycj, Yang:2019iqa}.

In this final chapter of the second part of this thesis, we will study hierarchical triples using \acs{BH} perturbation theory and investigate \acs{GW} emission from binaries around \acs{SMBH}s. Using the methods we already applied in previous chapters, we will be able to probe resonant excitation of \acs{QNM}s in triple systems, and capture for free all of the relativistic
effects which have so far been included only at a phenomenological level in the literature.

\section{Setup}
%
\begin{figure}[t]	
\centering
	\includegraphics[width=8.0cm,keepaspectratio]{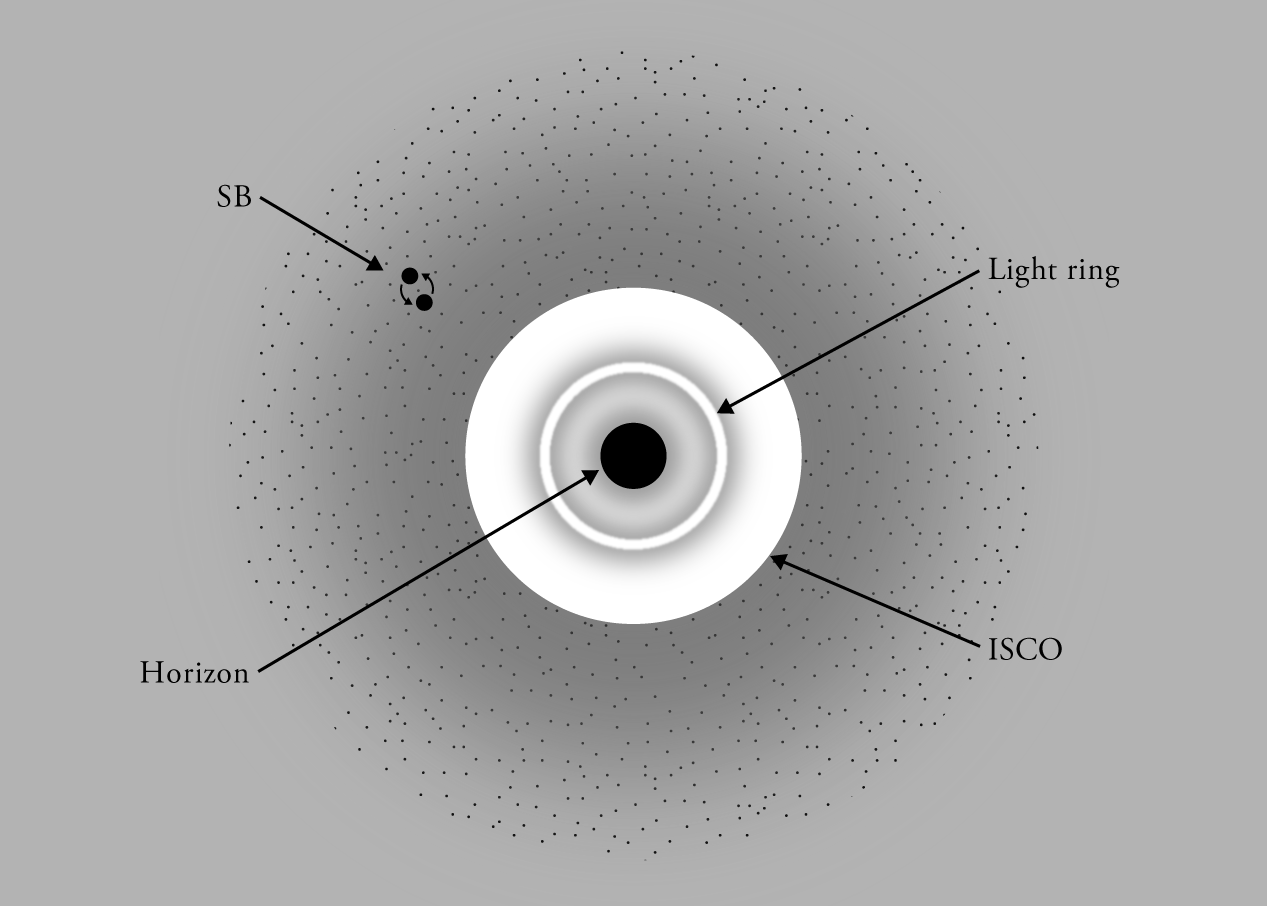} 
	\caption{Equatorial slice of a spacetime with a hierarchical triple system. In the center there is a \acs{SMBH}. At large distances away from the central region, physics is nearly Newtonian. The external gray area is the entire region where stable circular motion is possible. We place a small compact binary here, with characteristic frequency $\omega_0$. Its \acs{CM} orbits the \acs{SMBH} with angular frequency $\Omega_\text{CM}$. At the \acs{ISCO} ($r_\text{ISCO}=6M$ when the \acs{SMBH} is non-spinning), circular motion is marginally stable, and unstable for smaller radius. High-frequency waves can be trapped at the \acs{LR} ($r_\text{LR}=3M$ for non-spinning \acs{BH}s). Such motion is unstable, and as seen in previous chapter, can also be associated with the ``ringdown'' excited during \acs{BH} mergers.}\label{fig:anatomy}
\end{figure}

Our setup is similar to the one used in Chapter.~\ref{ch:LR}, where we had a small binary in the vicinity of a large \acs{BH}. An illustration of this sytem can be found in Fig.~\ref{fig:anatomy}. We consider the small binary is sourcing perturbations to a Kerr background~\eqref{eq:KerrLineElement} and model it as two point-particles. In Chapter~\ref{ch:LR} we considered the \acs{CM} of the small binary was plunging onto the central \acs{BH}, but here we will take it to be at some fixed radius, $r_\text{CM}$, while it describes a circular orbit around the central \acs{BH}. For the small binary, we again take elliptic orbits around its \acs{CM}
\beq
r^{\pm} &=& r_\text{CM} \, , \\
\varphi^{\pm}&=&\Omega_\text{CM}t\pm \epsilon_\varphi \sin{\omega_0 t}\quad ,\quad \theta^{\pm}=\pi/2 \pm \epsilon_\theta \cos{\omega_0 t} \,,
\eeq
where $\epsilon_\theta,\epsilon_\varphi \ll 1$ parametrize the two axis of the ellipse $\delta r_\theta = \epsilon_\theta r_\text{CM} $, $\delta r_\varphi = \epsilon_\varphi r_\text{CM} $ of the small binary, and $\Omega_{\text{CM}}$ is the angular velocity of the \acs{CM}. $\Omega_\text{CM}$ and $\omega_0$ are coordinate frequencies, while the proper oscillation frequency of the small binary, $\omega_0'$, is obtained by rescaling the time component of the $4$-velocity of the \acs{CM}, i.e. $\omega_0'=u_\text{CM}^t\omega_0$.
For concreteness, we focus exclusively on equal-mass binaries, $m_p^\pm=m_p$ and a highly eccentric orbit with $\epsilon_\theta=0$. We do not see any qualitatively new phenomena in the general case, and this particular choice could mimic high-eccentricity binaries driven by Kozai-Lidov resonances~\cite{1962AJ.....67..591K, doi:10.1146/annurev-astro-081915-023315, poisson_will_2014}. 

There is a physical relation between $\epsilon_\varphi$ and $\omega_0$. In the rest frame of the small binary, $\delta r'_\varphi \propto 1/(\omega'_0)^{2/3}$, where the prime refers to \textit{proper} quantities. For binaries on circular geodesics, for example, doing the appropriate rescaling $\omega_0'=u_\text{CM}^t\omega_0$ and $\delta r_\varphi = \Delta/ \rho^2 \,\cdot \delta r'_\varphi $, we find
\beq
\epsilon_\varphi \propto \frac{\Delta}{\rho^2}\frac{1}{r_\text{CM}(u_\text{CM}^t\omega_0)^{2/3}} \, .  
\eeq                           

We will be looking for possible resonances in this triple system, which may happen when the forcing frequency of the binary equals the natural frequencies of the background, i.e its \acs{QNM}s. 
There are three important frequencies in the problem: the angular frequency of the \acs{CM}, the frequency of the \acs{LR}, which as we saw in Chapter~\ref{ch:LR} controls the \acs{QNM}s~\cite{Berti:2009kk}), and the angular velocity of the \acs{BH} horizon $\Omega_\text{H}$~\eqref{eq:AngularBH}. Close to the central \acs{BH}, all of them are of order $\mathcal{O}(1/M)$. To have $M \omega_0 \sim 1$, we need to ensure $\delta r_\varphi / m_p \sim (M/m_p)^{2/3}$. For a \acs{SMBH} with $M\sim10^4-10^6 M_\odot$, like Sagittarius A*, and a small binary composed by stellar-mass \acs{BH}s with $m_p \sim 1-100 M_\odot$, this would correspond to $\delta r/ m_p \sim 10^2-10^4$, so well within the inspiral phase. 


As before, the small binary enters as a source term in the Teukolsky equation~\eqref{eq:TeukolskyMaster}, which governs perturbations to the background spacetime of the central Kerr \acs{BH}. We study both the case of \acs{GW}s ($s=-2$ in the Teukolsky equation) and also scalar radiation ($s=0$) where the source term is again given by the trace of the energy-momentum tensor of the point particles composing the binary, as presented in Sec.~\ref{sec:Scalars}.

Again, we solve this problem numerically in the time domain with the Lax-Wendroff method introduced in Sec.~\ref{sec:Numerics}.

\section{Resonant excitation of QNMs and Energy emission}

%
\begin{figure}[t]
\centering
\includegraphics[width=0.95\linewidth]{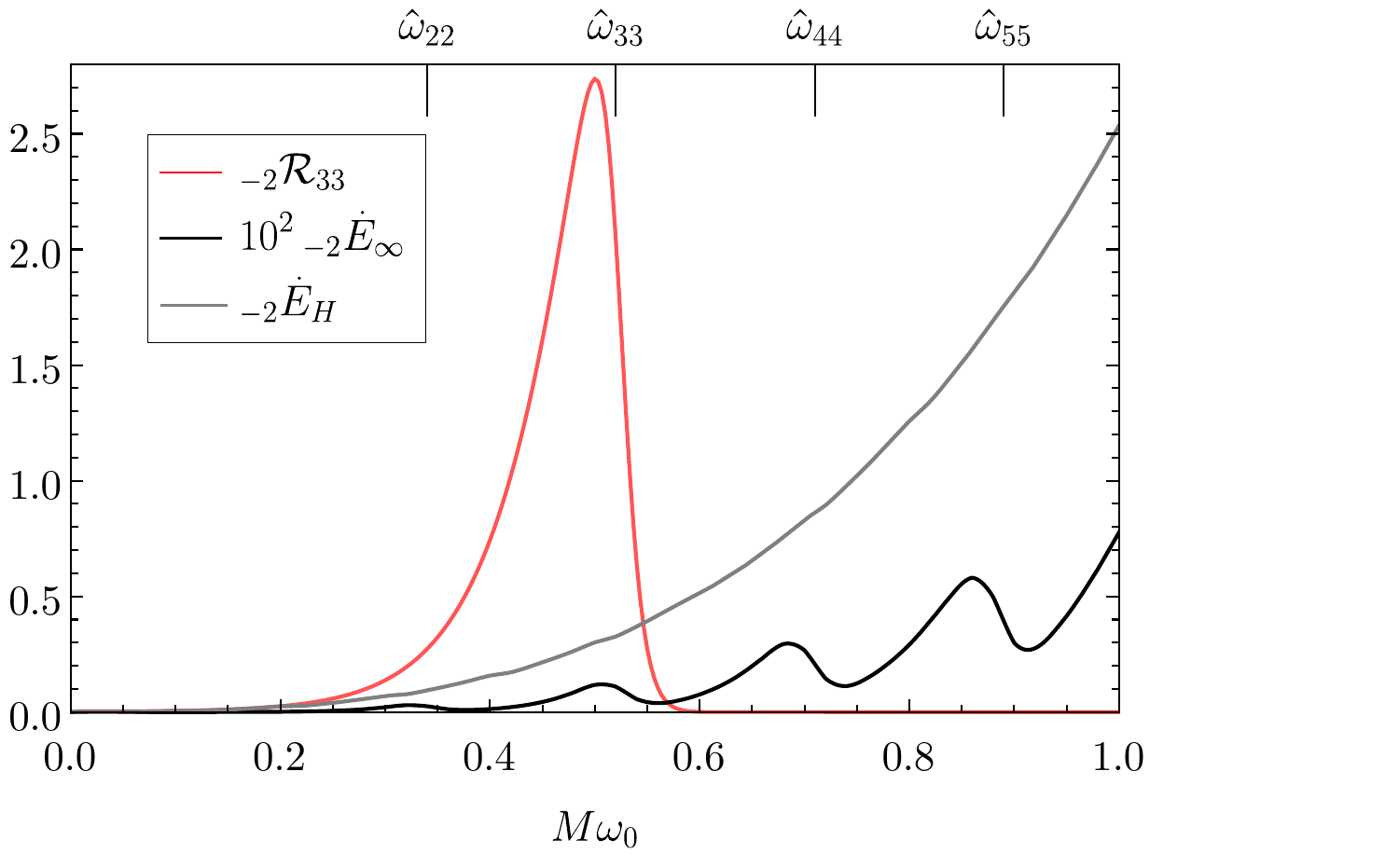}
\caption{
Energy output when a small binary stands at the \acs{ISCO} of a \acs{SMBH} with spin $a=0.9M$, as a function of the orbital frequency of the binary components, $\omega_0$. 
The modal energy output, as measured by the ratio with respect to flat spacetime $_{-2}\mathcal{R}$, peaks at a finite $\omega_0$ well described by the lowest \acs{QNM} (cf. Table~\ref{tab:MaxFlux}).
Also shown is the flux integrated over all modes: it has a substantial component going down the \acs{SMBH} horizon ($\dot{E}_H$), and the total flux at infinity is modulated by \acs{QNM} contributions ($\dot{E}_\infty$).
Here, $\hat{\omega}_{\ell m} = M\omega_\text{QNM}/2$.
}
\label{fig:Flux} 
\end{figure}
\begin{table}[t]
\adjustbox{width=0.9\width}{
\begin{tabular}{lccccccc} \hline\hline
$\ell$ &$s$&$a/M$& $M\omega_\text{QNM}/2$ & $M\omega_{0_\text{LR}}$ &  $M\omega_{0_\text{ISCO}}$ &$_s\mathcal{R}_{\text{LR}}$ &$_s\mathcal{R}_{\text{ISCO}}$\\ 
\hline   
$2$  &$0$&$0$& $0.242$ & $0.242$ & $0.189$  & $4.5$  & $2.0$ \\
$2$  &$-2$&$0$& $0.186$ & $0.175$ &$0.156$& $0.6$  & $1.5$ \\
$2$  &$-2$&$0.9$& $0.335$ & $0.332$ & $0.319$ & $88.0$  & $0.8$ \\
\hline
$3$  &$0$&$0$ & $0.338$ & $0.337$ &$0.255$  & $10.0$ & $2.5$ \\
$3$  &$-2$&$0$& $0.300$ & $0.289$ & $0.250$ & $2.0$  & $2.3$	 \\
$3$  &$-2$&$0.9$& $0.522$ & $0.520$ & $0.500$ & $515.8$  & $2.7$	\\
\hline
$4$  &$0$& $0$ & $0.434$ & $0.433$ & $0.317$  & $21.6$ & $3.0$ \\
$4$  & $-2$& $0$ & $0.405$ & $0.395$ & $0.326 $& $5.6$  & $3.0$ \\
$4$  & $-2$& $0.9$& $0.705$ & $0.704$ &$ 0.675$ & $1896.4$ &  $5.4$	 \\
%
%
\hline\hline
\end{tabular}
}
\caption{Frequency $M\omega_{0\,X}$ which maximizes the energy output of a small binary standing at location $X$ close to a \acs{SMBH}, in a given $(\ell, \ell)$ mode, as measured by the ratio $_s\mathcal{R}$ ($s=0,-2$ for scalar or gravitational perturbations, respectively). The binary's \acs{CM} is static, and sitting at the \acs{LR} or at the \acs{ISCO}. Notice the excellent agreement with the lowest \acs{QNM} frequency. The results for orbiting binaries are similar.}
\label{tab:MaxFlux}
\end{table}

We start by using the small binary as a tuning fork, placing it at some fixed radius, with its \acs{CM} fixed with respect to distant observers, and letting its frequency $\omega_0$ vary. We want to compare the (time-averaged) flux of energy in each $(\ell, \,m)$ mode with the corresponding value in flat space. We define this as the ratio
\beq
_s\mathcal{R}_{\ell\,m} =\, _s\dot{E}_{\ell\,m}/_s\dot{E}_{N\,\ell\,m}\,,
\eeq
where $N$ refers to the result in flat space, i.e setting the mass of the central \acs{BH} to zero. This can be computed numerically or through analytical estimates using the method of matched asymptotic expansions~\cite{Cardoso:2019nis}. When the binary is put at large distances from the central \acs{BH} this ratio tends to unity. 
%
%

Figure~\ref{fig:Flux} shows the behavior of $_{-2}\mathcal{R}_{33}$ as the small binary frequency $\omega_0$ changes, for a binary sitting at the \acs{ISCO} of a \acs{SMBH}.
The behavior is similar for other modes and fields. We observe a peak which we identify as a resonant excitation of the $\ell=m=3$ \acs{QNM}. As shown in Table~\ref{tab:MaxFlux}, the location of the peak is well described by the lowest \acs{QNM} frequency~\cite{Berti:2009kk}, for general binary locations. When the small binary is placed at the \acs{LR}, the agreement is excellent (better than $1$\% for scalars, and $4$\% for \acs{GW}s for the lowest modes). Recall from our discussion at the beginning of Chapter~\ref{ch:LR}, that \acs{QNM}s can be interpreted as waves marginally trapped in unstable orbits at the \acs{LR}~\cite{Cardoso:2008bp}. We conclude that a hierarchical triple system behaves as a driven harmonic oscillator~\cite{georgi1993physics}, where the small binary is the external harmonic force and the central \acs{BH} behaves as a damped oscillator.

As a side note, this behavior is analogous to the Purcell effect in quantum electrodynamics~\cite{PhysRev.69.37, PhysRevLett.110.237401}, describing the enhancement in the spontaneous decay of a quantum emitter inside a cavity, when its frequency matches those of the cavity modes. Our results are consistent with recent findings~\cite{PhysRevLett.110.237401}, namely that the contribution to the power spectrum independent of $r_\text{CM}$ is described by a Lorentzian curve $\mathcal{R}\propto \omega_{\text{QNM}}^2/(\omega_{\text{QNM}}^2+4 Q^2(\omega_0 - \omega_{\text{QNM}})^2 )$, where 
$Q=\text{Re}( \omega_{\text{QNM}} )/2 \text{Im}( \omega_{\text{QNM}})$ 
is the quality factor of the central \acs{BH}. Our results extend those of Ref.~\cite{Thornburg:2019ukt}, which also observed the resonant excitation of \acs{QNM}s in very eccentric \acs{EMRI}s, during passage on the periapsis. The effect is stronger the closer the particle can get to the \acs{LR}~\cite{Price:2015gia}.

As a rule of thumb, the flux peaks at lower frequencies the further the small binary is placed from the \acs{BH}, in agreement with blueshift/redshift corrections. Note that $\mathcal{R}$ smaller than unity does not imply that the system is emitting less energy than expected, since a portion of the radiation falls into the \acs{BH}. Also, a possible \acs{CM} orbital motion contributes to a shift in the resonant frequencies by $\pm m \Omega_\text{CM}$, fully consistent with our results. The maximum value of $\mathcal{R}$ in the entire $(r_\text{CM},\omega_0)$ parameter space does not occur precisely at the \acs{LR}, but close to it. The maximum is attained at locations $r_\text{CM}$ closer to the horizon for large $\ell$.
Finally, the magnitude of the resonance grows with $\ell$. For a fixed \acs{CM} location $r_\text{CM}$ and multipole $\ell$ we searched for $\omega_0$ for which $_s\mathcal{R}$ is a maximum $_s\mathcal{R}_\text{ peak}$.
We find an exponential dependence on $\ell$, $_s\mathcal{R}_\text{ peak}\sim a+b\exp(c\cdot \ell)$, at large $\ell$ with $a,\, b,\,c$ constants. 

Since we are using a mode decomposition centered at the \acs{SMBH}, radiation has support in higher modes as the binary is placed further away from it~\cite{Berti:2005gp, Gualtieri:2008ux}. Consequently, the lowest modes will not be dominant, and we need to sum over a sufficient amount of modes for the total fluxes to converge. Already for a small binary at the \acs{ISCO} of a non-rotating \acs{BH}, we find that the \acs{GW} flux at infinity is comparable to that at the horizon of the \acs{SMBH}. To compute it we use the Starobinsky identities that relate $\psi_0$, which controls the radiative degrees of freedom at the \acs{BH} horizon, with $\psi_4$ which controls them at infinity as seen in Eq.~\eqref{eq:RadiativeDegrees}~\cite{StarobinskyChurilov, 1974ApJ...193..443T}. At the \acs{BH} horizon, the ingoing solution of the $s=-2$ master function in the Teukolsky equation~\eqref{eq:TeukolskyMaster} behaves as
\beq
\Psi \sim \Delta^2 Z_\text{in} e^{-i(\omega t + k r_*)} e^{i m \varphi} \, _{-2}S_{\ell m} \left(\theta \right) \, ,
\eeq
with $k=\omega - m \Omega_\text{H}$ and $S_{\ell m} \left(\theta \right)$ the spin-weighted spheroidal harmonic obeying to~\cite{NIST:DLMF}
\beq
&&\frac{1}{\sin\theta}\frac{d}{d\theta}\left( \sin \theta \frac{dS}{d\theta} \right) \nonumber \\
&+& \left(a^2 \omega^2 \cos^2 \theta - \frac{m^2}{\sin^2 \theta} - \frac{2 a \omega s \cos \theta}{\sin^2 \theta} - \frac{2ms \cos \theta}{\sin^2 \theta} \right) S \nonumber \\ &= &- \left(\lambda- s^2 \right)S \, ,
\eeq
where the eingenvalues $\lambda$ are available online from the Black Hole Perturbation Theory Toolkit~\cite{BHPToolkit}. We can then compute the energy flux carried by \acs{GW}s through the \acs{BH} event horizon with
\beq
\dot{E}^H &=& \frac{128 \omega k \left(k^2 + 4 \epsilon^2_\text{BH} \right)\left(k^2+ 16 \epsilon_\text{BH}^2 \right)\left(2Mr_+\right)^5}{\left|C\right|^2}\left|Z_\text{in}\right|^2 \, , \\
\epsilon_\text{BH} &=& \frac{\sqrt{M^2-a^2}}{4Mr_+} \, , \\
\left|C \right|^2 &=& \left(Q^2 + 4 a \omega m - 4 a^2 \omega^2 \right) \left[ \left(Q- 2\right)^2 + 36 a \omega m -36 a^2 \omega^2 \right] \nonumber \\
&+& \left(2Q-1\right)\left(96a^2\omega^2 - 48a \omega m\right) + 144 \omega^2\left(M^2 - a^2\right) \ ,\\
Q &=& \lambda + a^2 \omega^2 - 2 a \omega m \, . 
\eeq
Formally, this equivalence is valid only in the frequency-domain but has been shown to yield correct results for circular orbits in the time-domain also.

As seen in Fig.~\ref{fig:Flux}, the effect is more dramatic when spin is included. This peculiar aspect is due to the similar length scales of the central \acs{BH} horizon and the radiation wavelength. \acs{GW}s are then efficiently absorbed by the \acs{BH}, in clear contrast with the inspiral phase of an \acs{EMRI}, whose 
wavelength is much larger than the \acs{BH} radius. This is our second result: hierarchical triple systems where the \acs{SMBH} occupies a large fraction of the small binary's sky will naturally probe strong-field physics, since the fraction of radiation that falls into the \acs{SMBH} is non-negligible. This property is essential for the dynamical evolutions of these systems.

For a fixed radius $r_\text{CM}$, the field has support on higher $\ell$ modes as the binary is vibrating at higher frequencies $\omega_0$. If we place it close enough to the \acs{SMBH}, it can resonantly excite its \acs{QNM}s, leading to characteristic peaks in the flux at infinity/horizon, as seen in Fig.~\ref{fig:Flux}. These structures correspond to the single multipolar excitations. 

\section{Waveforms: Doppler, aberration \& lensing}\label{sec:Waveform_tuningfork}

%
\begin{figure}[t]
\centering
\includegraphics[scale=0.65]{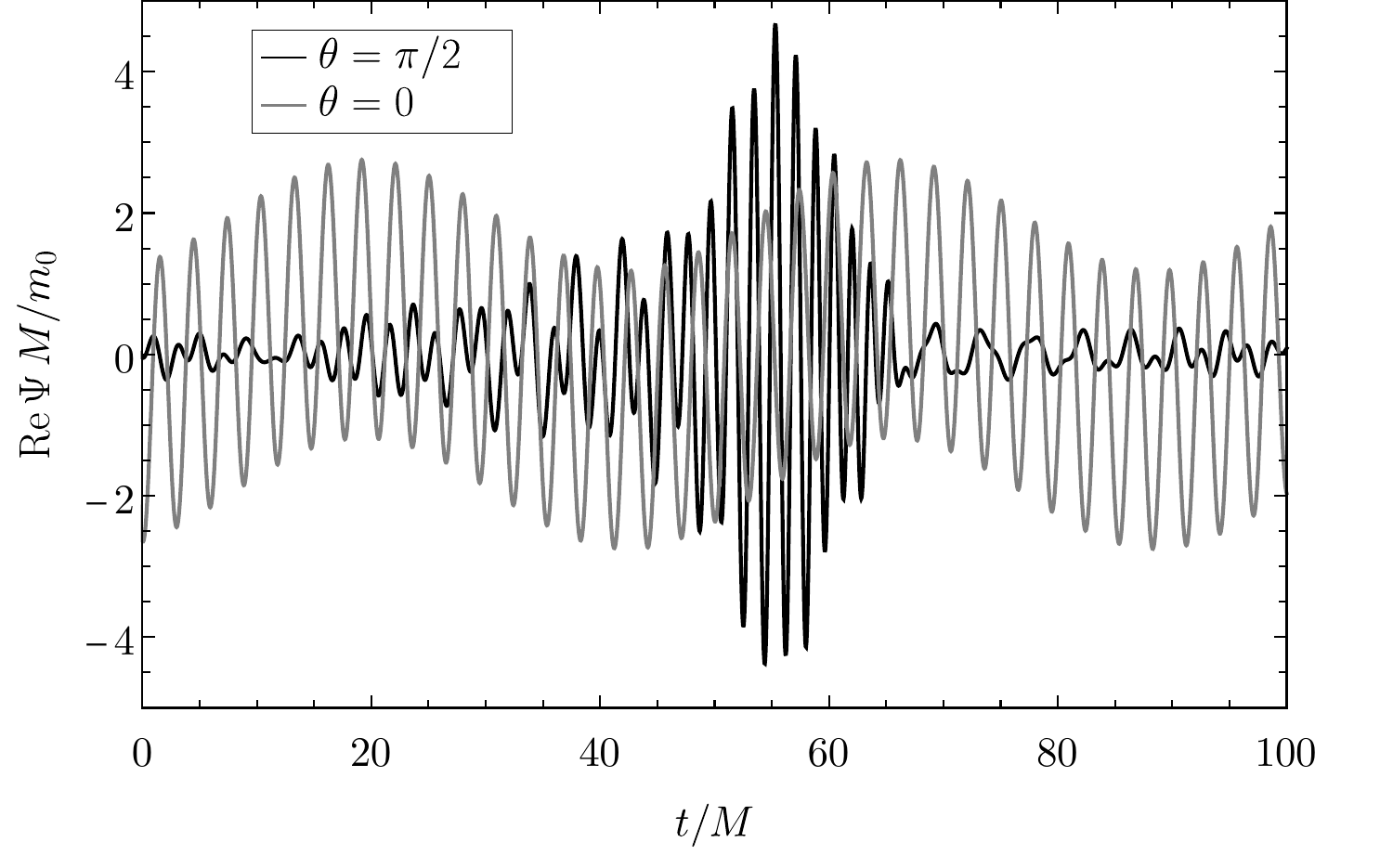}
\caption{Teukolsky function $\Psi$ measured by a stationary observer at large distances (either edge- or face-on, $\theta=\pi/2,\,0$ respectively), for a stellar binary around the \acs{ISCO} of a non-rotating \acs{BH} (we removed the \acs{CM} contribution, which just causes a low-frequency modulation). The orbital \acs{CM} period is $T_\text{CM}\approx 93 M$ and at $t=0M$ the observer is aligned with the small binary. Doppler effect induces frequency shifts, relativistic beaming and gravitational lensing modulations in the amplitude. The maximum blue-shift is well described by $\omega_\text{max}=\omega_0'\Upsilon (\,(\Upsilon+v_\text{CM})/(\Upsilon-v_\text{CM}) )^{1/2}$, with $\Upsilon=\sqrt{1-2M/r_\text{CM}}$, $M\omega_0'=1$ the proper frequency and $v_\text{CM}$ the \acs{CM} velocity~\cite{1972ApJ...173L.137C, 10.1093/mnras/stv172}. In this case, $\omega_\text{max}/\omega_0' \approx 1.4$.}
\label{fig:WaveCirc} 
\end{figure}
\begin{figure}[t]
\centering
	\includegraphics[scale=0.28]{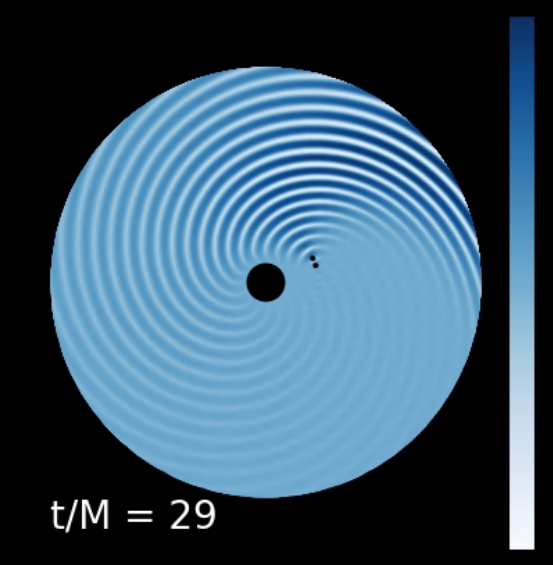}
	\includegraphics[scale=0.28]{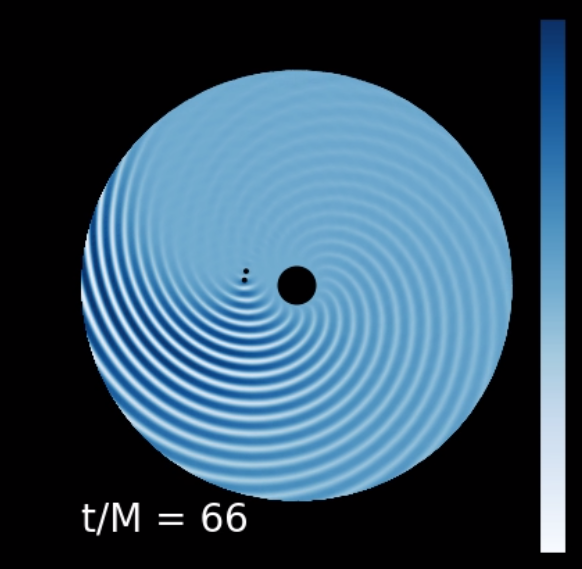} 
	\caption{Snapshots of the time-evolution of the hierarchical triple system with the \acs{CM} of the small binary orbiting the \acs{SMBH} at the \acs{ISCO} anti-clockwise. The shaded blues illustrate the wave emission by the system projected on the equatorial plane. There is a focus of radiation along the direction of motion, corresponding to relativistic beaming. We also observe the distortion of the path taken by waves due to gravitational pull exerted by the central \acs{BH}.}\label{fig:gifcirc}
\end{figure}

In addition to energy fluxes, we can also study qualitative strong-field effects in the waveforms from hierarchical triple systems. In Fig.~\ref{fig:WaveCirc} we show the Teukolsky function extracted at large distances for a small binary on circular motion at the \acs{ISCO} of a non-rotating \acs{BH}. We removed the (linear) \acs{CM} contribution, which only induces a low-frequency modulation. Observers on the equatorial plane see gravitational and Doppler-induced frequency shifts, consistent with analytical predictions when the \acs{CM} is moving towards the observer~\cite{1972ApJ...173L.137C, 10.1093/mnras/stv172}. The amplitude of the wave can vary by orders of magnitude because of relativistic beaming~\cite{Torres-Orjuela:2018ejx,Torres-Orjuela:2020cly, Gupta:2019unn} and gravitational lensing~\cite{Ezquiaga:2020spg, Ezquiaga:2020gdt}. Relativistic beaming focuses radiation along the direction of motion, and is significant for fast \acs{CM} motion (in this case $v_\text{CM}\approx 0.4$). Hence, when the binary is moving away from the observer, the radiation reaching it is small, and vice-versa. The maximum amplitude does not occur precisely when the stellar binary is moving towards the observer ($t\sim 70 M $ in Fig.~\ref{fig:WaveCirc}) but slightly before, when it is still behind the \acs{SMBH} with respect to the observer. This is due to lensing by the central \acs{SMBH}, which distorts the path taken by \acs{GW}s and concentrates radiation on certain directions, amplifying the signal~\cite{Nambu:2015aea,Nambu:2019sqn}. This effect is more relevant for larger frequencies, closer to the geometric optics limit when the radiation wavelength is much smaller than the \acs{SMBH} radius. These different effects are illustrated in Fig.~\ref{fig:gifcirc}, which corresponds to two snapshots of an animation representing the time evolution of wave emission of the hierarchical triple system.  

Observers facing the plane of motion ``face-on'' ($\theta = 0$) do not measure such modulations since the motion of the \acs{CM} is now transverse. The only feature is a modulation in amplitude coming from the \acs{CM} motion, which has also been reported in post-Newtonian studies of triple systems~\cite{Gupta:2019unn}.

\section{Discussion}\label{sec:Discussion_tuningfork}

This chapter highlights both the versatility of \acs{BH} perturbation theory and the robustness of our numerical framework to study \acs{GW} emission in more complicated systems than a standard binary. Effects like \acs{GW} lensing, aberration and amplitude modulations are naturally built-in, without the need to prescribe slow-motion approximations and force the orbital motion into the quadrupolar approximation. As a drawback, we need to sum over various multipoles in order to resolve the length scales of the small binary. 

One could question if it is physically possible for a stellar-mass binary to get so close to a \acs{SMBH} before being disrupted. While quantifying a detectability rate for the resonances we studied goes beyond the scope of our
work, we can make an estimate based on the Hills mechanism already mentioned in Chapter~\ref{ch:Cloud}~\cite{1988Natur.331..687H, Addison:2015bpa, Suzuki:2020vfw}. Disruption occurs if the tidal forces induced by the central \acs{BH} overcome the binary's self-gravity, which happens at a radius $r_\text{CM} \sim 2\delta r  \left(M/2m_p\right)^{1/3}$. The binary's frequency will be related to its separation by the Kepler's law $\omega_0 \sim \sqrt{2 m_p/\delta r_\text{CM}^3}$. We thus find $r_\text{CM} \lesssim 1/(M\omega_0)^{2/3} M$. Already for $M \omega_0 = 0.2$, tidal disruption happens at $r_\text{CM} \sim 5.84 M$, smaller than the \acs{ISCO} of a Schwarzschild \acs{BH}. Thus, small binaries very close to a central \acs{BH} and oscillating at relevant frequencies of the system have astrophysical interest. This is supported by more sophisticated numerical works~\cite{Brown:2018gar}.

We neglected spin-spin effects in the motion of the binary. The corrections are proportional to $\sigma=q J/m_0^2$, with $J$ the angular momentum of the binary~\cite{Jefremov:2015gza}. Again using Kepler's law, one finds that corrections to the motion scale like $\sigma\propto q^{2/3}$, which are extremely small for the systems we consider. 

We also did not consider a situation where the small binary is coalescing, or when its \acs{CM} describes a highly eccentric orbit around a spinning \acs{SMBH}~\cite{Cardoso:2020iji}. As the binary gets closer to the \acs{LR} in these orbits, the resonant excitation of the \acs{SMBH}'s modes is enhanced and can lead to the excitation of superradiant modes~\cite{Brito:2015oca}.
Another interesting hierarchical triple system is a pair of same-sized \acs{BH}s and a third lighter compact object orbiting around them. These spacetimes have been shown to have global properties not present in isolated BHs (e.g. global QNMs)~\cite{Bernard:2019nkv, Ikeda:2020xvt} and our results suggest that the lighter object can excite these global modes. 

\cleardoublepage
\part{Binaries in astrophysical environments}\label{pt:EMRIs}
\chapter{Galactic Black Hole}\label{ch:GBH}

In the Introduction, we went to great lengths to illustrate that the lack of relativistic models has arguably been the major flaw of studies on environmental effects for \acs{GW} astronomy. In the last part of this thesis we will use standard techniques of \acs{BH} perturbation theory to develop a generic, fully-relativistic formalism to handle environmental effects in \acs{EMRI}s in spherically-symmetric, but otherwise generic backgrounds~\cite{1967ApJ...149..591T,Detweiler:1985zz,Chandrasekhar:1991fi,Kojima:1992ie,Allen:1997xj,Sarbach:2001qq,Martel:2005ir,Martel:2003jj}. The main novelty in our approach is to treat perturbations to
matter on the same ground as those induced by the binary in the gravitational field, which allows to capture for free environmental effects such as dynamical friction, accretion, and halo feedback. This is inspired in studies of perturbations to relativistic stars~\cite{Allen:1997xj}, but here we want to apply it to extended distributions of matter surrounding \acs{BH}s.

Our starting point is therefore a static, spherically symmetric spacetime~\eqref{eq:SphericalLineElement} describing a \acs{BH} immersed in some environment, like an accretion disk or \acs{DM} halo. The environment is characterized in general by an anisotropic fluid with energy-momentum tensor
\beq
T_{\mu\nu}^{\text{env}}= \rho u_\mu u_\nu + p_r k_\mu k_\nu + p_t \Pi_{\mu\nu} \, , \label{eq:EnergyMomentumAnisotropic}
\eeq
where $\rho$ is the total energy density of the fluid, $p_r$ and $p_t$ are its radial and tangential pressure, respectively, $u^\mu$ the $4$-velocity of the fluid, $k^\mu$ a unit spacelike vector orthogonal to $u^\mu$, such that $k^\mu k_\mu = 1$ and $u^\mu k_\mu = 0$, and $\Pi_{\mu\nu}= g_{\mu\nu} + u_\mu u_\nu - k_\mu k_\nu$ is a projection operator orthogonal to $u^\mu$ and $k^\mu$ (environmental quantities are hereafter denoted with a superscript ``env''). We now envision a secondary object of mass $m_p$ (a star, asteroid or stellar-mass \acs{BH} for example) orbiting the primary \acs{BH} and causing perturbations to the geometry and matter energy-momentum tensor
\beq
g_{\mu\nu}=g^{(0)}_{\mu\nu}+ g^{(1)}_{\mu\nu},\,\,\,
T^\text{env}_{\mu\nu}&=&T^\text{env$(0)$}_{\mu\nu}+T^\text{env$(1)$}_{\mu\nu},
\eeq
where a superscript ``($1$)'' denotes perturbations to the background ``($0$)''. The $0$-th order Einstein's equations give
\beq
\frac{A'}{A} &=& \frac{2}{r}\frac{m +4\pi r^3 p_r^{(0)}}{r-2m } \, , \\
\rho^{(0)} &=& \frac{m'}{4\pi r^2} \, , \\
\frac{dp_r^{(0)}}{dr} &=& -p_r^{(0)} \frac{2r - 2m + 4\pi r^3 p_r^{(0)} - r m'}{r(r-2m)}  \, , \\
p_t^{(0)} &=& \frac{m\,m'}{8\pi r^2 (r-2m)} + \frac{2r m' - m}{2(r-2m)} p_r^{(0)} 
\eeq
where the prime denotes a derivative with respect to $r$, and $m=m(r)$ is the mass function
\beq
m(r) = \frac{r}{2}\left(1 - B(r)\right) \, .
\eeq

In Appendix~\ref{app:GBH}, we derive a set of ``wave-like'' partial differential equations governing the time evolution of gravitational and matter perturbations for this setup. We extend the framework introduced in Sec.~\ref{sec:BHPT} to include matter perturbations. Also, recall that spherical symmetry allows us to separate this problem in the axial and polar sectors. For the axial sector, gravitational perturbations decouple from matter ones, and we obtain a single master equation with an effective potential~
\beq
{\cal L}_1 \Psi^\text{ax}- V^\text{ax}\Psi^\text{ax} = S^\text{ax} \, ,
\eeq
where $ {\cal L}_v= v^2 \partial^2 /\partial r_*^2 -\partial^2/\partial t^2$ denotes the wave operator with characteristic speed of propagation $v$. Axial perturbations
propagate with the speed of light $v = 1$, and in vacuum the equation above corresponds to the Regge-Wheeler equation~\eqref{eq:MasterZRW}. The source depends on the motion of the secondary object (which we explicitly compute for circular motion in the Appendix~\ref{app:GBH}).

The polar sector is more involved because matter perturbations source gravitational ones and vice-versa. We re-expressed the problem as a set of $3$ equations for $\vec{\Psi}=(S,K,\delta\rho)$ 
\beq
\hat {\cal L} \vec{\Psi}= \hat {\textbf{B}} \partial_{r_*}\vec{\Psi} + \hat{\textbf{A}}\vec{\Psi}+\vec{S}\,,
\eeq
with $S = A/r \,( H_0 - K) $, and $\hat{\cal L} \vec{\Psi}=\left({\cal L}_1 S,{\cal L}_1 K ,{\cal L}_{c_{s_r}}\delta \rho\right)$, i.e., $S,\, K$ have characteristic velocity $v=1$, and $\delta \rho$ has $v=c_{s_r}$. We recall $K$ and $H_0$ are the gravitational perturbations given by Eq.~\eqref{eq:ExpansionMetric} and $\delta \rho$ is the perturbation of the energy density of the fluid. $c_{s_r}$ is the speed of sound along the radial direction and depends on the internal properties of the environment, i.e. its equation of state. This relates the pressure with density perturbations via
\beq
\delta p_{t,r}^{\ell m}(t,r)=c_{s_{t,r}}^2(r) \, \delta \rho^{\ell m}(t,r)\,. \label{eq:BarotropicEq}
\eeq
where $c_{s_{r}}(r)$ and $c_{s_{t}}(r)$ are, respectively, the radial and transverse sound speeds.

\section{Black Holes in Dark Matter Halos}

\subsection{The Hernquist Profile}

We now want to apply our framework to a particular background describing a \acs{BH} surrounded by a galactic \acs{DM} halo. Observations and large-scale simulations guide the profile of the halo matter distribution. Here, we pick Hernquist-type distributions appropriate to describe central bulges of galaxies and elliptical galaxies at the Newtonian level~\cite{1990ApJ...356..359H}. Their matter density is 
\beq
\rho = \frac{M a_0}{2\pi r \left(r + a_0\right)^3} \label{eq_rho:hernquist} \, ,
\eeq
where $M$ is the total mass of the ``halo'' and $a_0$ a typical lengthscale, where for astrophysical solutions $a_0 \gg M$. Recall that the Milky Way halo has 
$\sim 10^{12} M_\odot$ extended along $\sim 10^2\, \text{kpc}$, which in geometric units gives $M/a_0 \sim 10^{-6}$. Our choice is mainly practical since the Hernquist profile has ``well-behaved'' mathematical properties in comparison with other popular profiles such as the Navarro-Frenk-White~\cite{Navarro:1995iw}, Jaffe~\cite{1983MNRAS.202..995J} or King~\cite{1962AJ.....67..471K} which are known to have pathological behavior at large distances. 

The Hernquist model -- as well as others in the same ``family''~\cite{1990ApJ...356..359H,Navarro:1995iw,1983MNRAS.202..995J,1962AJ.....67..471K} --
have an increased density in the cores of the galaxies. However, in the presence of a \acs{BH} at the core, the density profile
is zero close to the horizon~\cite{Gondolo:1999ef,Sadeghian:2013laa, Speeney:2022ryg}. The \acs{DM} profile develops a cusp close to the horizon, with a length scale dictated by the \acs{BH}. The density profile vanishes at the horizon. The precise details of the profile depend on the equation of state, but eventually give way to Hernquist-like profiles.

\subsection{The Einstein Cluster}

Now let us place a \acs{BH} at the center of the Hernquist profile~\eqref{eq_rho:hernquist} and find a spherically symmetric \acs{GR} geometry which on small scales describes a \acs{BH} and on large scales describes matter distributed according to~\eqref{eq_rho:hernquist}. We can follow Einstein in his construction of a stationary system of many gravitating masses, an ``Einstein Cluster''~\cite{Einstein:1939ms,Geralico:2012jt}, and generalize it to include a central \acs{BH}. This recipe takes particles in all possible circular geodesics, and deals with an ``average'' stress-energy tensor~\cite{Einstein:1939ms,Geralico:2012jt}, characterized by the matter density $\rho$. 

The Einstein construction assumes then an effective energy momentum tensor $\langle T^{\mu\nu}\rangle=\frac{n}{m_p}\langle p^{\mu}p^{\nu}\rangle$, with $n$ the number density of particles with mass $m_p$, and $p^{\mu}$ the four-momentum satisfying the geodesic equations. This construction is equivalent to having an anisotropic material with energy-momentum tensor as in Eq.~\eqref{eq:EnergyMomentumAnisotropic} with only tangential pressure $p_t^{(0)}$, and vanishing radial pressure ($p_r^{(0)}=0$).

The second step in the Einstein Cluster construction is to assign a mass function $m(r)$ to the system. W explore the following choice inspired by the Hernquist profile
\beq
m(r)=M_\text{BH}+\frac{M r^2}{(a_0+r)^2}\left(1-\frac{2M_\text{BH}}{r}\right)^2\,.
\eeq
At small distances, this profile describes a source of mass $M_\text{BH}$ and at large distances recovers the mass function of the Hernquist distribution. Note that for astrophysical systems, the mass of the halo is much bigger than the mass of the \acs{BH}, so we have the hierarchy $a_0 \gg M \gg M_\text{BH}$. We experimented with other mass functions and the qualitative conclusions below do not change. 

Plugging this mass function in the Einstein's equations, we obtain a simple analytic solution for the background spacetime
\beq
A(r)&=&\left(1-\frac{2M_\text{BH}}{r}\right)e^{\Upsilon(r)}\,,\label{eq_fhairy}\\
\Upsilon(r)&=&-\pi\sqrt{\frac{M}{\xi}}+2\sqrt{\frac{M}{\xi}}\arctan{\frac{r+a_0-M}{\sqrt{M\xi}}}\,,\\
\xi&=&2a_0-M+4M_\text{BH}\,. \\
4\pi \rho(r)&=&=\frac{2M(a_0+2M_\text{BH})(1-2M_\text{BH}/r)}{r(r+a_0)^3}\,.
\label{eq:hernquist_GR}
\eeq

Let us dissect some of its properties. It has a horizon at $r_\text{H} = 2M_\text{BH}$, as Schwarzschild, and a curvature singularity at $r=0$, thus describing a \acs{BH} spacetime. At large distances away from the \acs{BH}, the Newtonian potential recovers that of the Hernquist profile, and the spacetime is asymptotically flat. The ADM mass is $M + M_\text{BH}$. For astrophysical relevant solutions, the gravitational potential at far away distances is dominated by the halo and the Ricci scalar behaves as $R\sim4Ma_0 / r^4$, while as it approaches the horizon it goes as $R\sim M/(a_0^2 M_\text{BH})$, which can be made small in a controlled way. For non-astrophysical parameters it is possible to have pathological solutions. For example if $M > 2\left(a_0 + 2M_\text{BH}\right)$, there are curvature singularities at $r=M - a_0 \pm \sqrt{M^2 - 2Ma_0 - 4M\,M_\text{BH}}$.

The weak and strong energy conditions are satisfied everywhere. However, since the matter density vanishes at the horizon but the tangential pressure remains finite, the dominant energy condition will be violated close to it. As we discussed above, the near-horizon region is very depleted so this violation should have little impact on the dynamics, as the pressure and density can be made arbitrarily small in this region.

Close to the central \acs{BH} we find that the redshift factor $e^\Upsilon\sim 1-2M/a_0$, a property that is generic for other configurations~\cite{1911MNRAS..71..460P, Liebling:2012fv,Annulli:2020lyc}.

\subsection{Redshift and Light Rings}

The spacetime above can be used as a proxy to explore the phenomenology of \acs{BH}s and \acs{GW}s from objects deep in the galactic potential.
Let us start by doing the simplest thing one can do when faced with a new spacetime geometry, which is to study geodesic motion. The location of the \acs{LR} can be computed as in Section~\ref{sec:LRKey} and is given by the roots of $r(r_\text{LR}) = 3m(r_\text{LR}) $. With this radius determined, we can then compute the \acs{LR} frequency $\Omega_\text{LR}$, the Lyapunov exponent $\lambda_\text{L}$ and the critical impact parameter $b_\text{c}$ for the capture of high-frequency waves which were introduced previously in Chapter~\ref{ch:LR}. Expanding these quantities in power of $M/a_0$ (which we recall needs to be small for astrophysical setups), we find up to order ${\mathcal O}(1/a_0^3)$
\beq
r_\text{LR} &\approx& 3M_\text{BH}\left(1 + \frac{M M_\text{BH} }{a_0^2} \right) \, , \\
M_\text{BH} \Omega_\text{LR}  &\approx& \frac{1}{3\sqrt{3}}\left(1 - \frac{M}{a_0} + \frac{M \left(M + 18 M_\text{BH}\right)}{6 a_0^2} \right)  \, , \label{eq:LRFreqHalo}\\
M_\text{BH} \lambda_\text{L} &\approx& \frac{1}{3 \sqrt{3}} \left(1 - \frac{M}{a_0} + \frac{M^2}{a_0^2} \right) \, , \\
b_\text{c}&=&3\sqrt{3}M_\text{BH}\left(1+\frac{M}{a_0}+\frac{M(5M-18M_\text{BH})}{6a_0^2}\right)\,, 
\eeq
The first order correction linear in $M/a_0$ is simply signaling a redshift of the mass scale of the system $M \rightarrow M\left(1 + M/a_0\right)$. The first ``non-linear" corrections in $M/a_0$, that would indicate the presence of new physics, appear at orders of magnitude which are too small to be detectable by \acs{BH} imaging experiments. We therefore conclude that tests on the nature of \acs{BH}s based on \acs{LR} physics like the ones performed by the Event Horizon Telescope and the GRAVITY collaboration~\cite{EventHorizonTelescope:2019dse,GRAVITY:2018ofz} are not going to be spoiled by the presence of a non-vacuum environment. \textit{A priori}, this is expected because the matter around the \acs{LR} region is very depleted but it is reassuring to see it be derived from the formalism without the need to rely on Newtonian-like approximations. Results are similar for other orbits of interest, like the \acs{ISCO}.

\section{Axial sector}

Having understood geodesic motion in our playground model for a \acs{BH} surrounded by a \acs{DM} halo, we move on to the study of its dynamics at a perturbative level, as introduced in Sec.~\ref{sec:BHPT}. We start with the more straightforward axial sector, where gravitational perturbations completely decouple from matter ones. Consequently, it is governed by the single master wave equation in Eq.~\eqref{eq:master_axial}.

\subsection{QNMs and Love Numbers}
%
\begin{figure}[htb]
	\centering
	\includegraphics[scale=0.28]{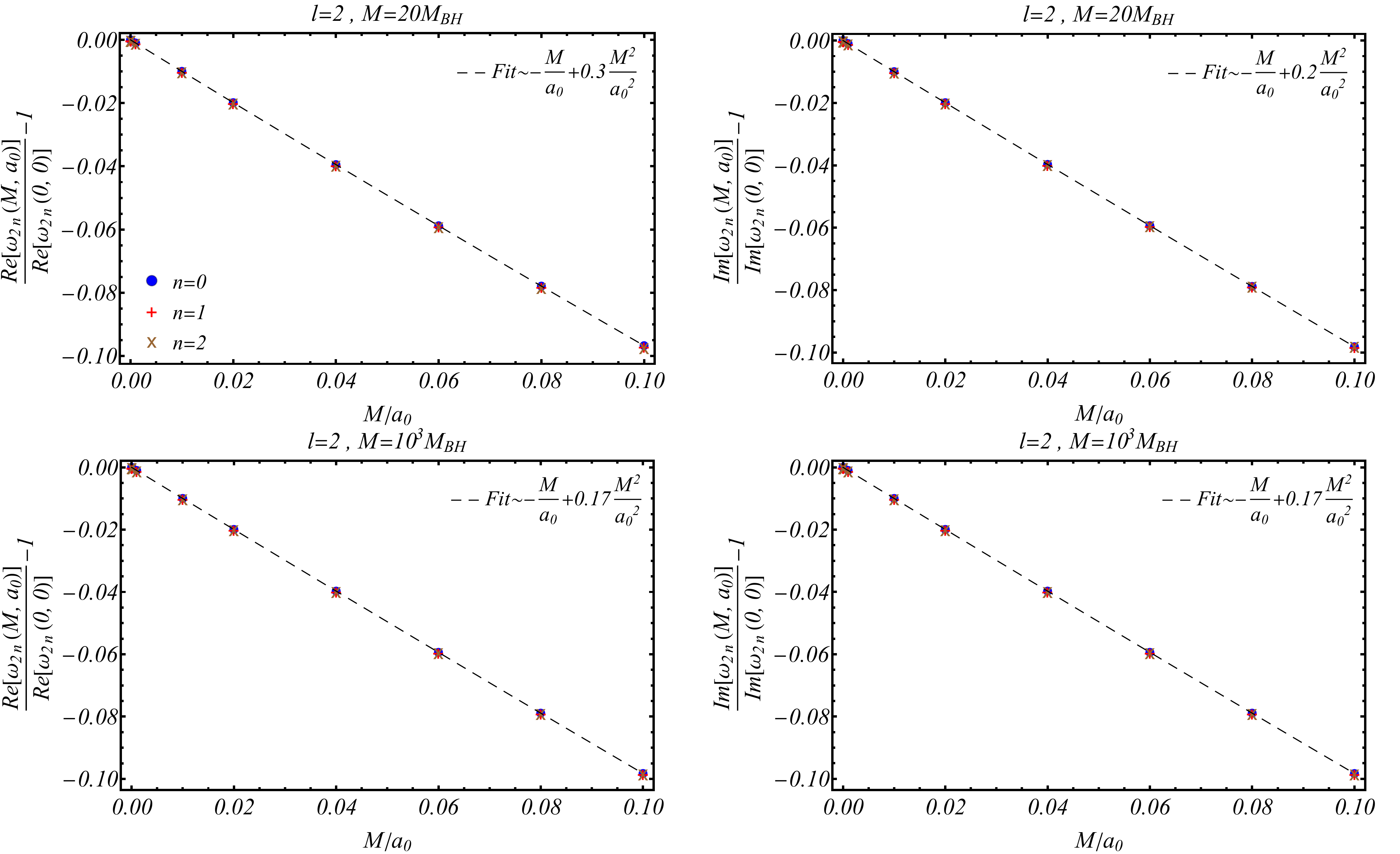}
	\caption{\textit{Top panel:} Real and imaginary part deviations of galactic axial \acs{QNM}s $\omega_{\ell n}(M,a_0)$, with $M=20 M_\text{BH}$, from the vacuum Scharzschild \acs{QNM}s $\omega_{\ell n}(0,0)$ for $\ell=2$ as a function of the compactness $M/a_0$. Different plot markers denote different overtones. The dashed black line represents a fit to the relative difference for powers of $M/a_0$. \textit{Bottom panel:} Same as the top panel with $M=10^3 M_\text{BH}$. The fit agrees with the \acs{LR} corrections obtained in Eq.~\eqref{eq:LRFreqHalo}.}\label{fig:axial_QNM_fit_l2}
\end{figure}

Following the same steps as in Chapter~\ref{ch:TLNs}, we can solve the axial problem in the static limit and compute the Love numbers of this configuration. We obtain a closed-form analytical expression in the small $M$ limit
\beq
k_{\ell=2}^B=\frac{Ma_0^4(5+12\log{(a_0+2M_\text{BH})})}{3(M+M_\text{BH})^5}\,.
\eeq
The scaling with $M$ and $a_0$ agrees with our results in Chapter~\ref{ch:TLNs} (cf. Eq~\eqref{eq:ShellAxialFarAway}).

We also computed the \acs{QNM}s of the axial sector using standard direct integration and spectral routines~\cite{Berti:2009kk,Jansen:2017oag,Cardoso:2017soq}~\footnote{The computation of the axial \acs{QNM} spectrum was conducted by Kyriakos Destounis and Rodrigo Panosso Macedo}. Our results are summarized in Fig.~\ref{fig:axial_QNM_fit_l2} for the quadrupole. 
We have accurately computed the fundamental mode and the first two overtones for various halo masses. We find that \acs{QNM}s depend solely on the compactness $M/a_0$ and only change slightly when the halo mass $M$ is increased by two orders of magnitude. We fitted the relative differences to the \acs{QNM}s with respect to vacuum to powers of $M/a_0$. In the astrophysical limit $M \gg M_\text{BH}$
\beq
\frac{\omega_{\ell n}(M, a_0)}{\omega_{\ell n}(0, 0)} \sim 1 - \frac{M}{a_0} + 0.17 \frac{M^2}{a_0^2}+ \mathcal{O}\left(\frac{M^3}{a_0^3}\right).
\eeq
The first-order correction in $M/a_0$ is again giving the redshift of the dynamics. Remarkably, the second-order agrees with the correction on the \acs{LR} frequency as in Eq.~\eqref{eq:LRFreqHalo}. This behavior of the \acs{QNM}s is consistent with the interpretation discussed on Chapter~\ref{ch:LR} that they correspond to high-frequency waves trapped at the \acs{LR} that are slowly escaping out. 

\subsection{Extreme-mass-ratio systems}
%
\begin{table}[ht]
	\centering
	\begin{tabular}{c c c c c c} 
		\hline\hline
		$\ell$ & $m$ & $\dot{E}^t_\infty$ & $\dot{E}^f_\infty$ & $\dot{E}^{\text{BHPT}}_\infty$ \\
		\hline\hline
		\multirow{2}*{2} & \multirow{2}*{$1$} & $8.1629$e$-7$ & $8.1631$e$-7$ & $8.1631$e$-7$ \\ 
						 &	  			    & $6.9156$e$-7$ & $6.9158$e$-7$ &		    \\  
		\hline
		\multirow{2}*{2} & \multirow{2}*{$2$} & $1.7068$e$-4$ & $1.7062$e$-4$ & $1.7062$e$-4$ \\     			   &	 & $1.6077$e$-4$ & $1.6208$e$-4$ &  \\
		\hline  
		\multirow{2}*{3} & \multirow{2}*{$2$} & $2.5198$e$-7$ & $2.5199$e$-7$ & $2.5198$e$-7$ \\
					     &	  			    & $2.1611$e$-7$ & $2.1612$e$-7$ &			\\
		\hline
		\multirow{2}*{3} & \multirow{2}*{$3$} & $2.5490$e$-5$ & $2.5473$e$-5$ & $2.5471$e$-5$ \\
						 &	  			    & $2.3163$e$-5$ & $2.3140$e$-5$ &			\\
		\hline
		\multirow{2}*{4} & \multirow{2}*{$3$} & $5.7750$e$-8$ & $5.7749$e$-8$ & $5.7749$e$-8$	\\
						 &	  			    & $5.0252$e$-8$ & $5.0252$e$-8$ &			\\
		\hline
		\multirow{2}*{4} & \multirow{2}*{$4$} & $4.7352$e$-6$ & $4.7260$e$-6$ & $4.7253$e$-6$ \\
						 &	  			    & $4.0458$e$-6$ & $4.0823$e$-6$ &	        \\
		\hline\hline
	\end{tabular}
\caption{Energy flux (in units of $m_p^2/M_\text{BH}^2$) emitted in \acs{GW}s to infinity in different modes by a particle in a circular orbit at radius $r_p = 7.9456 M_{BH}$ around a \acs{BH} surrounded by a Hernquist-type \acs{DM} halo. We show results for vacuum (first line of each mode) and for a halo with $\left(c_{s_{r}}, \, c_{s_{t}}\right)=(0.9,0)$, $M=10M_\text{BH}$ and $a_0=10M$ (second line of each mode). $\dot{E}^t_\infty$ is computed with the time-domain framework, whereas $\dot{E}^f_\infty$ are computed in the frequency domain and $\dot{E}^{\text{BHPT}}_\infty$ corresponds to results available online from the \acs{BH} perturbation toolkit, available only in vacuum. $\ell=m$ modes correspond to polar excitations whereas $\ell=m+1$ correspond to axial ones.
}\label{tab:Fluxes_Comparison}
\end{table}
\begin{figure}[ht]
	\centering
	\begin{tabular}{c}
		\includegraphics[width=0.8\linewidth]{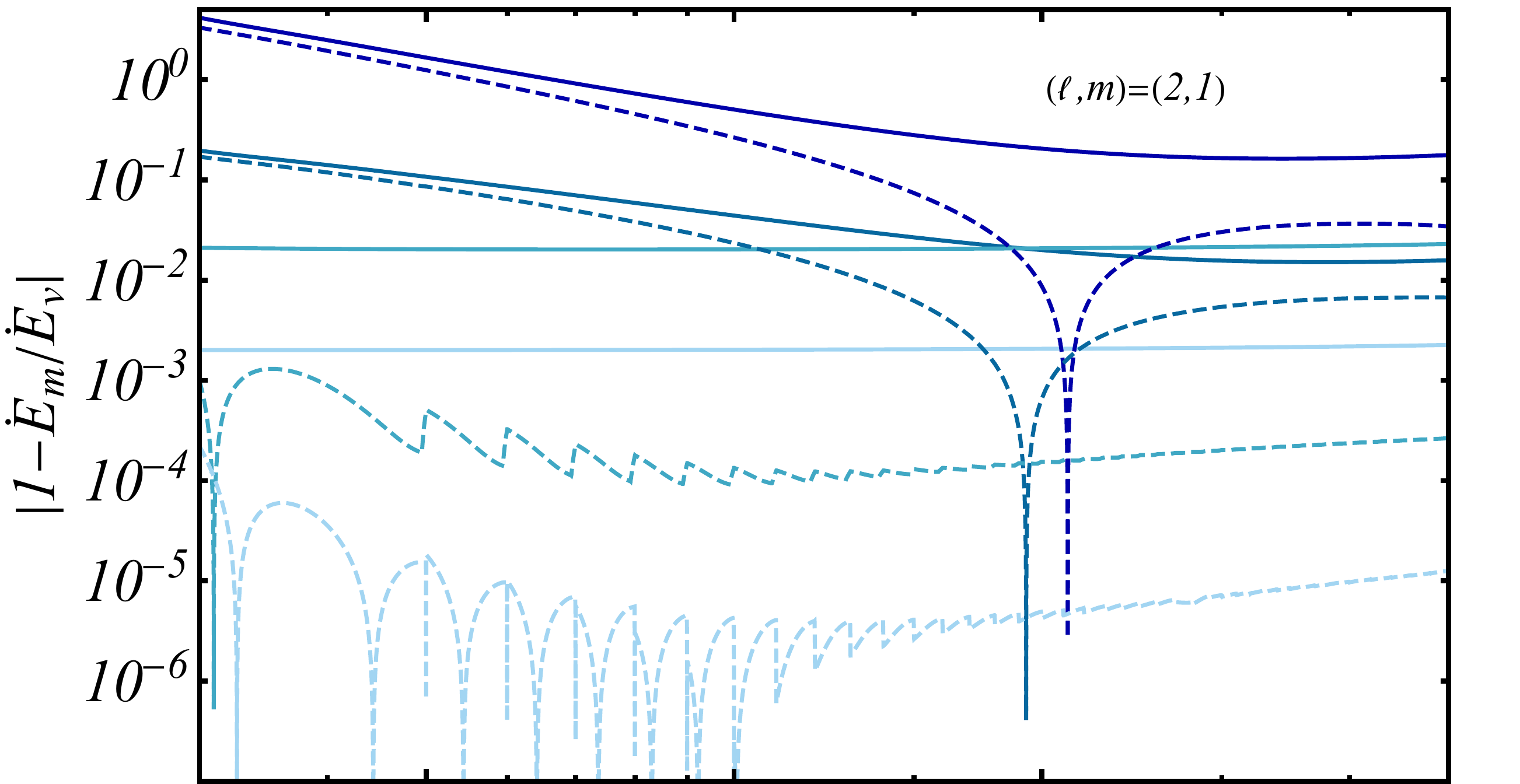} \\
		\includegraphics[width=0.8\linewidth]{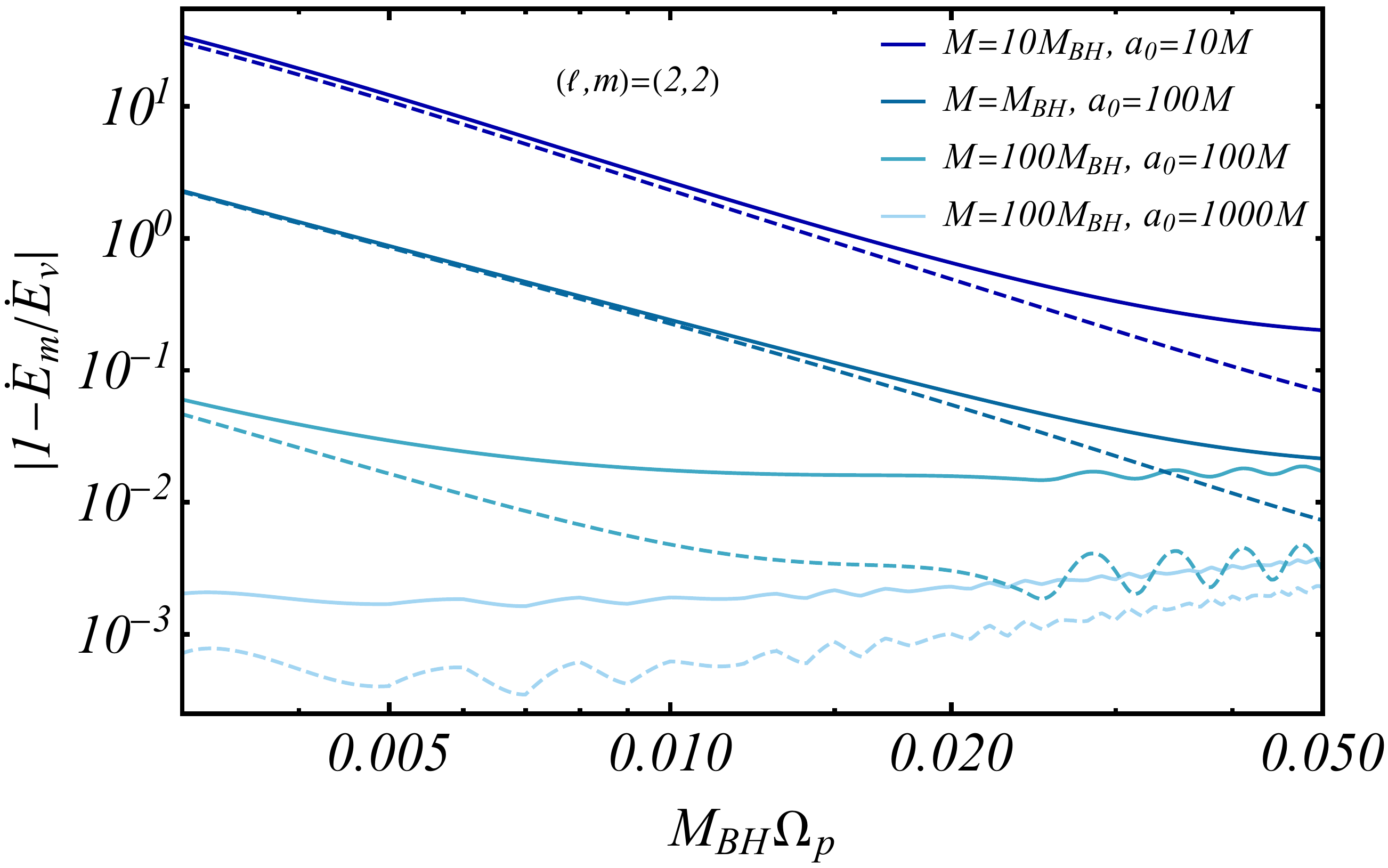} 
	\end{tabular}
	\caption{\textit{Top panel}: Relative difference between the energy flux emitted by a circular \acs{EMRI} in the dominant axial mode $\ell=2 \,, \, m=1$ as a function of the orbital frequency $\Omega_p$, for different halo configurations ($\dot{E}_m$) and in vacuum ($\dot{E}_v$) (solid lines). Dashed lines represent the same comparison but with the vacuum fluxes redshifted according to Eq.~\eqref{eq:RescalingOmega} . The frequencies correspond to a secondary at radius ranging from $r_p=50M_\text{BH}$ down to $r_p=6M_\text{BH}$; \textit{Bottom panel}: Same as the top panel but for the dominant polar mode $\ell=2 \, , \, m=2$. In this case, the redshift correction correctly captures the variation of the flux with the orbital frequency.}
	\label{fig:FluxHalo}
\end{figure}

Finally, we can study the \acs{EMRI} problem, where we put a point particle orbiting the central \acs{BH} (details for the source term are presented in Appendix~\ref{app:GBH}). We solved the problem numerically using the time-domain framework introduced in Sec.~\ref{sec:Numerics}, but also in the frequency-domain following Ref.~\cite{Cardoso:2016olt}~\footnote{The frequency-domain computations were led by Prof. Andrea Maselli. As in the time-domain, this code uses a smoothed distribution to approximate the point particle, $\sqrt{2\pi}\sigma\delta(r-r_p)=\exp{\left(-(r-r_p)^2/(2\sigma^2)\right)}$ where the width $\sigma$ is varied to ensure numerical convergence.}. We have tested our procedure and routines in the vacuum limit, i.e. using a geometry~\cite{Cardoso:2021wlq} with a low value of the halo mass $M=10^{-6}M_\text{BH}$, comparing the \acs{GW} fluxes with those available online from the Black Hole Perturbation Toolkit~\cite{BHPToolkit}. Results are summarized in Table~\ref{tab:Fluxes_Comparison} (for now let us focus the discussion on the axial modes with $\ell=m+1$), and compare favorably both between different implementations and with the Black Hole Perturbation Toolkit in vacuum. 

It is clear from Table~\ref{tab:Fluxes_Comparison} that, for a fixed \acs{BH} mass, the fluxes are smaller in the presence of a halo. One could then wonder that when accumulated over many orbits, this would significantly change the inspiral. Typically, relative differences of $\sim 1 \%$ in the energy fluxes may result in observable differences in the waveform of an \acs{EMRI}. However, we know that the \acs{EMRI} is evolving in a nontrivial gravitational potential of the \acs{DM} halo, so a decreasing flux might just signal some kind of redshift effect. Let us focus on realistic astrophysical systems, for which there is the scale hierarchy $a_0 \gg M \gg M_\text{BH}$. To linear order in $M/a_0$, 
\beq
\frac{dr}{dr_*} \approx \left(1-\frac{M}{a_0} \right)\frac{dr}{dr_*^\text{vac}} \, , 
\eeq
where $r_*^\text{vac}$ is the tortoise coordinate in a Schwarzschild geometry~\eqref{eq:SphericalLineElement}. For compact \acs{EMRI}s ($r_p \sim 10M_\text{BH}$), $S^\text{ax} \approx \left(1-3M/a_0 \right) S^\text{ax}_\text{vac}$. Combining these, expanding Eq.~\eqref{eq:master_axial} to linear order in $M/a_0$ we find (in the frequency domain)
\beq
\frac{d^2 \Psi^\text{ ax}}{d(r_{*}^\text{ vac})^2}  +
\left( \frac{\omega^2}{\gamma^2} -V^\text{ ax}_\text{Schw}\right)\Psi^\text{ ax}= \gamma  S^\text{ax}_\text{Schw} \,,
\label{eq:MasterAxialExpanded}
\eeq
where $\gamma=1-M/a_0$ is a redshift factor. Thus, to linear order in $\gamma$ the axial signal from the \acs{EMRI} immersed in the Hernquist halo is identical to that from a Schwarzschild \acs{BH}, with redshifted frequency and mass; in other words, the two setups are equivalent with the identification
\beq
\left(\Omega^\text{ vac}_p,\omega^\text{ vac},m^\text{ vac}_p\right) \rightarrow \left(
\frac{\Omega_p}{\gamma},\frac{\omega}{\gamma},\gamma m_p\right) \, . \label{eq:RescalingOmega}
\eeq

In the top panel of Fig.~\ref{fig:FluxHalo}, we present numerical results that confirm this. We show fluxes as a function of the frequency of the \acs{GW}s being measured by a distant stationary observer. \textit{A priori}, this direct comparison gives differences between a vacuum and a non-vacuum environment that are seemingly large. However, when we apply the redshift correction above, the fluxes in the presence of the halo are indeed well described by redshifted fluxes in vacuum. The agreement is better for larger halo mass $M$, and smaller compactness $M/a_0$. For realistic galactic configurations, it leads to relative differences that are extremely small and not expected to be observable. 
Note that for small scales, $a_0 \omega\lesssim 1$, the radiation wavelength is larger than the halo itself, and redshift is suppressed.

\section{Polar sector}

We now to move the polar sector, which dominates \acs{GW} emission. 

\subsection{Boundary conditions and sound speed}\label{sec:BCSoundSpeed}

%
\begin{figure}[t]
	\centering
	\begin{tabular}{c}
		\includegraphics[width=0.9\linewidth]{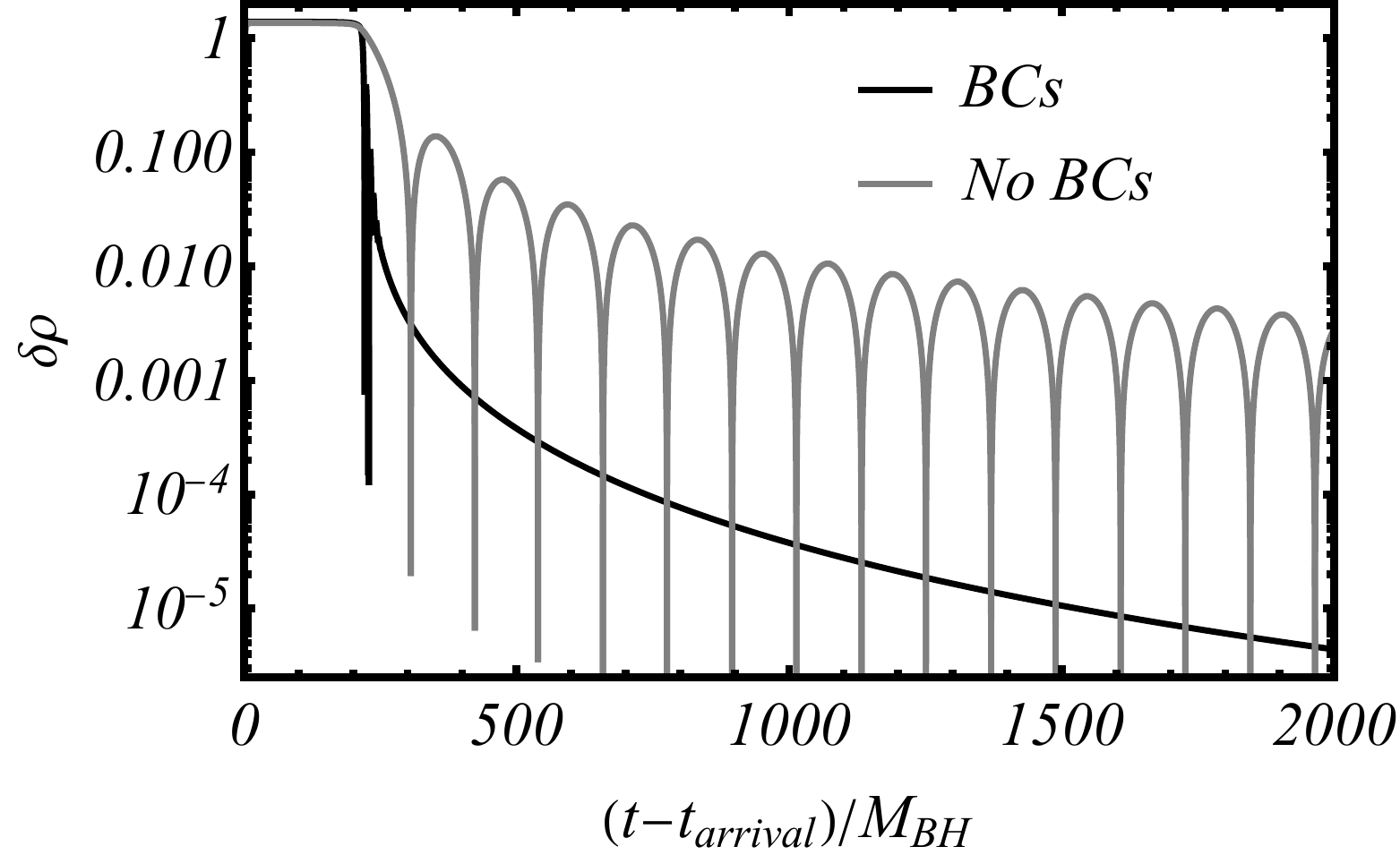} 
	\end{tabular}
	\caption{Evolution of $\delta \rho$ in a Schwarzschild background with $c_{s_r} = 0.9$, $c_{s_t} = 0.0$ when different boundary conditions are imposed. $t_{arrival}$ is the time of arrival of the first direct signal coming from the initial data prescribed. When no boundary conditions are imposed (\textit{No BCs}), and $\delta \rho$ is left free at the~\acs{BH} horizon, an oscillatory tail of the form $\delta \rho \propto t^{-5/6} \sin \left(\mu_\text{eff} \, c_{s_r}\right)$ sets in at late times, with $\mu_\text{eff} = \left(1- c_{s_r}^2\right)/8c_{s_r}^2 M_\text{BH}$. This is analogous to having a scalar field of mass $\mu_\text{eff}\, c_{s_r}$. On the other hand, when Dirichlet boundary conditions (\textit{BCs}), $\delta \rho = 0$, are imposed at a cutoff radius $r_\text{cut}$ (here $r_\text{cut}=3M_\text{BH}$), we find a universal power-law decay independent of $r_\text{cut}$ and $c_{s_r}$, consistent with $\delta \rho \propto t^{-3}$. }
	\label{fig:deltarho_inf}
\end{figure}

One of the major technical problems associated with the evolution of polar perturbations in non-vacuum backgrounds is that gravitational perturbations are coupled to the matter sector. However, in our setup,
at far away distances and very close to the \acs{BH} horizon, the matter density $\rho$ vanishes, and the field equations reduce to the case of a vacuum \acs{BH}. When this happens, the gravitational sector decouples from matter (and vice-versa). 

Considering first the case where the sound speeds introduced in Eq.~\eqref{eq:BarotropicEq} are \textit{constant} and taking the following ansatz for the density perturbation
\beq
\delta \rho = r^{\alpha} \left(r- 2M_\text{BH} \right)^\beta \Psi_\rho \, , 
\eeq
we obtain that $\Psi_\rho$ obeys to 
\beq
\mathcal{L}_{c_{s_r}} \Psi_\rho &=& V_\rho \Psi_\rho \, , \\
\alpha &=& \frac{1}{4} \left(-5 + \frac{1+4c_{s_t}^2}{c_{s_r}^2} \right) \, , \\
\beta &=& - \frac{3}{4} - \frac{1}{4c_{s_r}^2}\, , \\
V_\rho &\approx& \mathcal{O}\left(r^{-2} \right) \quad , \quad r \rightarrow \infty \\
V_\rho &\approx& \left( \frac{1-c_{s_r}^2}{8 c_{s_r}^2 M_\text{BH}} \right) \quad , \quad r \rightarrow 2M_\text{BH} \, .
\eeq

These properties are similar to the ones of a massive scalar field with effective mass $\mu_\text{eff} = \frac{1-c_{s_r}^2}{8 c_{s_r}^2 M_\text{BH}}$. Based on previous results for this system~\cite{Ching:1995tj,Hod:1998ra,Koyama:2001ee}, the scattering of a Gaussian pulse should lead to an oscillatory power-law with $\Psi_\rho \propto t^{-5/6} \sin \left( \mu_\text{eff} c_{s_r} \right)$, due to wave backscattering near the horizon.

To confirm this, we evolved the homogeneous version of the evolution equation for $\rho$~\eqref{eq:WaveEquationRho} in a Schwarzschild background with initial data 
\beq
\delta \rho \Big|_{t=0} &=& 0 \, , \nonumber\\
\partial_t \delta \rho \Big|_{t=0} &=& \exp \left[-\left(r_* - 100 M_\text{BH}  \right)^2/2  \right]\, .
\label{eq:IDdeltarho}
\eeq
The results presented in Fig.~\ref{fig:deltarho_inf} for the $\ell =2$ mode (in gray) agree with the expected behavior, and this was verified for different initial data and $c_{s_r}$.

However, since the matter profile vanishes at the horizon and spatial infinity, physical configurations should have asymptotically vanishing sound speed at these boundaries. If we impose a power-law decay at the boundary for $c_{s_r}$, then $\delta \rho$ has to satisfy Dirichlet boundary conditions $\delta \rho =0$ in order to be regular at the boundaries. We repeated the same numerical experiment as above, but now keeping $c_{s_r}$ constant and imposing Dirichlet boundary conditions at a cutoff radius $r_{cut}$ close to the \acs{BH} horizon. In this case ($c_{s_t}=0$), the late-time behavior should be $\delta \rho = t^{-3}$ for all multipoles, but now due to backscattering at far-away distances~\cite{Ching:1995tj}. This decay is independent of $\ell$ because the homogeneous part of the evolution equation for $\delta \rho$  in Eq.~\eqref{eq:WaveEquationRho} is independent of $\ell$ for $c_{s_t}=0$. Once again, we verified this numerically (in black in Fig.~\ref{fig:deltarho_inf}), independently of the cutoff radius $r_\text{cut}$ and the sound speed $c_{s_r}$. 

\subsection{QNM spectral stability}

%
\begin{figure}[t]
	\centering
	\begin{tabular}{c}
		\includegraphics[width=0.9\linewidth]{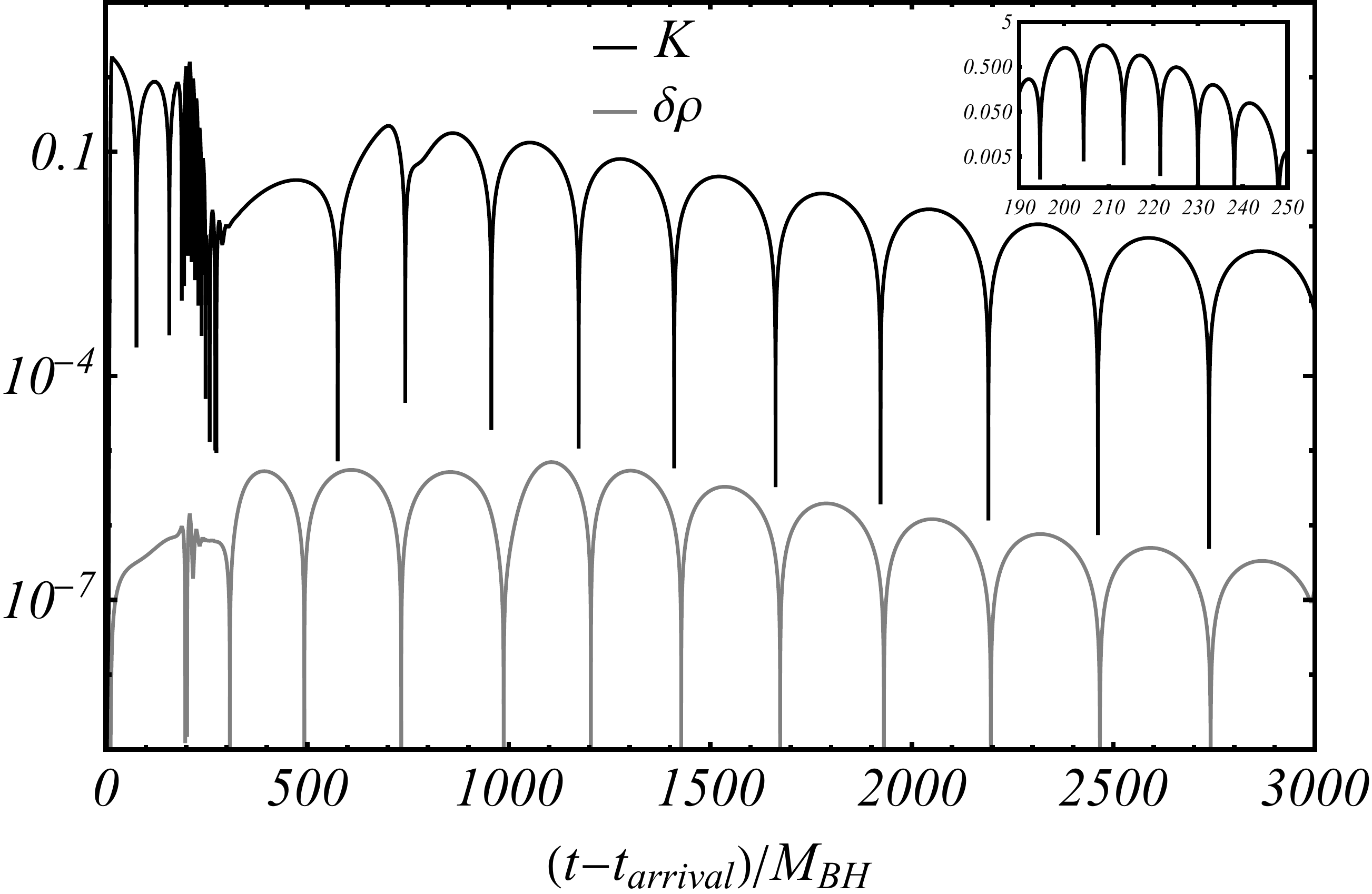} 
	\end{tabular}
	\caption{Evolution of the metric and density perturbation $K,\,\delta\rho$, with $M=10M_\text{BH}$, $a_0 = 10M$. We impose Dirichlet conditions at $r_{\text{cut}}=3M_\text{BH}$ and $c_{s_r}=\left[\left(2M_{\text{BH} } + a_0\right)/\left(r + a_0\right)\right]^4$, so that the radial speed of sound asymptotes to zero at large distances. At early times, a standard \acs{BH} ringdown is excited, as can be seen in the inset for $K$; at late times, the signal is dominated by a slowly-decaying, fluid-driven mode with period $\sim a_0$. There is a mutual conversion between gravitational and matter density waves, which is a manifestation of the spectral instability of \acs{BH} \acs{QNM}s. The qualitative conclusions are independent of the initial data given and the profile chosen for $c_{s_r}$.}
	\label{fig:K_halomodes}
\end{figure}

Taking into account the results of the previous section, from now on we impose vanishing sound speed $c_{s_r}$ at an interior cutoff radius close to the \acs{BH} horizon. Since in the polar sector, matter perturbations are coupled to gravitational ones we expect two families of modes, one traveling at the speed of light and determined by gravity, and the other traveling at the speed of sound $c_{s_r}$ and controlled by the matter distribution.

We repeated the scattering experiment of the previous section, with the same initial data given by Eq.~\eqref{eq:IDdeltarho} but for the metric function $K$, and for a very compact halo configuration with $M = 10M_\text{BH}$ and $a_0/M = 10$, to enhance the physical effects. We chose for the profile of the radial speed of sound 
\beq
c_{s_r} = \left(\frac{2M_\text{BH} + a_0}{r + a_0}\right)^4 \, ,
\eeq
but our qualitative conclusions are similar for other choices. In the polar sector, we do not use the hyperboloidal layers introduced in Section~\ref{sec:Hyperboloidal}, because matter waves do not travel at the speed of light and therefore do not reach null infinity. 

Our results are shown in Fig.~\ref{fig:K_halomodes} for the quadrupole $\ell=2$ mode. Initially, we observe a standard \acs{BH} ringdown, as highlighted in the inset, with a frequency close to the fundamental Schwarzschild \acs{QNM} with $\omega^\text{Schw}_{\ell=2} = 0.374 - 0.089i$. At late times, the signal is dominated by a long-lived fluid mode, whose period is of the order of the typical length scale of the halo $a_0$. We also observe the conversion between \acs{GW}s and matter waves, and vice-versa. In practice, this is a fluid mode and corresponds to a concrete manifestation of the \acs{QNM} spectral instability discussed in Chapter~\ref{ch:Elephant} for an astrophysically motivated system.  

\subsection{Extreme-mass-ratio systems}

We now revisit the \acs{EMRI} problem. Again, there is good agreement between the results from different codes and the ones from the Black Perturbation Toolkit in vacuum, as shown in Table~\ref{tab:Fluxes_Comparison} (polar modes are the ones with $\ell = m$). Since polar fluctuations couple to the fluid, it is harder to do the redshift analysis presented above for the axial sector. Nonetheless, we can directly apply the redshift corrections in Eq.~\eqref{eq:RescalingOmega} to the energy fluxes carried by \acs{GW}s and compare. This is illustrated in the bottom panel of Fig.~\ref{fig:FluxHalo}, where it is clear that polar perturbations are less prone to redshift effects, even in regions of the parameter space corresponding to large, near-galactic scales. 

Our results indicate the tantalizing possibility of using \acs{GW} astronomy to strongly constrain smaller scale distributions around \acs{BH}s. At $\omega M_\text{BH}=0.02$, the relative flux difference between a vacuum and a \acs{BH} immersed in a halo with $M=0.1M_\text{BH}$ and $a_0=10^2M,10^3M$ is $\sim 10\%,1\%$ respectively. These numbers are well within reach of next-generation detectors \cite{Bonga:2019ycj}.

\section{Discussion}\label{sec:Discussion_GBH}

This chapter served mainly as a proof-of-concept for the ability of our framework to study environmental effects in \acs{GW} physics at a full relativistic level. A natural next step is to apply it to other environments, for example by taking input from recent General Relativistic Magnetohydrodynamical simulations of accretion~\cite{Derdzinski:2020wlw, Zwick:2021dlg}, or to add rotation to the central \acs{BH}.

Our example shows that environments can easily destabilize the \acs{BH} spectra, as had recently been suggested with toy models \cite{Barausse:2014tra,Jaramillo:2021tmt,Destounis:2021lum,Cheung:2021bol,Berti:2022xfj} and discussed in Chapter~\ref{ch:LR}. It is unknown at this point if environmental resonances can be excited by supermassive \acs{BH}s, long-before merger. Also, we did not explore the effect of the environment in the late-time polynomial tails that follow the ringdown, nor in \acs{GW} memory~\cite{Favata:2010zu, PhysRevLett.67.1486}. Formally, these are zero frequency signals that should be sensitive to the full gravitational potential, possibly being more prone to changes due to the presence of a non-vacuum environment.  

The relative differences in the energy flux carried by \acs{GW}s appears to be more significant for lower frequencies. This behavior is particularly relevant in light of the recent announcement by the International Pulsar Timing Array of a stochastic \acs{GW} background produced by \acs{BH} binaries with billions of solar masses~\cite{Antoniadis:2023zhi, EuropeanPulsarTimingArray:2023qbc, Ghoshal:2023fhh, Ellis:2023dgf, NANOGrav:2023gor, NANOGrav:2023hvm, NANOGrav:2023hfp, Antoniadis:2023rey}. In fact, preliminary studies suggest that the inclusion of environmental effects describes better the observed spectrum in the lower-frequency end~\cite{Ghoshal:2023fhh, Ellis:2023dgf}.

Our works~\cite{Cardoso:2021wlq, Cardoso:2022whc} spurred a series of further explorations applying the Einstein Cluster construction to other \acs{DM} profiles~\cite{Konoplya:2022hbl, Destounis:2022obl, Figueiredo:2023gas, Zhao:2023tyo, Dai:2023cft, Daghigh:2022pcr, Igata:2022rcm, Zhang:2021bdr}. One worth highlighting is the exhaustive survey of the phase space of geodesics in Ref.~\cite{Destounis:2022obl}, including orbits with precession. Interestingly, there is a competition between the halo and \acs{GR} effects in the direction of precession, which can transition from prograde to retrograde for sufficiently compact configurations. The departure from vacuum typically results in non-integrable geodesics, which exhibit chaotic phenomena and resonant islands in the orbital phase space~\cite{MooreChaos, Destounis:2021rko, Destounis:2023khj}. 

The complicated intrinsic properties of the anisotropic halo constituting the \acs{DM} halo and the lack of a physical model for the sound speeds prevented us from studying matter fluxes, which should contain the information about accretion and dynamical friction on the \acs{EMRI}. On our final discussion in the next chapter we present a strategy on how to overcome these drawbacks and our future steps in the study of environmental effects in \acs{GW} astronomy.


\chapter{Concluding Remarks}\label{ch:Conclusion}

The formalism developed in the previous chapter has the potential to become a benchmarking tool for the systematic analysis of environmental effects in \acs{GW} astronomy. This can ultimately be fed into the analysis pipelines used by various \acs{GW} collaborations. For this reason, the codes we used are already publicly available~\cite{GRIT_REPO}. Nonetheless, as we discussed at the end of the previous chapter, there are still challenges to overcome. In the system we studied, we did not consider matter fluxes, so we could not retrieve information about accretion and dynamical friction. Moreover, at this stage we do not know how to correctly evaluate the backreaction on the orbit and evolve the inspiral. In a vacuum binary we can consider an adiabatic approximation where the smaller body ``flows through a sequence of geodesics"~\cite{Hughes:2021exa} of the background spacetime, determined by their respective conserved quantites: energy, angular momentum, and Carter constant (which is $0$ in Schwarzschild). However, in non-vacuum, the fluxes will contain contributions from the gravitational binding energy of the binary but also from the environment, which need to be disentagled.

\paragraph*{1. Improve modeling\\}

The first step of our plan is therefore to understand better the modeling aspects just described. To do that, we will apply our formalism to simpler environments, starting with \acs{EMRI}s surrounded by scalar fields~\cite{Baumann:2021fkf, Annulli:2020lyc, Annulli:2020ilw, Cardoso:2022nzc, Delgado:2023wnj, Fell:2023mtf}, for which we already have preliminary results. 




In the long run, we would like to study \acs{EMRI}s in thin accretion disks describing \acs{AGN}s. In this case, our techniques can be combined with state-of-the-art General Relativistic Magnetohydrodynamic simulations, which are currently only using Newtonian physics~\cite{Derdzinski:2020wlw, Zwick:2021dlg, Speeney:2022ryg}. This could also be relevant for computing the electromagnetic counterpart of binary coalescences in \acs{AGN}s. 

Moreover, accretion disks exhibit turbulence due to various phenomena, such as magnetorotational instability~\cite{1991ApJ...376..214B}. The \acs{GW} signature associated with these turbulent flows has yet to be studied. Even though the typical matter densities are too low to expect individual events, they could contribute to a putative non-negligible stochastic \acs{GW} background.

Finally, it will also be important to understand the most effective way to include rotation, which could be done either in terms of a small spin expansion or working in the tetrad formalism.

\paragraph*{2. Waveform implementation\\}

 After the modeling stage is completed, we can start implementing them in state-of-art waveform models, like the FastEMRIWaveform package~\cite{Katz:2021yft}. This is a modular code designed to facilitate custom changes. For example, one can simply add to the fluxes of energy and angular momentum the correction coming from a particular environment and a have a ready-to-use code which generates waveforms for the new system in seconds. This strategy has been used for gas torques in thin accretion disks~\cite{Speri:2022upm} and dynamical friction. Moreover, even though we have been focusing the discussion on \acs{EMRI}s, \acs{BH} perturbation theory agrees remarkably well with Numerical Relativity simulations up to mass-ratios of $\mathcal{O}(10)$~\cite{Albertini:2022rfe}. Consequently, it can also be useful to try to implement our results in Effective-One-Body~\cite{Taracchini:2013rva, Bohe:2016gbl} and Inspiral-Merger-Ringdown~\cite{Husa:2015iqa, Khan:2015jqa} models.

\paragraph*{3. Detectability/Measurability forecasts\\}

Once implemented, we can use the updated models to perform an exhaustive survey of their impact on the detectability and measurability of \acs{GW} events. Following previous works~\cite{Speri:2022upm, Speeney:2022ryg, LISA:2022kgy}, we want to answer the following questions:

\begin{itemize}

\item Considering mismodeling, statistical and instrumental systematic errors (which are currently unstudied), what is the expected level of accuracy of future waveforms and how strong environmental effects have to be to become distinguishable?

\item If the strength of the environmental effects surpasses this threshold, with what precision can we infer the properties of the environment? 

\item What bias will we incur in parameter estimation if environmental effects are ignored and how will that hinder tests of \acs{GR}?

\item What is the degeneracy between corrections introduced by the environment and those of modified gravity~\cite{Maselli:2021men}?

\item What is the impact of relativistic modeling of environmental in \acs{EMRI} population rates detectable by \acs{LISA}?

\end{itemize} 

This is certainly an ambitious plan which will require years of collaborative effort. Nonetheless, it is one that can bring precious dividends for a research topic that has been gaining widespread interest in recent years. This is exhibited by the increasing number of papers appearing every week on this topic, and by the organization of conferences/workshops entirely dedicated to it. \acs{LISA} and third-generation ground-based detectors will be operating by the end of next decade and therefore there is no better time to do the necessary theoretical work for them than \textit{now}. I would like to end with a word of appreciation for all the generations of physicists who have dedicated their careers to \acs{GW} astronomy. Their work has offered a once-in-a-lifetime opportunity for young scientists like me to be in the \textit{right} place, at the \textit{right} time, doing the \textit{right} science, and be part of what most likely will be a period marked by incredible discoveries in gravity and astrophysics.

\appendix
\cleardoublepage
\part{Appendix}
\chapter{Perturbation theory in Quantum Mechanics}\label{app:Cloud_PT}

In this appendix we explain how the problem studied in Chapter~\ref{ch:Cloud} can be treated analytically using standard perturbation theory in Quantum Mechanics.

In the non-relativistic limit, the scalar cloud obeys an equation which is formally equivalent to Schr\"{o}dinger's equation with a Coulomb potential governed by a single parameter
\beq
\alpha = M\mu \, .
\eeq
This can be seen by making the standard \textit{ansatz} for the dynamical evolution of $\Phi$~\cite{Page:2003rd, Mendes:2016vdr,Baumann:2018vus} 
\beq
\Phi\left(t,r^j\right)=\frac{1}{\sqrt{2\mu}}\left(\psi\left(t,r^j\right)e^{-i\mu t}+\psi^*\left(t,r^j\right)e^{i\mu t}\right)\,,
\eeq
where $\psi$ is a complex field which varies on timescales much larger than $1/\mu$. Then, one can expand the Klein-Gordon equation to first order in $\alpha$ and arrive at
\beq
i \partial_t \psi&=&\left(-\frac{1}{2\mu}\nabla^2-\frac{\alpha}{r}\right)\psi \, ,
\eeq
where we also kept only terms of order $\mathcal{O}\left(r^{-1}\right)$. 

The normalized eigenstates of the system are hydrogenic-like, with an adapted ``fine structure constant'' $\alpha$ and ``reduced Bohr radius'' $a_0$~\cite{Detweiler:1980uk,Brito:2015oca},
\beq
\psi_{n\ell m}&=&\,e^{-i\left(\omega_{n\ell m}-\mu\right)t}\,R_{n\ell}\left(r\right)Y_{\ell m}\left(\theta,\varphi\right) \, ,\\
R_{n \ell}\left(r\right)&=&C\left(\frac{2r}{n a_0}\right)^\ell L_{n-\ell-1}^{2\ell+1}\left(\frac{2r}{na_0}\right)e^{-\frac{r}{na_0}} \, , \nonumber\\
a_0&=&\frac{1}{\mu\alpha} \, ,\quad C=\sqrt{\left(\frac{2}{na_0}\right)^3\frac{\left(n-\ell-1\right)!}{2n\left(n+\ell\right)!}} \,,\label{eq:FrequencySpec}
\eeq
where $L_{n-\ell-1}^{2\ell+1}$ is the generalized Laguerre polynomial~\cite{NIST:DLMF}. 
We are adopting the convention for the quantum numbers used in Refs.~\cite{Baumann:2018vus,Berti:2019wnn}, where states are labeled by $n= \ell +1 , \, \ell +2, \,...\,$.
The eigenvalue is, up to terms of order $\alpha^5$~\cite{Baumann:2019eav}
\beq
\omega_{n\ell m}=\mu\left(1-\frac{\alpha^2}{2n^2}-\frac{\alpha^4}{8n^4}+\frac{\left(2\ell-3n+1\right)\alpha^4}{n^4\left(\ell+1/2\right)}\right)+\mathcal{O}\left(\alpha^5\right)\,.
\label{eq:FrequencyEigen}
\eeq
We can estimate the size of the scalar cloud by computing the expectation value of the radius on a given state
\be
\left< r \right> = \int_0^{\infty}dr\,r^3 R_{n\ell}^2\left(r\right) = \frac{a_0}{2}\left(3n^2-\ell(\ell+1)\right) \, .\label{cloud_radius}
\ee

When the binary companion is included, the tidal perturbation can be treated in the framework of perturbation theory in Quantum Mechanics. The tidal potential $\delta V$ entering in Schr\"{o}dinger's equation due to $\delta ds^2_{\text{tidal}}$ \eqref{eq:MetricPerturbation} is represented by a step function 
\beq
\delta V=-\Theta\left(t-t_0\right)\frac{M_c\, \mu}{R}\sum_{m = -2}^{2}\frac{4\pi}{5}\left(\frac{r}{R}\right)^2 Y_{\ell m}^*\left(\theta_c,\varphi_c\right)Y_{\ell m}\left(\theta, \varphi \right) \, , \nonumber \\
\eeq
where $t_0$ is the instant when we turn it on and $\Theta\left(t\right)$ is the Heaviside function. Though there is an implicit time dependence, if one lets the system evolve for sufficient time, it will end in a final stationary state (ignoring the loss of energy at the \acs{BH} horizon). To describe the final picture, time-independent perturbation theory is enough. 

Let us recall its standard procedure. 
We are solving Schr\"{o}dinger's equation 
\beq
\mathcal{H}\left|\psi_i\right> &=& \omega_i\left|\psi_i \right>\, , \label{eq:SchrodingerPert}\\
\mathcal{H}&=&\mathcal{H}_0+\lambda \, \delta V \label{eq:FullHamiltonian}\, ,
\eeq
where $\mathcal{H}_0$ is the Hamiltonian of the unperturbed problem, $\delta V$ is the potential corresponding to the perturbation, and $\lambda$ is a dimensionless expansion parameter varying between $0$ (no perturbation) and $1$ (full perturbation). Since we are now referring to a generic problem, we have dropped the triple indices of the ``hydrogenic" spectrum and instead label different eigenstates $\left|\psi_i\right>$ of the Hamiltonian (and the respective eigenvalue frequencies $\omega_i$) by a single index~\footnote{In Quantum Mechanics literature, it is common to use $E$ for the (energy) eigenvalues, but since we are working in natural units, $\hbar=1$, there is no distinction between them}. 

When the system is non-degenerate, the eigenstates $\left|\psi_k^{(0)}\right>$ of the unperturbed problem - which are assumed to be known and in our case are given by Eq.~\eqref{eq:FrequencySpec} - are in one-to-one correspondence with the eingenvalues, $\omega_k^{(0)}$,
\beq
\mathcal{H}_0\left|\psi_i^{(0)}\right>=\omega_i^{(0)}\left|\psi_i^{(0)}\right> \, ,
\eeq
and $\{\psi_n^{(0)}\}$ form a complete orthonormal basis 
\beq
\left<\psi_m^{\left(0\right)}|\psi_n^{\left(0\right)}\right>=\delta_{mn}\, .
\eeq

Now, we expand the eigenstates of the perturbed system, $\psi_i$, in terms of the basis $\{\psi_k^{\left(0\right)}\}$
\beq
\left|\psi_i\right>=\sum_k c_{ki}\left|\psi_k^{(0)}\right> \, , \label{eq:Eigenstates}
\eeq
and plugging this \textit{ansatz} in \eqref{eq:SchrodingerPert}, the coefficients $c_{ki}$ and the eigenvalues $\omega_i$ can be obtained as a power series in $\lambda$. If the perturbation is small enough, we expect the first-order expansions to be a good approximation~\cite{griffiths2017introduction}
\beq
\omega_i&=&\omega_i^{(0)}+\lambda \, \omega_i^{(1)} \, , \\
c_{ki}&=&c_{ki}^{(0)}+\lambda \, c_{ki}^{(1)} \, , \\
\omega_i^{(1)}&=&\left<\psi_i^{(0)}\right|\delta V\left|\psi_i^{(0)}\right> \, , \label{eq:EnCorr}\\
c_{ki}^{\left(1\right)}&=&\frac{\left<\psi_k^{(0)}\right|\delta V\left|\psi_i^{(0)}\right>}{\omega_i^{\left(0\right)}-\omega_k^{\left(0\right)}} \, , \, k\neq i \, , \label{eq:CoefCorr}
\eeq
where we omitted terms of order $\mathcal{O}\left(\lambda^2\right)$. In the end, we set $\lambda=1$, which is the same as reabsorbing it in $\delta V$. 

The timescales for the transitions between two modes can be estimated using time-dependent perturbation theory. This involves introducing the interaction picture and performing a Dyson series on the time-evolution operator. Since the eigenstates remain the same as in the time-independent unperturbed case, we will skip details on this procedure and directly import the result for the first-order correction on the coefficients $c_{ki}$ for a step-function perturbation \cite{sakurai2011modern}
\beq
c_{ki}^{\left(1\right)}&=&\frac{\left<\psi_k\right|\delta V\left|\psi_i\right>}{\omega_i-\omega_k}\left(1-e^{-i\left(\omega_i-\omega_k\right)t}\right) \, .
\eeq
Both the states $\left|\psi_i\right>$ and frequencies $\omega_i$ should be understood as the ones for the unperturbed system, but we ommit subscripts to avoid cluttering. Then, the probability  of the transition $\left|i\right> \rightarrow \left|k\right>$ is
\beq
\left|c_{ki}^{\left(1\right)}\right|^2&=&4\left|\frac{\left<\psi_k\right|\delta V\left|\psi_i\right>}{\omega_i-\omega_k}\right|^2\sin^2\left(\frac{(\omega_k-\omega_i)t}{2}\right) \, .
\eeq

Although we do not have a continuum spectrum, for large timescales we can take this limit. Then, at fixed $t$, we can treat the probabilities $\left|c_{ki}^{\left(1\right)}\right|^2$ as functions of
\beq
\Delta \omega_{ki}=\left|\omega_k-\omega_i\right|\, .
\eeq

Plotting it for different instants of time, one can verify this function becomes increasingly peaked around $\Delta \omega_{ki}=0$ as $t$ increases (check Fig.~$5.8$ of Ref.~\cite{sakurai2011modern}). This central peak scales with $t^2$ and has a typical width of $1/t$. If we wait enough time $\Delta t$ since the perturbation is introduced, the only transitions with appreciable probability are those satisfying
\beq
\Delta t=2\pi/\Delta \omega_{ki} \, .
\eeq

The final conclusion is that the typical timescale $\Delta t$ for the transition $\left|i\right> \rightarrow \left|k\right>$ to happen is
\beq
\Delta \omega_{ki} \, \Delta t \, \sim \, 1 \, ,
\label{eq:Timescales}
\eeq
which, if we momentarily insert factors of $\hbar$, can be seen as a manifestation of the energy-time uncertainty principle \cite{sakurai2011modern}
\beq
\Delta E \Delta t \sim \hbar \, .
\eeq

Returning to our problem, the initial data presented in Eq.~\eqref{Eq.axion cloud initial data} corresponds to the stationary state (reintroducing the triple ``hydrogenic"" indices) 
\beq
\left|i \right> \propto \left(\left|\psi^{(0)}_{211}\right> - \left|\psi^{(0)}_{21-1}\right> \right) \, , 
\label{eq:InitialState}
\eeq
up to a proportionality constant reflecting the renormalization done for numerical purposes. The final state should correspond to a stationary state $\left|f\right>$ which we can compute using the machinery developed before.
%
%
There is still a \textit{caveat}, which is the degeneracy between states with the same quantum number $m$ \eqref{eq:FrequencySpec}. Though a rotating \acs{BH} will lift this degeneracy, the energy shifts due to the perturbations considered are orders of magnitude higher than the energy scale associated with the rotation. Thus, for non-degenerate perturbation theory to be controlled, we would have to perform it at higher orders then what we presented. 

In the degenerate scenario, the equations presented are invalid (for example \eqref{eq:CoefCorr} diverges when $\omega_k^{\left(0\right)}=\omega_i^{\left(0\right)}$). Instead, we use the freedom in making a linear combination of unperturbed degenerate eigenstates, so that in every degenerate subspace, we pick a basis of the Hilbert space that diagonalizes the full Hamiltonian $\mathcal{H}$ \eqref{eq:FullHamiltonian}. After this step, we can apply non-degenerate perturbation theory, namely Eqs.~\eqref{eq:EnCorr} and \eqref{eq:CoefCorr}, using the new ``good'' basis.

Finally, the numerical data we present in the main text corresponds to multipole expansions of the field $\Phi$ and not to the coefficients $c_{ki}$ \eqref{eq:Eigenstates} describing the mix of the unperturbed states. To obtain these multipoles we have to select them from the space representation of the final state. The amplitude coefficients of the mode $\left|\psi_{n\ell m}\right>$ are obtained via
\beq
c_{n\ell m}&\propto& \frac{\left<\psi_{n\ell m}\left|\delta V\right|i\right>}{{\omega_{21}^{\left(0\right)}-\omega_{n\ell}^{\left(0\right)}}} \, , \\
\phi_{n\ell m}\left(r\right) &\propto& c_{n\ell m} R_{n\ell}\left(r\right) \, .
\eeq

In the end, we are interested in the ratio between amplitudes so the constant of proportionality is irrelevant. The matrix elements appearing here are explicitly presented in Eqs.$(3.7)-(3.9)$ of Ref.~\cite{Baumann:2018vus}. Notice that the relative amplitude between modes with the same quantum numbers $n$ and $\ell$ is completely determined by the angular integrals, and since these are (quasi)degenerate, they will also follow similar time evolutions. As a consequence, their relative amplitude  is independent of time and the value of $\alpha$, even at higher orders in perturbation theory. This is illustrated in Fig.~\ref{fig:a0_Mc001_mu01} for $\phi_{n33}/\phi_{n31}$.

A summary of the time-independent perturbation theory for transitions between overtones is shown in Table~\ref{tab:overtone_excitation} for $M\mu=0.1$, $\epsilon=10^{-8}$. The relative amplitudes $c_{n\ell m}/c_{211}$ indicate that the perturbation is not that small. 
This is even more obvious if we compute the first order corrections to the frequency eigenvalues, which for this configuration are of $\mathcal{O}\left(10^{-3}\right)$ for overtones $n>3$, as illustrated in Table~\ref{tab:CorrectedFreq}. For this reason, when computing the timescales of the transitions \eqref{eq:Timescales}, we used the first order corrected  $\omega_{n\ell m}$.

\begin{table}[h]
\centering
\caption{First order corrected frequencies $\omega_{n\ell m}$ predicted by time-independent theory, for a non-rotating \acs{BH} and a companion with the configuration $M\mu=0.1$, $\epsilon=10^{-8}$. A spinning \acs{BH} would break the degeneracy between states with the same $\ell$ but different $m$ quantum number. However, these corrections enter the frequency spectrum \eqref{eq:FrequencyEigen} only at order $\alpha^5$. For the above configuration, these would yield $\omega_{n33}-\omega_{n31} \sim 10^{-6}a/M\,n^3$.}
\begin{tabular}{c|c}
    ($n \, \ell $)   & $\omega_{n\ell m} \times 10^2$   \\
    \hline
    $2$ $1$    	   	   &        $9.9754$	      	\\
    $3$ $1$    	   	   &        $9.9224$  	    \\
    $4$ $1$    	   	   &     	$9.7570$          \\
    $5$ $1$    	   	   &     	$9.3980$          \\
    $4$ $3$ 	   	   &     	$9.9001$         	\\
\end{tabular}
\label{tab:CorrectedFreq}
\end{table}
%

\chapter{An isotropically-emitting star}\label{app:LR}

In this appendix, we provide some details on the calculation of the emission of isotropic stars in a Schwarzschild background spacetime. For that we need to describe the physics as seen by a freely-falling observer.
The following builds on Refs.~\cite{Misner:1974qy, 1975PhRvD..12..323C, Siwek:2015dqa}.

Consider two different observers: a static observer, i.e.  characterized by a wordline with $r=\theta=\varphi= \text{const}$; and a free-falling observer, who starts from rest at spatial infinity and has a purely radial motion. 

Let us start with the observer at rest at some radius $r_e$, which we take to be on the equatorial axis. Its proper reference frame has basis components
\beq
\omega^{\hat{t}} = \sqrt{1-\frac{2M}{r}} dt\,,\quad \omega^{\hat{r}} = \frac{1}{\sqrt{1-\frac{2M}{r}}} dr\,,\quad \omega^{\hat{\varphi}}=r d\varphi\,.\label{static_basis}	
\eeq

If we consider a photon emitted by a source at rest at infinity and received by the observer, its geodesic motion is fully determined by its energy $E$ and its impact parameter $b$.  The components of the photon's four momentum $p_\mu = dx_\mu / d\lambda$ read
\beq
p_t &=&-\left(1-\frac{2M}{r}\right)\frac{dt}{d\lambda} = -E \, , \nonumber \\
p_r &=& \frac{1}{\sqrt{1-\frac{2M}{r}}} \frac{dr}{d\lambda} = \frac{E}{1-\frac{2M}{r}}\sqrt{1-\left(1-\frac{2M}{r}\right)\frac{b^2}{r^2}} \, , \nonumber \\
p_\varphi &=& r^2 \frac{d\varphi}{d\lambda} = L = b\, E \, , \label{four_momentum}
\eeq
where we used the \acs{EOM} in Eqs.~\eqref{eq:EOMNullParticle}.

We must now compute the $p^{\hat{t}}$ component of the momentum in the observer's reference frame
%
%
%
%
%
\beq
p^{\hat{t}} = (\omega^{\hat{t}})_\mu p^\mu = \frac{-1}{\sqrt{1-\frac{2M}{r}}} p_t \,,\quad p^{\hat{r}} = ( \omega^{\hat{r}} )_\mu p^\mu =\sqrt{1-\frac{2M}{r}} p_r\,.
\eeq

The ratio of observed to emitted energy is then
\beq
\frac{p^{\hat{t}}}{E} = \frac{1}{\sqrt{1-\frac{2M}{r}}} \,, \label{blueshift}
\eeq
signaling a typical blueshift.

Moreover, the observer sees the photons come in at an angle $\alpha$ relative to its radial direction given by
\beq
\cos \alpha = -\frac{p^{\hat{r}}}{p^{\hat{t}}} = -\sqrt{1-\left(1-\frac{2M}{r}\right)\frac{b^2}{r^2}}\,.
\label{cos_alfa}
\eeq
%

Consider now free-falling observers. The basis one-forms of their proper reference frame are
\beq
\omega^{\hat{t}} = dt + \frac{\sqrt{2M/r_e}}{1-2M/r_e} dr\,,\quad \omega^{\hat{r}} = \sqrt{\frac{2M}{r_e}} dt + \frac{1}{1-2M/r_e}dr \,, \label{basis_freefall}
\eeq

When a photon with energy at infinity $E$ and impact parameter $b$ reaches the observer at $r=r_e$ and $\theta = \pi/2$, its four momentum is given by \eqref{four_momentum}. On the other hand, infalling observers will see the photon with an energy  $p^{\hat{t}} = \omega^{\hat{t}} \cdot \bm{p}$ and an angle $\alpha = \cos^{-1} (-p^{\hat{r}}/p^{\hat{t}})$ to the radial direction. 
%
%
Repeating the same steps as before we recover the results of Ref.~\cite{1975PhRvD..12..323C} 
%
\beq
\mathcal{P} &=& \frac{p_r}{p_t}\left(1-x_e^2\right) \quad , \quad x_e = \sqrt{\frac{2M}{r_e}} \\
\cos\alpha &=& - \left(\frac{p^{\hat{r}}}{p^{\hat{t}}} \right) =  - \frac{x_e+\mathcal{P}}{1+\mathcal{P} x_e}\,, \\
\frac{p^{\hat{t}}}{E} &=& - \frac{1+\mathcal{P} x_e}{x_e^2-1} = \frac{1}{1+ x_e \cos \alpha} \,,\label{shift_appendix}\\
%
\frac{b}{r_e} &=& \frac{\sin \alpha}{1 + x_e \cos \alpha}\,.
\eeq
%
\chapter{Unstable circular geodesics}\label{app:Greenhouse}

\begin{figure}[ht!]
\centering
\begin{tabular}{c}
\includegraphics[width=0.5\columnwidth]{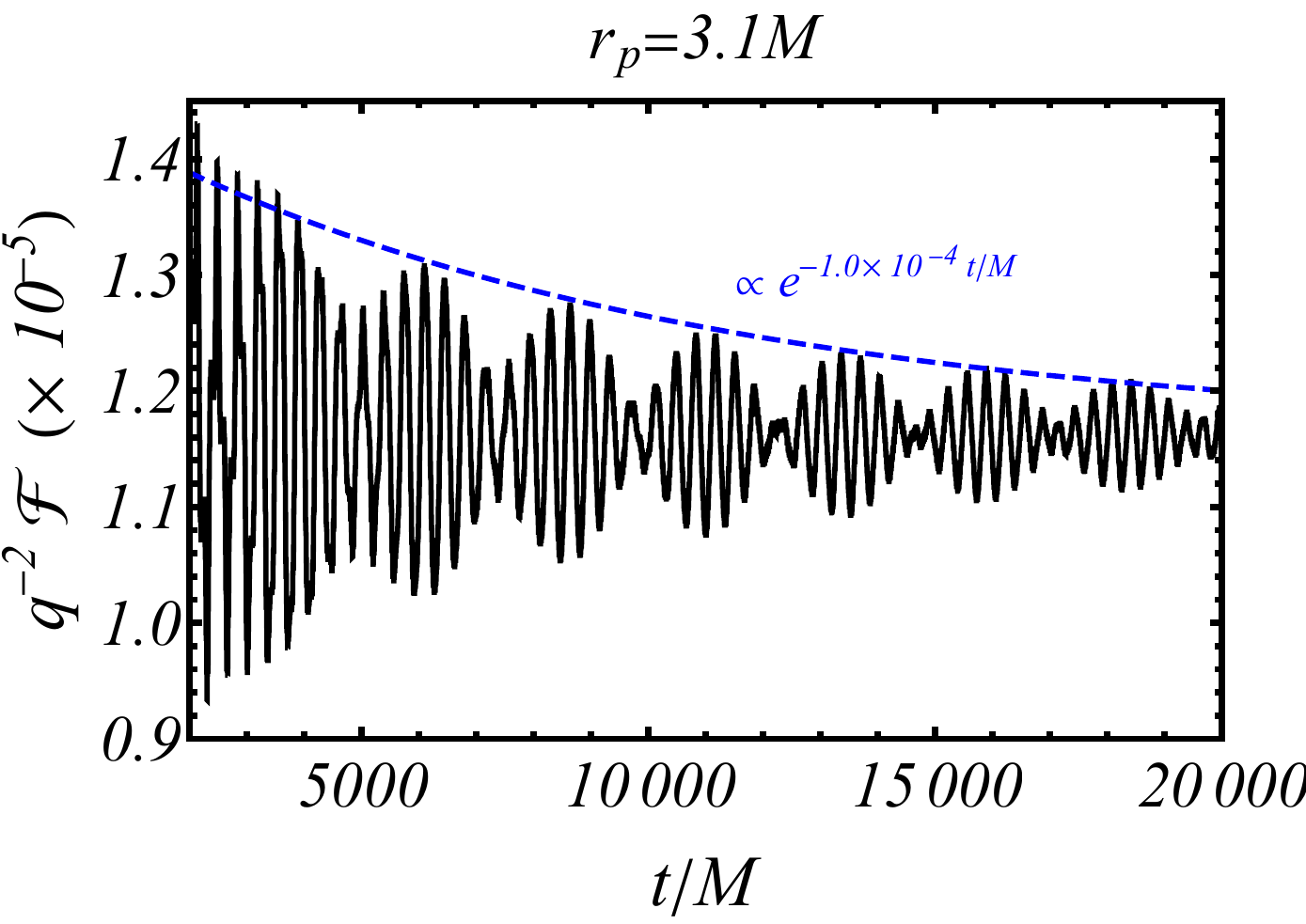} 
\includegraphics[width=0.5\columnwidth]{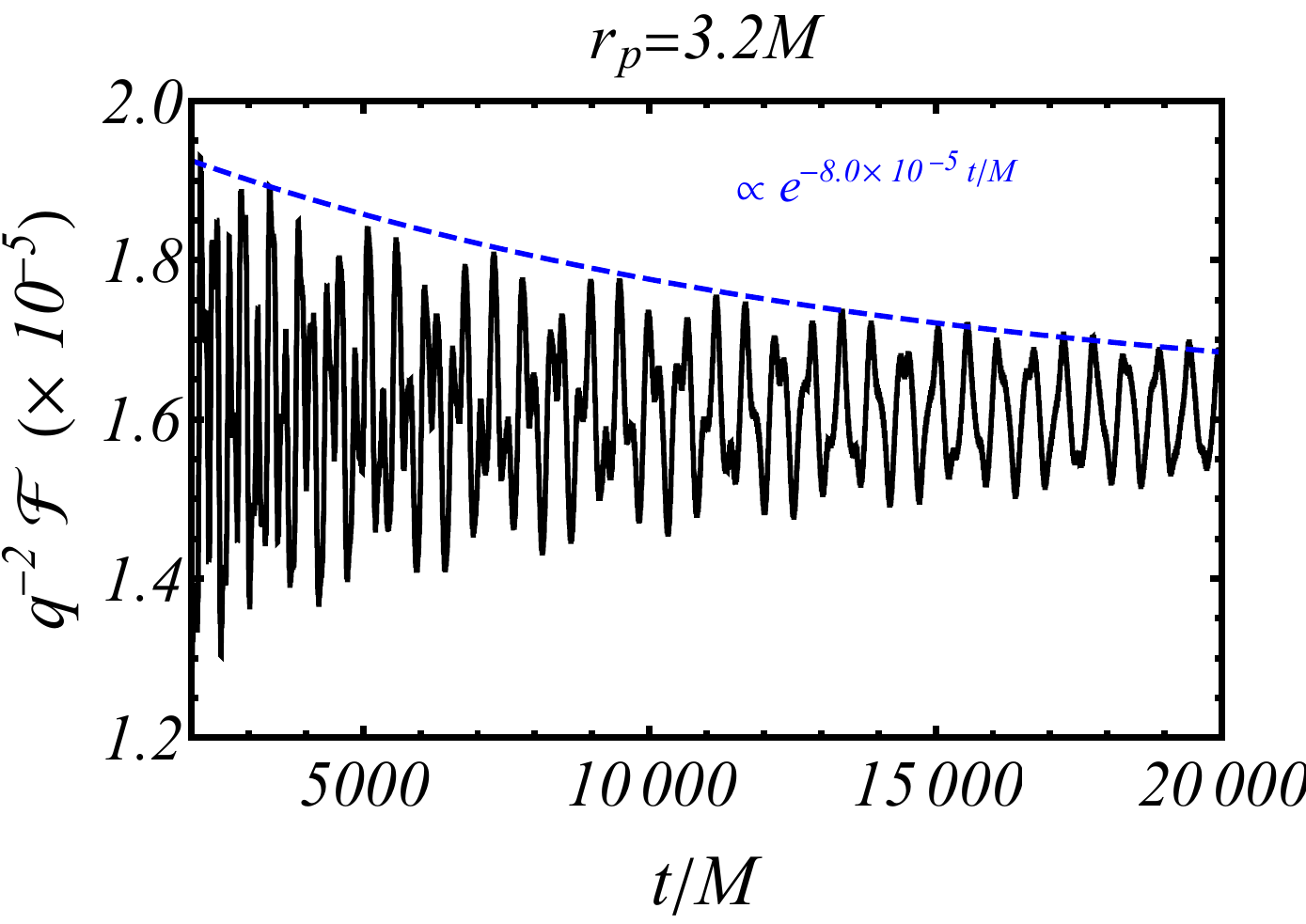} \\
\includegraphics[width=0.5\columnwidth]{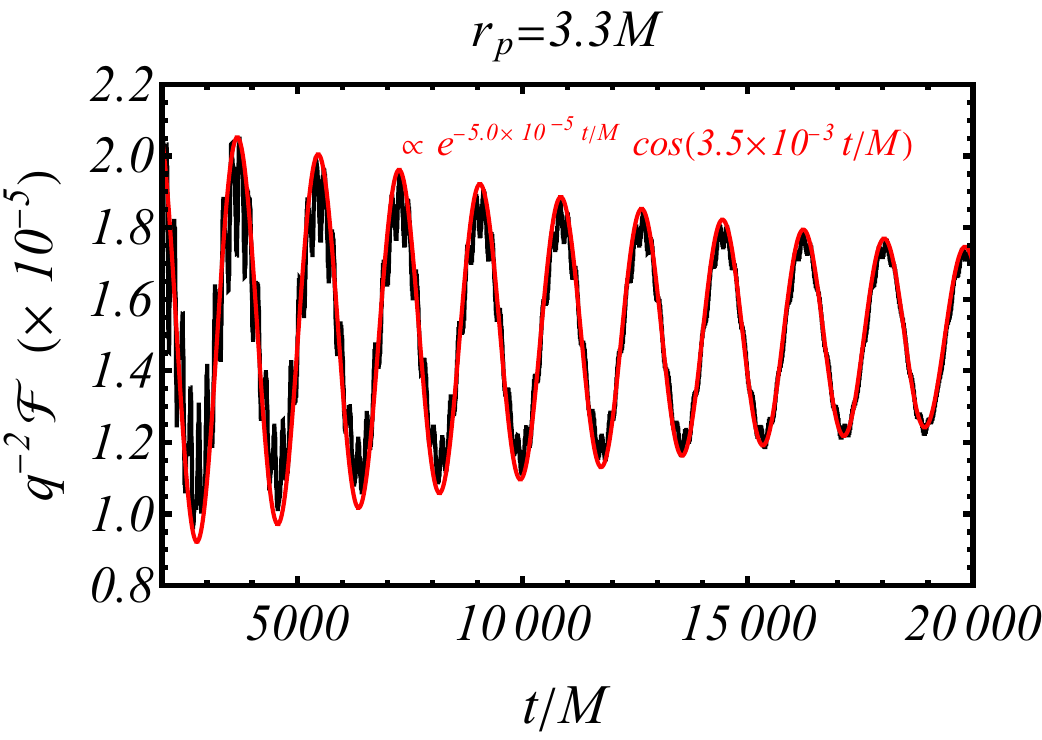} 
\includegraphics[width=0.5\columnwidth]{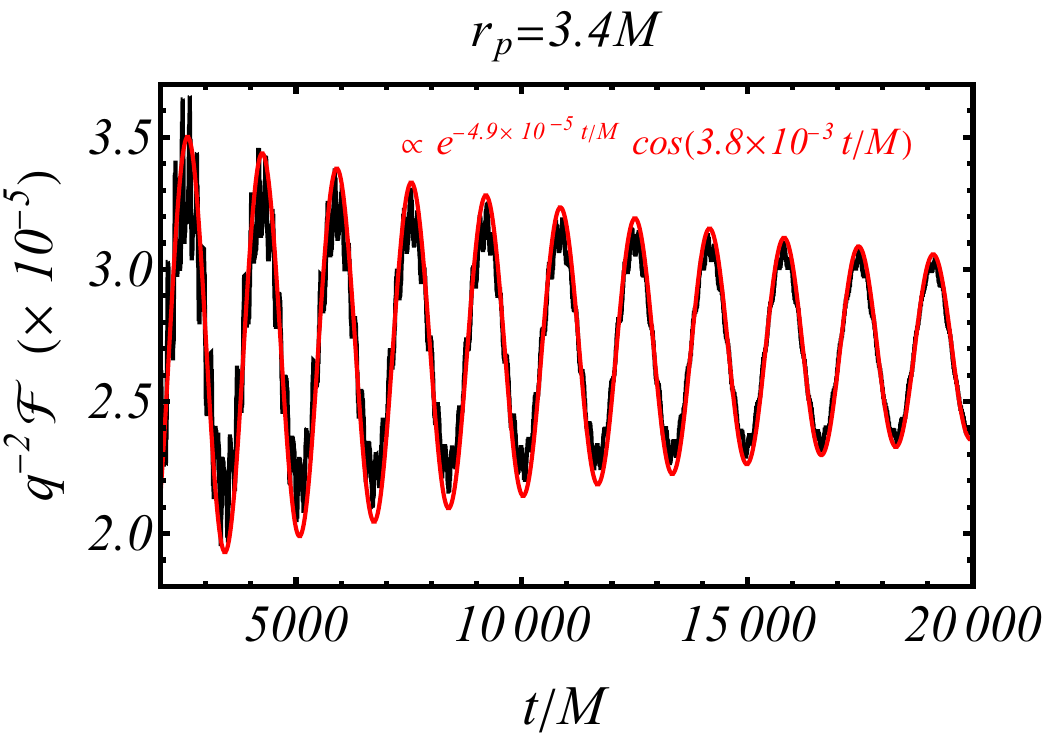} \\
\includegraphics[width=0.5\columnwidth]{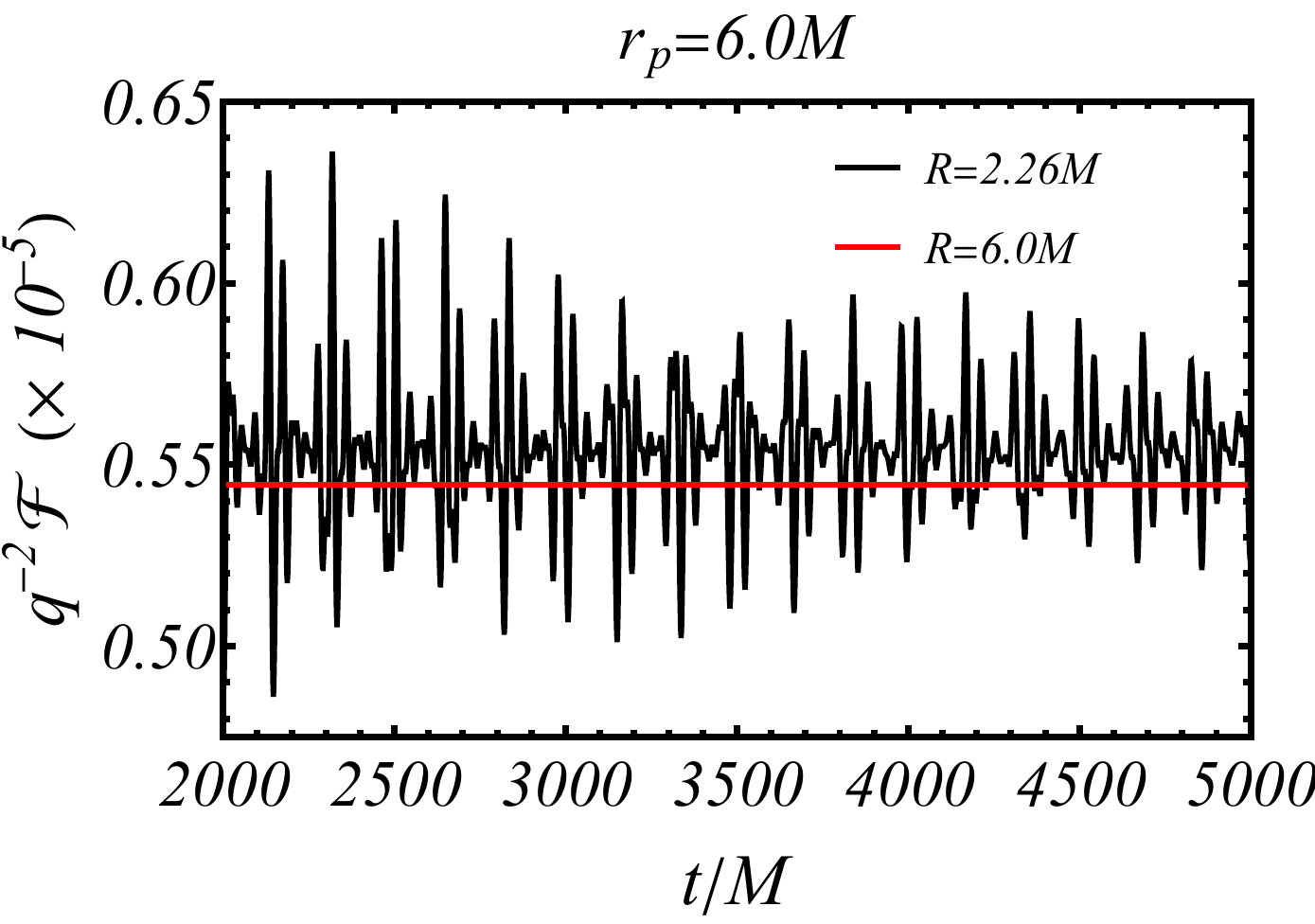} 
\end{tabular}
\caption{Same analysis as in Fig.~\ref{fig:l1} but considering circular geodesics around Schwarzschild, i.e. $a=0M$ in Eq.~\eqref{eq:AngularFreq}-\eqref{eq:Lz}. We do not show an analogous plot for the $M\omega=0.192$ since the particle would have to be put at the light-ring ($r_p=3.0M$) and we are only considering timelike motion. 
}
\label{fig:l1_a0} 
\end{figure}

In this Appendix, we present results similar to the ones discussed in Section~\ref{sec:build-up_numerical} but considering the point particle is on circular geodesics, i.e. with $a=0M$ in Eqs.~\eqref{eq:AngularFreq}-\eqref{eq:Lz}. Our results are summarized in Fig.~\ref{fig:l1_a0} (next page). We put the particle at a radius that yields the same frequency as those presented in Fig.~\ref{fig:l1}. Apart from a change in the absolute values for the flux, the relaxation timescales are in complete agreement with the ones obtained before, indicating the artificial motion we considered is irrelevant to the excitation of the constant-density star's \acs{QNM}s. 

%


\chapter{Perturbation equations in non-vacuum environments}\label{app:GBH}


In this appendix, we derive the full set of master wave-like partial differential equations that govern dynamical perturbations to spherically-symmetric but otherwise generic spacetimes given by the line element in Eq.~\eqref{eq:SphericalLineElement}. In Sec.~\ref{sec:BHPT} we already introduced the stepping stones of the formalism and now complement it with the perturbations to the matter sector. Recall that we are working in the Regge-Wheeler gauge~(cf. Eq.~\eqref{eq:RWgauge}). Since we will not be discussing matter fluxes, we focus our discussion on $\ell \geq 2$ modes, which represent the radiative degrees of freedom of the gravitational field.

%
%
Matter perturbations are encoded through the fluid perturbations in the density and pressure perturbations 
\beq
p^{(1)}_r&=&\sum_{\ell=2}^\infty \sum_{m=-\ell}^{\ell} \delta p_{r,\ell m}(t,r)Y_{\ell m}(\theta,\varphi)\label{math:preperttang}\ ,\\
p^{(1)}_t&=&\sum_{\ell=2}^\infty \sum_{m=-\ell}^{\ell} \delta p_{t,\ell m}(t,r)Y_{\ell m}(\theta,\varphi)\label{math:prepertrad}\ ,\\
\rho^{(1)}&=&\sum_{\ell=2}^\infty \sum_{m=-\ell}^{\ell} \delta \rho_{\ell m}(t,r)Y_{\ell m}(\theta,\varphi)\label{math:rhopert}\ ,
\eeq
In addition, we also need to perturb the fluid's $4$-velocity. These are described by three functions $\{U_{\ell m}(t,r),V_{\ell m}(t,r),W_{\ell m}(t,r)\}$ and the timelike condition $u^\mu u_\mu=-1$ (up to first order), which yields
\beq
u^{t}_{(1)}&=&\frac{1}{2A^{1/2}}\sum_{\ell=2}^\infty \sum_{m=-\ell}^{\ell}H^{\ell m}_0 Y_{\ell m}\,,\\
u^{r}_{(1)}&=&\frac{A^{1/2}}{B}\frac{1}{4\pi\left(p_r + \rho\right)}\sum_{\ell=2}^\infty \sum_{m=-\ell}^{\ell} W_{\ell m}Y_{\ell m}\,,\\
u^{\theta}_{(1)}&=&\frac{A^{1/2}}{4\pi\left(p_t + \rho\right) r^2}\sum_{\ell=2}^\infty \sum_{m=-\ell}^{\ell}\bigg[V_{\ell m}\partial_\theta -\frac{U_{\ell m}}{\sin\theta}\partial_\varphi\bigg]Y_{\ell m}\ ,\\
u^{\varphi}_{(1)}&=&\frac{A^{1/2}}{4\pi\left(p_t + \rho\right) r^2\sin^2\theta}\sum_{\ell=2}^\infty \sum_{m=-\ell}^{\ell}\bigg[V_{\ell m}\partial_\varphi +\frac{U_{\ell m}}{\sin\theta}\partial_\theta\bigg]Y_{\ell m}\ . \nonumber \\
\eeq
where we are now suppressing all coordinate dependences and the superscript $(0)$ to improve readability. The perturbed unit spacelike vector $k^\mu$ orthogonal to $u^\mu$ ($k^\mu k_\mu = 1$ and $k^\mu u_\mu = 0$) is then
\beq
k_{t}^{(1)}&=&-\frac{A}{B^{3/2}}\frac{1}{4\pi\left(p_r + \rho \right)}\sum_{\ell=2}^\infty \sum_{m=-\ell}^{\ell} W_{\ell m} Y_{\ell m}\,,\\
k_{r}^{(1)}&=&\frac{1}{2\sqrt{B}}\sum_{\ell=2}^\infty \sum_{m=-\ell}^{\ell} H_2^{\ell m}Y_{\ell m}\,,\\
k_{\theta}^{(1)}&=&k_{\varphi}^{(1)} =0\ ,
\eeq
from which we can then build the perturbed projection operator $\Pi^{(1)} = g_{\mu\nu}^{(1)} + u_\mu^{(0)}u_\nu^{(1)} + u_\mu^{(1)} u_\nu^{(0)} -k_\mu^{(0)}k_\nu^{(1)} - k_\mu^{(1)} k_\nu^{(0)} $
\beq
\Pi_{t\theta}^{(1)} &=& \frac{A}{4\pi \left(p_t + \rho \right) }\sum_{\ell=2}^\infty \sum_{m=-\ell}^{\ell} \left[\csc\theta\, U_{\ell m}\partial_\varphi + V_{\ell m} \partial_\theta \right] Y_{\ell m}\, , \\ 
\Pi_{t\varphi}^{(1)} &=&-\frac{A}{4\pi \left(p_t+ \rho \right) } \sum_{\ell=2}^\infty \sum_{m=-\ell}^{\ell} \left[\sin\theta\, U_{\ell m}\partial_\theta + V_{\ell m} \partial_\varphi \right] Y_{\ell m}\,,\\
\Pi_{\theta\theta}^{(1)} &=&r^2\sum_{\ell=2}^\infty \sum_{m=-\ell}^{\ell} K_{\ell m} Y_{\ell m}\,,\\
\Pi_{\varphi\varphi}^{(1)} &=&\sin^2\theta \, \Pi_{\theta\theta}^{(1)}\ .
\eeq

Finally, we also need to assume a (barotropic) equation of state that relates the density and pressure perturbations
\beq
\delta p_{r,\ell m}&=&c_{s_r}^2\delta \rho_{\ell m}\ , \\
\delta p_{t,\ell m}&=&c_{s_t}^2\delta \rho_{\ell m}\ ,
\eeq
where $c_{s_r}=c_{s_r}(r)$ and $c_{s_t}=c_{s_t}(r)$ are the sound speed along the radial and tangential directions. We can finally plug these in the energy-momentum tensor~\eqref{eq:EnergyMomentumAnisotropic} and arrive at the components 
\beq
T_{tt}^{\text{env}(1)}&=&A\sum_{\ell=2}^\infty \sum_{m=-\ell}^{\ell}(\delta\rho_{\ell m}-H^0_{\ell m}\rho)Y_{\ell m}\ ,\nonumber\\
T_{tr}^{\text{env}(1)}&=&-\sum_{\ell=2}^\infty \sum_{m=-\ell}^{\ell}\left[\frac{A}{4\pi B^2}W_{\ell m}+H^{1}_{\ell m} \rho\right] Y_{\ell m}\ ,\nonumber\\
T_{t\theta}^{\text{env}(1)}&=&
\frac{A}{4\pi}\sum_{\ell=2}^\infty \sum_{m=-\ell}^{\ell}\left[\csc\theta  U_{\ell m} \partial_{\varphi}-V_{\ell m}\partial_{\theta}\right]Y_{\ell m}\ ,\nonumber\\
T_{t\varphi}^{\text{env}(1)} &=&
\frac{-A}{4\pi}\sum_{\ell=2}^\infty \sum_{m=-\ell}^{\ell}\left[V_{\ell m} \partial_{\varphi}+U_{\ell m}\sin\theta\partial_{\theta}\right]Y_{\ell m}\ ,\nonumber\\
T_{rr}^{\text{env}(1)} &=& \frac{1}{B}\sum_{\ell=2}^\infty \sum_{m=-\ell}^{\ell} \left( p_r H^2_{\ell m}+
\delta p_{r,\ell m} \right)Y_{\ell m}\nonumber\ ,\\ 
T_{\theta\theta}^{\text{env}(1)} &=&r^2 \sum_{\ell=2}^\infty \sum_{m=-\ell}^{\ell}(p_t K_{\ell m}+
\delta p_{t,\ell m})Y_{\ell m}\nonumber\ ,\\ 
T_{\varphi\varphi}^{\text{env}(1)} &=&T_{\theta\theta}^{\text{halo}(1)}\sin^2\theta\ .
\eeq
%

\section{Energy-momentum tensor of the secondary for circular orbits}

We are interested in studying \acs{EMRI}s, so the source of our perturbations is a point particle orbiting the background spacetime, with energy-momentum tensor given by Eq.~\eqref{eq:PPTensor}. We will focus on circular orbits at some fixed radius $r_p$ and spherical symmetry allows us to consider motion only on the equatorial plane, i.e. $\theta_p=\pi/2$. The $4$-velocity of the source is~\cite{Cardoso:2008bp}
\beq
u_p = \left(\frac{E_p}{A_p} , 0,0, \frac{L_p}{r_p^2} \right) \, , 
\eeq
where $A_p = A(r_p)$ and $E_p$ and $L_p$ are the energy and angular momentum per unit rest mass of the orbiting body~\cite{Cardoso:2008bp}
\beq
E_p = \dfrac{A_p}{\sqrt{A_p - r_p^2 \Omega_p^2}},  \quad 
L_p = \dfrac{\Omega_p r_p^2}{\sqrt{A_p - r_p^2 \Omega_p^2}}.
\eeq
with the angular orbital frequency $\varphi_p(t) = \Omega_p \, t$ given by
\beq
\Omega_p = \sqrt{\dfrac{A'_p}{2r_p}}\, ,
\eeq
where recall that a prime denotes a derivative with respect to $r$.

For this orbital configuration, the tensor harmonics expansion~\eqref{harmonicexp} for the energy-momentum tensor of the secondary greatly simplifies. We have ${\cal A}_{\ell m}={\cal A}_{\ell m}^{1}={\cal B}_{\ell m}={\cal Q}_{\ell m}=0$, while the non-vanishing coefficients can be expressed in terms of the orbital parameters as follows
\beq
{\cal A}_{\ell m }^{0}&=&\frac{m_p\sqrt{A B}E_p}{r^2} Y^\star_{\ell m }\,\delta_{r_p}\ ,\nonumber\\
{\cal B}_{\ell m }^{0}&=&\frac{m_pi\sqrt{AB}L_p}{r^3\sqrt{(n+1)}}\,\delta_{r_p} \,\partial_\phi Y^\star_{\ell m }\ ,\nonumber\\
{\cal Q}_{\ell m }^{0}&=&-\frac{m_p\sqrt{AB}L_p}{r^3\sqrt{(n+1)}}\,\delta_{r_p} \,\partial_\theta Y^\star_{\ell m }\ ,\nonumber\\
{\cal G}_{\ell m }&=&\frac{m_pL_p^2\sqrt{AB}}{r^4\sqrt{2}E_p}\,\delta_{r_p} \, Y^\star_{\ell m}\ ,\nonumber\\
{\cal D}_{\ell m }&=&\frac{m_piL_p^2\sqrt{AB}}{E_pr^4\sqrt{2n(n+1)}}\,\delta_{r_p} \, \partial_{\theta\phi}Y^\star_{\ell m}
\ ,\nonumber\\
{\cal F}_{\ell m }&=&\frac{m_pL_p^2\sqrt{A B}}{r^42E_p\sqrt{2n(n+1)}}\, \delta_{r_p} \, (\partial_{\phi\phi}-\partial_{\theta\theta}) Y^\star_{\ell m}\ , \nonumber \\
\eeq
where 
$n=\ell(\ell+1)/2-1$, $Y^\star_{\ell m}=Y^\star_{\ell m}(\pi/2,\Omega_p t)$
and $\delta_{r_p}=\delta(r-r_p)$.

\section{Master Equations}

We now have all the ingredients necessary to derive a set of partial differential equations for the axial and polar  sectors. For the sake of clarity, hereafter, we will drop the sum over the multipolar indices.

\subsection{Axial Sector}

We start off with the simpler axial sector where gravitational perturbations decouple from the matter sector. We need to find a solution for $h^{\ell m}_0$, $h_1^{\ell m}$ and $U^{\ell m}$. Start by rewriting $\partial h^{\ell m}_0/\partial t$ using the combination of ${\cal E}_{\theta\theta}-{\cal E}_{\phi\phi}/\sin^2\theta$, where
\beq
\mathcal{E}_{\mu\nu}=G^{(1)}_{\mu\nu}-8\pi(T^{\text{env}(1)}_{\mu\nu}+T^p_{\mu\nu})\ ,
\eeq
and $G^{(1)}_{\mu\nu}$ is the perturbed Einstein tensor. We arrive at
\beq
	\frac{\partial h_0}{\partial t}=A B\frac{dh_1}{dr} +\frac{A(1-B+r B')}{2r}h_{1} -
	i \frac{4\sqrt{2}\pi r^2A}{\sqrt{n(1+n)}}{\cal D}\ ,
\eeq
where from now on we omit the angular indexes to avoid cluttering 

If we define the variable
\beq
\Psi^\text{ax} = \frac{\sqrt{A B}}{r} h_1 \, ,
\eeq
the ${\cal E}_{r\theta}$ component provides a second-order non-homogeneous 
differential equation for $\Psi^\text{ax}$. In terms of the generalized 
tortoise coordinate $dr_*/dr=(A\, B)^{-1/2}$, the master equation for the axial metric perturbations can be written as 
\beq
\left[ - \frac{\partial^2 }{\partial t^2} + \dfrac{\partial^2 }{\partial r_*^2} - V^\text{ax}\right]\Psi^\text{ax} =  S^\text{ax}\ ,\label{eq:master_axial} 
\eeq
where the potential reads 
\beq
V^\text{ax} = \dfrac{A}{r^2}\left[\ell(\ell+1) - \dfrac{6m(r)}{r} + m'(r)  \right]\ ,
\eeq
with $m(r) =  r(1-B(r))/2$ and the source term is
\beq
S^\text{ax} = i \frac{2\sqrt{2}\pi r^2}{\sqrt{n(n+1)}}\left( \frac{A'}{A}\mathcal{D} +\frac{\partial \mathcal{D}}{\partial r} \right) \, .
\eeq

In the vacuum limit, i.e. $M\rightarrow 0$, or alternatively $r\rightarrow \infty$, the equation above reduces to the Regge-Wheeler master equation~\eqref{eq:MasterZRW} which governs axial perturbations in Schwarzschild. Notice that both at the \acs{BH} horizon and at infinity the effective potential goes to zero. Therefore, the master function behaves as wave, with physical solutions corresponding to ingoing waves at the horizon and outgoing at large distances.

\subsection{Polar Sector}

We now move to the polar sector, which is more involved because matter perturbations couple to gravitational ones. First, we redefine the perturbations
\beq
H_0 &=& K + \frac{r}{A} S \, , \\
H_1 &=& \frac{r}{A} \tilde{H_1} \, .
\eeq
This choice is inspired in studies of perturbations around relativistic stars~\cite{Allen:1997xj} and the factoring of $r$ and $A$ captures the asymptotic behavior at, respectively, large distances and near the \acs{BH} horizon. 

\begin{itemize}
 
\item $\mathcal{E}_{\theta\theta} - \mathcal{E}_{\varphi\varphi}/\sin^2 \theta$ gives an algebraic relation for $H_2$
\beq
H_2 &=& K +  \frac{r}{A} S - \frac{16 \pi r^2}{\sqrt{2n(n+1)}} \,  \, \mathcal{F} \, . \label{eq:H2Relation}
\eeq

\item $\mathcal{E}_{r\theta}$ gives a relation between $\tilde{H}_1$ and the other perturbations which we use to substitute $\partial \tilde{H}_1/\partial t$ and $\partial^2 \tilde{H}_1/\left(\partial t\partial r_*\right)$ when necessary 
\beq
\frac{\partial \tilde{H}_1}{\partial t} &=& \sqrt{\frac{A}{B} } \frac{\partial S}{\partial r_*} + \frac{A}{r}S + \frac{A^2}{r^2B}\left(1-B+ 8\pi r^2 p_r \right)K \nonumber \\
& - &  \frac{8 \pi}{\sqrt{2n(n+1)}} \frac{A^2}{B}\left(1 + B + 8\pi r^2 p_r \right) \mathcal{F} \, .
\eeq

\item $\mathcal{E}_{tt} = 8 \pi T_{tt}$ yields a constraint between $K$, $S$ and $\delta \rho$
\beq \label{eq:HamiltonianConstraint}
\frac{\partial^2 K}{\partial r_*^2} &=&  \sqrt{\frac{B}{A}} \frac{\partial S}{\partial r_*} - \frac{1}{2r}\sqrt{\frac{A}{B}} \left( 1 + 3B - 8 \pi r^2 \rho \right)\frac{\partial K}{\partial r_*} \nonumber \\
&+& A \left(\frac{\ell(\ell +1)}{r^2} - 8 \pi \rho \right)K  \nonumber\\
&+& \left( \frac{\ell(\ell+1)+4B}{2r}  -8\pi r \left(\rho + p_r\right) \right)S   \nonumber \\
&-& 8 \pi A \,\delta \rho - 8 \pi A^0 -  \frac{16\pi A B r}{\sqrt{2n(n+1)}} \frac{\partial \mathcal{F}}{\partial r}  \nonumber \\
&+& \frac{8\pi\, A}{\sqrt{2n(n+1)}}  \left( 16 \pi r^2 \rho -4B -2 - \ell \left(\ell +1 \right) \right)  \mathcal{F}  \, . \nonumber \\
\eeq

\item $\mathcal{E}_{tt} - AB \mathcal{E}_{rr}$ gives the first second-order ``wavelike'' equation for $K$
\beq \label{eq:WaveEquationK}
&-&\frac{\partial^2 K}{\partial^2 t} + \frac{\partial^2 K}{\partial^2 r_*} + \frac{2\sqrt{AB}}{r}\frac{\partial K}{\partial r_*} \nonumber \\
&+& \frac{A}{r^2}\left[8\pi r^2 \, \left(\rho + p_r  \right) + 2-2B - \ell(\ell+1) \right] K \nonumber \\
&=& -8 \pi A \left(1 - c_{s_r}^2 \right) \, \delta \rho + \frac{2}{r}\left(B -  4 \pi r^2 \left(\rho + p_r\right) \right)S \nonumber \\
&-& \frac{8 \pi  \,A}{{\sqrt{2n(n+1)}}} \left( 2 + 2B + \ell \left(\ell +1 \right) - 16 \pi r^2 \rho \right) \mathcal{F}   \nonumber \\
&-& \frac{16\pi  A B r}{\sqrt{2n(n+1)}} \frac{\partial \mathcal{F}}{\partial r} - 8 \pi \mathcal{A}^0	 \, . \nonumber \\
\eeq

\item $\mathcal{E}_{\theta\theta} + \mathcal{E}_{\varphi\varphi}/\sin^2 \theta$  gives another second-order ``wavelike'' equation for $S$, where we used the previous equations to substitute the necessary derivatives
\beq
&-&\frac{\partial^2 S}{\partial^2 t} + \frac{\partial^2 S}{\partial^2 r_*} \nonumber \\
&+& \frac{A}{r^2}\left(4\pi r^2 \, \left(\rho + 3 p_r \right) + B -1  - \ell(\ell+1) \right) S \nonumber \\
&=& -\frac{A^2}{r^3 B} \Bigg[ 7B^2 + \left(1 + 8\pi r^2 p_r \right)^2 \nonumber \\
&-& 8B\left(1 + 4\pi r^2 p_r - 2\pi \left(p_t + \rho \right) \right) \Bigg] K \nonumber \\
&-& 16\pi \frac{a^2}{r}\left(c_{s_r}^2 - c_{s_t}^2  \right) \delta \rho + 8\pi \sqrt{2} \, \frac{A^2}{r} \mathcal{G}_{\ell m} \nonumber \\
&-& \frac{8 \pi}{\sqrt{2n(n+1)}} \frac{A^2}{r B}\Bigg[ 5 B^2 + B\left(\ell\left(\ell+1\right) -4 - 32\pi r^2 p_r \right) \nonumber \\
&& \qquad- \left(1+ 8\pi r^2 p_r \right)^2 \Big] \mathcal{F} \nonumber \\
&-& \frac{16 \pi \, A}{\sqrt{2n(n+1)}} \left ( \frac{A}{2}\left(1 +B + 8 \pi r^2 p_r \right) \frac{\partial \mathcal{F}_{\ell m}}{\partial r}   + r \frac{\partial^2 \mathcal{F}}{\partial t^2}\right ) \nonumber \, . \\
\label{eq:WaveEquationS}
\eeq

\item The final ``wavelike'' equation for $\delta \rho$ is obtained from the conservation of the perturbed energy-momentum tensor of the surrounding fluid. $\nabla_\mu T^{\mu \theta}_{\text{env}(1)} = 0$ gives  
\beq
\frac{\partial V }{\partial t} &=& \left(\rho + p_r\right) \left(K + \frac{r}{A}S \right) - 4 \pi c_{s_t}^2 \delta \rho \nonumber \\
&+& \frac{32 \pi^2 \, r^2 \, }{{\sqrt{2n(n+1)}}} \left(p_t -  p_r\right) \mathcal{F} \, . \nonumber \\
\eeq

\item whereas $\nabla_\mu T^{\mu r}_{\text{env}(1)} = 0$
\beq
&&\frac{\partial W}{\partial t} = -4\pi r \frac{B^2}{A^2}\left(p_r + \rho\right) \frac{\partial \tilde{H}_1}{\partial t} - 4 \pi B c_{s_r}^2 \sqrt{\frac{B}{A}}\frac{\partial \delta \rho}{\partial r_*} \nonumber \\
&-& 2\pi r \frac{B}{A} \sqrt{\frac{B}{A}} \left( p_r + \rho  \right) \frac{\partial S }{\partial r_*}  + 2\pi B \sqrt{\frac{B}{A}} \left( 2p_t - p_r+ \rho  \right) \frac{\partial K }{\partial r_*} \nonumber \\
&+& 2\pi \frac{B}{A}\left(1 - 2B + 8 \pi r^2 p_r \right) \left( p_r + \rho\right)S \nonumber \\
&+& 4\pi \frac{B}{r}\left(B - 1 + 8 \pi r^2 p_r \right) \left( p_r + \rho\right)K \nonumber \\
&+& \frac{2\pi B}{r}\Bigg[\left(1 + c_{s_r}^2 \right)\left(1+ 8 \pi r^2  p_r\right) -B\left( 1 + 5 c_{s_r}^2 - 4  c_{s_t}^2 + 4 r  c_{s_r} c'_{s_r} \right) \Bigg]\delta \rho \nonumber \\
&-& \frac{64\pi^2 r B}{\sqrt{2n(n+1)}} \left(1 - B + 8 \pi r^2 p_r\right)  \left( p_r + \rho\right) \mathcal{F} \, . 
\eeq

\item Doing $\nabla_\mu T^{\mu r}_{\text{env}(1)} = 0$ and using the previous relations we finally arrive at 
\beq\label{eq:WaveEquationRho}
&-&\frac{\partial^2 \delta \rho}{ \partial t^2} + c_{s_r}^2 \frac{\partial^2 \delta \rho}{\partial r_*^2} \nonumber \\
&+& \frac{1}{2r}\sqrt{\frac{A}{B}} \Bigg[ \left(c_{s_r}^2 - 1 \right)\left(1 + 8 \pi r^2 p_r \right) \nonumber \\ 
&+& B \left(1 + 7 c_{s_r}^2 - 4 c_{s_t}^2 + 8 r c_{s_r} c_{s_r}' \right) \Bigg] \frac{\partial \delta \rho}{\partial r_*}  \nonumber \\
&-& \frac{A}{2r^2 B} \Bigg[ \left(1+ c_{s_r}^2\right)\left(1+ 8 \pi r^2 p_r \right)^2 \nonumber \\
&+& 2B \Big[ c_{s_t}^2 \left(4 + \ell + \ell^2 + 16 \pi r^2\left(p_r -  \rho \right) \right)\nonumber \\ 
&-&1 + c_{s_r}^2\left( 8 \pi r^2 \left(\rho  -3 p_r \right) \right) -5 + r c_{s_r}c_{s_r}' \left(8\pi r^2 \left(\rho -2p_r \right) -3 \right)\Big] \nonumber \\
&+& B^2\left(1 + 5c_{s_r}^2	- 4c_{s_t}^2 - 4r^2\left(c'_{s_r}\right)^2 + 8 r c_{s_t} c_{s_t}' - 2 r c_{s_r}\left(5 c_{s_r}' + 2 r c_{s_r}'' \right) \right) \Bigg]\delta \rho \nonumber \\
&=&\mathcal{S}_\rho \, ,
\eeq
with the source term
\beq
&&\mathcal{S}_\rho = - \frac{r}{A}\left(p_r + \rho \right) \frac{\partial^2 S}{\partial r_*^2}  -\frac{1}{4r}\sqrt{\frac{A}{B}}\Big[ 72 \pi r^2 p_r^2 + 4 (3B - 2)p_t + \rho \nonumber \\
&+& p_r \left(9 - 13B + 8 \pi r^2 \left(\rho - 8 p_t\right) \right) - B\left(\rho + 2 r(2p_t' + \rho') \right) \Big] \frac{\partial K}{\partial r_*} \nonumber \\
&-&\frac{1}{4A\sqrt{AB}}\Bigg[ -2rB\left(p_r+ \rho\right )A'-A\Big[24\pi r^2 p_r^2 + \rho + 2r\rho\left(8\pi r \rho + B'\right)\nonumber \\
&+& p_r\left(1 - 13B + 2r \left(20\pi r \rho + B' \right)\right) -B\left(4 p_t + 17 \rho + 6 r \rho'\right)  \Big] \Bigg] \frac{\partial S}{\partial r_*} \nonumber \\
&-&\frac{A}{2r^2 B}\Bigg[64 \pi^2 r^4 p_r^3 + \rho + 8 \pi r^2 p_r^2 \left(2 - 5 B + 8 \pi r^2 \rho\right)\nonumber \\
&+& B\Big[\rho \left( B -2 + 24\pi r^2 \rho \right) - 2p_t\left(\ell^2 + \ell - 2 + 2B - 32\pi r^2 \rho \right) \nonumber \\
&+& 4 r \left(B-1\right) \rho'  \Big] + p_r \Big[1 + 5B^2 + 16\pi r^2 \rho^2 \nonumber \\
&+&	 2B (\ell^2 + \ell -3 + 8 \pi r^2 (6p_t+\rho + 2 r\rho') )  \Big] \Bigg]K \nonumber \\ 	
&-&\frac{1}{4r B}\Bigg[64 \pi^2 r^4 p_r^3 + \rho + 8 \pi r^2 p_r^2 \left(2 - 7 B + 8 \pi r^2 \rho\right)\nonumber \\
&+& B\Big[ 4p_t\left(1 - \ell - \ell^2 - 2B  \right) + \rho \left(10B -3  - 32\pi r^2 \rho \right) \nonumber \\ 
&+& 2 r \left(4B-1\right) \rho'  \Big]+ p_r \Big[1 + 2B^2 + 16\pi r^2 \rho + B ( (1+2\ell)^2 \nonumber \\
&-& 8 \pi r^2 (4p_t+15\rho + 2 r\rho') )  \Big] \Bigg]S \nonumber \\
&-& 4 \pi \left(\rho - p_r + 2p_t \right) \mathcal{A}^0 - 4 \pi \sqrt{2} A \left(p_r + \rho \right) \mathcal{G} \nonumber \\
&-& \frac{4\pi A r}{\sqrt{2(2n+1)}}\Bigg[40 \pi r^2 p_r^2 - 4 p_t - 5 \left(B-1 \right) \rho\nonumber \\
&+& p_r\left(5 - B + 40\pi r^2 \rho \right)\Bigg]\frac{\partial\mathcal{F}}{\partial r} \nonumber \\
&+&\frac{8\pi}{\sqrt{2n(n+1)}}\frac{A}{B} \Bigg[64\pi r^4 p_r^3 + \rho + 16\pi r^2p_r^2 \left(1 + 4 \pi r^2 \rho \right) \nonumber \\
&+&B \Big[\rho \left(3 - 4B + 24\pi r^2 \rho \right) -3 \left(B -3\right)\rho'\nonumber \\
&-& 2p_t \left(4B + \ell^2 + \ell -2 -32 \pi r^2 \rho \right) \Big]  \nonumber \\
&+& p_r \Big[1+ 4B^2 + 16 \pi r^2 \rho + B( 2 \ell (1+ \ell) \nonumber \\
&-& 1 + 8\pi r^2 (12p_t + 7\rho + 3 r \rho') ) \Big] \Bigg]\mathcal{F} \, .
\eeq

\end{itemize}

There are many points worth highlighting. First, if we set the mass of the halo to zero, $M=0$, the evolution equation for $\delta \rho$ decouples from the gravitational perturbations and becomes sourceless. This means if no initial data is given to $\delta \rho$, it remains $0$ for the whole evolution, as it should for an \acs{EMRI} in vacuum. Moreover, one can check explicitly by direct substitution that the vacuum Eqs.~\eqref{eq:WaveEquationK} and~\eqref{eq:WaveEquationS}, together with the constraint in Eq.~\eqref{eq:HamiltonianConstraint}, are completely equivalent to the Zerilli equation in Eq.~\eqref{eq:MasterZRW}. Asymptotic flatness then guarantees this equivalence as $r \rightarrow \infty$. 

We already discussed the asymptotic behavior of $\delta \rho$ in the main text, in Section~\ref{sec:BCSoundSpeed}. It is clear that $S$ also behaves as wave propagating at the speed of light both at infinity and at the \acs{BH} horizon, similarly to the master Zerilli function. From the homogeneous part of the evolution equation for $K$, one could naively assume that it would decay as $1/r$ at large distances. However, note that $K$ is also being sourced by $S$. If as $r\rightarrow \infty$, $S$ is represented by an outgoing wave, $S\sim S_0^\infty e^{-i \omega (t-r_*)}$, then writing $K \sim \left(K_0^\infty + K_1^\infty / r \right)e^{-i \omega (t-r_*)}$ and expanding Eq.~\eqref{eq:WaveEquationK} in powers of $r$ for large distances we find
\beq
K_0^\infty = -\frac{i}{\omega}S^\infty_0 \, , 
\eeq
so $K$ also behaves as a wave asymptotically. In fact, at far-away distances the relation between $K$ and the Zerilli master function is~\cite{Berti:2009kk}
\beq
K \underset{r \rightarrow \infty}{=} \frac{d\Psi_Z}{dr_*}\, .
\eeq
Moreover, for monochromatic waves and at large distances $d\Psi_\text{Z}/dr_* = i \omega \Psi_\text{Z} = - d \Psi_\text{Z} / dt$. Thus, for circular orbits we can use the absolute value of $K$ to evaluate the flux of energy carried by \acs{GW}s to infinity as determined by Eq.~\eqref{eq:FluxNonRot}.

We have also mentioned that the $\ell = 0$ and $\ell = 1$ modes are ``non-physical'' in \acs{GR} and can be removed in a region of spacetime by a gauge transformation. In fact, for $\ell \leq 1$ the Regge-Wheeler gauge is not completely fixed because there are less independent spherical harmonics (recall the discussion below Eq.~\eqref{eq:FluxNonRot}). In fact, Eq.~\eqref{eq:H2Relation} is actually identically $0$ for $\ell \leq 1$ (i.e. it becomes $0=0$). 

However, what is the actual meaning of gravitational perturbations for the lower multipoles, since they do not contribute to radiative degrees of freedom? The answer to this question is explained in detail in Appendix G of Ref.~\cite{Zerilli:1970wzz} (see also Refs.~\cite{Martel:2005ir, Detweiler:2003ci}) . The $\ell=0$ axial mode is identically $0$, while the respective $\ell=1$ case gives a perturbation to the angular momentum of the configuration. In the case of an orbiting particle, it would correspond to its conserved angular momentum. In a similar fashion, the $\ell=0$ polar mode leads to a correction in the mass of the system, e.g. by adding the mass of the particle to the mass of the central object. Finally, for the Regge-Wheeler gauge in which we adopt coordinates centered at the unperturbed spacetime, the $\ell=1$ polar mode yields the transformation to the center-of-momentum of the total system. In a Newtonian analysis, this would correspond to the appearance of a fictitious force because the reference frame of the central \acs{BH} is not inertial.


\cleardoublepage
\defbibheading{bibintoc}[\bibname]{%
  \phantomsection
  \manualmark
  \markboth{\spacedlowsmallcaps{#1}}{\spacedlowsmallcaps{#1}}%
  \addtocontents{toc}{\protect\vspace{\beforebibskip}}%
  \addcontentsline{toc}{chapter}{\tocEntry{#1}}%
  \chapter*{#1}%
}
\printbibliography[heading=bibintoc]


\cleardoublepage
\pdfbookmark[0]{Declaration}{declaration}
\chapter*{Declaration}
\thispagestyle{empty}

I declare that this document is an original work of my own and that it fulfills all the requirements of the Code of Conduct and Good Practices of the Universidade de Lisboa.
\bigskip
\vspace{5mm}

\noindent Name: Francisco Miguel Batista Duque\\
\noindent Student number: 84383\\
\bigskip
\vspace{5mm}

\noindent\textit{\myLocation, \myTime}

\smallskip

\begin{flushright}
    \begin{tabular}{m{6cm}}
        \\ \hline
        \centering\myName \\
    \end{tabular}
\end{flushright}

\cleardoublepage\pagestyle{empty}

\hfill

\vfill

\pdfbookmark[0]{Colophon}{colophon}
\section*{Colophon}
%
%
This document was typeset using the typographical look-and-feel \texttt{classicthesis} developed by Andr\'e Miede and Ivo Pletikosić.
The style was inspired by Robert Bringhurst's seminal book on typography ``\emph{The Elements of Typographic Style}''.

\bigskip

\noindent\finalVersionString

\end{document}